\documentclass[aps,aps,prb,twocolumn]{revtex4}
\usepackage{amsmath}
\usepackage{graphicx}
\usepackage{dcolumn}
\usepackage{color}
\newcommand{\be}{\begin{equation}}
\newcommand{\ee}{\end{equation}}
\newcommand{\bea}{\begin{eqnarray}}
\newcommand{\eea}{\end{eqnarray}}
\newcommand{\bse}{\begin{subequations}}
\newcommand{\ese}{\end{subequations}}

\begin{document}
\title{Magnetic Dipole Interactions in Crystals}
\author{David C. Johnston}
\altaffiliation{johnston@ameslab.gov}
\affiliation {Ames Laboratory and Department of Physics and Astronomy, Iowa State University, Ames, Iowa 50011}

\date{\today}

\begin{abstract}

The influence of magnetic dipole interactions (MDIs) on the magnetic properties of local-moment Heisenberg spin systems is investigated.  A general formulation is presented for calculating the eigenvalues $\lambda$ and eigenvectors $\hat{\mu}$ of the MDI tensor of the magnetic dipoles in a line (one dimension 1D), within a circle (2D) or a sphere (3D) of radius $r$ surrounding a given moment $\vec{\mu}_i$ for given magnetic propagation vectors~${\bf k}$ for collinear and coplanar noncollinear magnetic structures on both Bravais and non-Bravais spin lattices.  Results are calculated for collinear ordering on 1D chains, 2D square and simple-hexagonal (triangular) Bravais lattices, 2D honeycomb and kagom\'e non-Bravais lattices and 3D cubic Bravais lattices.  The $\lambda$ and $\hat{\mu}$ values are compared with previously reported results.  Calculations for collinear ordering on 3D simple tetragonal, body-centered tetragonal, and stacked triangular and honeycomb lattices are presented for $c/a$ ratios from 0.5 to 3 in both graphical and tabular form to facilitate comparison of experimentally determined easy axes of ordering on these Bravais lattices with the predictions for MDIs.  Comparisons with the easy axes measured for several illustrative collinear antiferromagnets (AFMs) are given.  The calculations are extended to the cycloidal noncollinear 120$^\circ$ AFM ordering on the triangular lattice where $\lambda$ is found to be the same as for collinear AFM ordering with the same~{\bf k}.  The angular orientation of the ordered moments in the noncollinear coplanar AFM structure of GdB$_4$ with a distorted stacked 3D Shastry-Sutherland spin-lattice geometry is calculated and found to be in disagreement with experimental observations, indicating the presence of another source of anisotropy.  Similar calculations for the undistorted 2D and stacked 3D Shastry-Sutherland lattices are reported.  The thermodynamics of dipolar magnets are calculated using the Weiss molecular field theory for quantum spins, including the magnetic transition temperature $T_{\rm m}$ and the ordered moment, magnetic heat capacity and anisotropic magnetic susceptibility $\chi$ versus temperature~$T$\@.  The anisotropic Weiss temperature $\theta_{\rm p}$ in the Curie-Weiss law for $T>T_{\rm m}$ is calculated.  A quantitative study of the competition between FM and AFM ordering on cubic Bravais lattices versus the demagnetization factor in the absence of FM domain effects is presented.  The contributions of Heisenberg exchange interactions and of the MDIs to $T_{\rm m}$ and to $\theta_{\rm p}$ are found to be additive, which simplifies analysis of experimental data. Some properties in the magnetically-ordered state versus~$T$ are presented, including the ordered moment and magnetic heat capacity, and for AFMs the dipolar anisotropy of the free energy and the perpendicular critical field. The anisotropic $\chi$ for dipolar AFMs is calculated both above and below the N\'eel temperature $T_{\rm N}$ and the results are illustrated for a simple tetragonal lattice with $c/a > 1$, $c/a=1$ (cubic) and $c/a<1$, where a change in sign of the $\chi$ anisotropy is found at $c/a=1$.  Finally, following the early work of Keffer [Phys.\ Rev.\ {\bf 87}, 608 (1952)], the dipolar anisotropy of $\chi$ above $T_{\rm N} = 69$~K of the prototype collinear Heisenberg-exchange-coupled tetragonal compound MnF$_2$ is calculated and found to be in excellent agreement with experimental single-crystal literature data above 130~K, where the smoothly increasing deviation of the experimental data from the theory on cooling from 130~K to $T_{\rm N}$ is deduced to arise from dynamic short-range collinear $c$-axis AFM ordering in this temperature range driven by the exchange interactions.

\end{abstract}


\maketitle

\section{\label{Intro} Introduction}

Local magnetic moments generate magnetic dipole fields around them.  In local-moment spin systems, the long-range magnetic dipole interaction between the local magnetic moments (spins) is always present.  However its strength is usually small compared to other interactions such as exchange and RKKY interactions between the spins.  The thermal-average magnitude of the interaction energy is of order $E \sim \mu^2/r^3$, where $\mu$ is the thermal-average value of the magnetic moment and $r$ is the distance between nearest-neighbor spins.  Taking, e.g., $\mu = 7~\mu_{\rm B}$ for Gd$^{+3}$ or Eu$^{+2}$ ($\mu_{\rm B}$ is the Bohr magneton) and $r = 3$~\AA\ gives $E/k_{\rm B} \sim 0.02$~K ($k_{\rm B}$ is Boltzmann's constant), which is usually very small compared to the other interactions between the spins.  However, even when the dipole interactions are weak, these interactions can be decisive in determining the orientations of the ordered moments in magnetic structures of local-moment ferromagnets (FMs) or antiferromagnets (AFMs).

If the distance between local moments is large enough, the magnetic dipole interaction can dominate the exchange interactions in local-moment insulators and result in either FM or AFM dipolar ordering.  Examples include FM ordering between Mn$_6^{+3}$ clusters with spin $S = 12$ at the Curie temperature $T_{\rm C} = 0.161(2)$~K in monoclinic ${\rm Mn_6O_4Br_4(Et_2dbm)_6}$,\cite{Morello2003,Morello2006} and AFM ordering in the face-centered cubic (fcc) diamond lattice of rare-earth $R$ atoms ($R$ = Gd, Dy, Er) in $R{\rm PO_4(MoO_3)_{12}\cdot30 H_2O}$ with N\'eel temperatures $T_{\rm N} = 0.01$~K to 0.04~K.\cite{White1993}

The theoretical study of magnetic dipole interactions and associated magnetic structures in crystals has a long history.  In 1946 Luttinger and Tisza solved for the possible magnetic structures of simple cubic (sc), body-centered cubic (bcc) and fcc Bravais spin lattices arising solely from classical magnetic dipole interactions, where the ordered moments all had the same magnitude (equal-moment magnetic structures).\cite{Luttinger1946}  They found that the ground state for the sc lattice is an AFM state with propagation vector ${\bf k} = (\frac{1}{2},\frac{1}{2},0)$~r.l.u., whereas a FM state with {\bf k} = 0 is the ground state for the bcc and fcc lattices if the samples are in the shape of long thin needles, but AFM structures are the most stable structures otherwise with ${\bf k} = (\frac{1}{2},\frac{1}{2},0)$~r.l.u.\ and ${\bf k} = (\frac{1}{2},\frac{1}{2},\frac{1}{2})$~r.l.u., respectively.  The abbreviation r.l.u.\ means reciprocal lattice unit, where 1~r.l.u. = $2\pi/a$ for cubic lattices and $a$ is the cubic lattice parameter.  Cohen and Keffer confirmed using spin-wave theory that FM cannot be the ground state at $T=0$ for pure magnetic dipole interactions in thin needles of a sc spin lattice but can be the ground state for bcc and fcc lattices.\cite{Cohen1955}  The magnetic structures of two-dimensional (2D) Bravais spin lattices induced by magnetic dipole interactions have also been investigated.\cite{Belobrov1985,Brankov1987,Rozenbaum1991,Feldman2008,Bolcal2012}

Luttinger and Tisza also showed that in a classical cubic dipolar AFM in the magnetically-ordered state at temperature $T=0$ with a magnetic field $H_z$ applied perpendicular to the easy axis of ordering, the component $\mu_z$ of the ordered moment per spin in the direction of $H_z$ is proportional to $H_z$ for $0\leq H_z\leq H_{\rm c}$ and is equal to the saturation moment $\mu_{\rm sat}$ for $H_z>H_{\rm c}$, where $H_{\rm c}$ is termed the critical field.\cite{Luttinger1946} An expression for the magnetic susceptibility $\chi_z=\mu_z/H_z$ for $0\leq H_z\leq H_{\rm c}$ was given. The high-field state with $H\geq H_{\rm c}$ is a field-induced paramagnetic (PM) state in which the magnetic moments are ferromagnetically aligned in the direction of $H_z$ with $\mu_z=\mu_{\rm sat}$.  According to the Weiss molecular field theory (MFT), precisely the same type of $\mu_z(H_z)$ behavior for the perpendicular magnetization occurs for both collinear and coplanar noncollinear AFMs with the spins interacting only by Heisenberg exchange.\cite{Johnston2015}  The susceptibility parallel to the easy axis at $T<T_{\rm N}$ for dipolar AFM ordering in a uniaxial (tetragonal or hexagonal) crystal has not been calculated before to our knowledge.

The so-called pyrochlore spin lattice has attracted much attention over the past two decades in the context of spin-ice compounds.\cite{Bramwell2001}  This non-Bravais fcc spin lattice with 16 spins per fcc unit cell consists of a 3D network of corner-sharing tetrahedra formed by either the $A$ or $B$ sublattices of a pyrochlore-structure compound $A_2B_2X_7$ or by the $B$ sublattice of a spinel-structure compound $AB_2X_4$.  An example is the Ho sublattice in the pyrochlore compound ${\rm Ho_2Ti_2O_7}$, where due to crystalline electric field effects the Ho cations behave at low~$T$ like Ising spins that can only point along the [111] and equivalent crystal directions (the Ti$^{+4}$ cations are nonmagnetic).  The spin-ice arrangement of the Ho moments at low~$T$ gives rise to a macroscopic degeneracy and a nonzero spin entropy at $T=0$, as occurs in water ice.  Magnetic dipole interactions between the Ho moments have been determined to be important to this magnetic behavior,\cite{Bramwell2001} and hence these compounds are sometimes referred to as dipolar spin ices.

On another front, dynamic magnetic fluctuations in long-range ordered 3D AFMs mediated by magnetic dipole interactions are stronger than for exchange interactions on the same lattice,\cite{White1993,Corruccini1993} contrary to what might have been anticipated from the classical origin of the magnetic dipole interaction.  In particular, in the cubic diamond lattice dipolar AFMs $R{\rm PO_4(MoO_3)_{12}\cdot30H_2O}$ ($R$ = Gd, Dy, Er), White {\it et al.}\ found that the suppression of the $T\to0$ susceptibility versus $1/S$ ($S$ is the effective spin quantum number) due to quantum fluctuations was a factor of two stronger than predicted for the nearest-neighbor Heisenberg model on the diamond lattice.\cite{White1993}  Corruccini and White found that within spin-wave theory, the 3D sc lattice exhibits quantum corrections to the N\'eel state that are also a factor of two larger than those of the nearest-neighbor Heisenberg AFM on the same lattice, indicating that dipolar magnets are more quantum mechanical than generally suspected,  whereas the 2D dipolar square lattice does not exhibit long-range order at finite temperature.\cite{Corruccini1993}  On the other hand, several authors have found that dipolar interactions in conjunction with Heisenberg interactions can induce long-range order at finite temperatures on 2D spin lattices.\cite{Maleev1976,Bruno1991,Ivanov1996}

The influence of magnetic dipole interactions on the magnetic properties of 3D Bravais spin lattices with unit cell symmetries lower than cubic has been discussed for particular cases.  Rotter has discussed the predictions of dipolar interactions for the easy axis of collinear AFMs with AFM propagation vectors {\bf k} that are determined by isotropic Heisenberg exchange interactions in a variety of collinear AFM compounds containing sc, fcc, hexagonal and body-centered tetragonal (bct) Gd sublattices.\cite{Rotter2003}  He found that in most cases the easy axis is consistent with that predicted for magnetic dipole interactions. Several authors calculated the local dipolar fields at a lattice site for general simple tetragonal and bct Bravais spin lattices versus a parameter not proportional to the $c/a$ ratio.\cite{Lo2001a,Lo2001b,Huang2004}  Maurya {\it et al.}\ calculated the influence of magnetic dipole interactions on the magnetization versus field isotherms of three AFMs containing Eu$^{+2}$ spins-7/2 below their N\'eel temperatures of 12~K to 15~K.\cite{Maurya2014, Maurya2015}

Classical Monte Carlo (MC) simulations on Heisenberg spin systems have been carried out on a variety of spin lattices to examine the influence of magnetic dipole interactions on the properties with either dipolar interactions only or in combination with other spin interactions.  For purely dipole interactions, Bouchaud and Z\'erah studied FM on the fcc lattice and determined the  Curie temperature $T_{\rm C}$.\cite{Bouchaud1993} They studied the critical exponents at $T_{\rm C}$ and determined the anisotropy constants $K_1$ and $K_2$, where they found that the ordered moment direction in the collinear FM state at the lowest $T$ was along [100], with a crossover from [111] at higher~$T$\@.  Tomita reported MC simulations on 2D triangular, square, honeycomb and kagom\'e spin lattices with only dipolar interactions and studied the ground state magnetic structures and critical phenomena.\cite{Tomita2009} One result was that the kagom\'e lattice has a FM ground state with 1/3 of the spins disordered at $T=0$ [an amplitude-modulated (AM) state] with residual entropy (``missing entropy'') at $T=0$ resulting from macroscopic degeneracy of the ground state.  [An AM magnetic structure is one where the magnitude of the {\it ordered} moment is not the same for all (identical) spins in the spin lattice.] Very recent MC simulations on the kagom\'e lattice by Holden {\it et al.}\cite{Holden2015} and Maksymendo {\it et al.}\cite{Maksymenko2015} instead found a noncollinear coplanar equal-moment ground-state magnetic structure on the kagom\'e latice.  Thus when an AM magnetic structure is obtained theoretically for a particular spin lattice, this may indicate that a lower-energy equal-moment magnetic structure exists in which the moments have their maximum (saturation) value.\cite{Maksymenko2015}  Other MC simulations examined the influence of dipolar interactions on the properties in combination with other interactions.\cite{Maksymenko2015,Suzuki1983, Fernandez2006}  

In this paper previous work on the effects of magnetic dipolar interactions on the magnetic and thermal properties of magnetic systems is significantly extended.  Usually exchange and/or RKKY interactions are stronger than dipole interactions and determine the nature (FM or AFM) and {\bf k} of the magnetic structure.  However, when the exchange interactions are Heisenberglike (isotropic), some sort of anisotropy is needed to determine the directions of the ordered moments in the ordered state as discussed above, even if very weak compared to the exchange interactions.  The present work was initially motivated by the lack of systematic studies of this topic for uniaxial tetragonal and hexagonal Bravais spin lattices versus the $c/a$ ratio to compare with experimental results such as for the Eu$^{+2}$ spins~$S=7/2$ in ${\rm EuCu_2Sb_2}$ on a bct sublattice that exhibit collinear AFM ordering below $T_{\rm N} = 5.1$~K.\cite{Anand2015,Ryan2015}

We study the influence of dipolar interactions on the magnetic ordering temperature $T_{\rm m}$, on the collinear ordered moment directions and the temperature $T$ dependence of the ordered moment and other properties at $T \leq T_{\rm m}$, and on the Weiss temperature in the Curie-Weiss law at $T \geq T_{\rm m}$ in a systematic way for a variety of spin lattices including 1D, 2D and 3D spin lattices using our recent formulation of the Weiss MFT.\cite{Johnston2015}  All spins in a given system are assumed to be identical and crystallographically equivalent.  The 3D spin lattices studied here include sc, bcc, fcc, simple tetragonal, bct and simple hexagonal (triangular) Bravais lattices.  Non-Bravais spin lattices are also studied which include the honeycomb (chickenwire) lattice, the kagom\'e lattice and the Shastry-Sutherland lattice.  For the uniaxial stacked lattices, the eigenvalues and eigenvectors of the magnetic dipole interaction (MDI) tensor are calculated for $c/a$ ratios from 0.5 to~3.  We utilize an appropriately modified theory to calculate the properties of noncollinear AFM structures and compare the results with calculations assuming collinear AFM structures for the same {\bf k}.  Within MFT, the contributions of different sources of molecular fields to the Weiss temperatures and the magnetic ordering temperatures are additive.  Therefore, for example, when dipolar and exchange interactions are simultaneously present, one can calculate the dipole contributions to good accuracy and then subtract them from the observed values to obtain the contributions from the exchange interactions.  Then with a model for the exchange interactions one can estimate their values.  In addition to calculating the magnetic and thermal properties of pure dipolar magnets, the anisotropy in the susceptibility of Heisenberg AFMs in the PM regime with $T \geq T_{\rm m}$ is also computed.  We compare our predictions to the magnetic properties measured for illustrative real materials.  In this paper we do not consider critical phenomena, domain formation and similar effects in FMs or other potential sources of magnetic anisotropy in a spin system such as single-ion effects.

Our theoretical framework allows easy extensions to calculate the dipolar contributions to the magnetic properties of spin lattices not discussed here such as collinear or noncollinear ordering on orthorhombic, monoclinic and triclinic Bravais or other non-Bravais spin lattices.  

In Sec.~\ref{Sec:Theory} we first write down the expressions relating the macroscopic magnetic induction, applied magnetic field and magnetization including shape (demagnetizing) effects. The expression for the local field seen by a spin is discussed in Sec.~\ref{Sec:LocalField}.  The part of that local field (the near field) due to discrete moments inside a macroscopic Lorentz sphere is discussed in Sec.~\ref{Sec:NearField}, together with the energy of a spin interacting with the near field.  Applications of the general theory in Sec.~\ref{Sec:NearField} to magnetically-ordered states in collinear magnets, non-Bravais spin lattices and coplanar noncollinear helical or cycloidal AFMs are presented in Secs.~\ref{Sec:CollinearThy}, \ref{Sec:NonBravais} and~\ref{Sec:HelCyc}, respectively.  The expression for the near field due to moments within a Lorentz line (1D), circle (2D) or sphere (3D) is discussed in Sec.~\ref{Eq:NearField}.  Some details about calculations of the MDI tensor are given in Sec.~\ref{Sec:Gcalcs}.  In Appendix~\ref{Methods} some information useful for implementing the theory in Sec.~\ref{Sec:Theory} is discussed.

The calculations of the eigenvalues and eigenvectors of the MDI tensor for collinear magnetic structures with specific magnetic propagation vectors for 1D and 2D spin lattices are given in Secs.~\ref{Sec:1D2D}, where the 2D spin lattices include the square, triangular, honeycomb and kagom\'e lattices.  Three-dimensional spin lattices are considered in Sec.~\ref{Sec:3DLattices}, where results are given for the three cubic Bravais lattices, the two tetragonal Bravais lattices, the simple hexagonal lattice and the honeyomb lattice.  For the 3D tetragonal and hexagonal lattices the eigenvalues and eigenvectors are obtained versus the $c/a$ ratio from $c/a=0.5$ to 3 in 0.1 increments.  For all spin lattices, we carry out calculations of the MDI tensor of a central spin with its neighbors by direct summation with increasing radius away from the central moment until convergence is achieved within at least 0.001\%. The convergence of the dipolar sums is discussed in the corresponding section, and representative convergence plots are given in Appendix~\ref{App:RtoInfty}.

The predictions of the easy axis for collinear AFM ordering are compared with experimental results for the simple-tetragonal Mn and Fe sublattices in ${\rm BaMn_2As_2}$ and ${\rm BaFe_2As_2}$, and for the bct spin lattices in ${\rm GdCu_2Si_2}$, ${\rm EuCu_2Sb_2}$ and MnF$_2$.  For these cases we compare the results of the eigenvalues and eigenvectors versus the $c/a$ ratio in graphical format with the experimental data, and the graphical results for other cases are placed in Appendix~\ref{App:lambdaFigs}.  The treatment of noncollinear AFMs is presented in Sec.~\ref{Sec:NoncollearAFMs}, with application to the 120$^\circ$ ordering on the triangular lattice, to the 90$^\circ$ ordering on the distorted Shastry-Sutherland GdB$_4$ compound and to the undistorted 2D and 3D Shastry-Sutherland lattices.

Section~\ref{Tm} presents the calculation of the FM ordering temperature $T_{\rm C}$ and AFM ordering temperature $T_{\rm N}$ arising from dipolar interactions within our recent formulation of MFT.\cite{Johnston2015}  A quantitative discussion of the competition between FM and AFM ordering on cubic Bravais lattices versus the demagnetization factor of a sample in the absence of FM domain formation is given in Sec.~\ref{cubicFMAFM}.  The properties of dipolar magnets in the magnetically-ordered state are derived in Sec.~\ref{Eq:PropsBelowTm}.  The ordered moment and heat capacity of dipolar magnets in zero magnetic field versus temperature are presented in Sec.~\ref{Sec:mu0Cmag}, where the results are the same within MFT for both FMs and AFMs.  The dipolar anisotropy parameter $K_1$  for uniaxial dipolar AFMs versus temperature is derived in Sec.~\ref{Sec:AFMAnisotropy}.  Calculations of the perpendicular susceptibility below $T_{\rm N}$ and the associated critical field for uniaxial AFMs are presented in Secs.~\ref{Sec:ChiPerp} and~\ref{Sec:Hc}, respectively.

The Curie-Weiss law for dipolar magnets in the PM state is derived in Sec.~\ref{Sec:CWLaw}, where the Weiss temperature is found to be anisotropic in general.  In Sec.~\ref{Sec:AnisChiAFM} we specialize to spherical samples of collinear AFMs, where the anisotropic susceptibilities $\chi$ for temperatures above $T_{\rm N}$ as well as both the parallel and perpendicular susceptibilities below $T_{\rm N}$ are presented and discussed.  Examples of these anisotropic $\chi(T)$ behaviors are given in Sec.~\ref{Sec:AFMChiExample} for simple tetragonal lattices with $c/a<1$, $c/a=1$~(sc) and $c/a>1$.

The anisotropic $\chi(T)$ of a Heisenberg-exchange AFM at $T > T_{\rm N}$ due to MDIs is derived in Sec.~\ref{Sec:AnisHeisAFM} and applied to fit the experimental data for single-crystal MnF$_2$.  The paper concludes with a short summary in Sec.~\ref{Sec:Summary}.

Tables of values of the dipolar eigenvalues and eigenvectors versus $c/a$ plotted in the text and Appendix~\ref{App:lambdaFigs} are available in the Supplemental Material.\cite{SupplInfo}

\section{\label{Sec:Theory} Theory}

The magnetization per unit volume of magnetic materials can be significant compared to the applied field and results in a demagnetizing field and an internal field smaller than the applied field.  In the following the theory for this important demagnetizing correction is discussed within the Gaussian cgs system of units\cite{Jackson1962,Johnston2010} that is used throughout this paper.

\subsection{\label{Sec:MacroFields} Macroscopic Fields}

We initially assume that a sample has the shape of an ellipsoid of revolution and that the applied field is along one of the three principal axes~$\alpha$.  Then  the volume magnetization (net magnetic moment per unit volume) {\bf M} (units: G) is uniform in the sample and the magnetic induction {\bf B} (units: G), the magnetic field {\bf H} (units: Oe = G) and {\bf M} are collinear with components $M_\alpha$, $H_\alpha$ and $B_\alpha$ for the external field ${\bf H}_\alpha$ applied along the $\alpha$ axis.  For each point in space one has
\bse
\label{Eqs:B}
\be
B_\alpha = H_\alpha + 4\pi M_\alpha.
\ee
Thus internal to the sample one has 
\be
B_{{\rm int}\,\alpha} = H_{{\rm int}\,\alpha} + 4\pi M_\alpha.
\label{Eq:Bin}
\ee
The demagnetizing field internal to the sample due to $M_\alpha$ is
\be
H_{{\rm d}\alpha} = -4\pi N_{{\rm d}\alpha}M_\alpha,
\label{Eq:Hdemag}
\ee
\ese
where here the demagnetizing factor $N_{{\rm d}\alpha}$ is defined as in the {\it Syst\`eme International} system of units for which $0 \leq N_{{\rm d}\alpha} \leq 1$ and $\sum_{\alpha=1}^3 N_{{\rm d}\alpha} = 1$.  Thus the internal magnetic field $H_{{\rm int}\,\alpha}$ and the magnetic induction $B_{{\rm int}\,\alpha}^{\rm shape}$ due to sample shape effects and including the applied field $H_{\alpha}$ are
\bse
\label{Eqs:HinBin}
\bea
H_{{\rm int}\,\alpha} &=& H_{\alpha} -4\pi N_{{\rm d}\alpha}M_\alpha,\label{Eq:Hin}\\*
B_{{\rm int}\,\alpha}^{\rm shape} &=& H_{{\rm int}\alpha} + 4\pi M_\alpha \nonumber\\*
&=& H_{\alpha} + (1-N_{{\rm d}\alpha}) 4\pi M_\alpha.\label{Eq:BintShape}
\eea
\ese
For a given $M_\alpha$, the internal field is $H_{{\rm int}\alpha}$ in Eq.~(\ref{Eq:Hin}).  Thus in  descriptions of the magnetic behavior of a sample in terms of $M_\alpha$ and $H_{\alpha}$, one can correct for the demagnetizing field by retaining the measured value of $M_\alpha$ but replacing $H_{\alpha}$ by $H_{\alpha} -4\pi N_{{\rm d}\alpha}M_\alpha$, where $N_{{\rm d}\alpha}$ is estimated from the sample shape and the field orientation with respect to the sample (see below). 

The magnetic susceptibility of a material is often defined as $\chi = M(H)/H$, which in general is field-dependent. In the present discussion, $M$ is the volume magnetization, so $\chi$ is the susceptibility per unit volume and is dimensionless.  The observed susceptibility is then $\chi^{\rm obs}_\alpha = M_\alpha/H_{\alpha}$ and the intrisic susceptibility is $\chi_\alpha = M_\alpha/H_{{\rm int}\,\alpha}$.  Utilizing Eq.~(\ref{Eq:Hin}) one obtains $\chi_\alpha$ from $\chi_\alpha^{\rm obs}$ according to
\bea
\chi_\alpha &=& \frac{M_\alpha}{H_{{\rm int}\,\alpha}}  = \frac{M_\alpha}{H_{\alpha} -4\pi N_{{\rm d}\alpha}M_\alpha} = \frac{M_\alpha/H_{\alpha}}{1 -4\pi N_{{\rm d}\alpha}M_\alpha/H_{\alpha}}\nonumber\\*
&=& \frac{\chi^{\rm obs}_\alpha} {1-4\pi N_{{\rm d}\alpha}\chi^{\rm obs}_\alpha}.
\label{Eq:ChiCorr}
\eea
At each temperature one can correct the observed susceptibility for the demagnetizing field using Eq.~(\ref{Eq:ChiCorr}).

Alternatively, using Eq.~(\ref{Eq:ChiCorr}) one can write the observed susceptibility in terms of the intrinsic one as
\be
\chi_\alpha^{\rm obs} = \frac{\chi_\alpha} {1+4\pi N_{{\rm d}\alpha}\chi_\alpha}.
\ee
Thus when $4\pi N_{{\rm d}\alpha}\chi_\alpha \gg1$, one obtains the field-independent susceptibility and linear $M_\alpha(H_{\alpha})$ behavior
\be
\chi_\alpha^{\rm obs} = \frac{1} {4\pi N_{{\rm d}\alpha}}, \qquad M_\alpha = \frac{1} {4\pi N_{{\rm d}\alpha}}H_{\alpha}.
\label{Eq:chiObsLimit}
\ee
The latter behavior holds until $M_\alpha$ reaches it saturation (maximum) value $M_{{\rm sat}\,\alpha}$; at higher fields $M_\alpha$ is of course equal to its constant saturation value.  In practice, the limiting behaviors in Eqs.~(\ref{Eq:chiObsLimit}) are realized only when a material is approaching its FM transition temperature from above.

An expression for the demagnetizing factors $N_{{\rm d}\alpha}$ for the general ellipsoid of revolution was calculated long ago.\cite{Osborn1945}  For sample shapes other than ellipsoids, $M$ is not uniform within the sample except for limiting cases.  What is then relevant in the present context is the demagnetizing field averaged over the sample volume as expressed in the associated ``magnetometric'' demagnetizing factor.  Such sample shapes include the cylinder and the rectangular parallelepiped (rectangular prism) for which the magnetometric $N_{\rm d\alpha}$ values have been calculated for arbitrary sample dimensions in Refs.~\onlinecite{Chen1991} and~\onlinecite{Aharoni1998}, respectively.

\subsection{\label{Sec:LocalField} Local Magnetic Induction from Magnetic Dipole Interactions}

Theoretical predictions of magnetic properties for local magnetic moments are often cast in terms of the local magnetic induction ${\bf B}_{{\rm int}\,i}^{\rm local}$ seen by a local moment $\vec{\mu}_i$ at position ${\bf r}_i$.  This local magnetic induction along a given principal axis~$\alpha$ is traditionally written for a 3D spin lattice in terms of the four contributions\cite{Kittel2005}
\be
B_{{\rm int}\,\alpha\,i}^{\rm local} = B_{\alpha} + B_{{\rm int}\,\alpha}^{\rm shape} + B_{{\rm int}\,\alpha}^{\rm Lorentz} + B_{{\rm int}\,\alpha\,i}^{\rm near},
\label{Eq:Bintalphai}
\ee
where $B_{\alpha} = H_{\alpha}$ is the applied magnetic induction arising from currents outside the sample and $B_{{\rm int}\,\alpha}^{\rm shape}$ is the contribution in Eq.~(\ref{Eq:Bin}) due to the sample shape.  The contribution
\be
B_{{\rm int}\,\alpha}^{\rm Lorentz} = \frac{4\pi}{3}M_\alpha
\label{Eq:BintLorentz}
\ee
is the Lorentz cavity field inside a spherical cavity of radius $R$ surrounding the point at its center at position ${\bf r}_i$ at which $B_{{\rm int}\,\alpha\,i}^{\rm local}$ is to be calculated.  The fourth contribution $B_{{\rm int}\,\alpha\,i}^{\rm near}$ is the sum of the dipolar fields at position ${\bf r}_i$ arising from the other magnetic dipoles inside the Lorentz cavity at positions ${\bf r}_j$.  This is the only term that depends on the crystal structure of the material.  The Lorentz cavity radius $R$ is much larger than the distance between magnetic moments in a sample and is large enough so that the calculated $B_{{\rm int}\,\alpha\,i}^{\rm near}$ becomes independent of $R$ to within some specified precision.  Substituting Eqs.~(\ref{Eq:BintShape}) and~(\ref{Eq:BintLorentz}) into~(\ref{Eq:Bintalphai}) gives
\bse
\be
B_{{\rm int}\,\alpha\,i}^{\rm local} = H_{\alpha} + \left(\frac{1}{3}-N_{{\rm d}\alpha}\right) 4\pi M_\alpha + B_{{\rm int}\,\alpha\,i}^{\rm near}.
\label{Eq:Bintalphai2}
\ee
This is an important fundamental equation for calculating the local field.

Two special cases of Eq.~(\ref{Eq:Bintalphai2}) are of use. In the first, one  corrects the applied field for the demagnetizing field in the measurements as described above which is equivalent to removing $N_{\rm d\alpha}$ from Eq.~(\ref{Eq:Bintalphai2}), yielding
\be
B_{{\rm int}\,\alpha\,i}^{\rm local} = H_{\alpha} + \frac{4\pi}{3} M_\alpha + B_{{\rm int}\,\alpha\,i}^{\rm near}.
\label{Eq:Bintalphai3}
\ee
This equation is sometimes favored for comparison of theoretical predictions of the dipolar magnetic properties with experimental data because it is independent of sample shape.  Here $M_\alpha$ is the total magnetic moment per unit volume.  If all spins are identical and crystallographically equivalent as assumed throughout this paper, one can write $M_\alpha = \mu_\alpha/V_{\rm spin}$ where $\mu_\alpha$ is the net average ordered and/or induced moment per spin in the $\alpha$ direction and $V_{\rm spin}$ is the volume per spin, so an equivalent form of Eq.~(\ref{Eq:Bintalphai3}) is
\be
B_{{\rm int}\,\alpha\,i}^{\rm local} = H_\alpha + \frac{4\pi}{3V_{\rm spin}} \mu_\alpha + B_{{\rm int}\,\alpha\,i}^{\rm near}.
\label{Eq:Bintalphai4}
\ee
Note that $M_\alpha=\mu_\alpha=0$ for an AFM in $H=0$.

Alternatively, one can shape a sample into a sphere, giving $N_{\rm d\alpha}=1/3$ for all three principal directions $\alpha$, and then Eq.~(\ref{Eq:Bintalphai2}) becomes
\be
B_{{\rm int}\,\alpha\,i}^{\rm local} = H_{\alpha}+ B_{{\rm int}\,\alpha\,i}^{\rm near},
\label{Eq:Bintalphai5}
\ee
\ese
which eliminates the effect of the Lorentz field but only applies to a spherical sample.  This formulation is desirable if one wishes to ameliorate the tendency of the Lorentz field to enhance dipolar FM ordering with respect to AFM ordering, as illustrated in Fig.~\ref{Fig:TmagVsNda_sc} below where FM is favored for small values of $N_{\rm d\alpha}$ for bcc and fcc Bravais lattices.

\subsection{\label{Sec:NearField} Magnetic Induction Due to Collinear Alignment of Magnetic Dipoles Inside Lorentz Cavity}

The magnetic induction ${\bf B}_{ij}$ seen by a central moment $\vec{\mu}_i$ at a position ${\bf r}_i$ due to a point magnetic dipole moment $\vec{\mu}_j$ at position ${\bf r}_j$ is 
\bse
\be
{\bf B}_{ij} = \frac{1}{r_{ji}^5}[3(\vec{\mu}_j\cdot {\bf r}_{ji}){\bf r}_{ji} - r_{ji}^2\vec{\mu}_j],
\ee
where
\be
{\bf r}_{ji} = {\bf r}_{j} - {\bf r}_{i},\qquad r_{ji}=|{\bf r}_{ji}|.
\ee
\ese

The energy of interaction $E_i$ of $\vec{\mu}_i$ at position ${\bf r}_i$ due to the magnetic induction ${\bf B}_{ij}$ is
\be
E_i =-\frac{1}{2} \vec{\mu}_i\cdot {\bf B}_{ij} = -\frac{1}{2r_{ji}^5}[3(\vec{\mu}_i\cdot {\bf r}_{ji})(\vec{\mu}_j\cdot {\bf r}_{ji}) - r_{ji}^2\vec{\mu}_i\cdot \vec{\mu}_j],
\label{Eq:Eij}
\ee
where the factor of 1/2 in the first equality recognizes that the interaction energy of the $\vec{\mu}_i$ with ${\bf B}_{ij}$ from $\vec{\mu}_j$ is equally shared between $\vec{\mu}_i$ and $\vec{\mu}_j$.  Expanding the first term on the right side of Eq.~(\ref{Eq:Eij}) in Cartesian coordinates, one can write the term in matrix form as
\begin{widetext}
\be
(\vec{\mu}_i\cdot {\bf r}_{ji})(\vec{\mu}_j\cdot {\bf r}_{ji}) = (\mu_{ix}\ \mu_{jy}\ \mu_{jz}) \left(
\begin{array}{ccc}
r_{jix}^2 & r_{jix}r_{jiy} & r_{jix}r_{jiz}\\
r_{jix}r_{jiy} & r_{jiy}^2 & r_{jiy}r_{jiz}\\
r_{jix}r_{jiz} & r_{jiy}r_{jiz} & r_{jiz}^2\\
\end{array}
\right)
\left(
\begin{array}{c}
\mu_{jx}\\
\mu_{jy}\\
\mu_{jz}\\
\end{array}
\right) = \vec{\mu}_i^{\rm T}{\bf r}_{ji}{\bf r}_{ji} \vec{\mu}_j,
\label{Eq:1stterm}
\ee
\end{widetext}
where $\vec{\mu}_i^{\rm T}$ is the transpose of the column vector $\vec{\mu}_i$, $\vec{\mu}_j$ is a column vector and ${\bf r}_{ji}{\bf r}_{ji}$ is a $3\times3$ dyadic.  Similarly, the scalar product in the second term on the right side of Eq.~(\ref{Eq:Eij}) can be written in matrix form as 
\be
\vec{\mu}_i\cdot \vec{\mu}_j = \vec{\mu}_i^{\rm T}{\bf 1}\vec{\mu}_j,
\label{Eq:2ndterm}
\ee
where {\bf 1} is the $3\times3$ identity matrix.  Using Eqs.~(\ref{Eq:1stterm}) and~(\ref{Eq:2ndterm}), Eq.~(\ref{Eq:Eij}) can be summed over all neighbors~$\vec{\mu}_j$ within a length of chain (1D), a circle of specified radius (2D) or Lorentz sphere (3D), all centered on $\vec{\mu}_i$, and then can be succinctly written in matrix form as
\bse
\label{Eqs:Ei}
\be
E_{i} = -\frac{1}{2}\vec{\mu}_i^{\rm T}{\bf G}_{i}\vec{\mu}_j.
\label{Eq:Ei}
\ee
where the $3\times3$ symmetric tensor ${\bf G}_{i}$ is 
\be
{\bf G}_{i} =  \sum_{j\neq i}\frac{1}{r_{ji}^5}(3{\bf r}_{ji}{\bf r}_{ji} - r_{ji}^2{\bf 1}).
\ee
\ese
In order to solve Eq.~(\ref{Eq:Ei}) for the eigenenergies $E_i$ and eigenvectors $\hat{\mu}_i$ of the tensor~${\bf G}_{i}$, one must first express each $\vec{\mu}_j$ in terms of $\vec{\mu}_i$.  In the following three sections we discuss our methods for doing so for collinear magnetic structures on Bravais and non-Bravais spin lattices and  coplanar noncollinear AFM structures, respectively.

\subsection{\label{Sec:CollinearThy} Collinear Magnetic Structures}

In this section we consider collinear magnetic structures with magnetic wavevector {\bf k} where
\be
\vec{\mu}_j = \cos({\bf k}\cdot{\bf r}_{ji})\vec{\mu}_i.
\label{Eq:muj}
\ee
Since the cosine function is a scalar with a value between $\pm1$, Eq.~(\ref{Eq:muj}) expresses that $\vec{\mu}_j$ can be either parallel or antiparallel to $\vec{\mu}_i$. For $\cos({\bf k}\cdot{\bf r}_{ji}) = \pm1$ for each $\vec{\mu}_j$ the magnetic structure is an ``equal-moment'' (EM) structure where the ordered moments all have the same magnitude $\mu$ (which depends on $T$).  For $\cos({\bf k}\cdot{\bf r}_{ji}) \neq \pm1$ for some $\vec{\mu}_j$, $\mu$ depends on~$j$ and the structure is a collinear AM AFM structure.  Collinear magnetic structures include both FM ({\bf k} = 0) and AFM structures below the magnetic ordering temperature $T_{\rm m}$ and the FM-aligned magnetic structure induced above $T_{\rm m}$ by an external magnetic field applied along one of the three principal axes of the MDI in Eq.~(\ref{Eq:Gk2}) below.  From Eq.~(\ref{Eq:muj}) one obtains the ``extinction condition''
\be
\vec{\mu}_j = 0\quad {\rm if}\quad {\bf k}\cdot{\bf r}_{ji} = {\rm odd~multiple~of}\ \frac{\pi}{2}~{\rm rad},
\label{Eq:Extinction}
\ee
as in Eq.~(\ref{Eq:kExtinction}) for AM AFM structures associated with specific {\bf k} values and spin lattices.  A general ${\bf k}$ corresponds to either an EM or AM collinear AFM structure.  

All simple Bravais lattices have EM magnetic structures.  Amplitude-modulated AFM structures occur when the simple Bravais lattices have more than one spin in the unit cell such as for bcc, bct and fcc spin lattices.  With the $\cos({\bf k}\cdot{\bf r}_{ji})$ term as given in Eq.~(\ref{Eq:muj}), EM structures occur for bcc and bct lattices with ${\bf k} = \left(\frac{1}{2},\frac{1}{2},0\right)$, $\left(\frac{1}{2},0,\frac{1}{2}\right)$ and (001)~r.l.u., and AM structures for ${\bf k} = \left(\frac{1}{2},0,0\right)$ and $\left(\frac{1}{2},\frac{1}{2},\frac{1}{2}\right)$~r.l.u.  For the fcc lattice, EM structures occur for ${\bf k} = \left(\frac{1}{2},\frac{1}{2},\frac{1}{2}\right)$ and~(0,0,1)~r.l.u, whereas AM structures occur for ${\bf k} = \left(\frac{1}{2},0,0\right)$, $\left(\frac{1}{2},\frac{1}{2},0\right)$, $\left(\frac{1}{2},0,1\right)$ and $\left(\frac{1}{3},\frac{1}{3},\frac{1}{3}\right)$.  With the exception of the last one, all AM structures considered can be converted into EM structures by inserting an additive phase in the cosine term: $\cos({\bf k}\cdot{\bf r}_{ji})\to \cos({\bf k}\cdot{\bf r}_{ji}+\phi)$, where $\phi=\pi/4$~rad.  In that case, all eigenvalues are reduced in magnitude by the factor $\cos(\pi/4)= 1/\sqrt{2}$, which corresponds to a reduction in the ordered moment by a factor of $1/2^{1/4}$.  All eigenvalues plotted or listed in this paper were obtained for $\phi = 0$.

In pure magnetic dipole AFMs, the above discussion shows that the AFM ground state can be an AM state, depending on the AFM wavevector.  However, even in systems in which the magnetic dipole interaction is not expected to play an important role, this interaction can still cause a small modulation of the ordered moment versus position in the magnetic unit cell. Furthermore, large-amplitude AM AFM structures are observed in geometrically-frustrated systems such as in ${\rm Gd_2Ti_2O_7}$.\cite{Paddison2015}  Because AM structures contain at least some fraction of spins with ordered moments less than the saturation moment and hence show strong quantum fluctuations in the ground state, the entropy increase on heating from low temperatures would be less than the value $R\ln(2S+1)$ per mole of spins.  This can be checked by calorimetry.

The discussion throughout this paper applies to identical crystallographically-equivalent spins with identical saturation moments $\mu_{\rm sat}$  and with thermal-average (ordered) magnetic moments $\vec{\mu}_i = \mu\hat{\mu}_i$, where $\mu$ can be different for different spins in AM structures.  We express $r_{ji}$ in units of the lattice parameter $a$ of the respective crystal structure. The crystallographic unit cell often contains more than one spin per unit cell in the examples described.  Then using Eq.~(\ref{Eq:muj}), Eqs.~(\ref{Eqs:Ei}) become\cite{Lax1952}
\bse
\label{Eqs:EiSoln2}
\be
E_{i} = -\epsilon\ \hat{\mu}_i^{\rm T}\widehat{{\bf G}}_i({\bf k})\hat{\mu}_i,
\label{Eq:Ei3}
\ee
where
\be
\epsilon = \frac{\mu^2}{2a^3}
\label{Eq:eps}
\ee
has dimensions of energy and the dimensionless symmetric MDI tensor is
\be
\widehat{{\bf G}}_{i} = \sum_{j\neq i}\frac{1}{(r_{ji}/a)^5}\bigg(3\frac{{\bf r}_{ji}{\bf r}_{ji}}{a^2} - \frac{r_{ji}^2}{a^2}{\bf 1}\bigg)\cos({\bf k}\cdot{\bf r}_{ji}).
\label{Eq:Gk2}
\ee 
Labeling the eigenvalues of $\widehat{{\bf G}}_{i}({\bf k})$ as $\lambda_{{\bf k}\alpha}$, Eq.~(\ref{Eq:Ei3}) gives the eigenenergies as
\be
E_{i\alpha} = -\epsilon\ \lambda_{{\bf k}\alpha},
\label{Eq:Eilambda}
\ee
\ese
where the subscript $\alpha$ refers to a Cartesian principal ordering axis eigenvector of the collinear magnetic structure, where the three principal axes are orthogonal to each other.  Thus the ground state energy and ordering axis for a given {\bf k} due to the MDI corresponds to the largest of the three $\lambda_{{\bf k}\alpha}$ eigenvalues.  The MDI energy scale is set by the value of $\epsilon$ in Eq.~(\ref{Eq:eps}) which is system-dependent.  The value of $\epsilon/k_{\rm B}$ is typically of order 0.01--0.1~K\@.

The magnetic propagation vector {\bf k} must be specified in terms of the reciprocal lattice translation vectors in Cartesian coordinates in advance of computing $\widehat{{\bf G}}_{ia}({\bf k})$.  One can calculate the $\lambda_{{\bf k}\alpha}$ eigenvalues and corresponding eigenvectors (ordered moment axes $\hat{\mu}_i$) for various {\bf k} vectors, including {\bf k} = 0 for FM-aligned moments which may occur due to FM ordering in applied field $H=0$ or to $H > 0$ in the PM state.  Usually the magnetic {\bf k} vector observed by, e.g., neutron diffraction measurements, is determined by exchange or RKKY interactions rather than dipole interactions.  In that case one can still test whether the easy axis predicted by the MDI is consistent with the observed one.  A negative answer would indicate that the MDI does not contribute to determining the easy axis, and hence some stronger source of magnetocrystalline anisotropy must be present that overcomes the preference of MDIs.  A positive answer would mean that the MDI at least contributes to ordering along the observed easy axis; however, this does not rule out other sources of anisotropy that may also contribute.

A general feature of the eigenvalues $\lambda_{{\bf k}\alpha}$ of the MDI tensor $\widehat{{\bf G}}_{i}$ for a given {\bf k} and spin lattice is that their sum over the three eigenvectors $\alpha$ is identically zero when no {\it a priori} constraint is placed on the ordering axis of $\vec{\mu}_i$.  This sum rule is violated when such a constraint is imposed such as for coplanar noncollinear helical or cycloidal AFM order as discussed in Secs.~\ref{Sec:HelCyc} and~\ref{TriangAFMs}.  In those cases, one of the $\lambda_{{\bf k}\alpha}$ is the eigenvalue for FM ordering ({\bf k} = 0) along the helix or cycloid axis.  The other two eigenvalues and corresponding eigenvectors are the ones associated with the actual AFM components of the helix or cycloid.

\subsection{\label{Sec:NonBravais} Non-Bravais Spin Lattices}

A crystal structure consists of a Bravais lattice plus a basis of atoms attached to each Bravais lattice point.  Non-Bravais spin lattices are Bravais lattices with more than one spin in the basis.  These include, e.g., the fcc diamond lattice and the 2D hexagonal honeycomb (or chickenwire) lattice, each with two spins in the basis, and the kagom\'e lattice with three spins in the basis.  In such cases one must modify Eq.~(\ref{Eq:Gk2}) to include a sum over the atoms in the basis, in addition to the sum over Bravais lattice points already included in Eq.~(\ref{Eq:Gk2}) via, e.g., Eqs.~(\ref{Eqs:spinPos}).  AFM structures in such non-Bravais spin lattices include those with AFM propagation vector {\bf k} = 0  for N\'eel-type ordering on the 2D honeycomb lattice, which is the same propagation vector as for FM ordering.  In such AFM structures where the magnetic and crystallographic unit cells are the same, in order to calculate $\widehat{{\bf G}}_{i}$ one must specify the orientations of the ordered moments within a unit cell with respect to the orientation of a central moment $\vec{\mu_i}$.  Thus Eq.~(\ref{Eq:Gk2}) is modified to read
\bse
\label{Eqs:NonBravais}
\be
\widehat{{\bf G}}_{i} = \sum_{j}\sum_k\frac{1}{(r_{jki}/a)^5}\bigg(3\frac{{\bf r}_{jki}{\bf r}_{jki}}{a^2} - \frac{r_{jki}^2}{a^2}{\bf 1}\bigg){\bf R}_{ki},
\label{Eq:G3}
\ee 
where the sum over $j$ again refers to the sum over the Bravais lattice positions, the sum over $k$ sums over all atoms in the basis, the position ${\bf r}_i$ of the central moment $\vec{\mu}_i$ is not necessarily at the origin of of a central unit cell, and the vector from $\vec{\mu}_i$ to a moment $\vec{\mu}_k$ is
\be
{\bf r}_{jki} = {\bf r}_{j} + {\bf r}_{k} - {\bf r}_{i},
\ee
where ${\bf r}_{k}$ is the position of moment $\vec{\mu}_k$ in the basis with respect to the position of the associated Bravais lattice point ${\bf r}_{j}$.  The term with ${\bf r}_{jki} = 0$ is omitted from the sum because that term corresponds the difference in position of moment $\vec{\mu}_i$ with itself.  The Cartesian rotation matrix ${\bf R}_{ki}$ in Eq.~(\ref{Eq:G3}) expresses the moment direction of $\vec{\mu}_k$ in the basis with respect to that of the central moment $\vec{\mu}_i$ via
\be
\vec{\mu}_k = {\bf R}_{ki}\vec{\mu}_i,
\label{Eq:RotMatDef}
\ee
\ese
similar to Eq.~(\ref{Eq:muj}) for collinear ordering associated with a magnetic propagation vector~{\bf k}.  Prior to calculating $\widehat{{\bf G}}_{i}$, the $3\times 3\ {\bf R}_{ki}$ rotation matrix must be specified for each spin in the basis via a model for the AFM structure.  For example, for the N\'eel AFM structure in Fig.~\ref{Fig:Honeycomb_Lattice} below, if $\vec{\mu}_i$ were at a red position ${\bf r}_i/a = \frac{1}{3}\hat{\bf a} + \frac{2}{3}\hat{\bf b}$, then ${\bf R}_{1i}$ for a spin at another red position would be ${\bf R}_{ki} = {\bf 1}$ and that for a black position would be $-{\bf 1}$, where again {\bf 1} is the $3\times3$ identity matrix.  This procedure is easily generalized to more than two spins per Bravais lattice point, as illustrated in Sec.~\ref{Sec:GdB4} below for calculating $\widehat{{\bf G}}_{i}$ for the known coplanar noncollinear AFM structure of tetragonal GdB$_4$ in Fig.~\ref{Fig:GdB4_mag_struct} containing four moments in the basis, each pointing in different directions, and for the related Shastry-Sutherland spin lattice.

\subsection{\label{Sec:HelCyc} Coplanar Noncollinear Helical or Cycloidal Antiferromagnets}

Here we extend the above discussion to coplanar noncollinear helical or cycloidal AFM ordering on tetragonal or hexagonal Bravais lattices.  For both types of AFM order, the ordered moments are defined to lie in the crystallographic $ab$~plane.  For helical AFM ordering, the ordered moments are FM-aligned in the $ab$~plane and the helix wavevector {\bf k} axis is the $c$~axis.  For cycloidal AFM ordering, {\bf k} lies in the $ab$~plane and the moments in planes perpendicular to both ${\bf k}$ and the $ab$~plane are aligned ferromagnetically.  The Cartesian $x$-axis is parallel to {\bf a}, the $y$-axis is perpendicular to {\bf a} in the $ab$~plane and the $z$~axis is perpendicular to the $ab$~plane along the $c$~axis.  Pictures of the helical and cycloidal structures are given in Refs.~\onlinecite{Johnston2012} and~\onlinecite{Goetsch2014}, respectively.  In either structure, the azimuthal angle $\phi_{ji}= \phi_{j}-\phi_{i}$ with respect to the positive {\bf a}~axis between moments $\vec{\mu}_j$ and $\vec{\mu}_i$ in the $ab$ plane is given by
\bse
\be
\phi_{ji} = {\bf k}\cdot{\bf r}_{ji}.
\ee
The relationship between the central moment direction $\hat{\mu}_i$ at position ${\bf r}_i$ and that of another moment at position ${\bf r}_j$ in either AFM structure is
\be
\hat{\mu}_j = \left(
\begin{array}{ccc}
\cos\phi_{ji} & 0 & 0\\
0&\sin\phi_{ji}&0\\
0&0&1\\
\end{array}
\right)\hat{\mu}_i,
\ee
which can be written
\be
\hat{\mu}_j = (\hat{\bf x}\hat{\bf x}\,\cos\phi_{ji} + \hat{\bf y}\hat{\bf y}\,\sin\phi_{ji} + \hat{\bf z}\hat{\bf z})\hat{\mu}_i,
\ee
\ese
where the Cartesian coordinate system is used throughout.  Then $\widehat{{\bf G}}_{i}$ in Eq.~(\ref{Eq:Gk2}) becomes
\bea
\widehat{{\bf G}}_{i} &=& \sum_{j\neq i}\frac{1}{(r_{ji}/a)^5}\bigg(3\frac{{\bf r}_{ji}{\bf r}_{ji}}{a^2} - \frac{r_{ji}^2}{a^2}{\bf 1}\bigg)\label{Eq:Gk3}\\*
&&\times (\hat{\bf x}\hat{\bf x}\,\cos\phi_{ji} + \hat{\bf y}\hat{\bf y}\,\sin\phi_{ji} + \hat{\bf z}\hat{\bf z}).\nonumber
\eea

As with collinear AFM ordering, one must specify {\bf k} in terms of the reciprocal lattice translation vectors in Cartesian coordinates in advance of computing $\widehat{{\bf G}}_i({\bf k})$.  Note that when $\widehat{{\bf G}}_i({\bf k})$ is diagonalized, one eigenvalue and corresponding eigenvector are for FM ordering along the $z$~axis and are not relevant to those for the helix, whereas the other two sets of eigenvalues and eigenvectors are for the helix. As a result, the sum of the three eigenvalues do not add to zero as they do for all other AFM structures discussed above.

\subsection{\label{Eq:NearField} Near Field}

The value of $B_{{\rm int}\,\alpha\,i}^{\rm near}$ in Eq.~(\ref{Eq:Bintalphai}) that is seen by a given moment~$\vec{\mu}_i$ at position ${\bf r}_i$ in a given magnetic structure with a given ordered moment configuration, due to the sum of the magnetic fields from the magnetic moments around it within the Lorentz cavity of radius $R$, is simply given as 
\be
B_{{\rm int}\,\alpha\,i}^{\rm near} = -\frac{2E_i}{\mu_\alpha} = \frac{\mu\lambda_{{\bf k}\alpha}}{a^3},
\label{Eq:Bi}
\ee
where the factor of 2 arises because the energy per pair is split evenly between each pair of moments, whereas the magnetic field arises only from the neighbor of each pair, the second equality was obtained using Eqs.~(\ref{Eq:eps}) and~(\ref{Eq:Eilambda}) and $B_{{\rm int}\,\alpha\,i}^{\rm near}$ can be either positive or negative, depending on the sign of $\lambda_{{\bf k}\alpha}$. If the MDI is the only source of anisotropy present, this field must be positive because then the ordered moment is parallel to the local magnetic induction, which minimizes the free energy of the moment.  The quantity $B_{{\rm int}\,\alpha\,i}^{\rm near}$ is needed to calculate the total local magnetic induction at the site of a local moment according to Eq.~(\ref{Eq:Bintalphai2}).  If $E_i$ is expressed in cgs units of erg and those of $\mu$ in cgs units of erg/G (=~G\,cm$^3$), then $B_{{\rm int}\,\alpha\,i}^{\rm near}$ has the correct cgs units of G\@.

Using Eq.~(\ref{Eq:Bi}), the total local field in Eq.~(\ref{Eq:Bintalphai3}) seen by central moment $\vec{\mu}_i$ becomes
\be
B_{{\rm int}\,\alpha\,i}^{\rm local} = H_\alpha + \left(\frac{4\pi\mu_\alpha/\mu}{3V_{\rm spin}/a^3} + \lambda_{{\bf k}\alpha}\right) \frac{\mu}{a^3},
\label{Eq:Blocal99}
\ee
where the first term in parentheses is the Lorentz field, where we distinguish the moment component $\mu_\alpha$ in the $\alpha$ direction per spin averaged over the sample and the magnitude $\mu$ of the average moment per spin.  A nonzero value of $\mu_\alpha$ only occurs in a FM or in an AFM in the presence of an external magnetic field.  The second term in parentheses arises from the near field.

\subsection{\label{Sec:Gcalcs} Calculation and Diagonalization of the Magnetic Dipole Interaction Tensor}

We chose to carry out the sums in the expressions for the dipole interaction tensor $\widehat{{\bf G}}_{i}$ in Eqs.~(\ref{Eq:Gk2}), (\ref{Eq:G3}) and~(\ref{Eq:Gk3}) directly instead of by using the Ewald-Kornfeld technique,\cite{Born1954}  because we wanted to study the convergence properties of the eigenvalues $\lambda_{\bf k\alpha}$ versus the radius $R$ of the circle or Lorentz sphere for 2D and 3D lattices, respectively.  The calculations and diagonalizations of $\widehat{{\bf G}}_{i}$ were carried out using standard {\tt Macintosh} laptop and desktop computers and {\tt Mathematica} software.  For the 1D chain with FM and N\'eel AFM states, the eigenvalues and eigenvectors of $\widehat{{\bf G}}_{i}$ are trivially determined exactly for the infinite chain as shown in Sec.~\ref{Sec:FM1D}.  For 2D lattices the sums were carried out within circles of radius up to $R/a = 1000$ containing up to $1\times10^7$ spins (for the kagom\'e lattice containing three spins per unit cell).  For the 3D lattices the sums were carried out within a Lorentz sphere, usually up to a radius $R/a = 50$ containing up to $6\times10^6$ spins.  Calculations were also done for two AFM structures out to a sphere radius of $R/a=100$ containing $1.7\times10^7$ spins to check convergence.

For the FM spin structures in 2D, the values of $\widehat{{\bf G}}_{i}$ versus $1/(R/a)$ were extrapolated to $1/(R/a)=0$.  As shown in Appendix~\ref{App:RtoInfty}, the calculations of $\widehat{{\bf G}}_{i}$ for AFM structures generally converge more rapidly with increasing $R/a$ than for FM structures.  These procedures determined $\lambda_{\bf k\alpha}$ to accuracies of $\lesssim \pm10^{-6}$ for 2D lattices and $\lesssim \pm0.001$ for 3D lattices, more than sufficient for our purposes.  The eigenvectors $\hat{\mu}_i$ usually converged very quickly with increasing~$R/a$.  For the various 3D tetragonal and hexagonal lattices, $\widehat{{\bf G}}_{i}$ was calculated for $c/a$ ratios from 0.5 to 3 in 0.1 increments.

Figures~\ref{Fig:SqLattk00} and~\ref{Fig:SqLattk11} in Appendix~\ref{App:RtoInfty} show the convergence of $\lambda_{\bf k\alpha}$ with increasing $R/a$ for FM and N\'eel AFM moment alignments along the $c$~axis in the 2D simple square lattice, respectively.  Figures~\ref{Fig:PTCA1_5_100Convergence}(a) and \ref{Fig:PTCA1_5_100Convergence}(b) show plots for a  simple tetragonal lattice with FM alignment of the moments along the $c$~axis for $c/a=1.5$ and~3, respectively.  Figures~\ref{Fig:AllPTk111ConvergCA1_5}(a) and \ref{Fig:AllPTk111ConvergCA1_5}(b) show analogous plots for the simple tetragonal lattice with N\'eel-type AFM ordering where {\bf k} = $(\frac{1}{2},\frac{1}{2},\frac{1}{2})$ and alignment of the moments along the $c$~axis for $c/a=1.5$ and~3, respectively.

For the 3D FM and AFM structures, the values of $\lambda_{\bf k\alpha}$ were typically obtained for $R/a=1$ to~50 in increments of~1 and the last 10 or~20 values were averaged to obtain the data in the figures in the text and Appendix~\ref{App:lambdaFigs} and in the tables in the Supplemental Material.\cite{SupplInfo}

\section{\label{Sec:1D2D} Eigenvalues and Eigenvectors for Magnetic Ordering on One- and Two-Dimensional Spin Lattices}

\subsection{\label{Sec:FM1D} Spin Chain}

We assume that the spin chain lattice is oriented along an axis desigated as the $a$~axis ($x$~axis) with spacing $a$ between adjacent spins, so 
\be
\frac{r_{ji}}{a} = n_a.
\ee
Ferromagnetic alignment corresponds to {\bf k} = 0.  This alignment can occur either in the FM-ordered state or in the PM state in the presence of an applied magnetic field.  The central spin is positioned at $n_a=0$, so $n_a$ of the neighbors runs from $-\infty$ to $\infty$, excluding $n_a = 0$.  Numerical diagonalization of $\widehat{{\bf G}}_{i}$ in Eq.~(\ref{Eq:Gk2}) with {\bf k} = 0 and $|n_x^{\rm max}| = 1000$ (2000 neighbors of the central moment) shows that the principal axes of the interaction tensor are parallel and perpendicular to the $a$~axis.   The lowest energy configuration with a calculated $\lambda_{(0,0,0)[100]}= 4.80823$ is with the ordered moments aligned along the $a$~(chain) axis.  This makes sense because the lowest energy configuration of a moment is when each moment points along the local field seen by the moment, which is along the axis of the chain.  The eigenvalues for the two higher-energy orthogonal directions are $\lambda_{(0,0,0)[010]} = \lambda_{(0,0,0)[001]} = -\lambda_{(0,0,0)[100]}/2$.  

For the present case of the 1D spin chain one can also evaluate $\lambda_{(0,0,0)[1,0,0]}$ exactly.  Equation~(\ref{Eq:Gk2}) yields the eigenvalue
\be
\lambda_{(0,0,0)[1,0,0]} = 4\sum_{n_a=1}^\infty\frac{1}{n_a^3}.
\label{Eq:EijChain}
\ee
The sum is $\sum_{n_x=1}^\infty=\zeta(3)$,\cite{Karetnikova2001} yielding
\be
\lambda_{(0,0,0)[1,0,0]} = 4\zeta(3) \approx 4.808\,228
\label{Eq:EiExact}
\ee
as shown in Table~\ref{Tab:EvecsEvals}, where $\zeta(z)$ is the Riemann zeta function with $\zeta(3)\approx 1.20206$.  The above numerical value of 4.80823 obtained for $\lambda_{(0,0,0)[1,0,0]}$ agrees with this exact value to six-place accuracy.  This shows that the value of $|n_a^{\rm max}| = 1000$ and a spin chain containing 2000 neighbors of the central moment used in the numerical calculation is sufficient to obtain this accuracy.

It is of interest to examine the approach to the infinite-chain limit of $\lambda_{(0,0,0)[1,0,0]}$ on increasing $|n_a|$.  For large $|n_a|$ one can replace the sum in Eq.~(\ref{Eq:EijChain}) in the region where $n_a$ is large  by an integral $\int n_a^{-3} dn_a \propto -1/n_a^2$.  Thus we expect for $n_a\gg1$ that
\be
\lambda_{(0,0,0)[1,0,0]} = 4\zeta(3) - \frac{A}{n_a^2} = 4\zeta(3)\left[1 - \frac{A}{4\zeta(3)n_a^2}\right],
\label{Eq:lamdaFMChainExpansion}
\ee
where $A$ is a positive constant.  An exact series expansion of the sum in Eq.~(\ref{Eq:EijChain}) about $n_a=\infty$ indeed gives $\lambda_{(0,0,0)[1,0,0]} = 4\zeta(3) - 2/n_a^2 + {\cal O}(n_a^{-3})$, yielding $A = 2$.  Equation~(\ref{Eq:lamdaFMChainExpansion}) then predicts six-place accuracy for $\lambda_{(0,0,0)[1,0,0]}$ for $|n_a|=1000$, consistent with the above comparison.

Here we also examine the N\'eel-type AFM wavevector ${\bf k} = (1/2,0,0)$~r.l.u., where $1~{\rm r.l.u.} = 2\pi/a$ is the reciprocal lattice unit for this spin lattice.  A numerical calculation using Eq.~(\ref{Eq:Gk2}) shows that the eigenvalues of Eq.~(\ref{Eq:Eilambda}) converge to six significant figures even with a small $|n_a|^{\rm max} = 70$.  These calculations also show that the most stable ordered moment direction is perpendicular to the chain with 
\be
\lambda_{(1/2,0,0)[0,1,0]} = \lambda_{(1/2,0,0)[0,0,1]} = 1.80309
\ee
and the unstable $x$-axis direction has
\be
\lambda_{(1/2,0,0)[1,0,0]} = -2\lambda_{(1/2,0,0)[0,1,0]}= -3.60617.
\ee

An exact calculation for the $a$-axis eigenvalue is obtained using the AFM version of Eq.~(\ref{Eq:EijChain}), yielding
\bse
\be
\lambda_{(1/2,0,0)[1,0,0]} = 4\sum_{n_a=1}^\infty\frac{(-1)^{n_a}}{n_a^3} = -3\zeta(3)\approx -3.60617,
\label{Eq:lambdaChainAF}
\ee
as listed in Table~\ref{Tab:EvecsEvals}.  This value is identical to within six places with the numerical result for $\lambda_{(1/2,0,0)[1,0,0]}$ obtained above using only a 141-spin chain (including the central spin).  Thus the dipole fields seen by a central moment converge much faster with increasing $n_a^{\rm max}$ for the AFM structure than for FM one for the spin chain.  The two lower-energy eigenvalues are
\bea
\lambda_{(1/2,0,0)[0,1,0]} &=& \lambda_{(1/2,0,0)[0,0,1]} = -\frac{1}{2}\lambda_{(1/2,0,0)[1,0,0]}\nonumber\\*
 &=& \frac{3}{2}\zeta(3)\approx 1.80309.
\label{Eq:lambda12ChainAF}
\eea
\ese

Comparing Eqs.~(\ref{Eq:EiExact}) and~(\ref{Eq:lambda12ChainAF}) shows that the eigenvalue for the FM-aligned state with the ordered/induced moments aligned along the $a$~axis is larger than the maximum AFM one, and hence the energy per spin is smaller according to Eq.~(\ref{Eq:Eilambda}) for the FM state than for the AFM state.  The FM state is thus expected to be the magnetic ground state of the linear spin chain for purely dipolar interactions provided the ordering is not destroyed by quantum fluctuations.

\begin{table*} 
\caption{\label{Tab:EvecsEvals} One- and two-dimensional spin lattices.  Eigenvalues $\lambda_{{\bf k}\alpha}$ and eigenvectors $\hat{\mu} = [\mu_x,\mu_y,\mu_z]$ in Cartesian coordinates of the MDI tensor $\widehat{{\bf G}}_i({\bf k})$ in Eq.~(\ref{Eq:Gk2}) for various values of the magnetic wavevector ${\bf k}$ in reciprocal lattice units with {\bf collinear} magnetic moment alignments.  The most positive $\lambda_{{\bf k}\alpha}$ value(s) corresponds to the lowest energy value according to Eq.~(\ref{Eq:Eilambda}).  The Cartesian $x$, $y$ and $z$ axes are along the $a$, $b$ and $c$ axes of orthogonal-axis lattices, respectively.  For the hexagonal lattice, the $x$~axis is parallel to the hexagonal $a$~axis and the $y$~axis is perpendicular the the $a$~axis, rather than along the $b$~axis.  The linear chain is aligned along the $a$~axis and the square and hexagonal lattices are aligned in the $ab$~plane.}
\begin{ruledtabular}
\begin{tabular}{cccc}
{\bf k} 							& $\lambda_{{\bf k}\hat{\mu}}$  			& $\hat{\mu}$ 		& $\hat{\mu}\cdot\hat{\bf k}$\\
\hline
\underline{1D linear chain} \\
(0,~0,~0)	(FM)						& $4\zeta(3)\approx 4.808\,228$			& [100]		\\
								& $-2\zeta(3)\approx -2.404\,114$			& [010], [001]		\\
$\left(\frac{1}{2},~0,~0\right)$ (N\'eel AFM)& $\frac{3}{2}\zeta(3)\approx 1.803\,085$ & [010], [001]	& 	0\\
								& $-3\zeta(3)\approx -3.606\,171$			& [100]	&	1\\
\underline{2D square lattice}\\
$\left(\frac{1}{2},~0,~0\right)$ (stripe AFM) & 5.098\,873						& [010]	&	0\\
								& 0.935\,462							& [001]	&	0\\
								& $-6.034\,335$						& [100]	&	1\\
(0,~0,~0)	(FM)						& 4.516\,811							& [100], [010]		\\
								& $-9.033\,622$						& [001]		\\
$\left(\frac{1}{2},~\frac{1}{2},~0\right)$ (N\'eel AFM)	& 2.645\,887			& [001]	&	0\\
						& $-1.322\,943$								& [100], [010]	&	$1/\sqrt{2}\approx 0.7071$\\
\underline{2D simple hexagonal (triangular) lattice}\\
(0,~0,~0)	 (FM)					& 5.517\,088						& [100], [010]	\\
								& $-11.034\,176$					& [001]		\\
$(1,~0,~0)$						& 5.517\,088						& [100], [010] & 1		\\
								& $-11.034\,176$					& [001]	& 0	\\
$(\frac{1}{2},~0,~0)$				& 4.094\,909						& $[\frac{1}{2},-\frac{\sqrt{3}}{2},0]$	&		1/2\\
								& 1.839\,029						& [001]		&	0\\
								& $-5.933\,939$				& $[-\frac{\sqrt{3}}{2},
\frac{1}{2},~0]$		& $-\sqrt{3}/2\approx -0.8660$\\
$(\frac{1}{2},~\frac{1}{2},~0)$		& 4.094\,909					& $[-\frac{1}{2},-\frac{\sqrt{3}}{2},0]$	&	$-(\sqrt{3}+1)/2^{3/2}\approx -0.9659$\\
								& 1.839\,029					& [001]	&	0	\\
								& $-5.933\,939$				& $[-\frac{\sqrt{3}}{2},\frac{1}{2},~0]$	& 	$1/\sqrt{2}\approx 0.7071$\\
$(\frac{1}{3},~\frac{1}{3},~0)$		& 2.331\,796					& [001]	&	0	\\
								& $-1.165\,898$				& [100], [010] &	$1/\sqrt{2}\approx 0.7071$		\\
\underline{2D hexagonal honeycomb lattice}\\
(0,~0,~0)	 (FM)					& 17.092\,359						& [100], [010]	\\
								& $-34.184\,718$					& [001]		\\
$(\frac{1}{2},~0,~0)$				& 12.827\,051					& $[-\frac{1}{2},\frac{\sqrt{3}}{2},0]$	 &		$-1/2$\\
								& $-0.090\,183$					& [001]						&	0\\
								& $-12.736\,868$					& $[\frac{\sqrt{3}}{2},\frac{1}{2},~0]$	 & $\sqrt{3}/2\approx 0.8660$\\
$(0,~0,~0)$ (N\'eel-type)			& 12.116\,366							& [001]	 &		\\
								& $-6.058\,183$					& [100], [010]						&	\\
\underline{2D hexagonal kagom\'e lattice}&								&				\\
(0,~0,~0)	 (FM)					& 51.321\,197						& [010]	\\
								& 11.205\,800						& [100]		\\
								& $-62.526\,996$					& [001]		\\
$(0,~1,~0)$ (ferrimagnet)			& 40.458\,644						& [001]	&   0 \\
								& $-0.171\,624$					& [100]	& 0	\\
								& $-40.287\,021$					& [010] 	& 1\\
$(\frac{2}{3},~\frac{2}{3},~0)$		& 13.213\,509						& [001]	& 0\\
								& 9.212\,253						& [100]	& $1/\sqrt{2}\approx 0.7071$	\\
								& $-22.425\,762$					& [010]	& $1/\sqrt{2}\approx 0.7071$	\\
$(0,~\frac{1}{2},~0)$ 				& 4.094\,910						& [100]	& 0\\
								& 1.839\,029						& [001] 	& 0		\\
								& $-5.933\,939$					& [010]	& 1	\\
\end{tabular}
\end{ruledtabular}
\end{table*}

\subsection{Two-Dimensional Spin Lattices}

For the simple square lattice, one has $a=b$ and the spin positions given by the first two terms in Eq.~(\ref{Eq:rji}).  The normalized wavevectors are the first two terms in Eq.~(\ref{Eq:ka}).  The largest (positive) eigenvector $\lambda_{{\bf k}\alpha}$ of $\widehat{{\bf G}}_{ia}({\bf k})$ in Eq.~(\ref{Eq:Gk2}) with {\bf k} = 0 for FM alignment occurs for the $a$ or~$b$ easy axes, with $\lambda_{(0,0,0)[1,0,0]} = \lambda_{(0,0,0)[0,1,0]}$ and $\lambda_{(0,0,0)[0,0,1]} = -2\lambda_{(0,0,0)[1,0,0]}$.  Shown in Fig.~\ref{Fig:SqLattk00}(a) in Appendix~\ref{App:RtoInfty} is the dependence of $\lambda_{(0,0,0)[0,0,1]}$ on the inverse radius $R^{-1}$ for the $c$-axis eigenvalue $\lambda_{(0,0,0)[0,0,1]}$.  According to the discussion in Sec.~\ref{Sec:FM1D}, in 2D one should have $\lambda_{(0,0,0)[0,0,1]}(R/a \gg1) = \lambda_{(0,0,0)[0,0,1]}(a/R =0) + A/R$, in agreement with the calculations in Fig.~\ref{Fig:SqLattk00}(a).  Fitting the data for $0.001\leq a/R \leq 0.002$ gives
\bea
\lambda_{(0,0,0)[0,0,1]}(a/R = 0) &=& -9.0336220(1),\\*
A &=& 6.28356(9).\nonumber
\eea
The deviations of the data from the fit are shown in Fig.~\ref{Fig:SqLattk00}(b), where it is seen that the deviations are of the order of 1 part in $10^7$ for $0.001\leq a/R \leq 0.0026$.  The graininess of the lattice becomes more apparent at larger values of $a/R$.  

The eigenvalues and eigenvectors for square-lattice AFM propagation vectors {\bf k} = $\left(\frac{1}{2},0,0\right)$ (stripe AFM) and $\left(\frac{1}{2},\frac{1}{2},0\right)$ (N\'eel-type AFM) were also computed as shown in Table~\ref{Tab:EvecsEvals}.  One sees that of these and the FM propagation vector, the stripe AFM propagation vector with the ordered moments aligned along the $b$~axis (perpendicular to {\bf k} as shown in the last column of Table~\ref{Tab:EvecsEvals}) has the lowest energy.  Our eigenvalues $\lambda_{(0,0,0)[100]}$ and $\lambda_{(\frac{1}{2},0,0)[100]}$ are in agreement with, but are more precise than, those previously reported in Ref.~\onlinecite{Rozenbaum1984}, and our $\lambda_{(0,0,0)[001]}$ and $\lambda_{(1/2,1/2,0)[001]}$ values are in precise agreement with the results  in Ref.~\onlinecite{Bishop2002}.

For the 2D simple-hexagonal (triangular) lattice the eigenvalues and eigenvectors were calculated for the FM state and four AFM states.  From Table~\ref{Tab:EvecsEvals}, the lowest-energy states are the FM state and the AFM state with {\bf k} = (1,0,0) (stripe-type), with the moments aligned within the $ab$~plane in both cases.  Data for the AM AFM state with {\bf k} = $\left(\frac{1}{3},\frac{1}{3},0\right)$  are included in Table~\ref{Tab:EvecsEvals} because the classical ground state of a triangular lattice of spins with Heisenberg interactions is the well-known coplanar noncollinear 120$^\circ$ state that can be described by {\bf k} = $\left(\frac{1}{3},\frac{1}{3},0\right)$ which we consider further in Sec.~\ref{TriangAFMs}.  Our eigenvalues $\lambda_{(0,0,0)[100]}$ and $\lambda_{(\frac{1}{2},\frac{1}{2},0)[-1/2,-\sqrt{3}/2,0]}$ are in agreement to seven significant figures with those previously reported in Ref.~\onlinecite{Rozenbaum1995}.

\begin{figure}
\includegraphics[width=2.5in]{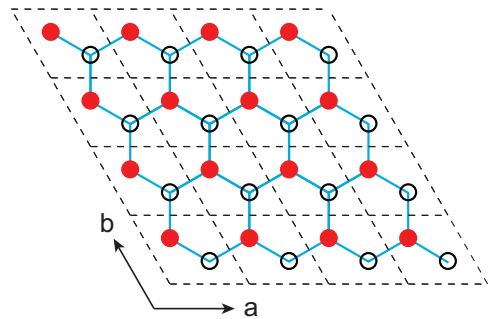}
\caption {(Color online) Honeycomb lattice.  Each honeycomb cell (not a unit cell) is bounded by solid blue lines.  The hexagonal unit cells with translation vectors {\bf a} and {\bf b} are outlined by dashed black lines.  The 2D space group is $p6m$ (No.~17) with two spins in Wyckoff positions $2b~\left(\frac{1}{3},\frac{2}{3}\right),\ \left(\frac{2}{3},\frac{1}{3}\right)$. Bipartite N\'eel ordering of the two spins per unit cell is shown.  The solid red circles represent  half the magnetic moments pointing in one direction and the open black circles correspond to half the moments pointing in the opposite direction. }
\label{Fig:Honeycomb_Lattice}
\end{figure}

\begin{figure}
\includegraphics[width=2.25in]{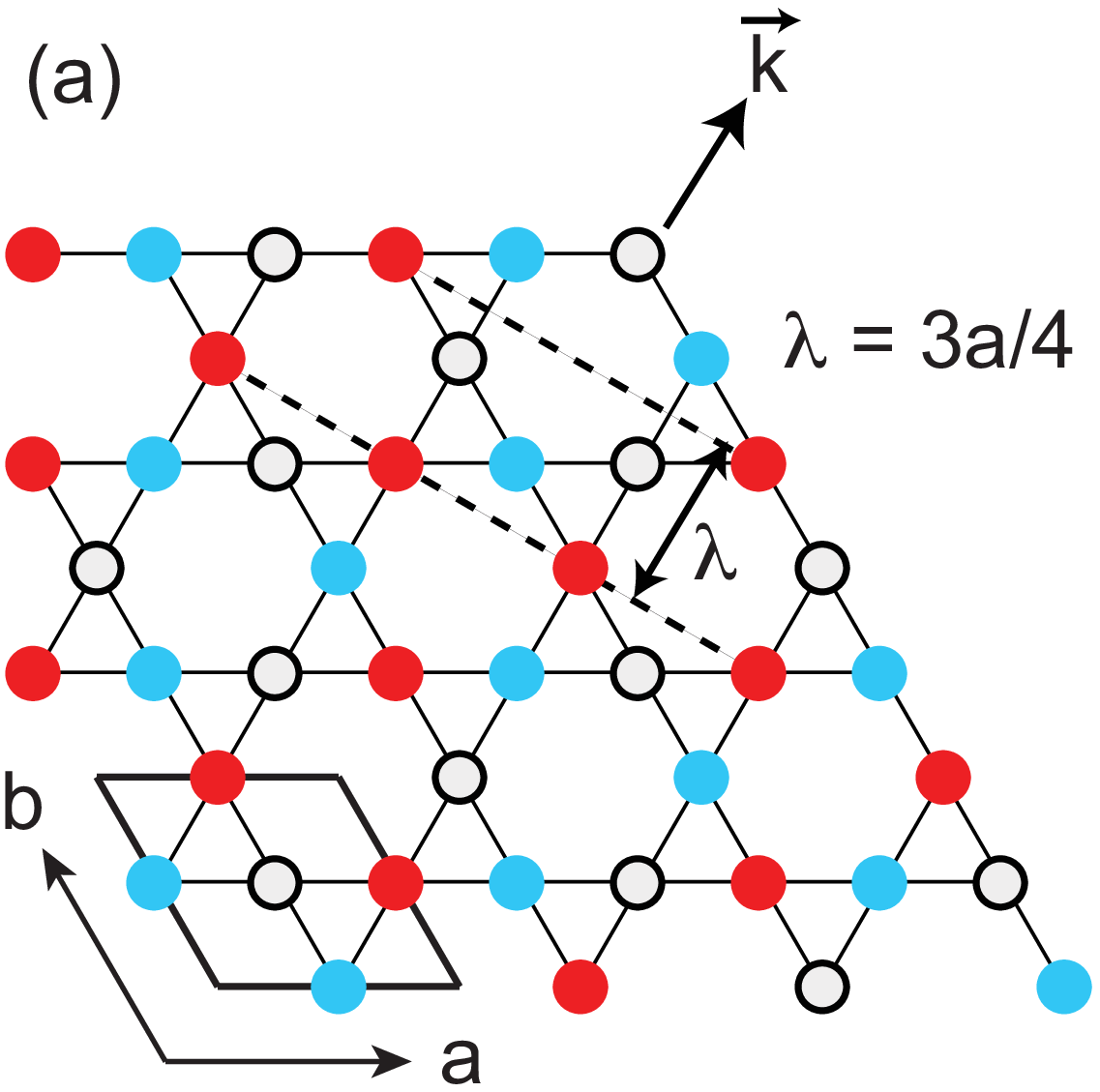}
\includegraphics[width=2.25in]{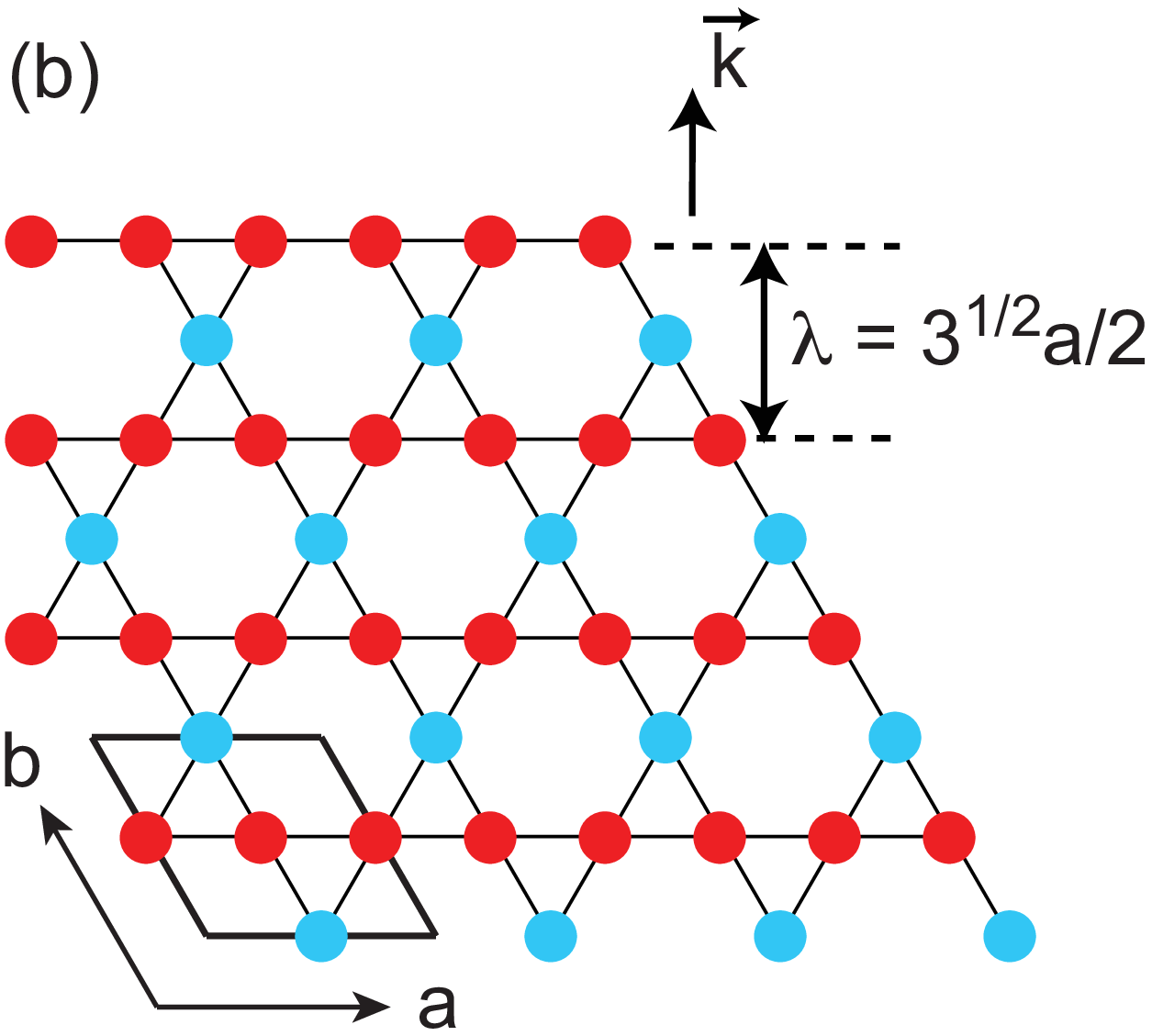}
\includegraphics[width=2.25in]{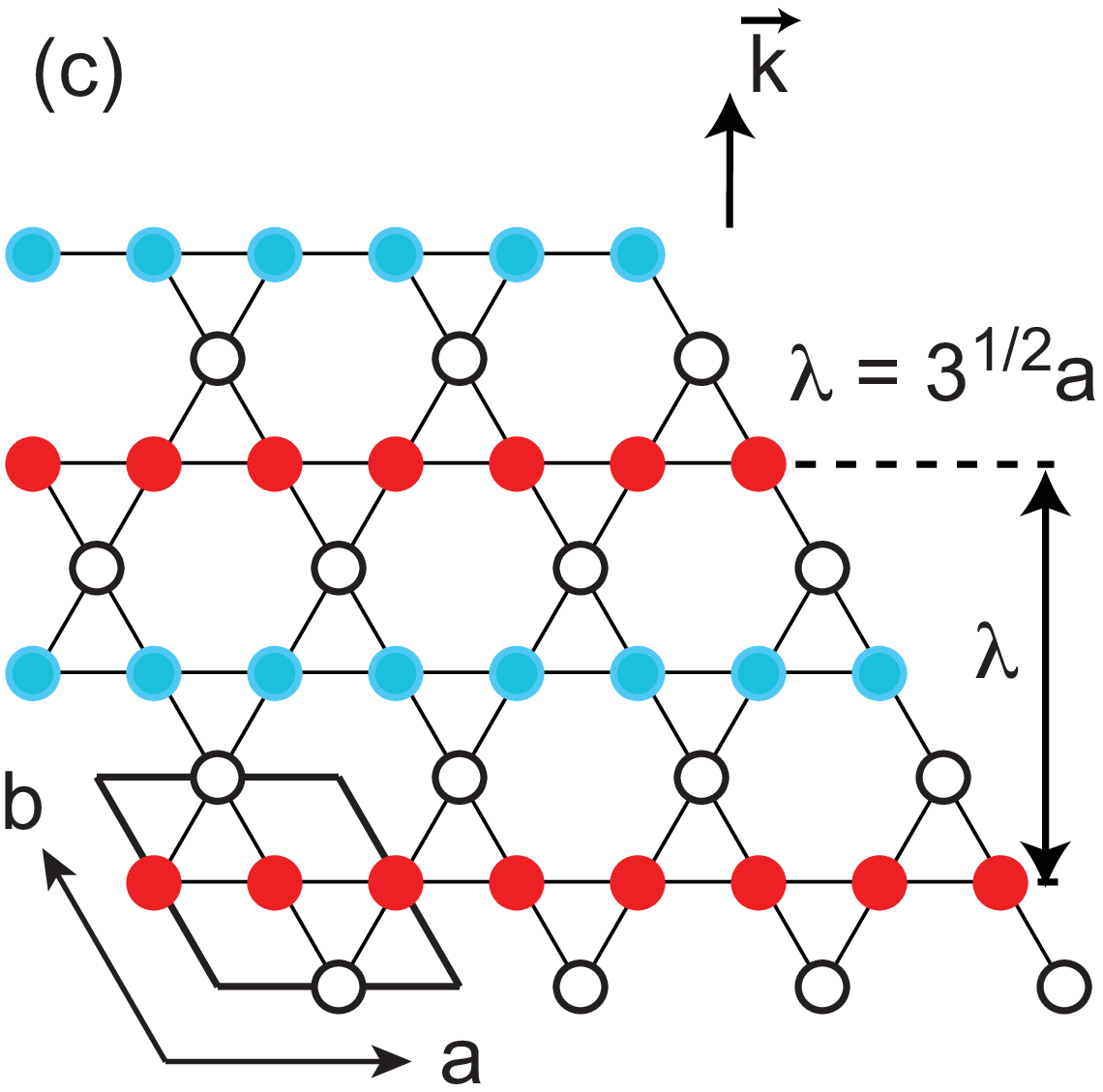}
\caption {(Color online)  Two-dimensional hexagonal kagom\'e lattices.  The hexagonal unit cell is shown at the lower left of each panel oulined in heavy black lines and contains three spins.  The unit cell edges $a$ and~$b$ are twice the length of the triangular-lattice unit cell edge.  Three magnetic structures are shown.  (a) The red, blue and open circles represent moments that are mutually at an angle of 120$^\circ$ to each other within the $ab$~plane, so a given moment has no nearest neighbors with the same orientation.  The cycloid spin configuration shown has a wavevector {\bf k} = $\left(\frac{2}{3},\frac{2}{3},0\right)$~r.l.u.  (b)~Collinear magnetic structure for {\bf k} = $\left(0,1,0\right)$~r.l.u.  The red circles represent moments in one direction and the blue circles represent moments in the opposite direction. Because there are twice as many red as blue circles, this magnetic structure is a ferrimagnet (a net FM).  (c)~Collinear AFM structure for {\bf k} = $\left(0,\frac{1}{2},0\right)$~r.l.u.  The red and blue circles have the same meanings as in~(b).  There are equal numbers of red and blue circles, but the open circles represent spins with zero ordered moment, so the magnetic structure is an AM AFM.}
\label{Fig:Kagome_Lattice}
\end{figure}

The 2D hexagonal honeycomb lattice is a non-Bravais spin lattice containing two spins per unit cell as shown in Fig.~\ref{Fig:Honeycomb_Lattice}.  The eigenvalues and eigenvectors of $\widehat{\bf G}_i$ for collinear magnetic ordering were calculated for this lattice according to the method of Sec.~\ref{Sec:NonBravais} and the results are listed in Table~\ref{Tab:EvecsEvals} for the FM and N\'eel-type (Fig.~\ref{Fig:Honeycomb_Lattice}) AFM states, where the magnetic propagation vector for both states is {\bf k} = (0,0,0).

Rozenbaum found that the ground state of the 2D honeycomb lattice is noncollinear with all spins aligned in the $ab$~plane,\cite{Rozenbaum1995} in contrast to the collinear FM and AFM structures assumed in the above calculations.  He gave the ground state energy per spin as $E/N = -\frac{\mu^2}{2a_{\rm nn}^3}(4.453\,809)$, where $a_{\rm nn}= a/\sqrt{3}$ is the nearest-neighbor spin-spin distance.  Converting to our notation for $N=2$ spins per unit cell according to Eqs.~(\ref{Eqs:EiSoln2}) gives the eigenvalue $\lambda = 3^{3/2}2(4.453809) = 46.2853$.  This eigenvalue is more than a factor of two larger (more stable) than the most stable collinear magnetic structure in Table~\ref{Tab:EvecsEvals} for the 2D honeycomb lattice.

The 2D hexagonal kagom\'e lattice is very popular for studying the effects of geometric frustration on the properties for AFM interactions.  This lattice is shown in Fig.~\ref{Fig:Kagome_Lattice}.  The lattice is generated from a doubled triangular lattice by removing the spin at the origin of the unit cell, which shows that a kagom\'e spin lattice is a $\frac{1}{4}$-depleted triangular lattice containing three spins per hexagonal unit cell.  The 2D hexagonal space group of the kagom\'e lattice is $p6m$ (No.~17), with three spins in Wyckoff positions $3c~\left(\frac{1}{2},0\right),\ \left(0, \frac{1}{2}\right),\,\ \left(\frac{1}{2},\frac{1}{2}\right)$.   For the kagom\'e lattice the cycloid wavevector in Fig.~\ref{Fig:Kagome_Lattice}(a) is {\bf k} = $\left(\frac{2}{3},\frac{2}{3},0\right)$~r.l.u.\  instead of {\bf k} = $\left(\frac{1}{3},\frac{1}{3},0\right)$~r.l.u.\ for the triangular lattice, due to the factor of two increase in the $a$- and $b$-axis lattice parameters compared to the triangular lattice.  The magnetic structure shown in the figure is the well-known classical $120^\circ$ structure for nearest-neighbor AFM Heisenberg interactions.  However, the ground state for collinear magnetic ordering arising from only dipole interactions is seen from Table~\ref{Tab:EvecsEvals} to be a FM structure with the moments pointing perpendicular to the plane of the lattice.  Also shown in the table are results for two AFM wavevectors directed along the $\hat{\bf b}^\ast$~($y$) direction discussed next.

A net FM (ferrimagnetic) collinear structure is shown in Fig.~\ref{Fig:Kagome_Lattice}(b) with magnetic wavevector ${\bf k} = \left(0,1,0\right)$~r.l.u.  There are twice as many moments pointing one way compared to the other way, so the net ordered FM moment is $\mu_{\rm sat}/3$, where $\mu_{\rm sat}$ is the saturation moment of each spin.  An AM collinear AFM structure is shown in Fig.~\ref{Fig:Kagome_Lattice}(c) with AFM propagation vector ${\bf k} = \left(0,\frac{1}{2},0\right)$~r.l.u.  The red and blue circles have the same meaning as in~(b), but the black open circles represent spins with no ordered moment.  Therefore the average AFM ordered moment per spin is $2\mu_{\rm sat}/3$.

According to Table~\ref{Tab:EvecsEvals}, the lowest-energy (largest eigenvalue) collinear magnetic structure for the 2D kagom\'e lattice is the FM structure with moments directed along the $y$ direction (vertically upwards in Fig.~\ref{Fig:Kagome_Lattice}) within the $ab$~plane.  The collinear structures shown in Figs.~\ref{Fig:Kagome_Lattice}(a)--(c) are significantly less stable.  Classical MC simulations determined that the ground state magnetic structure is an EM noncollinear ferrimagnetic structure with all ordered moments lying in the $ab$ plane.\cite{Holden2015,Maksymenko2015}  The ground state energy per spin is quoted as $E/{\rm spin}=-2.38895~\mu^2/a_{\rm nn}^3$, where $a_{\rm nn}=a/2$ and $a_{\rm nn}$ is the nearest-neighbor spin-spin distance.\cite{Holden2015}  In terms of our notation, $E/{\rm spin} = -\frac{\mu^2}{6a^3}\lambda$ which takes into account the three spins per unit cell.  Then also taking into account the relation $a=2a_{\rm nn}$, one obtains the ground-state eigenvalue $\lambda = 48(2.38895) = 114.670$, more than a factor of two larger (more stable) than the value of $\approx 51.3$ listed for $\lambda_{(0,0,0)[010]}$ for collinear FM in Table~\ref{Tab:EvecsEvals}.  Thus the classical MC simulations reveal a noncollinear ground state that is much more stable than the most stable classical collinear FM state.

The results for the 2D spin lattices in Table~\ref{Tab:EvecsEvals} provide very useful reference points for 3D lattices, where the 2D results correspond to the limit $c/a\to\infty$.  Indeed, in plots of $\lambda_{\bf k\alpha}$ versus $c/a$ for uniaxial 3D spin lattices below, we include horizontal dashed lines in the plots to observe the rate at which the 2D limits are approached with increasing~$c/a$ ratio within the calculated range $0.5\leq c/a\leq 3$.

\section{\label{Sec:3DLattices} Eigenvalues and Eigenvectors for Three-Dimensional Spin Lattices}

\begin{table}
\caption{\label{Tab:SCEvecsEvals} {\bf Simple cubic spin lattice.}  Eigenvalues $\lambda_{{\bf k}\alpha}$ and eigenvectors $\vec{\mu} = [\mu_x,\mu_y,\mu_z]$ in Cartesian coordinates of the MDI tensor $\widehat{{\bf G}}_i({\bf k})$ in Eq.~(\ref{Eq:Gk2}) are listed for various values of the magnetic wavevector ${\bf k}$ in reciprocal lattice units with {\bf collinear} magnetic moment alignments.  The most positive $\lambda_{{\bf k}\alpha}$ value(s) corresponds to the lowest energy value according to Eq.~(\ref{Eq:Eilambda}).  Also shown are the differences between the eigenvalues for the different eigenvectors for a given {\bf k} and spin lattice, which are proportional to the respective magnetic anisotropy energies and fields.  The Cartesian $x$, $y$ and $z$ axes are along the $a$, $b$ and $c$ axes of the cubic unit cell, respectively.  The labels A-, B-, C- and G-type for the different wavevectors are from Ref.~\onlinecite{Wollan1955}.  The $\lambda_{{\bf k}\alpha}$ values agree with the $f_2$--$f_7$ eigenvalues in Table~II of Ref.~\onlinecite{Luttinger1946}.}
\begin{ruledtabular}
\begin{tabular}{ccr}
$(k_x,k_y,k_z)$  &  $[\mu_x,\mu_y,\mu_z]$  &  $\lambda_{{\bf k}\alpha}$   \\
\hline
(0,0,0) (FM, B-type)  &  [100], [010], [001]  		&  0  				\\
$\left(\frac{1}{2},0,0\right)$ (A-type) &  [100]  	&  $-9.6874$  			\\
  &  [010], [001]  								&  4.8437  			\\
  &  $[001]-[100]$  							&  14.5311  			\\
$\left(\frac{1}{2},\frac{1}{2},0\right)$ (C-type)&  [100], [010]&  $-2.6767$  		\\
  &  [001]  									&  5.3535  			\\
  &  $[001]-[100]$  							&  8.0302  			\\
$\left(\frac{1}{2},\frac{1}{2},\frac{1}{2}\right)$ (N\'eel- or G-type)&[100], [010], [001] &  0  	\\
\end{tabular}
\end{ruledtabular}
\end{table}

\begin{table}
\caption{\label{Tab:BCCEvecsEvals} Body-centered cubic spin lattice.  Symbol definitions are the same as in Table~\ref{Tab:SCEvecsEvals}.  The $\lambda_{{\bf k}\alpha}$ values agree with the eigenvalues in Table~IV of Ref.~\onlinecite{Luttinger1946}.}
\begin{ruledtabular}
\begin{tabular}{ccr}
$(k_x,k_y,k_z)$  &  $[\mu_x,\mu_y,\mu_z]$  &  $\lambda_{{\bf k}\alpha}$  \\
\hline
(0,0,0) (FM)  &  [100], [010], [001]  				&   0			 \\
$\left(\frac{1}{2},0,0\right)$ &  [100]  			&  $-9.6874$  			 \\
  &  [010], [001]  								&  4.8437  			 \\
  &  $[001]-[100]$  							&  14.5311  			 \\
$\left(\frac{1}{2},\frac{1}{2},0\right)$   &  [001]  	&  5.3534      \\
  &  $[1\bar{1}0]$  							&  7.9437  	    \\
  &  [110]  									&  $-13.2971$      \\
  &  $[001]-[110]$  							&  18.6505  	   \\
  &  $[001]-[1\bar{1}0]$  						&  $-2.5903$  	    \\
$\left(\frac{1}{2},\frac{1}{2},\frac{1}{2}\right)$ &[100], [010], [001] &  0  	   \\
(1,0,0)  &  [100], [010], [001]  					&  0  		\\
\end{tabular}
\end{ruledtabular}
\end{table}

\begin{table}
\caption{\label{Tab:FCCEvecsEvals} Face-centered cubic spin lattice.  Symbol definitions are the same as in Table~\ref{Tab:SCEvecsEvals}.  The designations of the AFM type are from Ref.~\onlinecite{Chatterji2006}. The $\lambda_{{\bf k}\alpha}$ values agree with the eigenvalues in Table~V of Ref.~\onlinecite{Luttinger1946}.}
\begin{ruledtabular}
\begin{tabular}{ccc}
$(k_x,k_y,k_z)$  &  $[\mu_x,\mu_y,\mu_z]$  &  $\lambda_{{\bf k}\alpha}$  \\
\hline
(0,0,0) (FM)  &  [100], [010], [001]  				&  0  \\
$\left(\frac{1}{2},0,0\right)$ (Type-IA AFM) &  [100]  			&  $-25.679$  \\
  &  [010], [001]  								&  12.8393  \\
  &  $[001]-[100]$  							&  38.518  \\
$\left(\frac{1}{2},\frac{1}{2},0\right)$ (Type-IV AFM)  &  $[1\bar{1}0]$	&  14.383  \\
  &  $[\bar{1}\bar{1}0]$  						&  $-19.736$  \\
  &  [001]  									&  5.3535  \\
  &  $[1\bar{1}0]-[\bar{1}\bar{1}0]$  				&  34.119  \\
  &  $[1\bar{1}0] - [001]$ 						&  9.029  \\
(0,0,1)  (Type-I AFM)&  [100], [010]  				&  8.6687  \\
  &  [001]  									&  $-17.3374$  \\
  &  $[100]-[001]$  							&  26.0061  \\
$\left(\frac{1}{2},\frac{1}{2},\frac{1}{2}\right)$ (Type-II AFM)  &  $[\bar{1}\bar{1}\bar{1}]$  &  $-28.9204$  \\
  &  $[2\bar{1}\bar{1}]$, $[0\bar{1}1]$  			&  14.4602  \\
  &  $[2\bar{1}\bar{1}] - [\bar{1}\bar{1}\bar{1}]$  		&  43.381  \\
$\left(\frac{1}{3},\frac{1}{3},\frac{1}{3}\right)$ 	&  $[\bar{1}\bar{1}\bar{1}]$  & $-30.0587$  \\
	& $[2\bar{1}\bar{1}],\ [0\bar{1}1]$			&	15.0293	\\
	& $[2\bar{1}\bar{1}]-[\bar{1}\bar{1}\bar{1}]$		&	45.0881		\\
$\left(\frac{1}{2},0,1\right)$ (Type-III AFM) & [100]	& 6.3040	\\
	& [010], [001] 							& $-3.1520$ \\
	& $[100]-[010]$							& 9.4560	\\
\end{tabular}
\end{ruledtabular}
\end{table}

\subsection{Cubic Spin Lattices}

The eigenvalues and eigenvectors of the dipolar interaction tensor for the cubic Bravais lattices are well-known but are presented here in modern notation for completeness and as a check on our calculation methods.  Our parameters for sc, bcc and fcc lattices are found to agree with previous results\cite{Luttinger1946} and are listed in Tables~\ref{Tab:SCEvecsEvals}, \ref{Tab:BCCEvecsEvals} and~\ref{Tab:FCCEvecsEvals}, respectively, for various values of {\bf k} along with the common magnetic structure designations.\cite{Wollan1955}  Belobrov {\it et al.}\ carried out an exact calculation of the ground state spin configuration and energy of the sc lattice and found degenerate noncollinear and noncoplanar AFM ground states with energy per spin corresponding to eigenvalue $\lambda = 5.344$,\cite{Belobrov1983} which is essentially the same as our value $\lambda_{(1/2,1/2,0)[001]} = 5.3535$ for collinear AFM ordering with wavevector ${\bf k} = \left(\frac{1}{2},\frac{1}{2},0\right)$ in Table~\ref{Tab:SCEvecsEvals}.  The designations of the AFM type for fcc lattices in Table~\ref{Tab:FCCEvecsEvals} are from Ref.~\onlinecite{Chatterji2006}. 

\subsection{\label{Sec:stLattices} Simple Tetragonal Spin Lattices}

The eigenvalues and eigenvectors for FM moment alignments [{\bf k} = (0,0,0)] for simple tetragonal lattices with $c/a = 0.5$--3 are shown in Fig.~\ref{Fig:SummaryPTDataCA} in Appendix~\ref{App:lambdaFigs} and in a table in the Supplemental Material.\cite{SupplInfo}  For $c/a<1$, moment alignment along the $c$~axis is energetically favorable, whereas for $c/a>1$, $ab$-plane moment alignment is preferred.

The eigenvalues and eigenvectors for a number of 3D AFM structures for simple tetragonal spin lattices were determined versus $c/a$.  The 2D limits corresponding to $c/a\to\infty$ are shown as black horizontal dashed lines in the figures.  The data for {\bf k} = (1/2,0,0) are plotted in Fig.~\ref{Fig:AllPTDataCAk100Ave} in Appendix~\ref{App:lambdaFigs}.  In this case there are three distinct $\lambda_{\bf k\alpha}$ values for the three eigenvectors [100], [010] and [001] because this {\bf k} breaks the fourfold rotational symmetry about the $c$~axis, with the easy axis switching from [001] for $c/a<1$ to [010] for $c/a>1$.  One sees that the respective  2D limits in Table~\ref{Tab:EvecsEvals} are reached for $c/a\gtrsim 2$.  Similarly, data for {\bf k} = (1/2,1/2,0) and (0,0,1/2) are plotted in Fig.~\ref{Fig:AllPTDataCAk11Ave} in Appendix~\ref{App:lambdaFigs} and the data are listed in the Supplemental Material.\cite{SupplInfo}

\begin{figure}[t]
\includegraphics[width=3.3in]{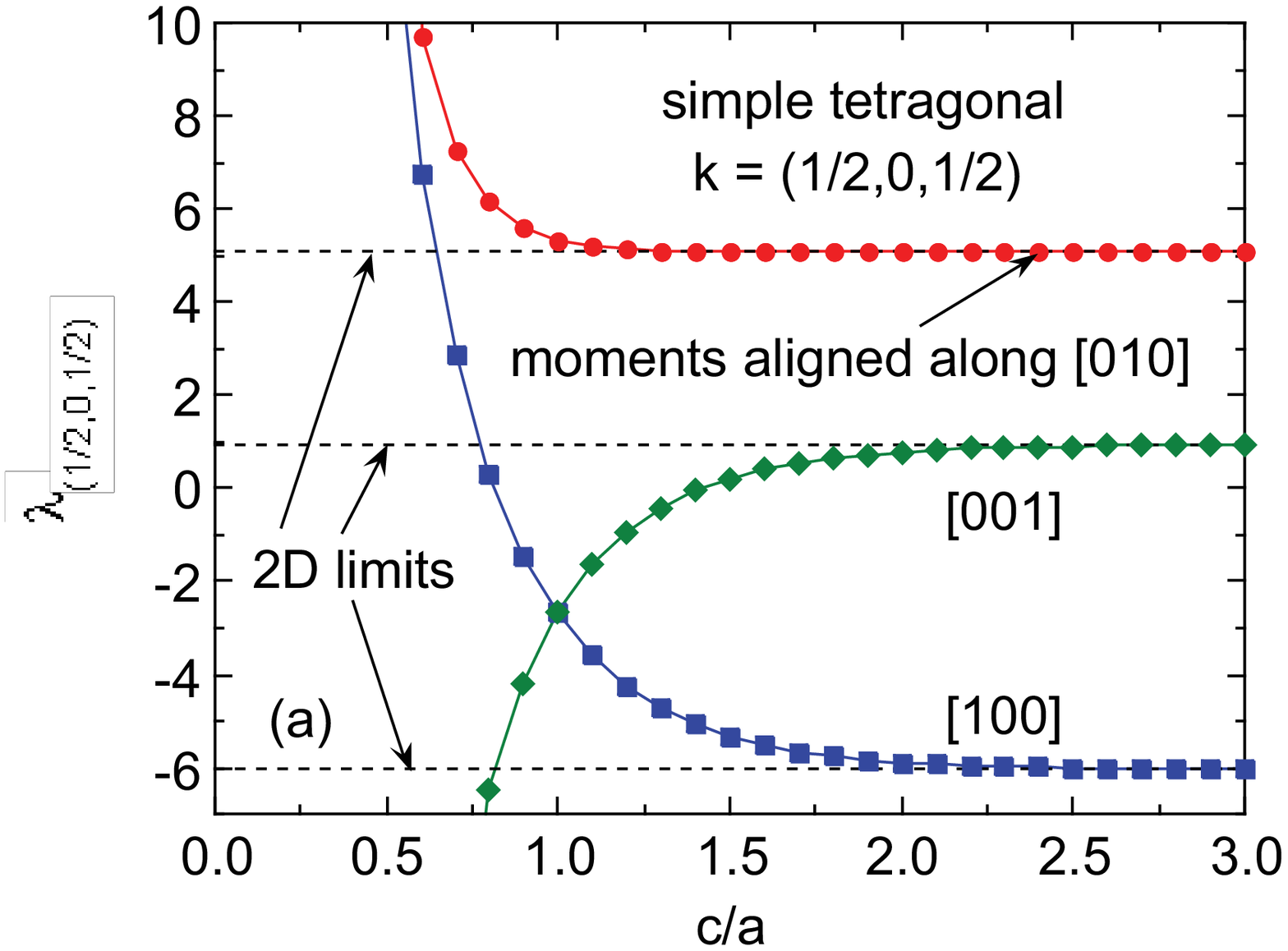}
\includegraphics[width=3.3in]{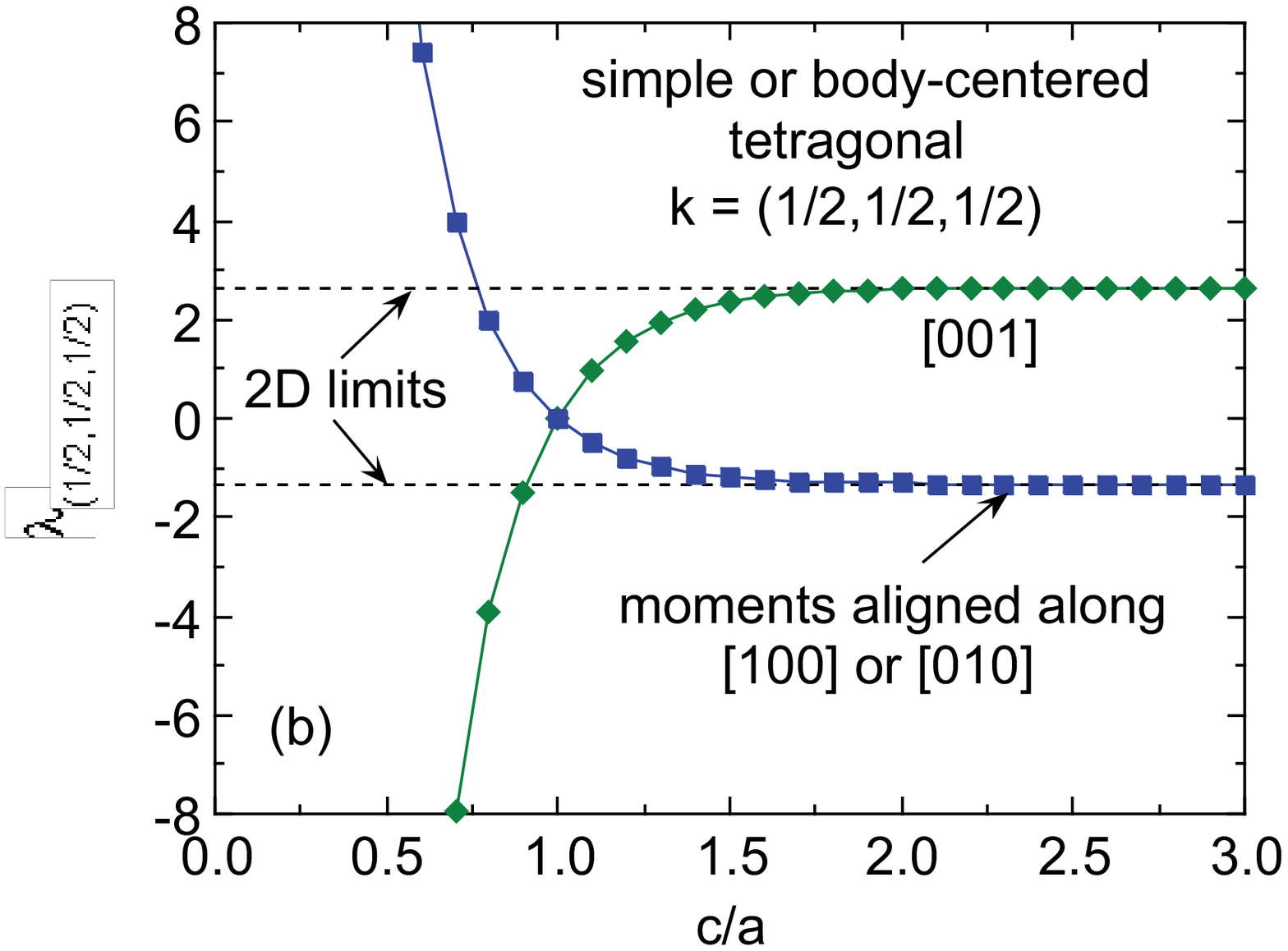}
\caption {(Color online) Eigenvalues (a) $\lambda_{(1/2,0,1/2)}$ for AFM wavevector {\bf k} = (1/2,0,1/2)\,r.l.u.\ and (b) $\lambda_{(1/2,1/2,1/2)}$ for AFM wavevector {\bf k} = (1/2,1/2,1/2)\,r.l.u.\ versus the $c/a$ ratio for a simple tetragonal lattice with the moments aligned along [010] ($b$~axis, solid red circles), [001] ($c$~axis, solid green diamonds) and [100] ($a$~axis, solid blue squares).}
\label{Fig:AllPTDataCAk101Ave}
\end{figure}

The eigenvalues for AFM wavevectors {\bf k} = (1/2,0,1/2) and~(1/2,1/2,1/2) are plotted for the respective eigenvectors versus the $c/a$ ratio for a simple tetragonal lattice in Figs.~\ref{Fig:AllPTDataCAk101Ave}(a) and~\ref{Fig:AllPTDataCAk101Ave}(b), respectively, with the numerical values listed in the Supplemental Material.\cite{SupplInfo}  Here again, the respective 2D limits in Table~\ref{Tab:EvecsEvals} are reached rather quickly with increasing $c/a$ in Fig.~\ref{Fig:AllPTDataCAk101Ave} at $c/a\sim2$.

\begin{figure}[t]
\includegraphics[width=2.75in]{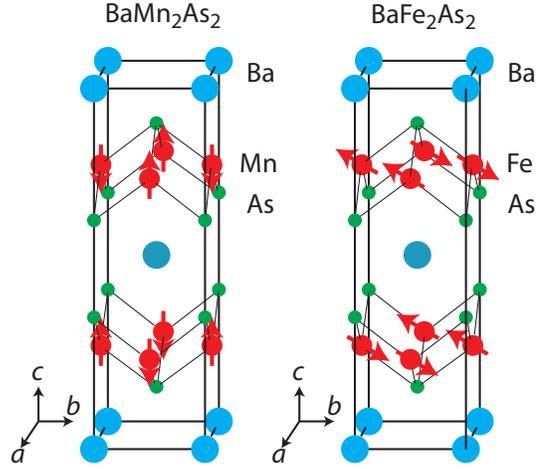}
\caption {(Color online) Crystallographic structures of bct ${\rm ThCr_2Si_2}$-type ${\rm BaMn_2As_2}$ and ${\rm BaFe_2As_2}$.\cite{Johnston2010}  The magnetic atoms Mn and Fe form simple-tetragonal sublattices.  The AFM structure of ${\rm BaMn_2As_2}$ is N\'eel-type (G-type) with AFM propagation vector ${\bf k} = \left(\frac{1}{2},\frac{1}{2},\frac{1}{2}\right)$ and with the ordered moments aligned along the $c$~axis, whereas the AFM structure of ${\rm BaFe_2As_2}$ is stripe-type with AFM propagation vector ${\bf k} = \left(\frac{1}{2},0,\frac{1}{2}\right)$ and with the ordered moments aligned along the $a$~axis of the simple-tetragonal Fe sublattice structure (due to an orthorhombic distortion, the $a$ and $b$ axes have slightly different lengths at $T\leq T_{\rm N}$ in ${\rm BaFe_2As_2}$).}
\label{Fig:BaMnFe2As2_structs}
\end{figure}

Shown in Fig.~\ref{Fig:BaMnFe2As2_structs} is the bct ${\rm ThCr_2Si_2}$-type crystal structure (space group $I4/mmm$) of ${\rm BaMn_2As_2}$ and ${\rm BaFe_2As_2}$.\cite{Johnston2010}  In both compounds the transition-metal atoms Mn and Fe form a simple tetragonal sublattice with lattice parameters $a_{\rm Mn/Fe}= a_{\rm bct}/\sqrt{2}$ and $c_{\rm Mn/Fe}= c_{\rm bct}/2$, yielding $c_{\rm Mn/Fe}/a_{\rm Mn/Fe} = (c_{\rm bct}/a_{\rm bct})/\sqrt{2} = 2.285$ for ${\rm BaMn_2As_2}$ and 2.32 for ${\rm BaFe_2As_2}$.  ${\rm BaMn_2As_2}$ has a G-type (N\'eel-type) AFM structure with {\bf k} = $\left(\frac{1}{2},\frac{1}{2},\frac{1}{2}\right)$ in the tetragonal lattice notation below $T_{\rm N} = 625$~K with the Mn ordered local moments aligned along the $c$~axis, whereas ${\rm BaFe_2As_2}$ has a stripe-type itinerant AFM structure with {\bf k} = $\left(\frac{1}{2},0,0\right)$ below $T_{\rm N} = 137$~K with the Fe ordered moments aligned along the $a$~axis of the simple-tetragonal sublattice in Fig.~\ref{Fig:BaMnFe2As2_structs}.  The ordered moment axis for ${\rm BaMn_2As_2}$ agrees with the prediction for the wavevector {\bf k} = $\left(\frac{1}{2},\frac{1}{2},\frac{1}{2}\right)$ in Fig.~\ref{Fig:AllPTDataCAk101Ave}(b).  However, as shown in Fig.~\ref{Fig:AllPTDataCAk101Ave}(a), for ${\rm BaFe_2As_2}$ MDIs favor the $b=[010]$ easy axis for {\bf k} = $\left(\frac{1}{2},0,\frac{1}{2}\right)$ and $c/a = 2.32$, perpendicular to the in-plane component ${\bf k}_{ab} = \left(\frac{1}{2},0\right)$ of the AFM propagation vector, whereas the easy axis is found to be the $a$~axis, parallel to ${\bf k}_{ab}$ (see Fig.~40 of Ref.~\onlinecite{Johnston2010}).  Therefore, there must be another source of anisotropy in ${\rm BaFe_2As_2}$ that overcomes that due to MDIs to determine the easy axis.

\subsection{Body-Centered Tetragonal Spin Lattices}

\begin{figure}[t]
\includegraphics[width=3.3in]{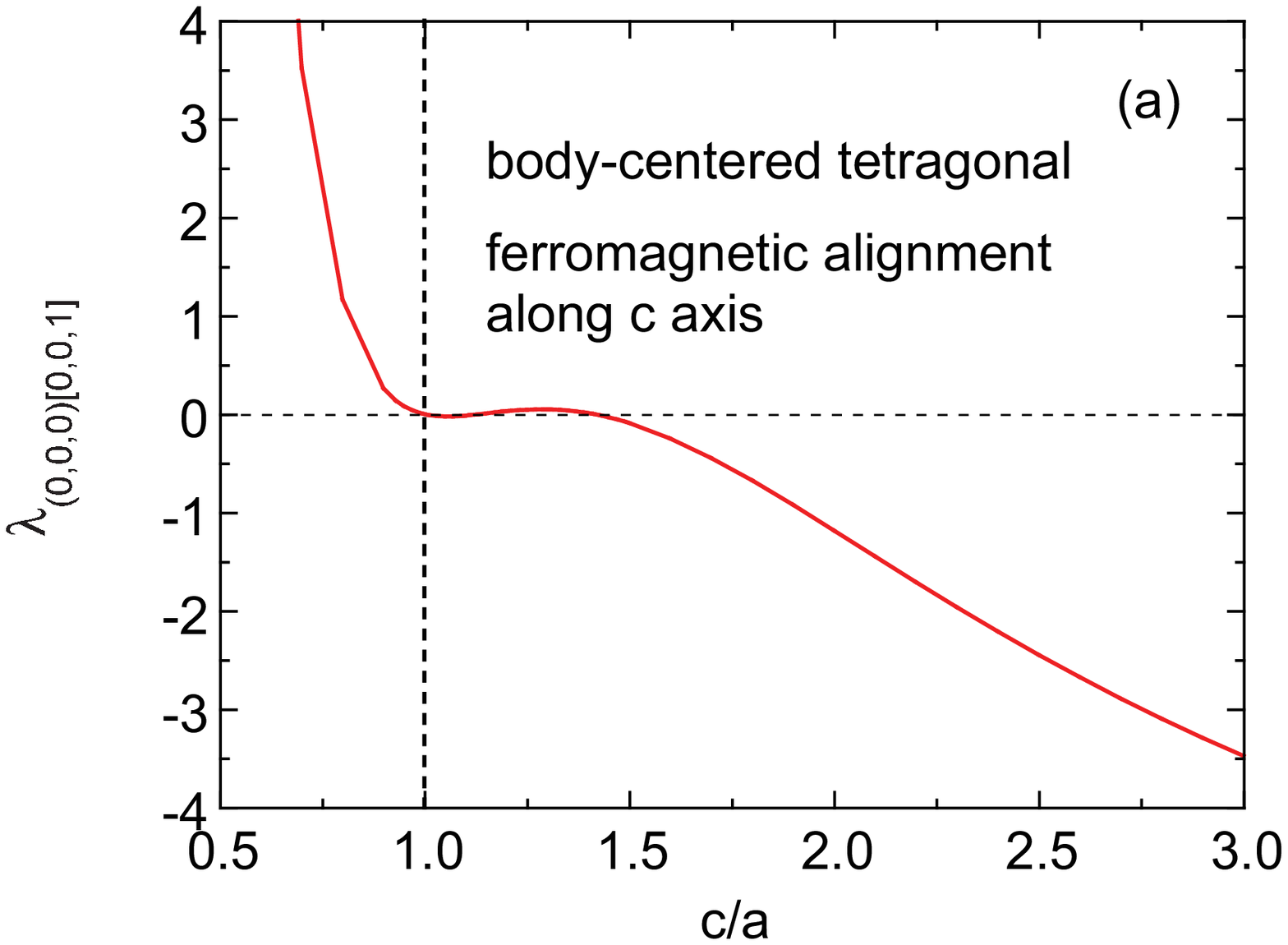}
\includegraphics[width=3.3in]{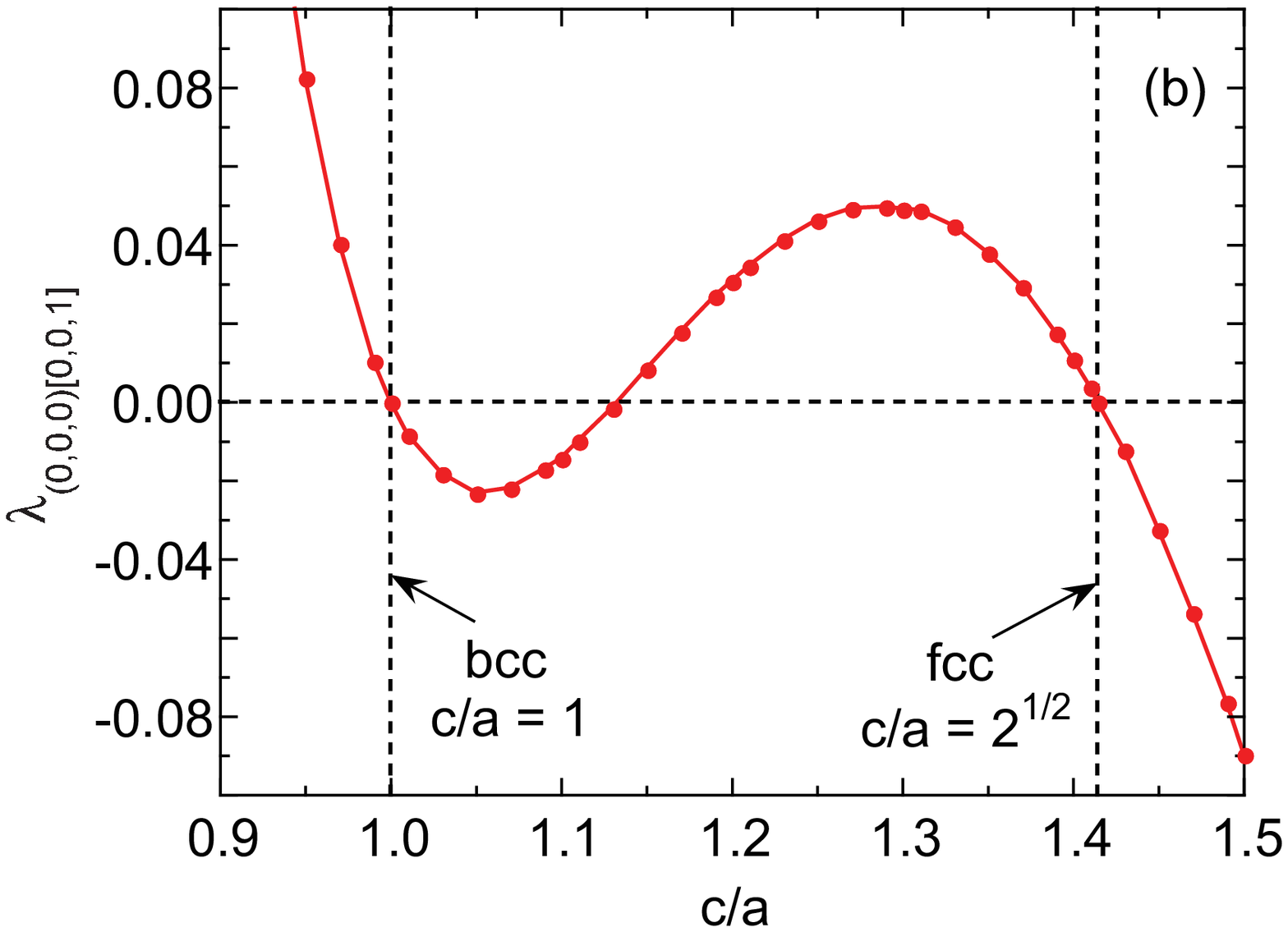}
\caption {(Color online) (a) Dependence of the eigenvalue $\lambda_{(0,0,0)[0,0,1]}$ on the $c/a$ ratio for a bct lattice with a FM alignment of the magnetic moments along the $c$~axis. (b) Expanded plot of the data in (a) for $0.95 \leq c/a\leq1.5$.  One sees that FM alignment along the $c$~axis is the most stable for $c/a < 1$, which from Fig.~\ref{Fig:SummaryPTDataCA} is also the case for the simple tetragonal lattice.  For $1<c/a\lesssim 1.3$ the easy axis for FM alignment is the $a$ or $b$ axis, for $1.3\lesssim c/a \leq \sqrt{2}$ the $c$~axis is favored, then for $c/a>\sqrt{2}$ the $a$ or $b$ axis is again favored.  For this magnetic structure, one has $\lambda_{(0,0,0)[100]} = \lambda_{(0,0,0)[010]}= -\lambda_{(0,0,0)[001]}/2$.}
\label{Fig:AllFMBCTDataAve}
\end{figure}

The behavior of the eigenvalue $\lambda_{(0,0,0)[001]}$ of the MDI tensor for FM ordering with {\bf k} = (0,0,0) and the ordered moment direction along the $c$~axis is shown versus $c/a$ in Fig.~\ref{Fig:AllFMBCTDataAve}(a).  A list of the numerical data is given in the Supplemental Material.\cite{SupplInfo}  An expanded plot of the data for $0.85 \leq c/a \leq 1.5$ is shown in Fig.~\ref{Fig:AllFMBCTDataAve}(b).  One sees an S-shaped oscillation in the latter range that was apparently first noticed by Lo {\it et al.}\cite{Lo2001b}  The first zero crossing occurs at $c/a=1$, corresponding to a bcc lattice, and the third zero crossing occurs at $c/a=\sqrt{2}$.  This latter $c/a$ value for the bct lattice corresponds to an fcc lattice within the bct lattice that is rotated by $45^\circ$ with respect to the bct lattice as shown in Fig.~15 of Ref.~\onlinecite{Johnston2010}.  The lattice parameters are related by $a_{\rm fcc} = \sqrt{2}a_{\rm bct} = c_{\rm bct}$, yielding $c_{\rm bct}/a_{\rm bct}=\sqrt{2}$.  Hence both values $c/a=0$ and~$\sqrt{2}$ correspond to cubic Bravais lattices, for which it is well known that $\lambda_{(0,0,0)}=0$ for all~$\hat{\mu}$.  We verified that our $\lambda_{(0,0,0)[001]}$ versus $c/a$ data in Fig.~\ref{Fig:AllFMBCTDataAve}(b) calculated by direct summation quantitatively agree with the corresponding eigenvalue data in Refs.~\onlinecite{Lo2001a,Lo2001b,Huang2004} that were calculated using the Ewald-Kornfeld method.

\begin{figure}
\includegraphics[width=3.3in]{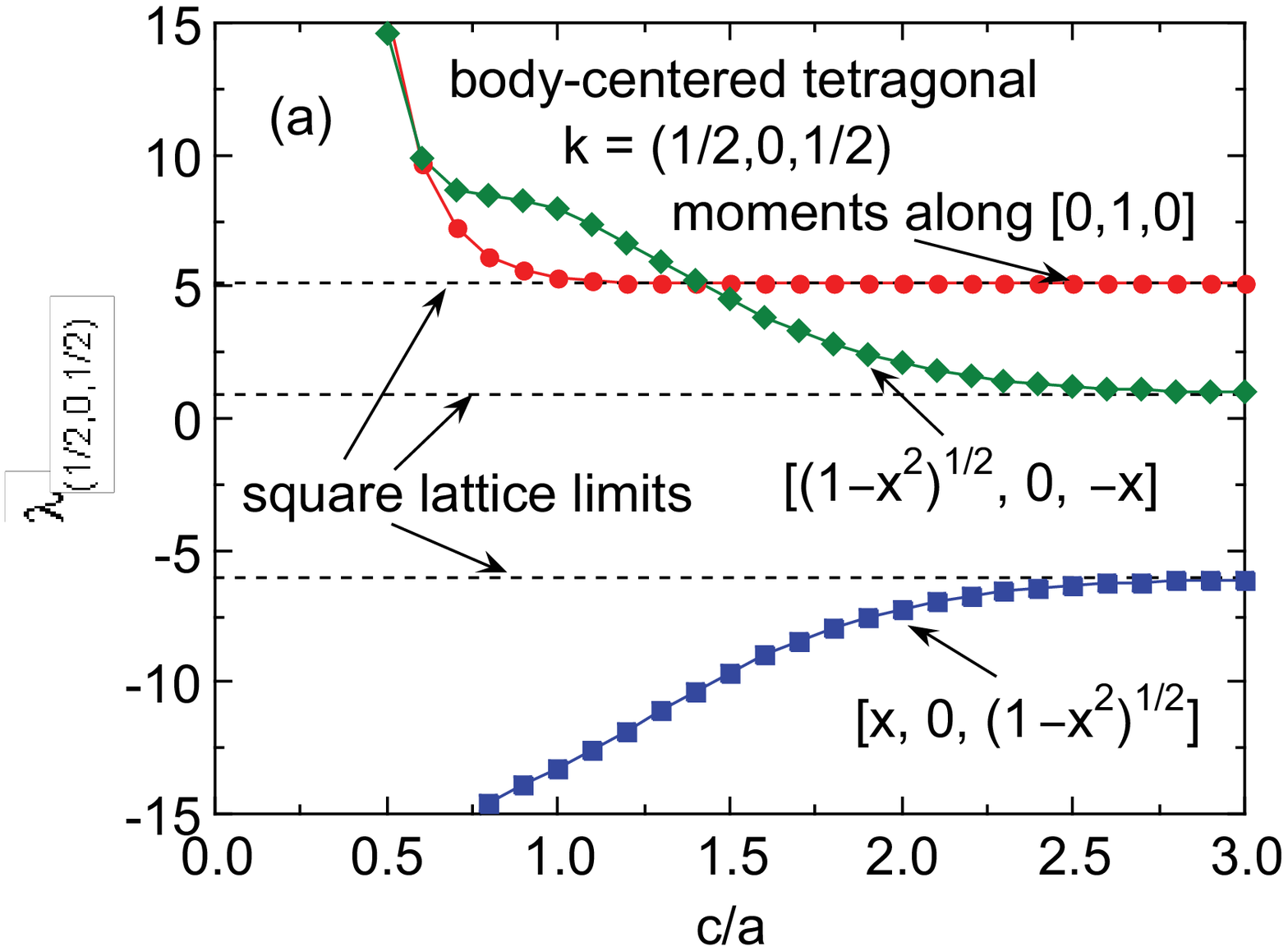}
\includegraphics[width=3.3in]{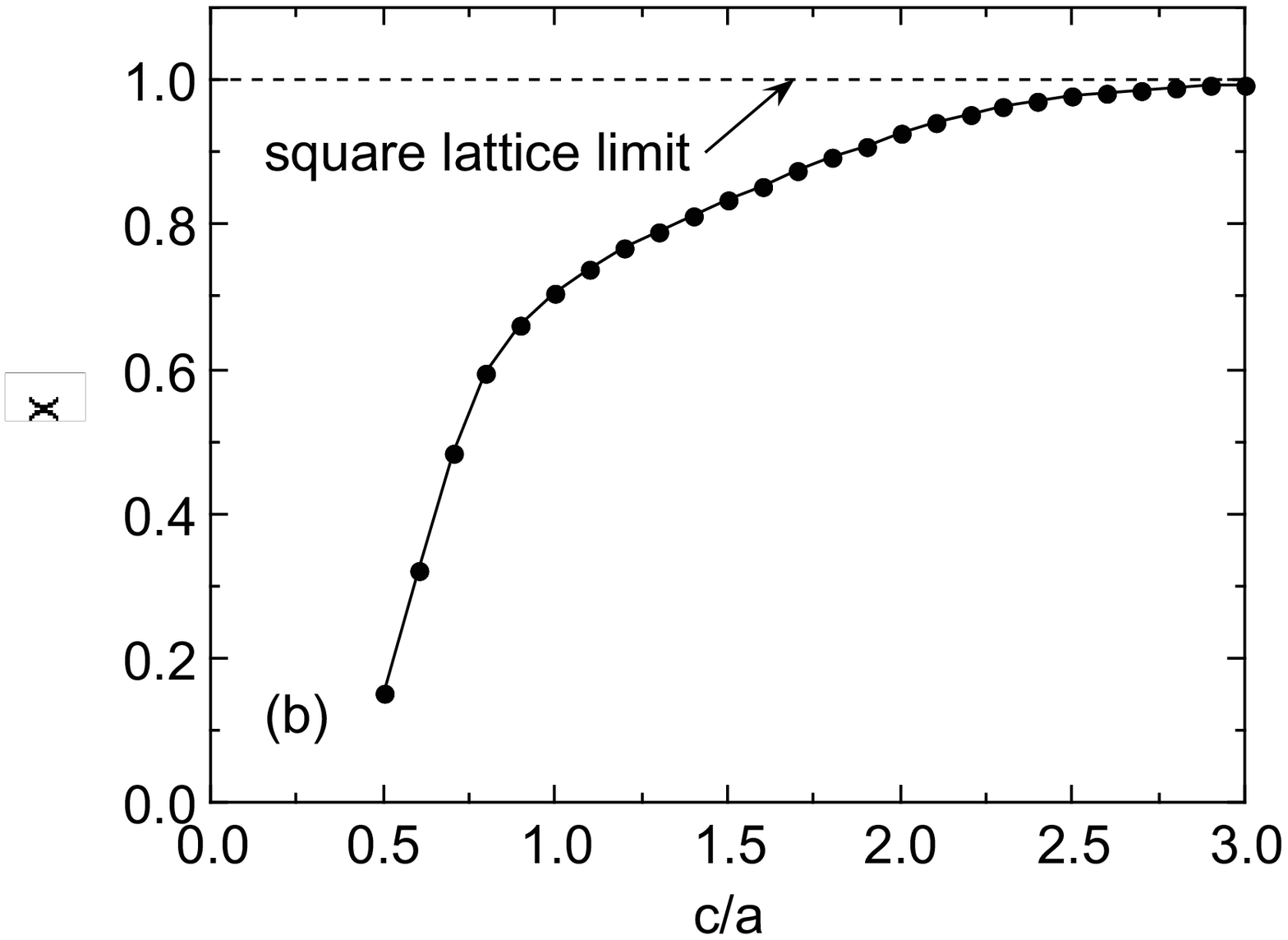}
\caption {(Color online) Eigenvalues for wavevector {\bf k} = $\left(\frac{1}{2},\frac{1}{2},0\right)$\,r.l.u.\ versus the $c/a$ ratio for a bct spin lattice with the moments aligned along (a) [0,~$\bar{1}$,~0] (solid red circles), $[-\sqrt{1-x^2},~0,~x]$ (solid green diamonds) or $[x,0,\sqrt{1-x^2}]$ (solid blue squares), where $x$ versus $c/a$ is shown in~(b). }
\label{Fig:AllBCTDataCAk101Ave}
\end{figure}

The eigenvectors and eigenvalues of $\widehat{\bf G}_i$ were calculated for several AFM propagation vectors.  The $\lambda_{(1/2,1/2,0)[001]}$ data for {\bf k} = $\left(\frac{1}{2},\frac{1}{2},0\right)$ are plotted versus $c/a$ in Fig.~\ref{Fig:AllBCTDataCAk110Ave} in Appendix~\ref{App:lambdaFigs} and a listing of the numerical data is given in the Supplemental Material.\cite{SupplInfo}  The eigenvalues for wave {\bf k} = $\left(\frac{1}{2},0,\frac{1}{2}\right)$ versus the $c/a$ ratio with the moments aligned in the [0,~$-1$,~0], $[-\sqrt{1-x^2},~0,~x]$ or $[x,0,\sqrt{1-x^2}]$ directions are plotted in Fig.~\ref{Fig:AllBCTDataCAk101Ave}(a), and $x$ versus $c/a$ is plotted in Fig.~\ref{Fig:AllBCTDataCAk101Ave}(b).  The numerical data in Fig.~\ref{Fig:AllBCTDataCAk101Ave} are listed in the Supplemental Material.\cite{SupplInfo}  

\begin{figure}
\includegraphics [width=1.5in]{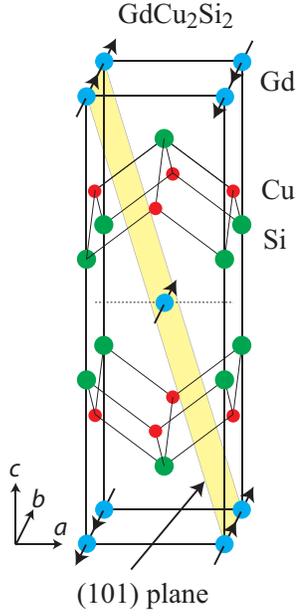}
\caption{(Color online) Crystal and magnetic structures of bct ${\rm GdCu_2Si_2}$ with the ${\rm ThCr_2Si_2}$-type crystal structure.  One crystallographic unit cell is shown.  The magnetic unit cell has dimensions $2a\times b\times 2c$ and contains four crystallographic unit cells.  The collinear magnetic structure has an AFM propagation vector $(\frac{1}{2},0,\frac{1}{2})$~r.l.u.\ perpendicular to the (101) plane shown, with the magnetic moments oriented along the $b$~axis.\cite{Barandiaran1988}  Within each such (101) plane, the Gd magnetic moments are FM aligned.}
\label{Fig:GdCu2Si2_Mag_struct}
\end{figure}

The compound ${\rm GdCu_2Si_2}$ has the bct ${\rm ThCr_2Si_2}$-type structure with space group $I4/mmm$ as shown in Fig.~\ref{Fig:GdCu2Si2_Mag_struct} and lattice parameters and $z$-axis Si positions $a=3.922$~\AA, $c=9.993$~\AA, $c/a = 2.548$ and $z_{\rm Si}=0.368$ at 24~K.\cite{Barandiaran1988}  The magnetic structure of ${\rm GdCu_2Si_2}$ is collinear, with the Gd ordered moments oriented along the tetragonal $b$~axis with an AFM propagation vector ${\bf k} = (\frac{1}{2},0,\frac{1}{2})$~r.l.u.,\cite{Barandiaran1988} as shown in Fig.~\ref{Fig:GdCu2Si2_Mag_struct}.  The ordered moment at 2~K is 7.2(4)~$\mu_{\rm B}$/Gd,\cite{Barandiaran1988} in agreement with the value of $7\,\mu_{\rm B}$/Gd obtained from the usual relation $\mu_{\rm sat} = gS\mu_{\rm B}$, where here $S=7/2$ and $g=2$.  Thus the Gd moments in (101) planes are FM aligned and are oriented perpendicular to ${\bf k}$.  From Fig.~\ref{Fig:AllBCTDataCAk101Ave}, dipolar interactions for ${\bf k} = (\frac{1}{2},0,\frac{1}{2})$ and $c/a = 2.548$ predict that the moment alignment should be along the $b$~axis, in agreement with the experimental AFM structure in Fig.~\ref{Fig:GdCu2Si2_Mag_struct}.

\begin{figure}
\includegraphics[width=3.in]{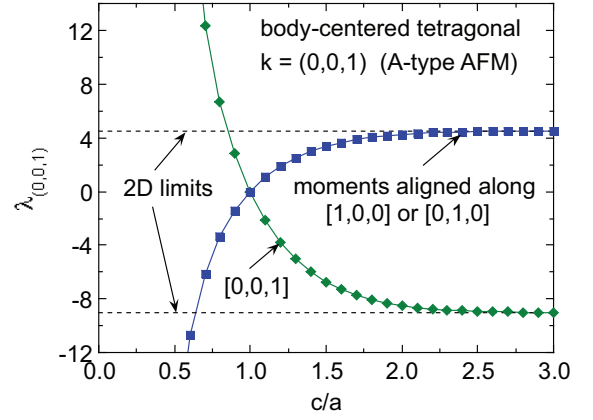}
\caption {(Color online) Eigenvalues for wavevector {\bf k} = (0,0,1)\,r.l.u.\ versus the $c/a$ ratio for a bct spin lattice with the moments aligned along [001] ($c$~axis, solid green diamonds) or [100] or [010]  ($a$ or $b$~axis, solid blue squares). The 2D limits for the square lattice for $c/a\to\infty$ are shown as horizontal black dashed lines.}
\label{Fig:AllBCTDataCAk002Ave}
\end{figure}

\begin{figure}
\includegraphics[width=1.5in]{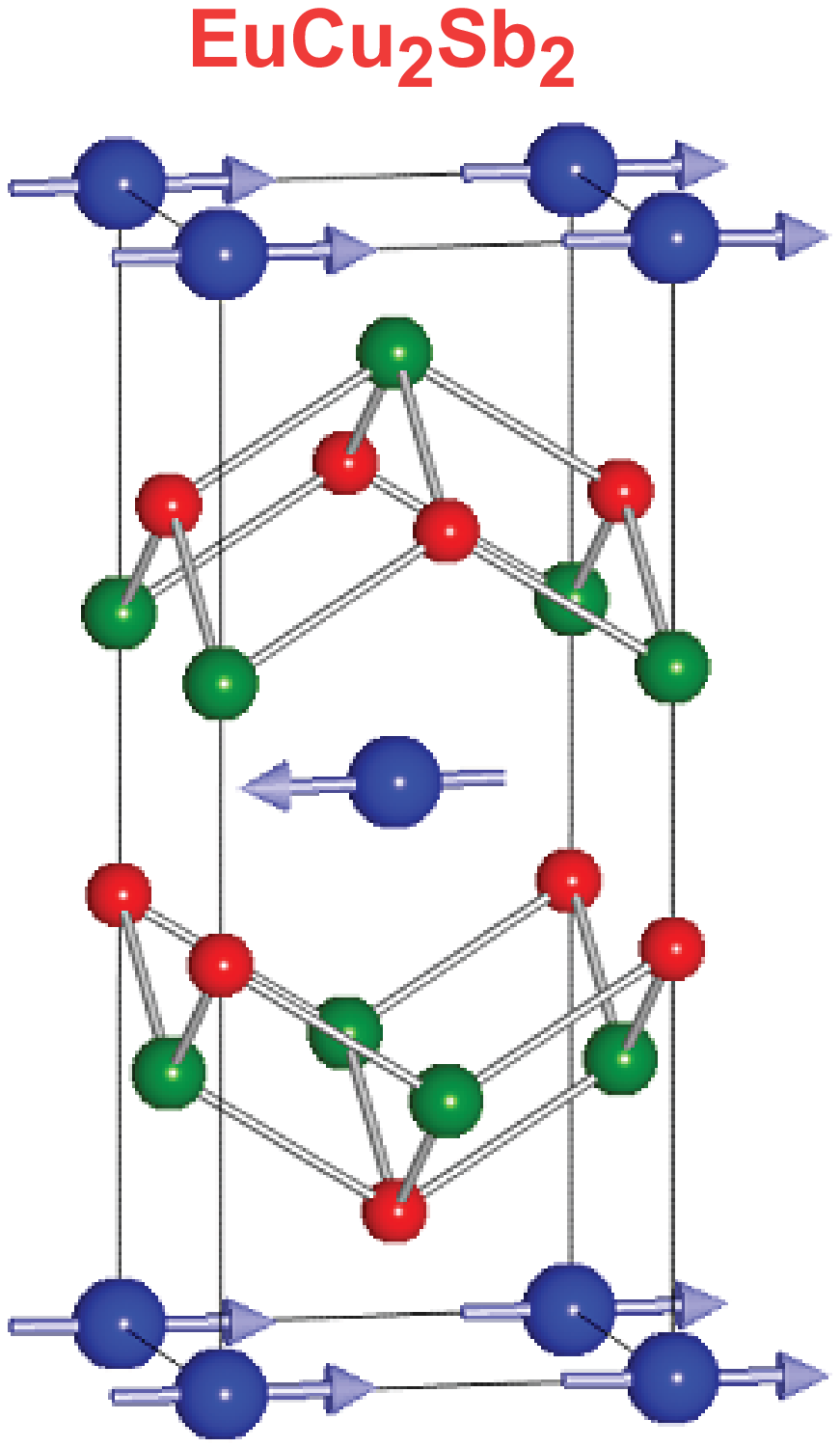}\vspace{0.2in}
\includegraphics[width=2in]{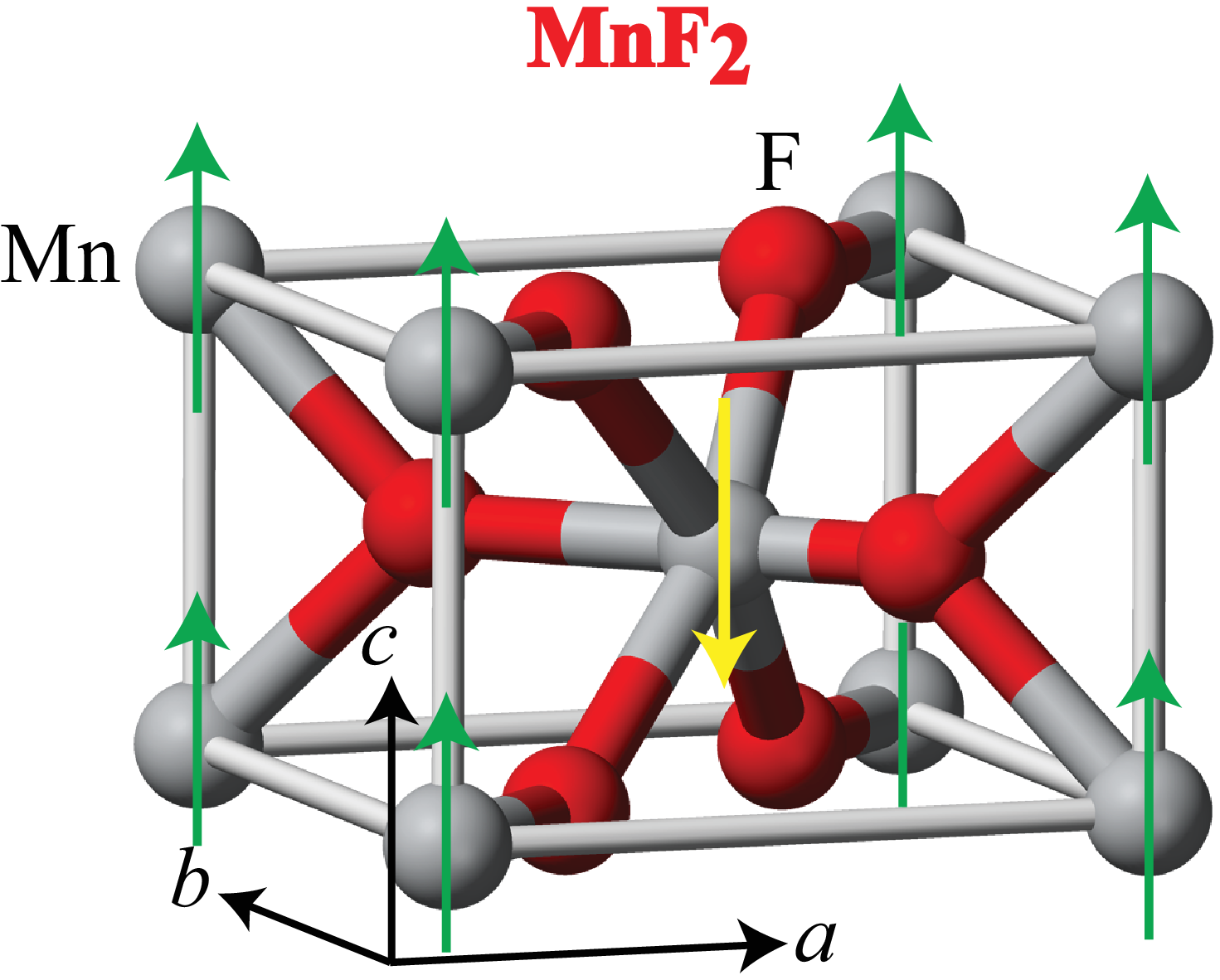}
\caption {(Color online) Crystallographic and AFM A-type structure with {\bf k} = (0,0,1) and $\hat{\mu} = [100]$ of ${\rm EuCu_2Sb_2}$ with $c/a = 2.401$ (Refs.~\onlinecite{Anand2015}, \onlinecite{Ryan2015}) and~MnF$_2$ with $c/a=0.6793$ and $\hat{\mu} = [001]$.\cite{Stout1954,Erickson1953,Goldman1987,Chatterji2010}  Each compound contains a bct sublattice of magnetic ions, but with $c/a < 1$ and $c/a > 1$, respectively, which is the crossover point between [001] and [100]- or [010]-axis ordering, respectively.}
\label{Fig:EuCu2Sb2_mag_Struct_A_type}
\end{figure}

The eigenvalues for AFM propagation vector {\bf k} = (0,0,1) in the bct spin lattice versus the $c/a$ ratio with the moments aligned along the $c$~axis or in the $ab$~plane are plotted in Fig.~\ref{Fig:AllBCTDataCAk002Ave} and listed in the Supplemental Material.\cite{SupplInfo}

The compound ${\rm EuCu_2Sb_2}$ has a primitive tetragonal ${\rm CaBe_2Ge_2}$-type crystal structure (space group $P4/nmm$) containing Eu$^{+2}$ ions in crystallographically-equivalent sites forming a bct sublattice as shown in one panel of Fig.~\ref{Fig:EuCu2Sb2_mag_Struct_A_type}.\cite{Anand2015}  Like Gd$^{+3}$, the Eu$^{+2}$ ions have spin~$S=7/2$, $g=2$, angular momentum $L=0$ and a saturation moment $\mu_{\rm sat} = gS\mu_{\rm B} = 7~\mu_{\rm B}$.  The compound orders antiferromagnetically below $T_{\rm N}=5.1$~K with an A-type structure, {\bf k} = (0,0,1), and with the Eu$^{+2}$ moments oriented in the $ab$~plane as shown in Fig.~\ref{Fig:EuCu2Sb2_mag_Struct_A_type}.\cite{Ryan2015,Anand2015}  The powder neutron diffraction measurements\cite{Ryan2015} can only determine that the ordered moments lie in the $ab$~plane and not their direction within this plane.\cite{Shirane1959}  ${\rm EuCu_2Sb_2}$ has lattice parameters $a = 4.488$~\AA, $c = 10.778$~\AA\ and $c/a = 2.401$.  From Fig.~\ref{Fig:AllBCTDataCAk002Ave}, for this $c/a$ value the ordering direction for {\bf k} = (0,0,1) is predicted for MDIs to be in the $ab$~plane, in agreement with the experimental data.

The compound MnF$_2$ has the primitive tetragonal rutile crystal structure with space group $P4_2/mnm$ and is widely considered to be the prototype for collinear AFM ordering.  The crystal and magnetic structures of ${\rm MnF_2}$ are shown in Fig.~\ref{Fig:EuCu2Sb2_mag_Struct_A_type}.  At $T =  298$~K, the lattice parameters are $a = 4.8734(2)$ and $c = 3.3099(5)$~\AA\ with $c/a=0.6792$ and the general F position parameter is $u = 0.310(3)$.\cite{Stout1954}  The Mn$^{+2}$ $d^5$ ions with expected high-spin $S = 5/2$ form a bct sublattice.  The Mn$^{+2}$ spins order in an A-type AFM structure\cite{Erickson1953} below the N\'eel temperature $T_{\rm N} = 67$~K (Ref.~\onlinecite{Goldman1987}) with an ordered moment at 5~K of 5.12(9)~$\mu_{\rm B}$/Mn.\cite{Chatterji2010} The ordered moment is in good agreement with the expected value $\mu_{\rm sat} = gS\mu_{\rm B} = 5\,\mu_{\rm B}$/Mn for $g=2$.  A fit to $\chi(T)$ measurements from 200 to 300~K by the Curie-Weiss law gave a molar Curie constant of ${\rm 4.47~cm^3\,K/mol}$ and a Weiss temperature $\theta=-97.0$~K\@.\cite{Corliss1950}  The Curie constant is close to the value of ${\rm 4.38~cm^3\,K/mol}$ expected for $S = 5/2$ and $g=2$.  From the $c/a$ ratio and Fig.~\ref{Fig:AllBCTDataCAk002Ave}, the MDI favors ordered moment alignment along the $c$~axis, in agreement with the easy axis observed in Fig.~\ref{Fig:EuCu2Sb2_mag_Struct_A_type}.  This ordering axis is perpendicular to the ordering axis for ${\rm EuCu_2Sb_2}$ with $c/a>1$ discussed above, as expected from MDIs.

\subsection{Simple Hexagonal (Triangular) and Honeycomb Spin Lattices}

The eigenvalues and eigenvectors of the MDI tensor $\widehat{\bf G}_i$ for stacked simple hexagonal lattices were calculated versus $c/a$ from 0.5 to 3 for FM alignment ({\bf k} = 0) and AFM wavevectors {\bf k} = (1,0,0), $\left(\frac{1}{2},\frac{1}{2},0\right)$, $\left(\frac{1}{3},\frac{1}{3},\frac{1}{3}\right)$, $\left(\frac{1}{3},\frac{1}{3},0\right)$ and $\left(\frac{1}{3},\frac{1}{3},\frac{1}{2}\right)$, and are plotted in Figs.~\ref{Fig:AllHexk000001DataAve}, \ref{Fig:AllHexk131313DataAve} and~\ref{Fig:AllHexk13130DataAve} in Appendix~\ref{App:lambdaFigs} and the numerical data are listed in the Supplemental Material.\cite{SupplInfo}  In contrast to the AFM cases, for the FM alignment the approach of the eignevalues to the asymptotic 2D ones with increasing $c/a$ is very slow as seen from comparison of the plots for FM alignments in Fig.~\ref{Fig:AllHexk000001DataAve}(a) with the AFM ones, which reach their 2D values by $c/a\sim2$.

The eigenvalues and eigenvectors of $\widehat{\bf G}_i$ for the honeycomb spin lattice in Fig.~\ref{Fig:Honeycomb_Lattice} calculated versus $c/a$ from 0.5 to 3 for {\bf k} = (0,0,0) (FM alignment) and AFM propagation vectors {\bf k} = $\left(\frac{1}{2},0,0\right)$, $\left(0,0,\frac{1}{2}\right)$ (N\'eel-type in all directions), (0,0,0)\,r.l.u.\ (N\'eel-type in $ab$~plane and FM alignment along $c$~axis) and $\left(0,0,\frac{1}{2}\right)$ (FM alignment intraplane and AFM alignment interplane) are plotted in Figs.~\ref{Fig:All3DHoneyCo000_100AveLaTeX}, \ref{Fig:All3DHoneyComb3DNeelAve} and~\ref{Fig:All3DHoneyCombk0012Ave} in Appendix~\ref{App:lambdaFigs}, respectively, and are listed in the Supplemental Material.\cite{SupplInfo}  Similar to the behavior of the eigenvalues for the simple hexagonal spin lattice, for FM alignment in the honeycomb lattice the approach of the eigenvalues to their 2D limits with increasing $c/a$ is very slow compared to behaviors for the AFM moment alignments.  For the N\'eel AFM alignments both just in the $ab$~plane and also along the $c$~axis, the approach with increasing $c/a$ to the infinite $c/a$ limits is very fast, being essentially complete by $c/a\sim1.5$.

\section{\label{Sec:NoncollearAFMs} Eigenvalues and Eigenvectors for Noncollinear Antiferromagnets}

The relationship between the ordered/induced central moment $\vec{\mu}_i$ and another moment $\vec{\mu}_j$ at position ${\bf r}_{ji}$ with respect to $\vec{\mu}_i$ in a collinear magnetic structure was given in Eq.~(\ref{Eq:muj}).  In noncollinear AFMs one must specify the directions of each of the moments in a crystal in order to calculate the net dipolar interaction of a given central moment $\vec{\mu}_i$ with its neighbors inside the Lorentz sphere.  There are two generic cases.  In the first, one can define a nonzero AFM propagation vector {\bf k} such that moments in a plane perpendicular to {\bf k} are FM-aligned and all change their directions from plane to plane along {\bf k}.  In the second, the spin lattice is a non-Bravais lattice and the magnetic and chemical unit cells are the same, where the AFM propagation vector is {\bf k} = (0,0,0) for such cases.  We consider the first type of AFM ordering in the 2D triangular lattice in the following section and then the second type of ordering in GdB$_4$ and the Shastry-Sutherland lattice.

\subsection{\label{TriangAFMs} 2D Triangular Lattice Antiferromagnets}

\begin{figure}
\includegraphics [width=1.75in]{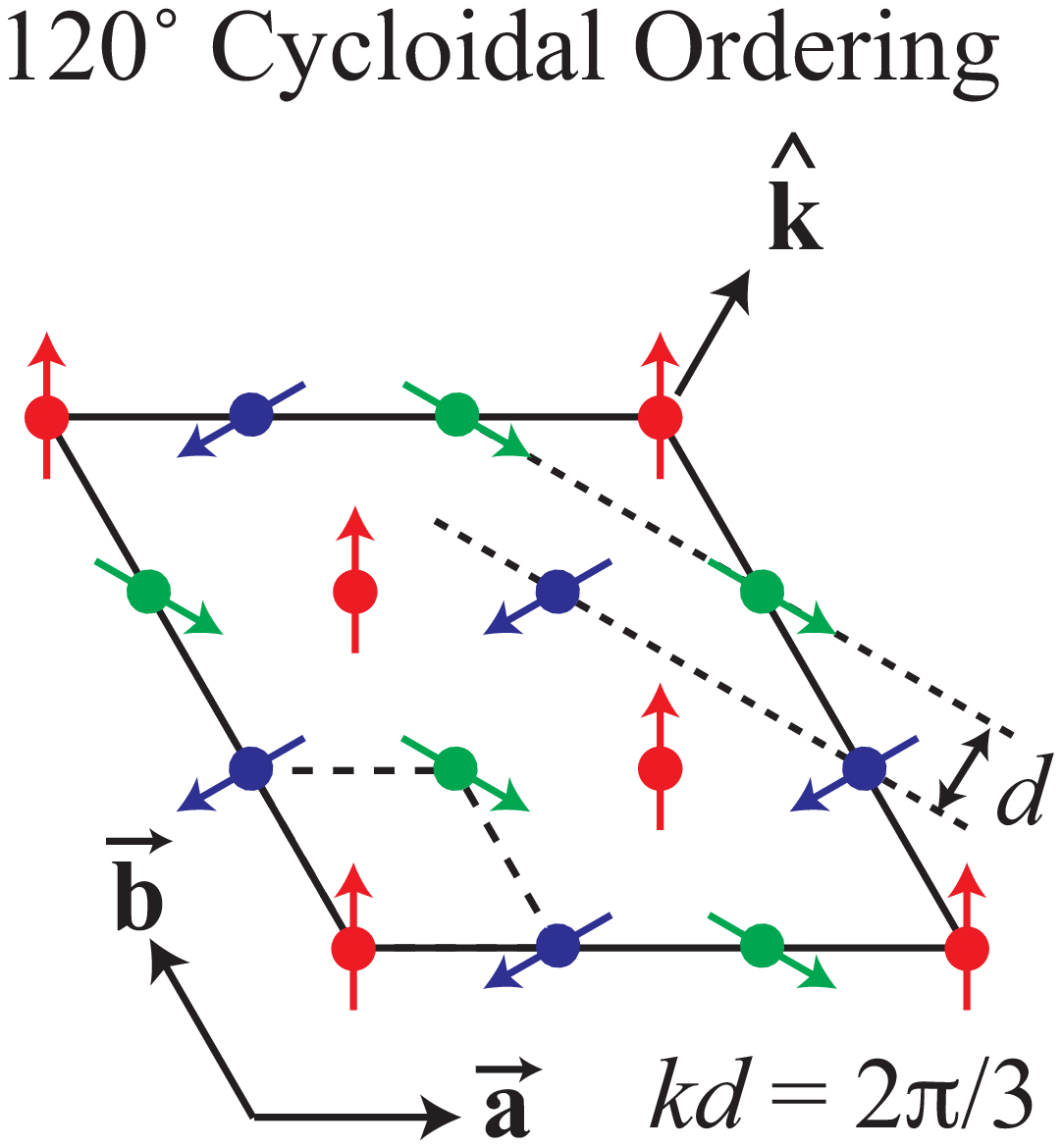}
\caption{(Color online) Coplanar noncollinear magnetic unit cell of classical $120^\circ$ ordering on the 2D simple hexagonal (triangular) spin lattice for cycloidal AFM ordering with a commensurate wavelength of 3$a$/2.  The hexagonal lattice translation vectors {\bf a} and {\bf b} ($a=b$) and the direction $\hat{\bf k}$ of the cycloid wavevector {\bf k} are indicated.  The long-dashed line is the outline of the hexagonal unit cell containing one spin and the solid line is the outline of the magnetic unit cell containing nine spins (nine unit cells).  The AFM propagation vector is ${\bf k} = \left(\frac{1}{3},\frac{1}{3}\right)$~r.l.u.  The quantity $d$ is the distance between lines of FM-aligned magnetic moments along the cycloid axis ($\hat{\bf k}$) direction.  The rotation angle of the magnetic moments between adjacent lattice lines in the $\hat{\bf k}$ direction is $\phi_{ji} = kd = \frac{2\pi}{3}$~rad.}
\label{Fig:TriangularLattWaveVector}
\end{figure}

It is well known that the classical ground state of a triangular lattice AFM interacting by isotropic Heisenberg exchange is the coplanar noncollinear $120^\circ$ structure, where each of the six neigbors of a given moment is at a $120^\circ$ angle with the given moment, as in the cycloidal AFM structure shown in Fig.~\ref{Fig:TriangularLattWaveVector} where the 2D AFM propagation vector is ${\bf k} = \left(\frac{1}{3},\frac{1}{3}\right)$~r.l.u.  In the absence of anisotropy, the energy of the spin lattice in Fig.~\ref{Fig:TriangularLattWaveVector} is invariant on rotating each spin by the same angle, thus retaining the $120^\circ$ angles between adjacent moments.  Here we examine whether the MDI can determine how the moments are oriented with respect to the hexagonal unit cell axes for the AFM structure in Fig.~\ref{Fig:TriangularLattWaveVector}, or indeed whether the MDI alone can stabilize this magnetic structure.

The approach we use is to first calculate the eigenvalues of the MDI tensor $\widehat{\bf G}_i$ for noncollinear moments and variable ${\bf k} = (x,x)$~r.l.u.\ and see whether the maximum eigenvalue is obtained for $x = 1/3$. If so, then we are done.  If not, we conclude that exchange interactions alone determine ${\bf k} = \left(\frac{1}{3},\frac{1}{3}\right)$~r.l.u.\ and then calculate for this {\bf k} what the moment orientations should be with respect to the crystal axes as predicted by the MDI.

\begin{figure}
\includegraphics [width=3.3in]{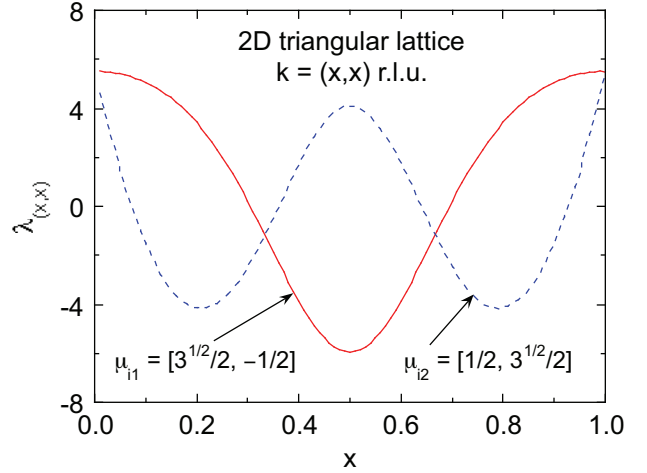}
\caption{(Color online) Variation in the eigenvalues $\lambda_{(x,x)}$ versus $x$ in the AFM propagation vector ${\bf k} = (x,x)$~r.l.u.\ for the two eigenvectors $\vec{\mu}_{i1}$ and $\vec{\mu}_{i2}$ of the MDI tensor for the orientation of a representative moment $\vec{\mu}_i$.  The first eigenvector is in the hexagonal ${\bf b}$ direction and the second is in the ${\bf b}^\ast$ direction, which is rotated clockwise by 90$^\circ$ from the first (see Fig.~\ref{Fig:Hexagonal_Recip_Latt} in Appendix~\ref{Methods}).  The two curves cross at $x=1/3$ and $x=2/3$.  For $x=1/3$ the eigenvectors are calculated as [100] (along the $a$~axis) and [010] (perpendicular to the $a$~axis). }
\label{Fig:Hex2DCycloidxx0Ave}
\end{figure}

The MDI tensor $\widehat{\bf G}_i$ was calculated using Eq.~(\ref{Eq:Gk3}).  Shown in Fig.~\ref{Fig:Hex2DCycloidxx0Ave} are plots of the two eigenvalues $\lambda_{(x,x)}$ versus~$x$ with ${\bf k} = (x,x)$~r.l.u.\ for the two eigenvectors $\hat{\mu}_{i1}$ and $\hat{\mu}_{i2}$ shown in the figure for the orientation of central moment $\vec{\mu}_i$ at the origin of the Cartesian coordinate system (the third eigenvalue is for FM ordering along the $c$~axis as discussed in Sec.~\ref{Sec:HelCyc} and is not relevant here). From Fig.~\ref{Fig:Hex2DCycloidxx0Ave}, there is no maximum in $\lambda_{(x,x)}$ at $x=1/3$ corresponding to the $120^\circ$ noncollinear structure.  Instead, the MDI favors {\bf k} = (1/2, 1/2)~r.l.u.  Setting $x=1/3$, we obtain $\lambda_{(1/3,1/3)} = -1.1659$ for the two degenerate eigenvectors $\hat{\mu}_i = [100]$ or [010]\@.  The AFM structure in Fig.~\ref{Fig:Hex2DCycloidxx0Ave} corresponds to $\hat{\mu}_i = [010]$.

Interestingly, the eigenvalue $\lambda_{(1/3,1/3)} = -1.1659$ is identical to the value in Table~\ref{Tab:EvecsEvals} obtained for {\it collinear} AM AFM ordering on the triangular lattice with {\bf k} = $\left(\frac{1}{3},\frac{1}{3}\right)$~r.l.u.\ with the same two eigenvectors.  This shows that the net energy of interaction of a moment with the magnetic fields of the other moments inside the Lorentz sphere only depends on the projections of those moments on the eigenvector axis.

The fact that $\lambda_{(1/3,1/3)}$ is negative, whereas the eigenvalue for collinear AM ordering along the easy $c$~axis for {\bf k} = $\left(\frac{1}{3},\frac{1}{3}\right)$ in Table~\ref{Tab:EvecsEvals} is positive, suggests that the MDI might tend to cant the moments in the classical 120$^\circ$ coplanar structure out of the $ab$~plane and also introduce an amplitude modulation of the ordered moments.

\subsection{\label{Sec:GdB4} GdB$_4$ and Shastry-Sutherland Antiferromagnets}

\begin{figure}
\includegraphics [width=2.25in]{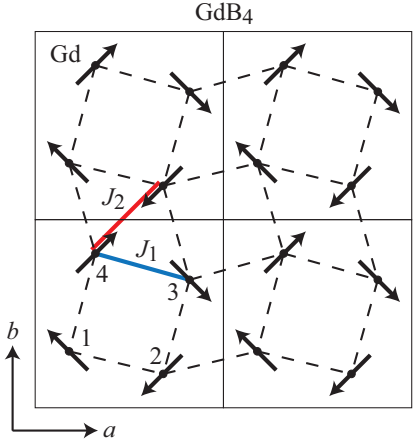}
\caption{(Color online) Four crystallographic and magnetic unit cells of the Gd sublattice of the tetragonal GdB$_4$ compound in the $ab$~plane.\cite{Johnston2012}  The Gd ordered moments all lie in the $ab$-plane in [110] and equivalent directions.\cite{Blanco2006} Also shown are the 2D in-plane Shastry-Sutherland\cite{Shastry1981} exchange interactions $J_1$ and $J_2$ between nearest- and next-nearest-neighbor Gd spins, respectively.  The four Gd spins in the lower-left unit cell are numbered counterclockwise as shown.  The spin interactions are topologically the same as in the undistorted Shastry-Sutherland square lattice model in which the ${\rm GdB_4}$ squares are not tilted with respect to the $a$ and $b$~axes.  Adjacent stacked layers along the $c$~axis are FM-aligned with FM (negative) nearest-neighbor exchange interaction $J_c$ (not shown).  Since the chemical and magnetic unit cell are the same, the AFM propagation vector is \mbox{{\bf k} = (0,0,0)}.}
\label{Fig:GdB4_mag_struct}
\end{figure}

The AFM structure for ${\rm GdB_4}$ shown in Fig.~\ref{Fig:GdB4_mag_struct} was deduced from neutron diffraction measurements.\cite{Blanco2006}  The configuration of the exchange interactions $J_1$ and $J_2$ shown in the figure is an example of a so-called Shastry-Sutherland Heisenberg exchange model in two dimensions.\cite{Shastry1981}  In ${\rm GdB_4}$, this AFM structure is stacked along the $c$~axis with FM alignments between nearest-neighbor layers and a corresponding FM interlayer interaction $J_c$ that is not included in the Shastry-Sutherland model.

Here we assume that the AFM structure is known, along with the \emph{relative} orientations of each of the ordered moments in a unit cell.  For noncollinear AFMs, Eq.~(\ref{Eq:muj}) cannot be used and instead one must express each $\vec{\mu}_k$ in a magnetic = crystallographic unit cell in terms of the central moment $\vec{\mu}_i$ around which the dipolar sum within the Lorentz sphere is calculated.  Thus we use the method described in Sec.~\ref{Sec:NonBravais} to obtain the orientation (eigenvector) of $\vec{\mu}_i$ with respect to the Cartesian coordinate system, together with the associated eigenvalue.    

${\rm GdB_4}$ has a primitive-tetragonal crystal structure with space group $P4/mbm$.\cite{Blanco2006}  The Gd atoms occupy the Wyckoff $4g$ positions (1)~$\left(\frac{1}{2}-x,x,0\right)$, (2)  $\left(1-x,\frac{1}{2}-1,0\right)$, (3) $\left(\frac{1}{2}+x,1-x,0\right)$ and (4)~$\left(x,\frac{1}{2}+x,0\right)$ with $x = 0.31746(2)$.  Thus from Eq.~(\ref{Eq:rji}) the absolute positions of the atoms within the unit cell normalized to the $a$-axis lattice parameter are
\bse
\bea
\frac{{\bf r}_{1}}{a} &=& \left(n_a + \frac{1}{2}-x,\ n_b+x,\ n_c\frac{c}{a}\right),\\*
\frac{{\bf r}_{2}}{a} &=& \left(n_a + 1-x,\ n_b+ \frac{1}{2}-x,\ n_c\frac{c}{a}\right),\\*
\frac{{\bf r}_{3}}{a} &=& \left(n_a + \frac{1}{2}+x,\ n_b+1-x,\ n_c\frac{c}{a}\right),\\*
\frac{{\bf r}_{4}}{a} &=& \left(n_a + x,\ n_b+\frac{1}{2}+x,\ n_c\frac{c}{a}\right),
\label{Eq:rk0}
\eea
\ese
where $n_a$, $n_b$ and $n_c$ are positive or negative integers or zero.  Taking the central moment $\vec{\mu}_i$ to be at position ${\bf r}_{1}$ with $n_a=n_b=n_c=0$, one obtains the ${\bf r}_{ki}={\bf r}_{k}-{\bf r}_{i}$ as
\bse
\bea
\frac{{\bf r}_{1i}}{a} &=& \left(n_a ,\ n_b,\ n_c\frac{c}{a}\right),\\*
\frac{{\bf r}_{2i}}{a} &=& \left(n_a + \frac{1}{2},\ n_b + \frac{1}{2} - 2x,\ n_c\frac{c}{a}\right),\\*
\frac{{\bf r}_{3i}}{a} &=& \left(n_a + 2x,\ n_b+1-2x,\ n_c\frac{c}{a}\right),\\*
\frac{{\bf r}_{4i}}{a} &=& \left(n_a -\frac{1}{2} + 2x,\ n_b+\frac{1}{2},\ n_c\frac{c}{a}\right).
\label{Eq:rk}
\eea
\ese
The $3\times3$ rotation matrices ${\bf R}_k$ for the four numbered moments in the lower-left unit cell in Fig.~\ref{Fig:GdB4_mag_struct} are
\bse
\bea
{\bf R}_1 &=& {\bf 1},\\*
{\bf R}_2 &=& {\bf y}{\bf x} - {\bf x}{\bf y},\\*
{\bf R}_3 &=& -{\bf 1},\\*
{\bf R}_4 &=& {\bf x}{\bf y} - {\bf y}{\bf x},
\eea
\ese
where ${\bf x}{\bf y}$ and~${\bf y}{\bf x}$ are $3\times3$ dyadics.

\begin{table}
\caption{\label{Tab:GdB4SSEvecsEvals} GdB$_4$ and Shastry-Sutherland lattice.  Eigenvectors $[\mu_x,\mu_y,\mu_z]$ for central spin $\vec{\mu}_i$ and eigenvalues $\lambda_{{\bf k}\alpha}$ for the 3D Gd sublattice in GdB$_4$ and for the 2D Shastry-Sutherland model. For both cases, the experimental 90$^\circ$ angles between adjacent spins was assumed with the order $\phi=\phi_0,\ \phi_0+90^\circ,\ \phi_0+100^\circ$ and $\phi_0+270^\circ$ on going clockwise around a Gd square as shown in Fig.~\ref{Fig:GdB4_mag_struct}, but with the value of $\phi_0$ undetermined for the moment in the lower left corner of each square. The experimental $x$ value and $c/a$ ratio for GdB$_4$ are 0.317\,46 and 0.567\,97, respectively. In the 2D Shastry-Sutherland model, $x=1/4$ and $c/a=\infty$. The symbol $\bar{1}$ means $-1$.}
\begin{ruledtabular}
\begin{tabular}{cccc}
System  	&  $x$ 			& $[\mu_x,\mu_y,\mu_z]$  &  $\lambda_{{\bf k}\alpha}$  \\
\hline
3D GdB$_4$ &  0.31746 		& [001]  				&  27.945  			\\
  		&  	(actual)		& $[1\bar{1}0]$ 		&  20.112  			\\
		&  				& $[110]$ 			&  $-48.055$  		\\
 		&				&  $[001]-[1\bar{1}0]$  	&  7.883 			 	\\
 		&				&  $[001]-[110]$  		&  76.000  			\\
		&  1/4   			&  $[1\bar{1}0]$  		&  40.013      		\\
  		&				&  [001]  			&  26.833  	    		\\
  		&				&  [110]  			&  $-66.845$      	\\
  		&				&  $[1\bar{1}0]-[001]$  	&  13.180  	   		\\
  		&				&  $[1\bar{1}0]-[110]$  	&  106.858  	   		\\
2D Shastry-  &  1/4  		&  $[1\bar{1}0]$		&  40.790\,982 		\\
Sutherland&				&  [001] 				&  7.483\,697		\\
		&  				&  $[\bar{1}\bar{1}0]$  	&  $-48.274\,678$  	\\
		&				&  $[1\bar{1}0]-[001]$	&  33.307\,285			\\
		&				&  $[1\bar{1}0]-[\bar{1}\bar{1}0]$	&  89.065\,660	\\
\end{tabular}
\end{ruledtabular}
\end{table}

The sums in Eq.~(\ref{Eq:G3}) were calculated out to a Lorentz sphere radius $R/a=50$ for 3D GdB$_4$.  Then diagonalizing $\widehat{{\bf G}}_i$ gave the eigenvectors and corresponding eigenvalues listed in Table~\ref{Tab:GdB4SSEvecsEvals}.  Recalling that the largest positive eigenvalue corresponds to the minimum energy according to Eq.~(\ref{Eq:Ei3}), the data in Table~\ref{Tab:GdB4SSEvecsEvals} show that the MDI favors moment alignment along the $c$~axis, contrary to the experimental result in Fig.~\ref{Fig:GdB4_mag_struct} which gives the alignment of the $k=0$ spin as the $[1,\bar{1},0]$ direction, corresponding to the second-highest $\lambda_{{\bf k}\alpha}$.  The highly unstable [110] direction for central moment~\#1 corresponds to all magnetic moments in Fig.~\ref{Fig:GdB4_mag_struct} rotating clockwise by 90$^\circ$ and hence all moments in each Gd$_4$ square pointing towards the center of the square.  The RKKY interaction between Gd spins and/or a high-order crystalline electric field effect  evidently give an anisotropic exchange interaction that is responsible for the observed ordered moment directions.

Calculations were also carried out for $x=1/4$, which corresponds to untilted Gd$_4$ squares in Fig.~\ref{Fig:GdB4_mag_struct}, as shown in Table~\ref{Tab:GdB4SSEvecsEvals}.  One sees significant differences in the eigenvalues compared to the results for the observed $x=0.31746$.  In particular, the Gd ordered moments are now predicted to have the experimental ordered-moment directions.  We also carried out calculations for the Shastry-Sutherland 2D lattice and the results are shown in Table~\ref{Tab:GdB4SSEvecsEvals}, where the favored ordered moment direction for Gd$_1$ is found to be the same as for $x=1/4$ and $c/a = 0.56797$, the observed $c/a$ ratio for GdB$_4$.  Thus the ground-state ordering direction predicted by the MDI is sensitive to the tilting angle of the Gd$_4$ squares.

\section{\label{Tm} Magnetic Ordering Temperature Due to Magnetic Dipole Interactions}

The MFT calculations in this and the following sections closely follow the development of the author detailed in Ref.~\onlinecite{Johnston2015}.  Therefore only an outline of the calculations associated with the MDI is given.

In this section, an AFM ordering (N\'eel) temperature arising from dipolar interactions only is denoted by $T_{\rm NA}$ and a FM ordering (Curie) temperature by $T_{\rm CA}$, where the subscript A refers to the quantity being the contribution from an anisotropic magnetic interaction.  Similarly, a N\'eel temperature arising from Heisenberg exchange interactions only is denoted by $T_{{\rm N}J}$ and a Curie temperature by $T_{{\rm C}J}$.  We use the Weiss MFT to calculate these transition temperatures where we assume that the spins are identical and crystallographically equivalent and we only treat EM (not AM) magnetic structures on Bravais lattices.  Within MFT, the contributions of the dipolar and exchange interactions to the actual ordering temperatures $T_{\rm N}$ and $T_{\rm C}$, respectively, are additive:
\be
T_{\rm N} = T_{\rm NA} + T_{{\rm N}J},\qquad T_{\rm C} = T_{\rm CA} + T_{{\rm C}J}.
\label{Eq:TNTNATNJ}
\ee

The magnetic ordering temperature $T_{{\rm m}J}$ (m = N, C) for both AFMs and FMs due to exchange interactions is given by the same expression\cite{Johnston2015}
\be
T_{{\rm m}J} = -\frac{S(S+1)}{3k_{\rm B}}\sum_j J_{ij}\cos\phi_{ji},
\label{Eq:TmJ}
\ee
where $\phi_{ji}$ is the angle between magnetic moments $j$ and~$i$ in the ordered state and $\phi_{ji} = \phi_{j}-\phi_{i} = 0$ for a FM\@.  We define the reduced ordered and/or applied magnetic field-induced average moment $\bar{\mu}$ for a spin~$S$ as
\be
\bar{\mu} \equiv \frac{\mu}{\mu_{\rm sat}} = \frac{\mu}{gS\mu_{\rm B}},
\label{Eq:barmuDef}
\ee
where $\mu_{\rm sat}=gS\mu_{\rm B}$ is the saturation moment of the spin and $g$ is the spectroscopic splitting factor.  Using Eq.~(\ref{Eq:TmJ}), one can  write the exchange field seen by a representative moment~$i$ in zero applied field~$H$ as
\be
H_{{\rm exch}i} = \frac{T_{{\rm m}J}}{C_1}\mu_0 = \frac{3k_{\rm B}T_{{\rm m}J}}{g\mu_{\rm B}(S+1)}\bar{\mu}_0,
\ee
where the subscript 0 in $\bar{\mu}_0$ signifies $H=0$, $C_1$ is the single-spin Curie constant (see below) and this expression applies to the ordered state.

The magnetic ordering temperature is determined within MFT by the criterion that $\bar{\mu}_0\rightarrow0$ for $T\to T_{\rm m}^-$.  For magnetic dipole ordering, the near-field contribution to the local magnetic induction is given by Eq.~(\ref{Eq:Bi}).  The magnetic moment $\mu$ in that equation is defined in general as either the ordered moment in a magnetic structure in $H=0$ ($\mu_0$) and/or an average moment induced by $H_\alpha > 0$ ($\mu$).  Using Eq.~(\ref{Eq:barmuDef}), Eq.~(\ref{Eq:Bi}) associated with MDIs becomes
\be
B_{{\rm int}\,\alpha\,i}^{\rm near} = \frac{g\mu_{\rm B}S\bar{\mu}\lambda_{{\bf k}\alpha}}{a^3}.
\label{Eq:Bnear}
\ee

\subsection{Antiferromagnetic Ordering (N\lowercase{\'eel}) Temperature}

Here we calculate $T_{\rm NA}$ in $H=0$ within MFT for a specified AFM wavevector {\bf k} and ordered moment axis $\hat{\mu}$ in the presence of MDIs but in the absence of exchange interactions.  The standard MFT prediction is obtained from\cite{Johnston2011,Johnston2012,Johnston2015}
\be
\bar{\mu}_0 = B_S\left(\frac{g\mu_{\rm B}B_{{\rm int}\,\alpha}^{\rm local}}{k_{\rm B}T}\right),
\label{Eq:TNACalc}
\ee
where we have dropped the subscript~$i$ because all moments are crystallographically equivalent in $H=0$, the subscript 0 in $\bar{\mu}_0$ signifies that $H=0$ as above, and $B_S(y)$ is the Brillouin function for spin~$S$ given by our unconventional expression
\be
B_S(y) = \frac{1}{2S} \left\{(2S+1)\coth\left[(2S+1)\frac{y}{2}\right]-\coth\left(\frac{y}{2}\right)\right\}
\label{Eq:BrillouinFunction22}.
\ee
There is no demagnetizing field for an AFM in $H=0$ because there is no net magnetization, so for AFM ordering in $H=0$ the local field is just the near field.  Inserting $B_{{\rm int}\,\alpha\,i}^{\rm near}$ from Eq.~(\ref{Eq:Bnear}) into~(\ref{Eq:TNACalc}) gives
\bse
\be
\bar{\mu}_{0} = B_S(y_0),
\label{Eq:SolveMu0}
\ee
where
\be
y_0 = \frac{g^2S\mu_{\rm B}^2\bar{\mu}_0\lambda_{{\bf k}\alpha}}{a^3k_{\rm B}T}.
\label{Eq:y0Def}
\ee
\ese
Then one obtains for a given {\bf k} and easy axis~$\alpha$ the N\'eel temperature\cite{Johnston2015}
\be
T_{{\rm NA}\alpha} = \frac{g^2S(S+1)\mu_{\rm B}^2\lambda_{{\bf k}\alpha}}{3a^3k_{\rm B}}.
\label{Eq:TNA1}
\ee
The relevant ordering axis $\alpha$ and hence $T_{{\rm NA}\alpha}$ is the one with the largest eigenvalue $\lambda_{{\bf k}\alpha}$ for the given AFM structure.

The single-spin Curie constant $C_1$ for spin~$S$ is given by\cite{Kittel2005}
\be
C_1 = \frac{g^2S(S+1)\mu_{\rm B}^2}{3k_{\rm B}},
\label{Eq:C1Def}
\ee
so Eq.~(\ref{Eq:TNA1}) can be written more succinctly as
\bse
\be
T_{{\rm NA}\alpha} = \frac{C_1\lambda_{{\bf k}\alpha}}{a^3}.
\label{Eq:TNA2}
\ee
Thus one can also write
\be
\frac{\lambda_{{\bf k}\alpha}}{a^3} = \frac{T_{{\rm NA}\alpha}}{C_1} = \frac{3k_{\rm B}T_{{\rm NA}\alpha}}{g^2S(S+1)\mu_{\rm B}^2}.
\ee
\ese
Then for $H=0$ and $T\leq T_{\rm NA}$ one can write the near field in Eq.~(\ref{Eq:Bnear}) in the direction of each ordered moment as
\bse
\label{Eqs:LocFieldSum}
\be
B_{{\rm int}\alpha}^{\rm near} = \frac{3k_{\rm B}T_{{\rm NA}\alpha}\bar{\mu}_0}{g(S+1)\mu_{\rm B}}.
\label{Eq:Bnear2}
\ee
The exchange field for $H=0$ seen by each moment in its ordering direction due to Heisenberg exchange interactions for either FM or AFM ordering can be written in the same form as\cite{Johnston2015}
\be
H_{\rm exch} = \frac{3k_{\rm B}T_{{\rm m}J}\bar{\mu}_0}{g(S+1)\mu_{\rm B}},
\label{Eq:HexchTN}
\ee
\ese
where $T_{{\rm m}J}$ is the contribution of Heisenberg exchange interactions to either a FM Curie temperature $T_{{\rm C}J}$ or an AFM N\'eel temperature $T_{{\rm N}J}$.  Using Eq.~(\ref{Eq:TNTNATNJ}), in the case of AFM ordering the sum of the two local fields in Eqs.~(\ref{Eqs:LocFieldSum}) can be written
\be
B_\alpha^{\rm local} = \frac{3k_{\rm B}(T_{{\rm N}J}+T_{\rm NA\alpha})\bar{\mu}_0}{g(S+1)\mu_{\rm B}} = \frac{3k_{\rm B}T_{\rm N}\bar{\mu}_0}{g(S+1)\mu_{\rm B}},
\label{Eq:BlocalGenAFM}
\ee
where $T_{\rm N}$ is the N\'eel temperature in the presence of both exchange and MDIs.

Because different sources of local fields are additive in their contributions to the observed $T_{\rm N}$ within MFT, if both exchange and dipolar interactions are present $T_{\rm NA}$ is the contribution of dipolar interactions to $T_{\rm N}$, which is usually but not always a small fraction of $T_{\rm N}$.

Quantum fluctuations generally increase as the dimensionality of a spin lattice decreases.  These quantum fluctuations can prevent long-range magnetic ordering from occurring.  Corruccini and White found from spin-wave calculations that AFM order cannot occur at finite temperature on the 2D square spin lattice due to dipolar interactions alone.\cite{Corruccini1993}  MFT does not take into account such quantum fluctuations associated with reduced dimensionality and hence predicts that AFM ordering can occur in 1D, 2D and 3D spin lattices.  

\subsection{\label{Sec:FMOrdering} Ferromagnetic Ordering (Curie) Temperature}

As is well-known, whether or not a particular sample exhibits FM ordering driven by the MDI  depends on the shape of the sample via the demagnetizing field as well as the competition with AFM states.  The former is evident from Eq.~(\ref{Eq:Bintalphai2}) which for $H_{\alpha}= 0$ becomes
\be
B_{{\rm int}\alpha i}^{\rm local} = \frac{g\mu_{\rm B}S}{a^3}\left[\lambda_{{\bf 0}\alpha} + \frac{4\pi}{V_{\rm spin}/a^3}\left(\frac{1}{3}-N_{\rm d\alpha}\right)\right]\bar{\mu},
\label{Eq:BlocFMAFM}
\ee 
where $a$ is the $a$-axis lattice parameter of the unit cell, $V_{\rm spin}$ is the volume per spin, $\lambda_{{\bf 0}\alpha}$ refers to the FM moment alignment, the magnetic moment per unit volume is $\mu/V_{\rm spin}=g\mu_{\rm B}S\bar{\mu}/V_{\rm spin}$, and we used Eqs.~(\ref{Eq:barmuDef}) and~(\ref{Eq:Bnear}).
Then following the same development as in the previous section gives the Curie temperature
\bea
T_{\rm CA\alpha} &=& \frac{g^2S(S+1)\mu_{\rm B}^2}{3k_{\rm B}a^3}\left[\lambda_{{\bf 0}\alpha} + \frac{4\pi}{V_{\rm spin}/a^3}\left(\frac{1}{3}-N_{\rm d\alpha}\right)\right]\nonumber\\*
&=& \frac{C_1}{a^3}\left[\lambda_{{\bf 0}\alpha} + \frac{4\pi}{V_{\rm spin}/a^3}\left(\frac{1}{3}-N_{\rm d\alpha}\right)\right] \ {\rm (FM)},\label{TCAa}
\eea
where $C_1$ was defined in Eq.~(\ref{Eq:C1Def}).  The system will choose the easy axis~$\alpha$ with the largest value of $\lambda_{{\bf 0}\alpha}$.  For a cubic Bravais lattice $\lambda_{{\bf 0}\alpha}=0$, so there is no preferred easy axis for FM ordering according to the present treatment.

Using Eq.~(\ref{TCAa}) one can write the local field in Eq.~(\ref{Eq:BlocFMAFM}) for FM moment alignments as
\be
B_{{\rm int}\alpha i}^{\rm local} = \frac{3k_{\rm B}T_{\rm CA\alpha}}{g\mu_{\rm B}(S+1)}\,\bar{\mu}.
\label{Eq:BlocFMAFM2}
\ee 
If Heisenberg exchange interactions are present, one adds the local exchange field in Eq.~(\ref{Eq:HexchTN}) to the dipolar contribution in Eq.~(\ref{Eq:BlocFMAFM2}) to obtain
\be
B_{{\rm int}\alpha i}^{\rm local} = \frac{3k_{\rm B}T_{\rm C\alpha}}{g\mu_{\rm B}(S+1)}\,\bar{\mu}.
\label{Eq:BlocFMAFM3}
\ee
where $T_{\rm C\alpha} = T_{\rm CA\alpha} + T_{{\rm C}J\alpha}$ according to Eq.~(\ref{Eq:TNTNATNJ}).

Comparing Eqs.~(\ref{Eq:BlocalGenAFM}) and~(\ref{Eq:BlocFMAFM3}) one sees that the same form of the local field in the direction of each ordered moment is obtained for both FM and AFM structures in the ordered states and one can therefore write the local magnetic induction seen by each moment in general for either FM or AFM moment alignments and dipolar and/or Heisenberg interactions as
\be
B_{{\rm int}\alpha i}^{\rm local} = \frac{3k_{\rm B}T_{\rm m\alpha}}{g\mu_{\rm B}(S+1)}\,\bar{\mu},
\label{Eq:BlocFMAFMGen}
\ee
where $T_{\rm m\alpha}$ is the Curie or N\'eel temperature for the collinear ordering axis $\alpha$.

\section{\label{cubicFMAFM} Competition Between Ferromagnetic and Antiferromagnetic Ordering}

One can have a crossover between FM and AFM ordering depending on the value of the  demagnetizing factor $N_{{\rm d}\alpha}$ and the possible AFM eigenvalues $\lambda_{{\bf k}\alpha}$ and FM eigenvalues $\lambda_{{\bf 0}\alpha}$.  The value of $N_{{\rm d}\alpha}$ depends on the shape of the sample.  For FM ordering, the field direction with the smallest value of $N_{{\rm d}\alpha}$ gives the lowest free energy and hence is the FM ordering direction provided that the calculated $T_{{\rm CA}\alpha}>0$ and that competing AFM states have a lower calculated $T_{{\rm NA}\alpha}>0$.

To examine this competition, we define the dimensionless reduced AFM and FM ordering temperatures obtained from Eqs.~(\ref{Eq:TNA2}) and~(\ref{TCAa}), respectively, as
\bse
\label{Eqs:TCARed}
\bea
\frac{T_{\rm NA\alpha}a^3}{C_1} &=& \lambda_{{\bf k}\alpha}\qquad ({\rm AFM}),\label{Eq:TNARed}\\*
\frac{T_{\rm CA\alpha}a^3}{C_1} &=& \lambda_{\bf0\alpha} +  \frac{4\pi}{V_{\rm spin}/a^3}\bigg(\frac{1}{3}-N_{{\rm d}\alpha}\bigg) \qquad ({\rm FM}).\nonumber\\*
\label{Eq:TCARed}
\eea
\ese

\begin{figure}
\includegraphics[width=3.3in]{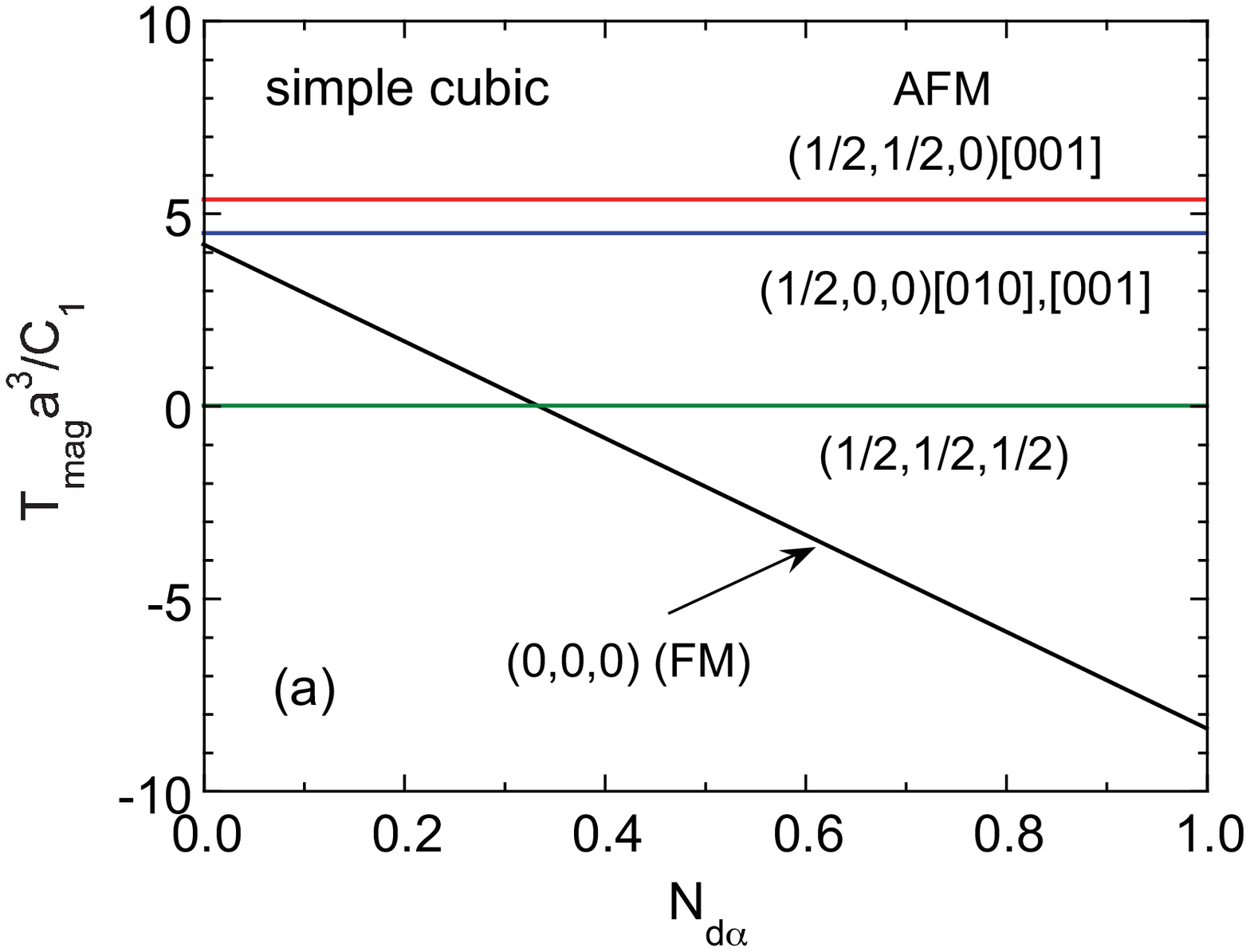}
\includegraphics[width=3.3in]{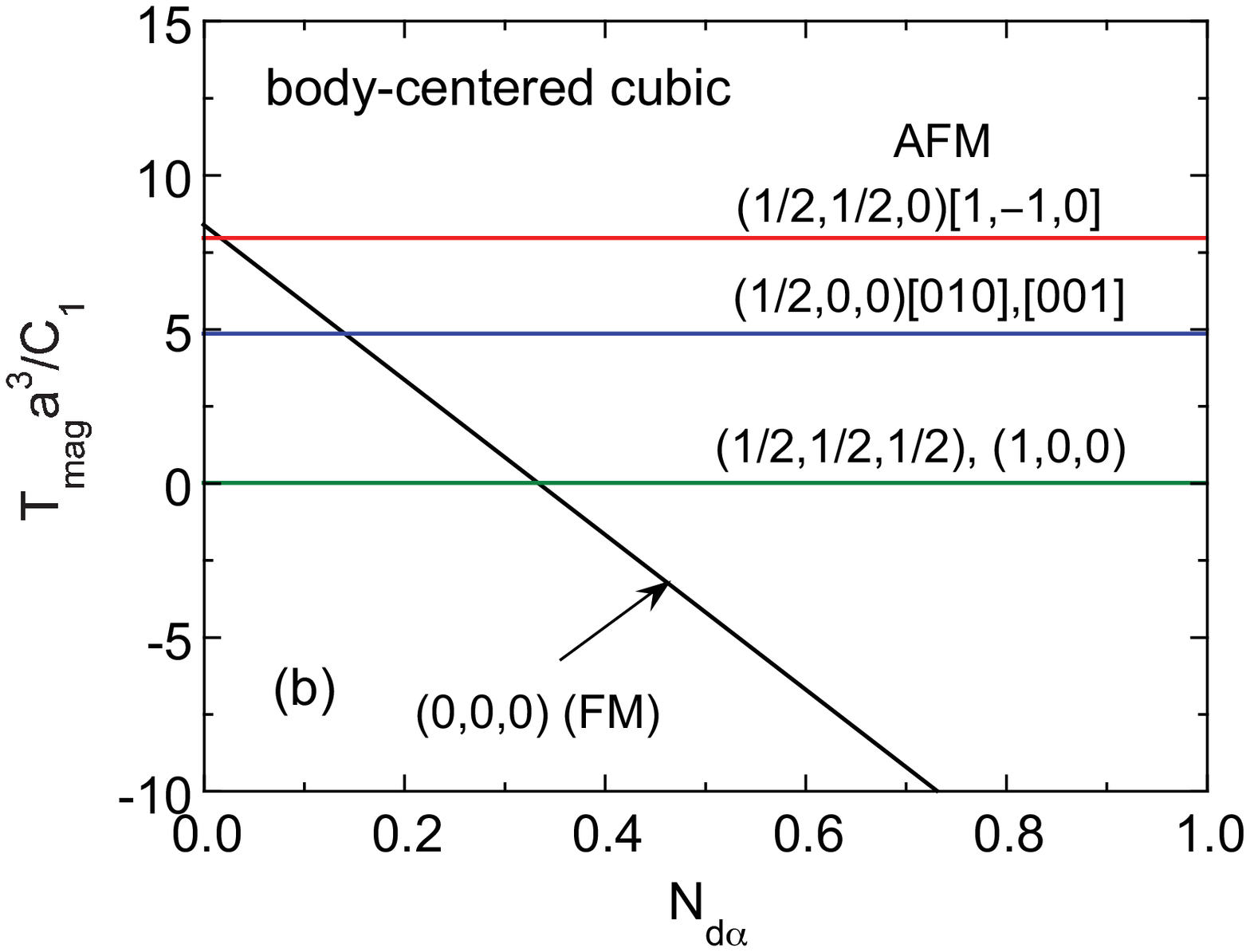}
\includegraphics[width=3.3in]{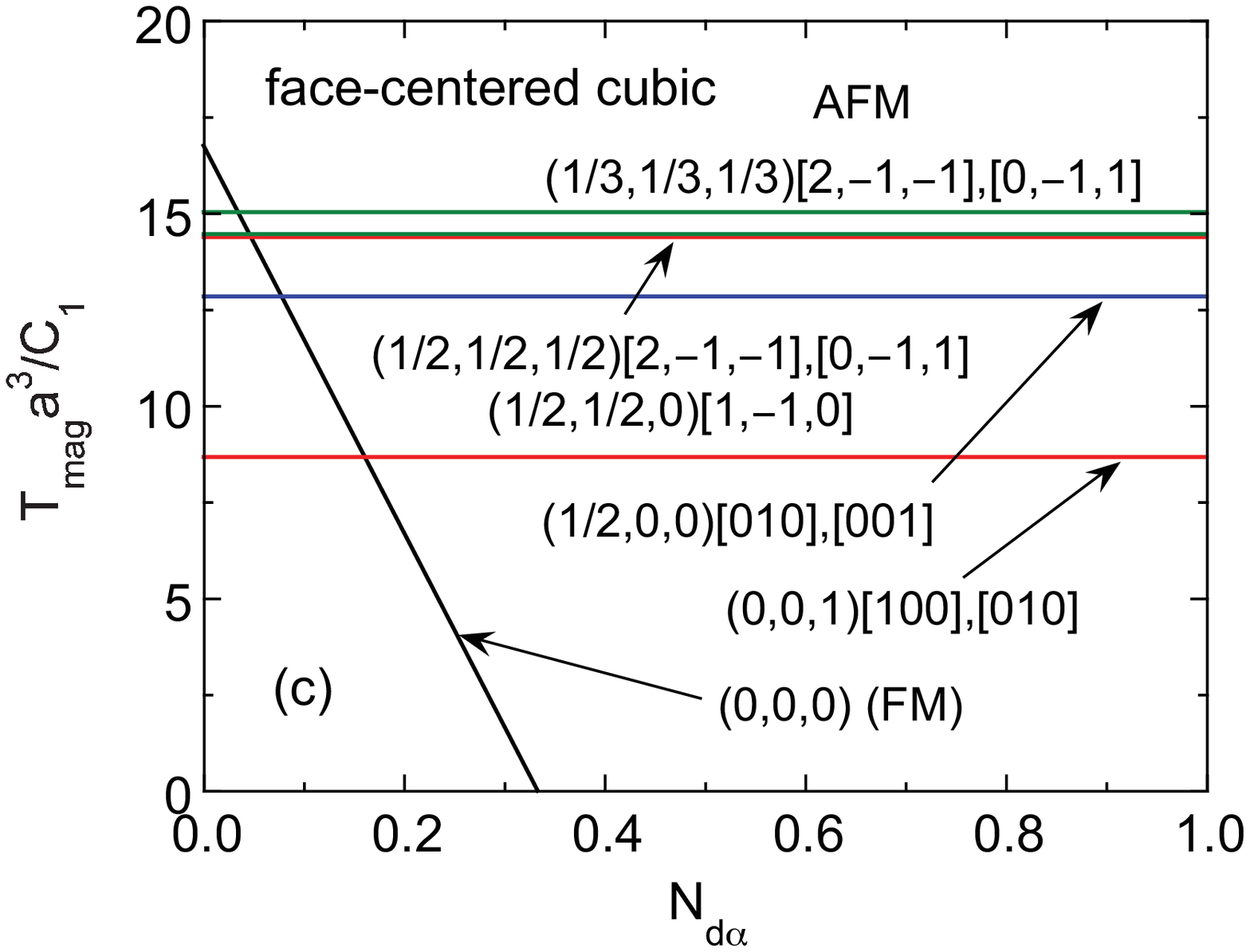}
\caption {(Color online) Reduced magnetic ordering temperature $T_{\rm mag}a^3/C_1$ versus the demagnetizing factor $N_{{\rm d}\alpha}$ with $0\leq N_{{\rm d}\alpha} \leq 1$ for pure magnetic dipolar ordering in (a) sc, (b) bcc and (c) fcc Bravais spin lattices as predicted for FM and AFM ordering by MFT via Eqs.~(\ref{Eqs:TCARed}). An ordering wavevector is labeled as $(m_1,m_2,m_3)$~r.l.u.\ and the ordered moment axis as $[\mu_x,\mu_y,\mu_z]$ in Cartesian coordinates. Values of $T_{\rm mag}<0$ are unphysical. For the fcc lattice, the most stable AFM wavevector shown is {\bf k} = (1/3,1/3,1/3)~r.l.u.}
\label{Fig:TmagVsNda_sc}
\end{figure}

As an example, we consider the competition between FM and AFM ordering due to dipolar interactions on sc, bcc and fcc Bravais lattices, which have $\lambda_{{\bf 0}\alpha} = 0$ and $V_{\rm spin}/a^3 = 1$, 1/2 and~1/4, respectively.  The reduced Curie temperature in Eq.~(\ref{Eq:TCARed}) is plotted versus $N_{\rm d\alpha}$ for sc, bcc and fcc Bravais spin lattices in Figs.~\ref{Fig:TmagVsNda_sc}(a), \ref{Fig:TmagVsNda_sc}(b) and \ref{Fig:TmagVsNda_sc}(c), respectively.  Using Eq.~(\ref{Eq:TNARed}) and the data in Tables~\ref{Tab:SCEvecsEvals}, \ref{Tab:BCCEvecsEvals} and~\ref{Tab:FCCEvecsEvals}, AFM $\lambda_{{\bf k}\alpha}$ values are plotted for the most stable (positive) $\lambda_{{\bf k}\alpha}$ value for each {\bf k} as horizontal lines for the sc, bcc and fcc lattices in Figs.~\ref{Fig:TmagVsNda_sc}(a), \ref{Fig:TmagVsNda_sc}(b) and \ref{Fig:TmagVsNda_sc}(c), respectively.  One sees from Fig.~\ref{Fig:TmagVsNda_sc} that for the magnetic structures considered, the ground state of the sc lattice is AFM-ordered with {\bf k} = $\left(\frac{1}{2},\frac{1}{2},0\right)$~r.l.u.\ and ordering axis $\hat{\mu}=[001]$ for all values of $N_{{\rm d}\alpha}$, the bcc lattice is unstable to FM ordering only for $N_{{\rm d}\alpha}\approx 0$ and the fcc lattice for $N_{{\rm d}\alpha}\lesssim0.03$.  These inferences are consistent with early results.\cite{Luttinger1946}  A sample with the shape of a long thin needle with the magnetization directed along the axis of the needle has a demagnetizing factor $N_{\rm d\alpha}\approx0$.

\section{\label{Eq:PropsBelowTm} Properties of the Magnetically-Ordered State}

\subsection{\label{Sec:mu0Cmag} Ordered Moment and Magnetic Heat Capacity}

For either an AFM or FM with Heisenberg and/or MDIs, Eq.~(\ref{Eq:BlocFMAFMGen}) gives the same form of the local magnetic induction seen by each spin in its ordering direction for $T\leq T_{\rm m}$.  Using Eq.~(\ref{Eq:BlocFMAFMGen}), the behavior of $\bar{\mu}$ versus $t$ is the same as for pure Heisenberg interactions and is shown for several values of the spin~$S$ in Fig.~10 of Ref.~\onlinecite{Johnston2011}.  

The magnetic energy per spin is given by 
\bse
\label{Eqs:EmagCmagCalc2}
\be
E_{{\rm mag}i} = -\frac{1}{2}\mu_i B_i^{\rm local},
\label{Eq:Ei2}
\ee
where the factor of 1/2 derives from the fact that $B_i^{\rm local}$ is attributed to the neighbors of $\mu_i$ whereas the energy is equally shared by pairs of interacting spins.  Inserting $B_i^{\rm local}$ from Eq.~(\ref{Eq:BlocFMAFMGen}) into~(\ref{Eq:Ei2}) for a mole of spins with $N=N_{\rm A}$ where $N_{\rm A}$ is Avogadro's number, one obtains
\be
E_{\rm mag} = -\frac{3RS}{2(S+1)}T_{\rm m}\bar{\mu}_0^2,
\ee
where $R=N_{\rm A}k_{\rm B}$ is the molar gas constant.  Then the magnetic heat capacity $C_{\rm mag}$ per mole of spins is obtained as\cite{Johnston2011,Johnston2015}
\be
\frac{C_{\rm mag}}{R} = -\frac{3S\bar{\mu}_0(t)}{S+1}\,\frac{d\bar{\mu}_0(t)}{dt},
\ee
\ese
where $t = T/T_{\rm mag}$ and the reduced ordered moment versus temperature $\bar{\mu}_0(t)$ in $H=0$ is obtained as described in Ref.~\onlinecite{Johnston2015}.  This equation is identical to that obtained for pure Heisenberg interactions, where plots of $C_{\rm mag}/R$ versus~$t$ for several values of $S$ are shown in Fig.~11 of Ref.~\onlinecite{Johnston2011}.  For quantum spins, $C_{\rm mag}$ decreases exponentially to zero for $t\to0$, whereas for classical spins $C_{\rm mag}/R\to1$ for $t\to0$.

\subsection{\label{Sec:AFMAnisotropy} Dipolar Anisotropy of Uniaxial Antiferromagnets in the Ordered State}

Here we calculate the dipolar anisotropy of the free energy between EM orthogonal principal collinear magnetic ordering axes denoted as the $\alpha$ and $\beta$ axes.  We consider collinear AFMs with noncubic spin lattices containing identical crystallographically-equivalent spins.  The lowest-order expression for the anisotropy free energy per spin $F_i$ is given by the usual expression
\be
F_i = K_1\sin^2\theta,
\label{FiDef}
\ee
where $\theta$ is defined as the angle between the ordered moment axis and the $\alpha$ axis.  We derive an expression for $K_1$ associated with the anisotropic MDI in terms of the eigenvalues and eigenvectors of the MDI tensor.

The orientation of a representative $T$-dependent ordered moment $\vec{\mu}_i$ in the $\alpha$-$\beta$ plane in $H=0$ with $\mu_0 = |\vec{\mu}_i|$ is
\be
\vec{\mu}_i = \mu_0(\cos\theta\,\hat{\alpha} + \sin\theta\,\hat{\beta}),
\label{Eq:VecMui}
\ee
where $\mu_0$ is the $T$-dependent ordered moment in $H=0$ and $\theta=0$ corresponds to $\vec{\mu}_i$ parallel to the $\alpha$~axis.  The corresponding $T$-dependent internal local field is
\be
{\bf B}_{{\rm int}\,i}^{\rm local} = B_{{\rm int}\,\alpha\,i}^{\rm local}\cos\theta\,\hat{\alpha} + B_{{\rm int}\,\beta\,i}^{\rm local}\sin\theta\,\hat{\beta}.
\label{Eq:VecBinti}
\ee
where the expression for ${\bf B}_{{\rm int}\,\alpha\,i}^{\rm local}$ is given in Eq.~(\ref{Eq:Bintalphai2}) with $H_{\alpha}=0$.  The differential $dF_i$ of the magnetic free energy of the moment is
\be
dF_i = -\frac{1}{2}\vec{\mu}_i\cdot d{\bf B}_{{\rm int}\,i}^{\rm local},
\label{Eq:dFi}
\ee
where the factor of 1/2 is present because ${\bf B}_{{\rm int}\,i}^{\rm local}$ arises from the neighboring moments of $\vec{\mu}_i$ whereas the free energy per moment is equally shared between each pair of moments.  Inserting Eqs.~(\ref{Eq:VecMui}) and~(\ref{Eq:VecBinti}) into~(\ref{Eq:dFi}) gives
\be
dF_i = \frac{\mu_0}{2}(B_{{\rm int}\,\alpha\,i}^{\rm local} - B_{{\rm int}\,\beta\,i}^{\rm local})\sin\theta\,\cos\theta\,d\theta.
\ee
Integrating $dF_i$ from $\theta=0$ to $\theta$ yields
\be
F_i = \frac{\mu_0}{4}(B_{{\rm int}\,\alpha\,i}^{\rm local} - B_{{\rm int}\,\beta\,i}^{\rm local})\sin^2\theta.
\label{Eq:Fi}
\ee
This expression for $F_i$ applies to moments along the collinear ordering axis with angles of either $\pm\theta$ to the $\alpha$~axis because the sine function is squared.  Comparing Eq.~(\ref{Eq:Fi}) with~(\ref{FiDef}) gives the anisotropy parameter~$K_1$ as
\be
K_1 = \frac{\mu_0}{4}(B_{{\rm int}\,\alpha\,i}^{\rm local} - B_{{\rm int}\,\beta\,i}^{\rm local}).
\label{Eq:K_1}
\ee

For an AFM in the ordered state, one has ${\bf B}_{{\rm int}\,\alpha\,i}^{\rm local} = {\bf B}_{{\rm int}\,\alpha\,i}^{\rm near}$.  Inserting ${\bf B}_{{\rm int}\,\alpha\,i}^{\rm near}$ in Eq.~(\ref{Eq:Bi}) into~(\ref{Eq:K_1}) gives
\be
K_1 = \frac{\mu_0^2}{4a^3}(\lambda_{{\bf k}\alpha} - \lambda_{{\bf k}\beta}).
\label{Eq:K_12}
\ee
From Eqs.~(\ref{FiDef}) and~(\ref{Eq:K_12}) one obtains
\be
F_i = \begin{cases}
0 & (\theta=0)\\
\frac{\mu_0^2(T)}{4a^3}(\lambda_{{\bf k}\alpha} - \lambda_{{\bf k}\beta}) & (\theta = \pi/2)
\end{cases}.
\label{Eq:FiCases}
\ee
Therefore if $\lambda_{{\bf k}\alpha} - \lambda_{{\bf k}\beta}>0$, the minimum free energy occurs if the moments are aligned along the $\alpha$~axis ($\theta=0$) and hence the easy axis is the $\alpha$~axis, whereas if $\lambda_{{\bf k}\alpha} - \lambda_{{\bf k}\beta}<0$, the $\beta$~axis ($\theta=\pi/2$) is favored for the ordering axis over the $\alpha$~axis.  These results are consistent with expectation because one expects a moment $\vec{\mu}_i$ to line up along the axis with the largest value of $B_{{\rm int}\,i}^{\rm near}$ in Eq.~(\ref{Eq:Bi}), i.e., with largest value of $\lambda_{\bf k}$.

\subsection{\label{Sec:ChiPerp} Perpendicular Magnetic Susceptibility of Collinear Antiferromagnets in the Ordered State}

The Heisenberg exchange Hamiltonian has no intrinsic magnetic anisotropy to determine the directions of the ordered moments in the ordered state with respect to the spin-lattice axes.  In this paper the only source of magnetic anisotropy is the MDI, and in this section we only consider collinear magnetic ordering.  The easy axis is the eigenvector of the interaction tensor $\widehat{{\bf G}}_{i}({\bf k})$ that corresponds to the largest eigenvalue for the given AFM propagation vector.

The single-spin magnetic susceptibility $\chi$ is rigorously defined as $\chi = \lim_{H\to0}\mu(H)/H$ where $\mu$ is the thermal-average moment of a spin in the direction of {\bf H} that is induced by {\bf H}\@.  Here we take the easy axis to be the $x$~axis and the applied infinitesimal field to be along a $z$~axis, perpendicular to the $x$~axis.  The magnitude of each ordered moment in zero field is $\mu_0$, which is $T$-dependent as shown in Ref.~\onlinecite{Johnston2011}.  In the presence of the perpendicular field, the magnitude of the moment does not change in the AFM phase\cite{Luttinger1946,Johnston2015} and the induced moment acquires a component along the $z$~axis.  Including the applied infinitesimal perpendicular field and both the exchange and dipolar fields and setting the net torque on a representative moment equal to zero following the procedure of Ref.~\onlinecite{Johnston2015} yields the perpendicular susceptibility

\be
\chi_\perp = \frac{C_1}{(T_{{\rm N}J} + T_{{\rm NA}x}-T_{{\rm CA}z})- \theta_{{\rm p}J} }.
\label{Eq:ChiPerpB}
\ee
The $T$-dependent ordered moment $\mu_0$ canceled out, so $\chi_\perp$ is independent of $T$ for $T\leq T_{\rm N}$, as also obtained for pure Heisenberg spin interactions.\cite{Johnston2015} 

Several special cases occur for Eq.~(\ref{Eq:ChiPerpB}).  If exchange interactions are negligible, the pure magnetic dipole prediction is obtained by setting $T_{{\rm N}J} = \theta_{{\rm p}J} = 0$, yielding
\bse
\label{Eqs:ChiPerp99}
\bea
\chi_\perp &=& \frac{C_1}{T_{{\rm NA}x}-T_{{\rm CA}z}  }\\*
&=& \frac{a^3}{\lambda_{{\bf k}x} - \lambda_{{\bf 0}z} - \frac{4\pi}{3V_{\rm spin}/a^3}}. \label{Eq:ChiPerpDipoles}
\eea
\ese
For cubic Bravais spin lattices for which $\lambda_{\bf 0\alpha}=0$ for all~$\alpha$, Eq.~(\ref{Eq:ChiPerpDipoles}) gives
\be
\chi_\perp = \frac{a^3}{\lambda_{{\bf k}x} - \frac{4\pi}{3V_{\rm spin}/a^3}}.
\label{Eq:ChiPerpSC}
\ee
This result agrees, e.g., with $\chi_\perp$ obtained from the equation between Eqs.~(29) and~(30) in Ref.~\onlinecite{Luttinger1946} which includes in the denominator of Eq.~(\ref{Eq:ChiPerpSC}) the ground state eigenvalue $\lambda_{{\bf k}x}= \lambda_{(1/2,1/2,0)[001]} = 5.351$ ($f_5$ in their notation) for the sc dipolar AFM, in good agreement with our value of 5.3535 in Table~\ref{Tab:SCEvecsEvals}.

When dipolar interactions are negligible, Eq.~(\ref{Eq:ChiPerpB}) gives for the pure Heisenberg exchange model
\bse
\be
\chi_\perp = \frac{C_1}{T_{{\rm N}J} - \theta_{{\rm p}J} }\qquad (T\leq T_{{\rm N}J}),
\label{Eq:ChiPerpC}
\ee
in agreement with Ref.~\onlinecite{Johnston2015}.  In the PM state at $T\geq T_{{\rm N}J}$, the isotropic susceptibility per spin is given by the Curie-Weiss law\cite{Johnston2015}
\be
\chi = \frac{C_1}{T - \theta_{{\rm p}J} }\qquad (T\geq T_{{\rm N}J}).
\label{Eq:chiHeisCW}
\ee
Comparing Eqs.~(\ref{Eq:ChiPerpC}) and~(\ref{Eq:chiHeisCW}) gives
\be
\chi_\perp = \chi(T_{{\rm N}J}) \qquad (T\leq T_{{\rm N}J}).
\label{Eq:ChiPerpD}
\ee
\ese

\subsection{\label{Sec:Hc} Perpendicular Critical Field}

As the perpendicular field is increased from zero at $T<T_{\rm N}$, the induced perpendicular moment $\mu_\perp$ increases as
\be
\mu_\perp = \chi_\perp H,
\ee
where $\chi_\perp$ is given by Eq.~(\ref{Eq:ChiPerpB}).  When $\mu_\perp$ reaches the ordered moment $\mu_0(T)$, the induced moments become parallel to {\bf H} and the system enters the PM state in a second-order transition.\cite{Luttinger1946,Johnston2015}  Setting $\mu_\perp=\mu_0$ with increasing $H$, the critical field $H_{\rm c}$ at which this happens is defined by $\mu_0=\chi_\perp H_{\rm c}$, yielding
\be
H_{\rm c}(T) = \frac{\mu_0(T)}{\chi_\perp}.
\ee  
Thus one obtains
\be
\frac{H_{\rm c}(T)}{H_{\rm c}(0)} = \frac{\mu_0(T)}{\mu_0(0)} = \frac{\mu_0(T)}{\mu_{\rm sat}} = \bar{\mu}_0(T),
\ee
where $\bar{\mu}_0$ is plotted versus $t\equiv T/T_{{\rm N}}$ in Ref.~\onlinecite{Johnston2011}.  Since within MFT $\mu_0(T)$ depends on the spin~$S$ of the moment, so does $\frac{H_{\rm c}(T)}{H_{\rm c}(0)}$. Near $t=1$, one obtains
\be
\frac{H_{\rm c}(T)}{H_{\rm c}(0)} \propto \sqrt{1-t}\qquad (t\to1^-).
\label{Eq:HcRed}
\ee

Previous classical calculations (not utilizing the Weiss MFT and hence not the Brillouin function for quantum spins) yielded the behavior in Eq.~(\ref{Eq:HcRed}) for the whole temperature range $0\leq t\leq 1$, with the proportionality replaced by an equality.\cite{Lax1952}  In that case, expanding the right-hand side of Eq.~(\ref{Eq:HcRed}) in a Taylor series about $t=0$ gives the linear dependence $\frac{H_{\rm c}(T)}{H_{\rm c}(0)} = 1-\frac{t}{2}$ $(t\ll1)$ instead of the exponential approach to unity for $t\to0$ obtained for quantum spins.

\section{\label{Sec:CWLaw} Curie-Weiss Law in Paramagnetic State}

In the PM state above the N\'eel or Curie temperature, all moments are aligned in the direction $\alpha$ of the magnetic field $H_{\alpha}$ applied along a principal axis of the spin lattice [the magnetic propagation vector is ${\bf k} = (0,0,0)\equiv {\bf 0}$].  For Heisenberg exchange  interactions, the exchange field in the PM state is isotropic and given by\cite{Johnston2012,Johnston2015}
\be
H_{\rm exch} = \frac{3k_{\rm B}\theta_{{\rm p}J}}{g\mu_{\rm B}(S+1)}\bar{\mu},
\ee
where $\bar{\mu}$ is the normalized moment induced by $H_{\alpha}$ and
\be
\theta_{{\rm p}J} = -\frac{S(S+1)}{3k_{\rm B}}\sum_jJ_{ij}
\label{Eq:ThetapJ}
\ee
is the contribution to the Weiss temperature in the Curie-Weiss law due to Heisenberg exchange interactions.  Then adding $H_{\rm exch}$ and $H_{\alpha}$ to the local dipolar field for $H_{\alpha}=0$ in Eq.~(\ref{Eq:BlocFMAFM}) gives the total local field seen by each moment as
\bea
B_{{\rm int}\alpha i}^{\rm local} &=& H_{\alpha} + \frac{3k_{\rm B}\theta_{{\rm p}J}}{g\mu_{\rm B}(S+1)}\bar{\mu}\label{Eq:BlocPM}\\*
&& +\ \frac{g\mu_{\rm B}S}{a^3}\left(\lambda_{{\bf 0}\alpha} + \frac{4\pi}{3V_{\rm spin}/a^3}\right)\bar{\mu},\nonumber
\eea
where we assume that the demagnetizing field has been corrected for in experimental data and hence the demagnetizing factor $N_{\rm d\alpha}$ does not appear in this expression.  To include it, replace the multiplicative factor $\frac{1}{3}$ in the last term by $\frac{1}{3}-N_{\rm d\alpha}$.

Analogous to Eq.~(\ref{Eq:TNACalc}) for $H_{\alpha}=0$, in the present case one has
\be
\bar{\mu} = B_S\left(\frac{g\mu_{\rm B}B_{{\rm int}\,\alpha\,i}^{\rm local}}{k_{\rm B}T}\right).
\label{Eq:muPara}
\ee
Inserting $B_{{\rm int}\,\alpha\,i}^{\rm local}$ from Eq.~(\ref{Eq:BlocPM}) into~(\ref{Eq:muPara}), Taylor expanding the Brillouin function $B_S(y)$ to first order in~$y$, solving for $\bar{\mu}$ and using Eq.~(\ref{Eq:barmuDef}) gives the Curie-Weiss law
\bse
\label{Eq:chiAlpha}
\bea
\chi_\alpha &=&  \frac{C_1}{T - \theta_{\rm p\alpha}},\label{Eq:CWLawGen}\\*
\theta_{\rm p\alpha} &=& \theta_{{\rm p}J} + \theta_{{\rm pA}\alpha},
\eea
where the single-spin Curie constant $C_1$ is given in Eq.~(\ref{Eq:C1Def}), $\theta_{{\rm p}J}$ is given in Eq.~(\ref{Eq:ThetapJ}) and the magnetic dipole contribution $\theta_{{\rm pA}\alpha}$ to the Weiss temperature is
\be
\theta_{{\rm pA}\alpha} = \frac{C_1}{a^3}\left(\lambda_{{\bf 0}\alpha} + \frac{4\pi}{3V_{\rm spin}/a^3}\right).
\label{Eq:thetapPM}
\ee

A comparison of Eq.~(\ref{Eq:thetapPM}) with~(\ref{TCAa}) shows that the contributions of dipolar interactions to the Weiss temperature and the Curie temperature of a FM are the same, i.e.,
\be
\theta_{{\rm pA}\alpha} = T_{{\rm CA}\alpha},
\ee
\ese
which is the same result as obtained from MFT for a system of local moments exhibiting a FM transition and interacting by Heisenberg exchange only.\cite{Johnston2015}

On the other hand, a comparison of Eqs.~(\ref{Eq:TNA2}) and ~(\ref{Eq:thetapPM}) shows that in general the contribution of dipolar interactions to the Weiss temperature for AFMs is not equal to the negative of the dipolar N\'eel temperature in Eq.~(\ref{Eq:TNA2}), as is also found in general for local-moment Heisenberg AFMs.\cite{Johnston2015}  Thus the ratio $f = \theta_{\rm p}/T_{\rm C}$ for a FM within MFT is
\bse
\be
f=1 \qquad ({\rm FM}),
\ee
whereas in general for an AFM it is
\be
f_\alpha = \frac{\theta_{\rm p\alpha}}{T_{\rm N\alpha}} = \frac{\theta_{{\rm pA}\alpha} + \theta_{{\rm p}J}}{T_{{\rm NA}\alpha} + T_{{\rm N}J}} < 1\qquad ({\rm AFM}).
\ee
\ese

\section{\label{Sec:AnisChiAFM} Anisotropic Magnetic Susceptibility of a Spherical Sample of a Pure Dipolar Antiferromagnet}

In the following, we assume that the sample is in the shape of a sphere, which cancels the Lorentz field within the Lorentz cavity according to Eq.~(\ref{Eq:Bintalphai2}) and hence ameliorates the competition of FM with AFM ordering.

\subsection{\label{Sec:PMSphere} Paramagnetic State}

For a dipolar collinear AFM at $T>T_{{\rm NA}x}$ where the easy axis is defined as the $x$~axis, the Curie-Weiss law in Eq.~(\ref{Eq:CWLawGen}) becomes
\be
\chi_\alpha = \frac{C_1}{T - \theta_{\rm pA\alpha}} \qquad (T > T_{{\rm NA}x}),
\label{Eq:CWLawDipole}
\ee
where $\theta_{\rm pA\alpha}$ is given by setting the second term in Eq.~(\ref{Eq:thetapPM}) to zero for a spherical sample, yielding
\be
\theta_{{\rm pA}\alpha} = \frac{C_1\lambda_{{\bf 0}\alpha}}{a^3}.
\label{Eq:thetapPMSphere}
\ee
This would be zero for a cubic Bravais spin lattice because in that case $\lambda_{{\bf 0}\alpha}=0$ for all~$\alpha$.  The N\'eel temperature in Eq.~(\ref{Eq:TNA2}) for the easy $x$~axis is
\be
T_{{\rm NA}x} = \frac{C_1\lambda_{{\bf k}x}}{a^3}
\label{Eq:TNAx}
\ee
and we define the ratio $f_{{\rm A}\alpha}$ as
\be
f_{{\rm A}\alpha} = \frac{\theta_{{\rm pA}\alpha}}{T_{{\rm NA}x}} = \frac{\lambda_{{\bf 0}\alpha}}{\lambda_{{\bf k}x}},
\label{Eq:FADef}
\ee
where the subscript~A in $f_{{\rm A}\alpha}$ signifies that the value of $f$ arises only from the anisotropic MDI and $\alpha$ can be any of the principal axes $x$, $y$ or~$z$.

Using Eqs.~(\ref{Eq:TNAx}) and~(\ref{Eq:FADef}), the Curie-Weiss law~(\ref{Eq:CWLawDipole}) for a single spin can be written in dimensionless form as
\bse
\label{Eqs:CW}
\be
\frac{\chi_\alpha \,T_{{\rm NA}x}}{C_1} = \frac{1}{t_{{\rm A}} -f_{{\rm A}\alpha}} \qquad (t_{\rm A} > 1),
\label{Eq:ChiRedPM}
\ee
where the reduced temperature $t_{{\rm A}}$ is defined as
\be
t_{{\rm A}} = \frac{T}{T_{{\rm NA}x}}.
\label{Eq:tADef}
\ee
\ese
Note that Eq.~(\ref{Eq:ChiRedPM}) is a law of corresponding states for all quantum spins~$S$, since $S$ only appears in $C_1$.

The reduced PM susceptibility at $T_{{\rm NA}x}$ from the Curie-Weiss law~(\ref{Eq:ChiRedPM}) is then
\be
\frac{\chi_\alpha(t_{\rm A} = 1^+) \,T_{{\rm NA}x}}{C_1}= \frac{1}{1 -f_{\rm A\alpha}} \qquad (t_{\rm A}= 1^+).
\label{Eq:ChiRedTNA}
\ee
From Eqs.~(\ref{Eq:ChiRedPM}) and~(\ref{Eq:ChiRedTNA}) one obtains
\be
\frac{\chi_\alpha(t_{\rm A})}{\chi_\alpha(t_{\rm A}=1^+)} = \frac{{1 -f_{\rm A\alpha}}}{t_{\rm A} -f_{\rm A\alpha}}  \qquad (t_{\rm A} > 1),
\label{Eq:ChiNormTgtrTN}
\ee
which yields the identity
\be
\frac{\chi_\alpha(t_{\rm A}= 1^+)}{\chi_\alpha(t_{\rm A}=1^+)} = 1,
\label{Eq:ChiNormTtoTN+}
\ee
as required.

\subsection{Perpendicular Susceptibility in the AFM-Ordered State}

In the AFM state at $T<T_{\rm NA}$ of a strictly dipolar AFM, one sets $T_{{\rm N}J}= \theta_{{\rm p}J} = 0$ and for spherical samples Eqs.~(\ref{Eqs:ChiPerp99}) yield
\bse
\bea
\chi_\perp(T\leq T_{{\rm NA}x}) &=& \frac{a^3}{\lambda_{{\bf k}x} - \lambda_{{\bf 0}z} }\\*
 &=& \frac{C_1}{T_{{\rm NA}x}-T_{{\rm CA}z}},
\label{Eq:ChiPerp}
\eea
\ese
where, as above, the $x$~axis is the easy axis for the collinear AFM ordering, $T_{{\rm NA}x}$ is the associated N\'eel temperature and the $z$~axis is perpendicular to the $x$~axis, i.e., $\chi_\perp=\chi_z$.  One can write Eq.~(\ref{Eq:ChiPerp}) in dimensionless form as
\bse
\label{Eqs:ChiPerp}
\be
\frac{\chi_\perp T_{{\rm NA}x}}{C_1} = \frac{1}{1-r_z}  \qquad (t_{\rm A} < 1,\ z\perp x),
\label{Eq:ChiPerpRed}
\ee
where according to Eq.~(\ref{Eq:TNA2}) and Eq.~(\ref{TCAa}) modified for a spherical sample one has
\be
r_z = \frac{T_{{\rm CA}z}}{T_{{\rm NA}x}} = \frac{\lambda_{{\bf 0}z}}{\lambda_{{\bf k}x}}.
\label{Eq:rDef}
\ee
\ese 

Using Eqs.~(\ref{Eq:ChiRedTNA}) and~(\ref{Eq:ChiPerpRed}) one obtains
\be
\frac{\chi_\perp(t_{\rm A}<1)}{\chi_\alpha(t_{\rm A}=1^+)} = \frac{1-f_{\rm A\alpha}}{1-r_z} .
\label{Eq:ChiNormTlessTN}
\ee
Comparing Eqs.~(\ref{Eq:ChiNormTtoTN+}) and~(\ref{Eq:ChiNormTlessTN}), one sees that in general the hard-axis $\chi_z$ is continuous on cooling below $T_{{\rm NA}x}$, where $\chi_\perp=\chi_z$ below $T_{{\rm NA}x}$.  If $\lambda_{0\alpha} = 0$ for all~$\alpha$ as in cubic Bravais lattices, $\chi_\perp$ is obtained for all axes below $T_{{\rm NA}x}$. 

\subsection{Parallel Susceptibility in the AFM-Ordered State}

When an infinitesimal field ${\bf H} = H\hat{\bf i}$ is applied in the positive $x$~direction along the collinear AFM ordering easy axis at a temperature $0<T<T_{{\rm NA}x}$,  an ordered moment initially pointing parallel (antiparallel) to {\bf H} increases (decreases) slightly in magnitude, where the vectorial change $d\vec{\mu}=d\mu\hat{\bf i}$ is the same for both moments.  Therefore in this section we only consider the change in the $x$-axis component of a representative moment $\vec{\mu}_i$ pointing towards the positive $x$~axis due to the applied field.

Following Ref.~\onlinecite{Johnston2015} we obtain the dimensionless equation
\bse
\label{Eqs:ChiPar}
\be
\frac{\chi_\parallel T_{{\rm NA}x}}{C_1} = \frac{1}{\tau^\ast(t_{\rm A}) - f_{{\rm A}x}},
\label{Eq:ChiParRed}
\ee
where $t_{\rm A}$ is defined in Eq.~(\ref{Eq:tADef}) and
\be
\tau^\ast(t_{\rm A}) = \frac{(S+1)t_{\rm A}}{3B_S^\prime(y_0)},\qquad f_{{\rm A}x} = \frac{\theta_{{\rm pA}x}}{T_{{\rm NA}x}} = \frac{\lambda_{{\bf 0}x}}{\lambda_{{\bf k}x}}.
\ee
\ese
$\mu_0(T)$ is obtained by numerically solving
\bse
\be
\mu_0 = g\mu_{\rm B}SB_S(y_0),
\ee
where
\be
y_0 = \frac{g\mu_{\rm B}}{k_{\rm B}T}\frac{\mu_0\lambda_{{\bf k}x}}{a^3}.
\ee
\ese
Here $B_S(y)$ is the Brillouin function in Eq.~(\ref{Eq:BrillouinFunction22}) and $B_S^\prime(y_0)\equiv [dB_S(y)/dy]|_{y=y_0}$.   Note that the parallel susceptibility in the dimensionless form in Eq.~(\ref{Eq:ChiParRed}) still depends on~$S$ since the Brillouin function on the right-hand side does.  This contrasts with the dimensionless forms of the Curie-Weiss and perpendicular susceptibilities above for dipolar interactions that do not depend on~$S$\@.

Useful limits are 
\be
\tau^\ast(t_{\rm A}\to0) = \infty,\qquad \tau^\ast(t_{\rm A}\to1) = 1,
\ee
yielding
\bse
\bea
\frac{\chi_\parallel T_{{\rm NA}x}}{C_1} &=& 0\qquad (t_{\rm A}\to0), \\*
\frac{\chi_\parallel T_{{\rm NA}x}}{C_1} &=& \frac{1}{1 - f_{{\rm A}x}}\qquad (t_{\rm A}\to1^-).
\eea
\ese
The latter $\chi_\parallel$ expression is identical with
\be
\frac{\chi_x\,T_{{\rm NA}x}}{C_1}= \frac{1}{1 -f_{{\rm A}x}} \qquad (t_{\rm A}= 1^+)
\ee
obtained from Eq.~(\ref{Eq:ChiRedTNA}) for the Curie-Weiss law at $t_{\rm A}= 1^+$ for the field applied along the $x$~axis.  Thus $\chi_\parallel = \chi_x$ for $t_{\rm A}<1$ joins continuously with $\chi_x$ for $t_{\rm A}>1$.

\subsection{\label{Sec:AFMChiExample} Example}

\begin{table*}
\caption{\label{Tab:AFMExample} Eigenvalues $\lambda$ and eigenvectors $\hat{\mu} = [\mu_a,\mu_b,\mu_c]$  of the dipolar interaction tensor for simple-tetragonal spin lattices with $c/a = 0.8$ and~1.2 and wavevectors $(m_1,m_2,m_3)$~r.l.u. The data were taken from tables in the Supplemental Material.\cite{SupplInfo}  The largest eigenvalue for ${\bf k} = 0$ is labeled as $\lambda_{\bf 0\alpha}$.  For ${\bf k} = \left(\frac{1}{2},\frac{1}{2},0\right)$~r.l.u.\ the maximum eigenvector is denoted as $\lambda_{{\bf k}x}$ and the value for the perpendicular direction as $\lambda_{{\bf k}z}$.  For each~{\bf k}, the values of $f_{\rm A}$ and~$r_z$ are listed as defined in Eqs.~(\ref{Eq:FADef}) and~(\ref{Eq:rDef}), respectively.  According to Eqs.~(\ref{Eq:ChiRedPM}) and~(\ref{Eq:ChiPerpRed}), the parameter $f_{\rm A\alpha}$ is relevant for the PM $T$ range and $r_z$ is relevant for the AFM-ordered $T$ range.  In the table, the assignments of the $x$ and~$z$ Cartesian axes to the $c$ and $a$ crystal axes, respectively, are shown in the subscripts to the parameters.}
\begin{ruledtabular}
\begin{tabular}{cccccc}
$c/a$ & {\bf k} (r.l.u.)  					& [100]		& [001] 		& $f_{\rm A}$ & $r_z$\\
\hline
0.8 & (0,0,0)  						&  	$\lambda_{{\bf 0}z,a}=-1.9691$	&	$\lambda_{{\bf 0}x,c}=3.9382$ & $f_{{\rm A}x,c} = 0.4104$, $f_{{\rm A}z,a} = -0.2052$	 		 \\
 & $\left(\frac{1}{2},\frac{1}{2},0\right)$  &  $\lambda_{{\bf k}z,a}= -4.7977$  	&  $\lambda_{{\bf k}x,c}= 9.5955$  & & $r_{z,a} = -0.2052$     \\
1.2 & (0,0,0)  						&    $\lambda_{{\bf 0}z,a}=0.9364$	& $\lambda_{{\bf 0}x,c}=-1.8728$	& $f_{{\rm A}x,c} = -0.5010$, $f_{{\rm A}z,a} = 0.2505$	 \\
 & $\left(\frac{1}{2},\frac{1}{2},0\right)$  &  	$\lambda_{{\bf k}z,a}= -1.8691$	&  $\lambda_{{\bf k}x,c}= 3.7381$   &&$r_{z,a} = 0.2505$   \\
\end{tabular}
\end{ruledtabular}
\end{table*}

\begin{figure}
\includegraphics[width=3.3in]{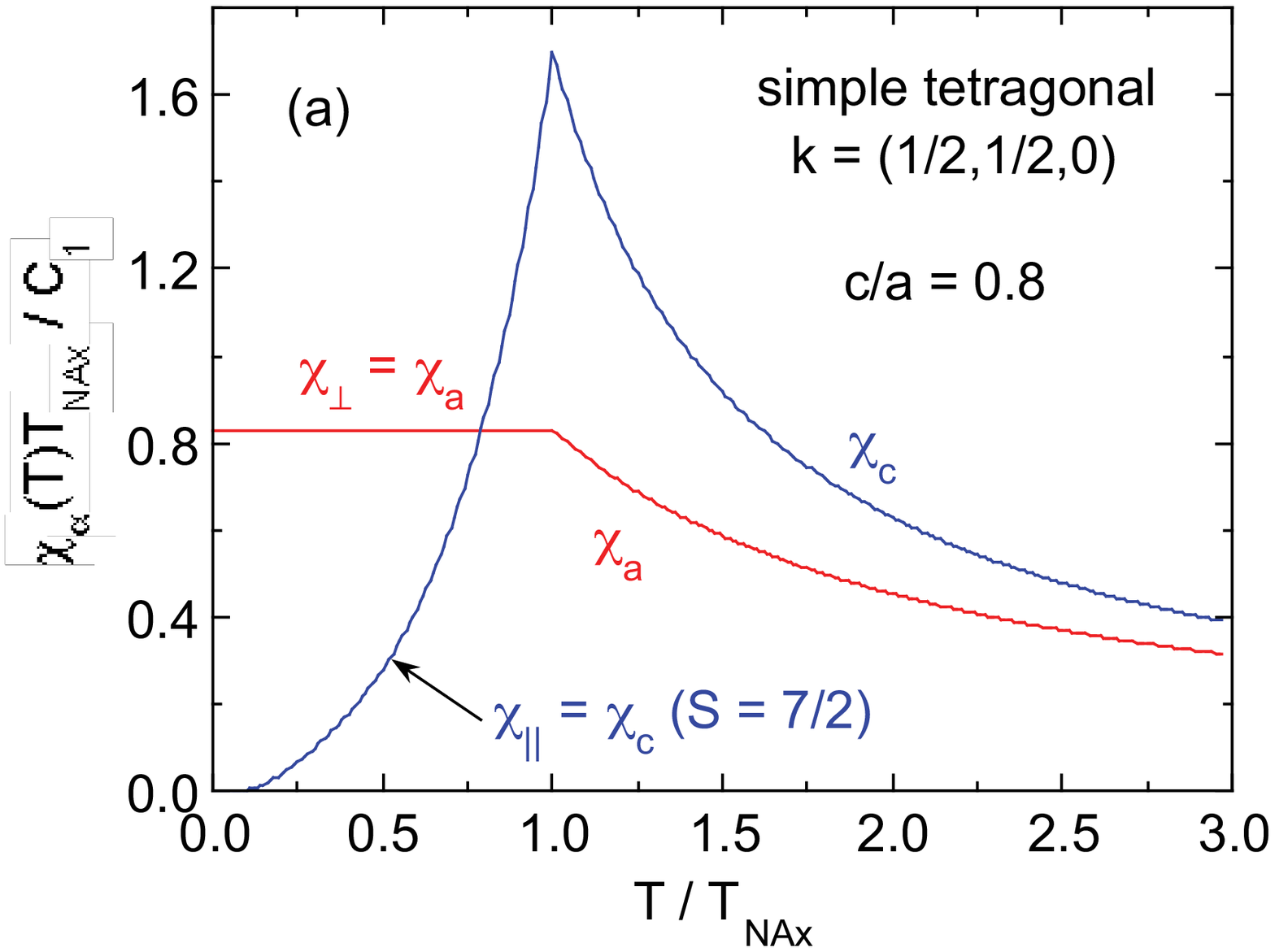}
\includegraphics[width=3.3in]{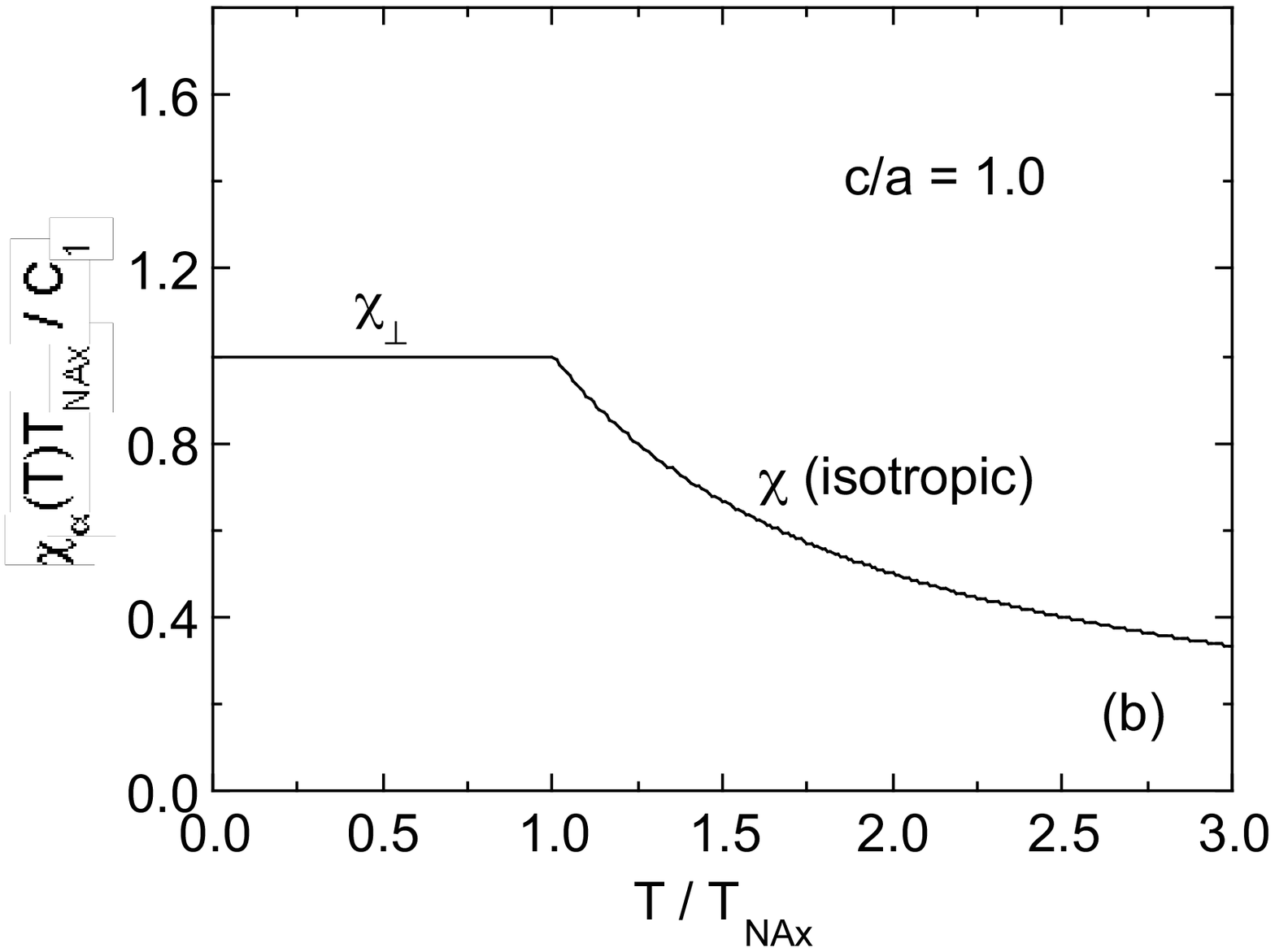}
\includegraphics[width=3.3in]{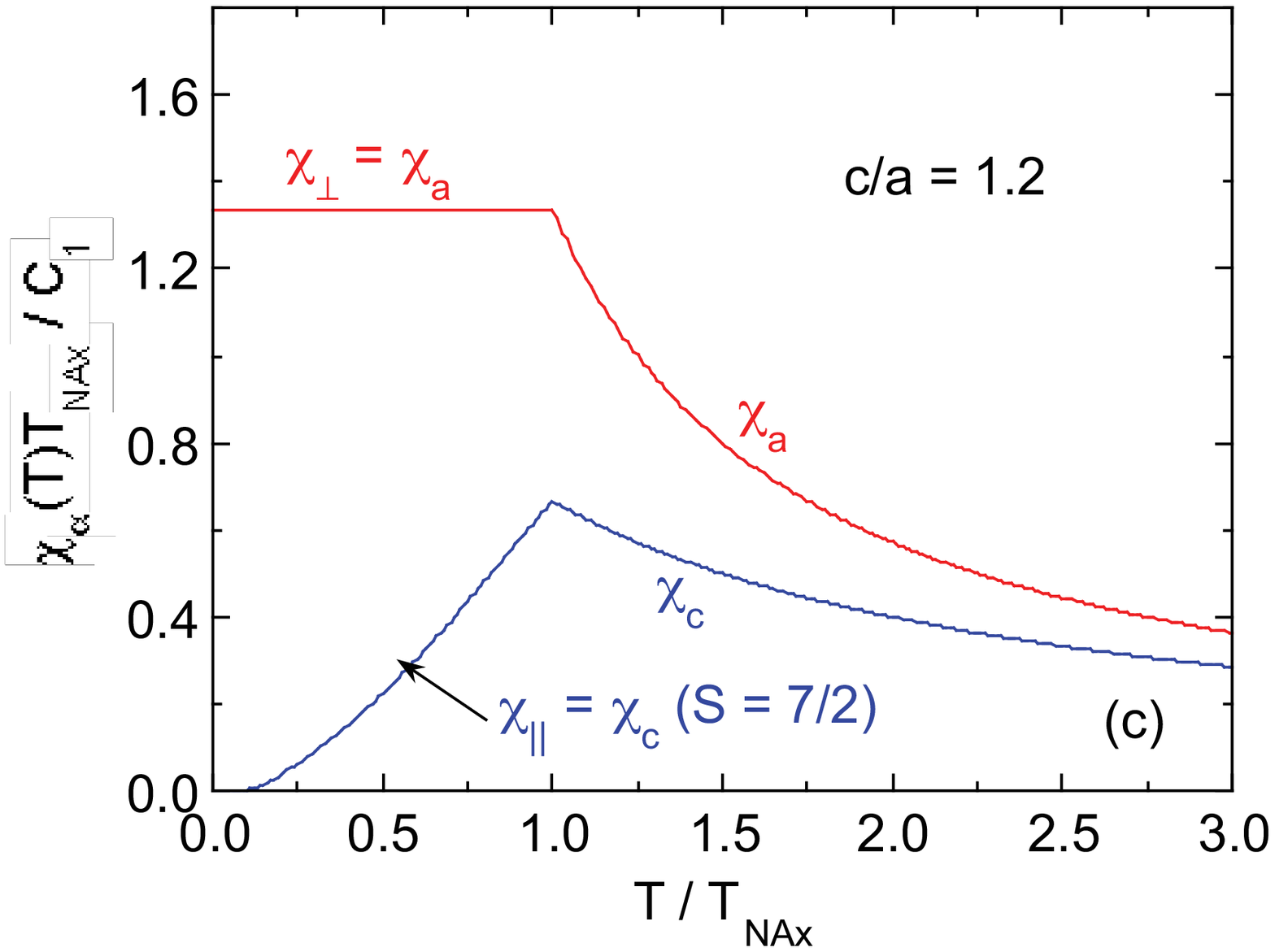}
\caption {(Color online) Anisotropy of the magnetic susceptibilities $\chi_a$ and $\chi_c$ due to MDIs versus reduced temperature $t_{\rm A} = T/T_{{\rm NA}x}$ for a simple tetragonal spin lattice with (a)~$c/a=0.8$, (b) 1.0 (sc lattice) and~(c)~$c/a=1.2$.  The AFM propagation vector in the ordered AFM state at $t_{\rm A}<1$ is ${\bf k} = \left(\frac{1}{2},\frac{1}{2},0\right)$~r.l.u.\ and the easy axis is the $c$~axis [001] for both $c/a=0.8$ and~1.2.  The data were plotted using Eqs.~(\ref{Eqs:CW}) (Curie-Weiss law) for $t_{\rm A}\geq 1$, and (\ref{Eqs:ChiPerp}) ($\chi_\perp$) and~(\ref{Eqs:ChiPar}) ($\chi_\parallel$) for $t_{\rm A}\leq 1$.}
\label{Fig:ChiExample}
\end{figure}

As an example, we consider the simple tetragonal Bravais spin lattice with $c/a = 0.8$, 1.0 and~1.2 and AFM propagation vector ${\bf k} = \left(\frac{1}{2},\frac{1}{2},0\right)$~r.l.u.\ for temperatures both above and below the N\'eel temperature. 
Recall that for $f_{{\rm A}\alpha}$, the $x$ and $\alpha$ axes are the easy principal axis for AFM ordering and any of the three principal axes, respectively, whereas for $r_z$, the $z$~axis must be an axis perpendicular to the $x$~axis.  In a real material, one must identify $x$, $z$ and~$\alpha$ with the appropriate crystal axes.

The eigenvalues and eigenvectors of the dipolar interaction tensor taken from tables in the Supplemental Material\cite{SupplInfo} are shown in Table~\ref{Tab:AFMExample} along with the respective values of $f_{\rm A\alpha}$ and~$r_z$ defined in Eqs.~(\ref{Eq:FADef}) and~(\ref{Eq:rDef}).  One sees that the AFM state is stable against the FM state below $T_{{\rm NA}x}$ for both $c/a$ values, but that the anisotropy in the PM state at $T > T_{\rm NA}$ changes sign between the two $c/a$ values.

Using the data in Table~\ref{Tab:AFMExample}, Eq.~(\ref{Eq:ChiNormTlessTN}) yields $\frac{\chi_\perp(t_{\rm A})}{\chi_a(t_{\rm A}=1^+)} = 1$ for the easy $a$ axis for both $c/a=0.8$ and~1.2.  For the sc lattice with $c/a=1$, one has $\lambda_{{\bf 0}\alpha}=f_{\rm A}=r_\alpha = \theta_{{\rm pa}\alpha} = 0$ for all $\alpha$.  Therefore $\chi(T)$ follows a Curie law for $t_{\rm A}\geq1$.  Also, there is no restoring force for keeping the easy axis parallel to the field, so the magnetization flops to the perpendicular orientation whenever this is attempted.  Thus only $\chi_\perp(T) = \chi(T_{{\rm NA}x})$ is measured for $t_{\rm A}\leq 1$.

Shown in Fig.~\ref{Fig:ChiExample} are plots of $\frac{\chi(t_{\rm A})T_{{\rm NA}x}}{C_1}$ versus $t_{\rm A}$ for $c/a = 0.8$, 1.0 and~1.2 illustrating the progression of the anisotropy in $\chi$ as $c/a$ traverses the sc of unity.  To our knowledge no theoretically-predicted behaviors such as in Figs.~\ref{Fig:ChiExample}(a) and~\ref{Fig:ChiExample}(c) have appeared before in the literature.

\section{\label{Sec:AnisHeisAFM} Anisotropy of Magnetic Susceptibility of a Heisenberg Paramagnet due to Magnetic Dipole Interactions}

In this section we assume that demagnetizing fields have been corrected for in experimental data and hence the demagnetizing factor $N_{\rm d\alpha}$ does not appear.  

In the PM state above $T_{\rm N}$, according to Eq.~(\ref{Eq:thetapPM}) the anisotropy in $\chi$ can only arise from a difference in the dipolar Weiss temperatures along different principal axis directions $\alpha$ and $\beta$, given by Eq.~(\ref{Eq:thetapPM}) as
\be
\theta_{{\rm pA}\alpha} - \theta_{{\rm pA}\beta} =\frac{C_1}{a^3}(\lambda_{{\bf 0}\alpha} - \lambda_{{\bf 0}\beta}).
\label{Eq:thetapPMAnis2}
\ee
For cubic Bravais lattices, one has no dipolar anisotropy in the PM state because $\lambda_{{\bf 0}\alpha}=0$ for all~$\alpha$.  Here we follow the approach of Keffer.\cite{Keffer1952}

For two susceptibilities $\chi_\alpha$ and $\chi_\beta$ measured along the $\alpha$ and $\beta$ principal axes, one has the identity
\bse
\be
\frac{1}{\chi_\beta} - \frac{1}{\chi_\alpha} = \frac{\chi_\alpha-\chi_\beta}{\chi_\alpha\chi_\beta},
\ee
or
\be
\chi_\alpha - \chi_\beta = \chi_\alpha\chi_\beta\left(\frac{1}{\chi_\beta} - \frac{1}{\chi_\alpha}\right).
\label{Eq:chiachib0}
\ee
\ese
Using Eqs.~(\ref{Eq:chiAlpha}), Eq.~(\ref{Eq:chiachib0}) yields
\be
\chi_\alpha - \chi_\beta = \frac{\chi_\alpha\chi_\beta}{C_1} (\theta_{{\rm p}\alpha} - \theta_{{\rm p}\beta}). 
\ee 
If the dipolar anisotropy in $\theta$ is small compared to the measured average Weiss temperature $\theta_{\rm p}$, one can define the geometric-mean susceptibility $\chi = \sqrt{\chi_\alpha\chi_\beta}$ and use Eq.~(\ref{Eq:thetapPMAnis2}) to obtain
\be
\chi_\alpha - \chi_\beta = \frac{\chi^2}{a^3} (\lambda_{0\alpha} - \lambda_{0\beta}).
\label{Eq:chia-chib}
\ee
Here the Curie-Weiss $\chi$'s are per spin and $a$ is the $a$-axis lattice parameter for the particular Bravais spin lattice considered.  The susceptibility difference per mole of spins is obtained by multiplying each $\chi$ on the left side of Eq.~(\ref{Eq:chia-chib}) and one $\chi$ on the right by Avogadro's number $N_{\rm A}$ and Eq.~(\ref{Eq:chia-chib}) yields the molar susceptibility difference
\be
\chi_{{\rm M}\alpha}(T) - \chi_{{\rm M}\beta}(T) = \frac{\chi_{\rm M}^2(T)}{N_{\rm A}a^3} (\lambda_{0\alpha} - \lambda_{0\beta}).
\label{Eq:chia-chib2}
\ee

\begin{figure}
\includegraphics[width=3.3in]{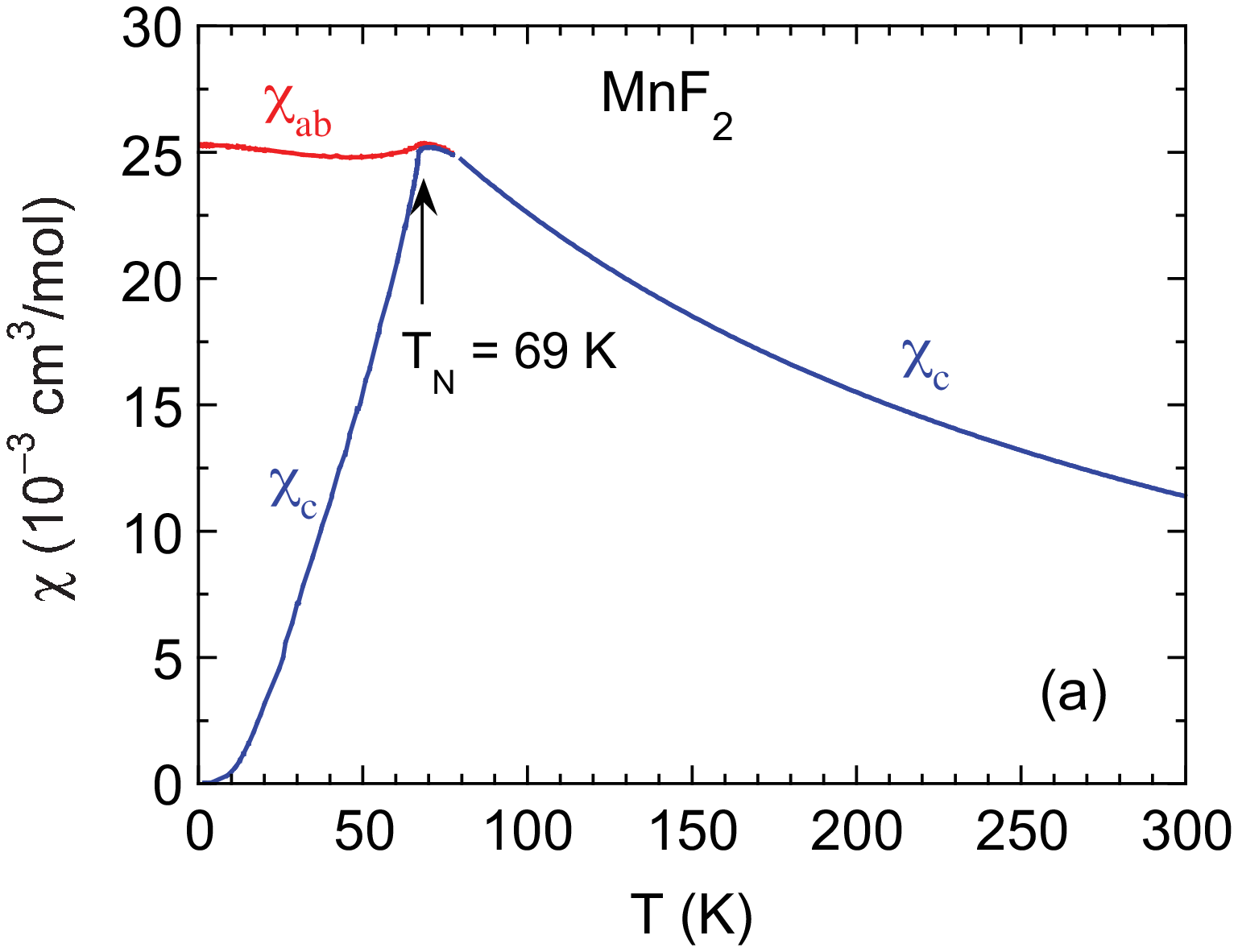}
\includegraphics[width=3.3in]{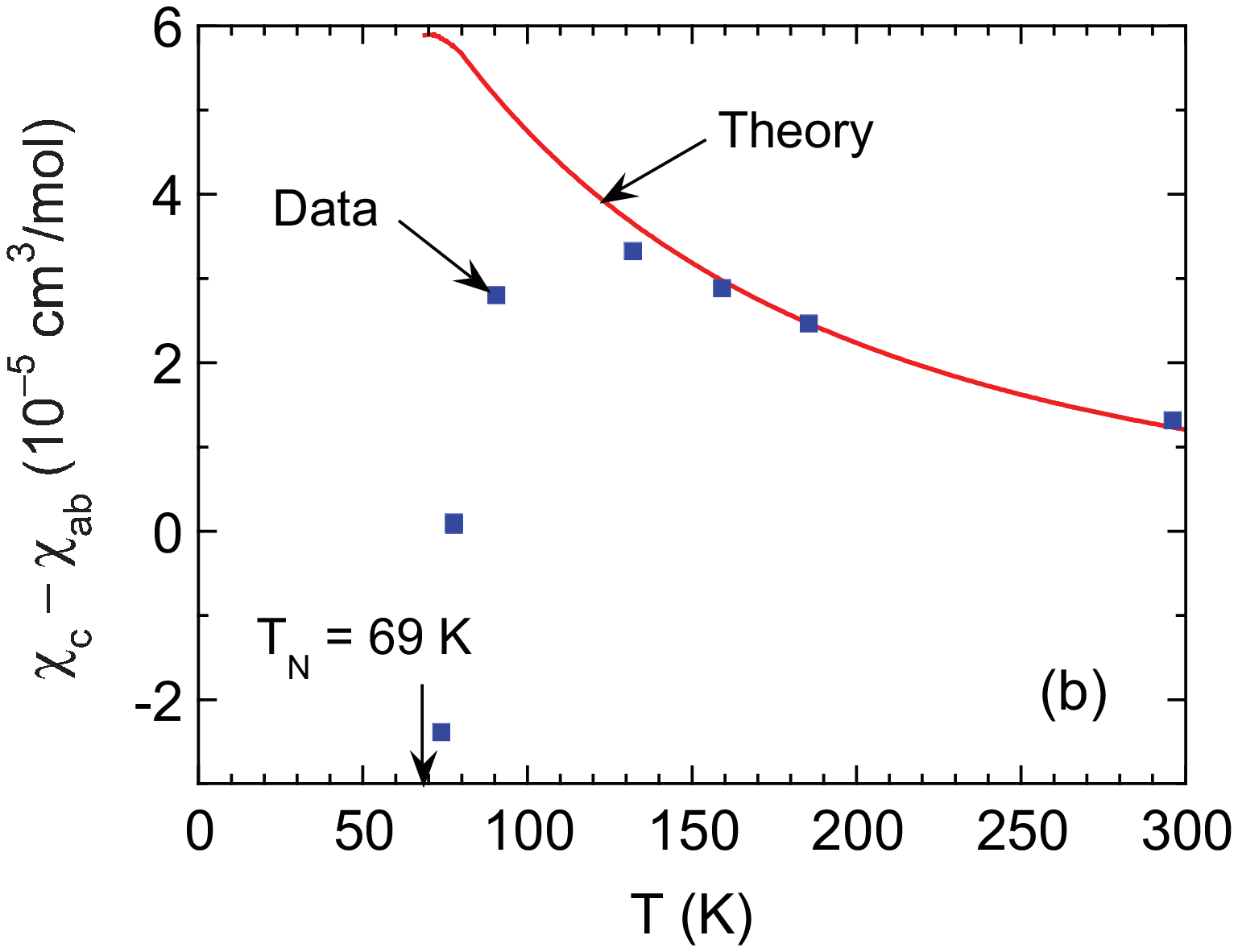}
\caption {(Color online) (a) Magnetic susceptibility $\chi$ versus temperature~$T$ of tetragonal MnF$_2$ crystals for applied fields along the $c$~axis ($\chi_c$) and in the $ab$~plane ($\chi_{ab}$).\cite{Trapp1963, Trapp1963a}  (b)~Anisotropy $\chi_c - \chi_{ab}$ versus $T$ (solid blue squares).\cite{Griffel1950}  Note the factor of 100 difference between the two ordinate scales in (a) and~(b).  The red solid curve is the MFT prediction for magnetic anisotropy arising from magnetic dipole interactions obtained using Eq.~(\ref{Eq:chia-chib2}).  }
\label{Fig:MnF2_Chi_Trapp_1963}
\end{figure}

Here we apply Eq.~(\ref{Eq:chia-chib2}) to the primitive tetragonal rutile-structure collinear AFM MnF$_2$ with $T_{\rm N} = 69$~K, which is often considered a prototype for collinear AFM ordering.  This compound contains a bct sublattice of Mn$^{+2}$ cations with spin $S = 5/2$ and an expected $g=2$ and orders into a A-type  AFM structure with AFM wavevector {\bf k} = (0,0,1) as shown in Fig.~\ref{Fig:EuCu2Sb2_mag_Struct_A_type}.  The lattice parameters are\cite{Stout1954, Griffel1950b}
\be
a = 4.8734(5)~{\rm \AA},\quad  c = 3.3103(10)~{\rm \AA},\quad \frac{c}{a} = 0.6793(3).
\label{Eq:LattPars}
\ee
For the given $c/a$ ratio and FM {\bf k} = {\bf0} we find $\lambda_{{\bf0}[001]} = 4.3219$ and $\lambda_{{\bf0}[100],[010]} = -2.1609$ yielding
\be
\lambda_{0[001]} - \lambda_{0[100]} = 6.4828,
\label{Eq:DeltaLambda(000)}
\ee
whereas for the ordering wavevector {\bf k} = (0,0,1)~r.l.u.\ we obtain $\lambda_{(001)[001]} = 13.8639$ and $\lambda_{(001)[100],[010]} = -6.9319$, with $\lambda_{(001)[001]} -\lambda_{(001)[100],[010]} = 20.7958$.  These values show that the [001] moment direction is energetically favored by the MDI both above and below $T_{\rm N}$, in agreement with experiment as follows.

The anisotropic $\chi(T)$ of MnF$_2$ crystals is shown in Fig.~\ref{Fig:MnF2_Chi_Trapp_1963}(a).\cite{Trapp1963, Trapp1963a}  Above $T_{\rm N}$, $\chi$ is found to be nearly isotropic.  Below $T_{\rm N}$, the data are a textbook example of the anisotropy expected for collinear AFM ordering, where in this case the easy axis is the $c$~axis.  According to MFT, $\chi_\perp = \chi_{ab}$ for $T \leq T_{\rm N}$ should be independent of $T$, which is well satisfied.  On the other hand, $\chi_\parallel = \chi_c$ should go to zero as $T \to 0$, as also observed.  We obtained a fairly good fit to $\chi_\parallel(T\leq T_{\rm N})$ using our MFT with no adjustable parameters.\cite{Johnston2012}  The fit function used was similar to the equation we obtained for $\chi_\parallel(T)$ for the pure dipole AFM in Eqs.~(\ref{Eqs:ChiPar}) and Fig.~\ref{Fig:ChiExample}.

The anisotropy $\Delta\chi(T) \equiv \chi_c(T)-\chi_{ab}(T)$ was measured with a torque magnetometer and the results are shown in Fig.~\ref{Fig:MnF2_Chi_Trapp_1963}(b).\cite{Griffel1950}  The $\Delta\chi$ data measured with the torque magnetometer\cite{Griffel1950} for $T<T_{\rm N}$ agree with the anisotropy calculated from the direct measurements\cite{Trapp1963, Trapp1963a} in Fig.~\ref{Fig:MnF2_Chi_Trapp_1963}.  For $T\gtrsim T_{\rm N}$, a comparison of the data in Figs.~\ref{Fig:MnF2_Chi_Trapp_1963}(a) and~\ref{Fig:MnF2_Chi_Trapp_1963}(b) shows that $|\Delta\chi|/\chi\sim 0.1$\% for $T>T_{\rm N}$.  From Eq.~(\ref{Eq:chia-chib2}), the anisotropy of $\chi$ is predicted to be
\be
\Delta\chi_{\rm M}(T)  = \frac{\chi_{\rm M}^2(T)}{N_{\rm A}a^3} (\lambda_{0[001]} - \lambda_{0[100]}).
\label{Eq:DeltaChiMnF2}
\ee
Using the values of $a$ and $\lambda_{0[001]} - \lambda_{0[100]}$ in Eqs.~(\ref{Eq:LattPars}) and~(\ref{Eq:DeltaLambda(000)}), respectively, and the $\chi_{\rm M}(T)$ data in Fig.~\ref{Fig:MnF2_Chi_Trapp_1963}(a), $\Delta\chi_{\rm M}(T)$ was calculated from Eq.~(\ref{Eq:DeltaChiMnF2}) and the result is shown as the solid red curve in Fig.~\ref{Fig:MnF2_Chi_Trapp_1963}(b) (see also Ref.~\onlinecite{Keffer1952}).  The calculation is in excellent agreement with the data for $T\gtrsim150$~K, suggesting that the MDI is responsible for the $\chi$ anisotropy in this $T$ range, or at least reinforces this anisotropy.  However, the data are increasingly suppressed to lower values below 130~K, which likely result from the onset of dynamic short-range collinear AFM correlations along the $c$~axis with a correlation length that eventually diverges at $T_{\rm N}=69$~K, where from Fig.~\ref{Fig:MnF2_Chi_Trapp_1963}(a), $\Delta\chi_{\rm M}$ grows to become large and even more negative below that temperature.  

\section{\label{Sec:Summary} Summary}

A detailed summary of the paper is given in the Abstract to the paper.  Here we provide a few additional comments.

The eigenvalues and eigenvectors of the MDI tensor were determined for specified magnetic wavevectors and spin lattices.  The eigenvalues give the energy of a spin in the magnetic fields of the local moments inside a Lorentz sphere of radius $R$ in units of the $a$-axis lattice parameter~$a$.  For 3D lattices, $R/a=50$ was usually used, for a 2D circle $R/a\leq 1000$ and for a spin chain with $R/a=\infty$ the eigenvalues were determined exactly.  The eigenvectors are the three orthogonal principal axis directions for collinear magnetic ordering.  For uniaxial 3D spin lattices, these were calculated for $c/a=0.5$ to~3 and the results presented in figures in the main text and Appendix~B and in tables in the Supplemental Material.\cite{SupplInfo}  We also calculated the eigenvalues and eigenvectors for noncollinear AFM structures including the 2D 120$^\circ$  triangular lattice and for the 2D and 3D coplanar noncollinear Shastry-Sutherland lattice and GdB$_4$ magnetic structure.  We compared the ordering-direction predictions with data for some Mn$^{+2},\ (S=5/2)$,  Gd$^{+3}$ and Eu$^{+2},\ (S=7/2)$  compounds and found good agreement.  Disagreement occurred for the itinerant AFM ${\rm BaFe_2As_2}$ and for the coplanar noncollinear AFM GdB$_4$, which indicates that a stronger anisotropy source must be present in these compounds that defeats the preferences of the MDI.  

A significant contribution of this paper was to apply our formulation of the Weiss MFT\cite{Johnston2012,Johnston2015} to predict many properties of the ordered and PM states arising from MDIs.  These include the magnetic ordering temperature $T_{\rm m}$, the ordered moment, the magnetic heat capacity, and for AFMs the perpendicular critical field, the anisotropic magnetic susceptibility versus temperature for $T\leq T_{\rm N}$, and the parameters of the Curie-Weiss law for the anisotropic susceptibility for both FMs and AFMs at $T\geq T_{\rm m}$.  Within MFT, the contributions of different molecular field sources to these properties are additive.  This means that the same theory can be used to treat purely magnetic dipole magnets or spin systems containing both exchange and dipole interactions.  We recently used the theory to separate the magnetic dipole and exchange contributions to the properties of the bct compound ${\rm EuCu_2Sb_2}$ with $c/a=2.4$ and $T_{\rm N}=5.1$~K, which then allowed estimates of the Eu--Eu exchange interactions to be made.\cite{Anand2015}

\acknowledgments

The author is grateful to Andreas Kreyssig for helpful discussions, to Vivek Anand for experimental collaborations relating to this work and to Jiping Huang for sending a list of published eigenvalue data.  This research was partially supported by the U.S.~Department of Energy, Office of Basic Energy Sciences, Division of Materials Sciences and Engineering.  Ames Laboratory is operated for the U.S.~Department of Energy by Iowa State University under Contract No.~DE-AC02-07CH11358.


\appendix

\section{\label{Methods} Direct and Reciprocal Lattices}

\subsection{Orthogonal Bravais Lattices}

In a Bravais spin lattice each spin position is a point of inversion symmetry with respect to the other spins.  For orthogonal lattices which include as special cases the linear chain, the simple square lattice,  sc, bcc, fcc, simple tetragonal and bct lattices, the unit cell origins are at
\bse
\label{Eqs:spinPos}
\be
\frac{{\bf r}_{ji}}{a} = n_a\hat{\bf a} + \frac{b}{a} n_b\hat{\bf b} + \frac{c}{a}n_c\hat{\bf c},
\label{Eq:rji}
\ee
where $n_a$, $n_b$ and $n_c$ are positive or negative integers or~0.  For all spin lattices, we normalize all spin positions and interspin distances by the $a$-axis lattice parameter~$a$.  For body-centered spin lattices one also has atoms at the body centers
\be
\frac{{\bf r}}{a} = \left(n_a+\frac{1}{2}\right)\hat{\bf a} + \left(n_b+\frac{1}{2}\right)\hat{\bf b} + \frac{c}{a}\left(n_c+\frac{1}{2}\right)\hat{\bf c},
\label{Eq:bcpositions}
\ee
\ese
where $c/a=1$ for the bcc lattice.  The central magnetic moment $\vec{\mu}_i$ is placed at ${\bf r}_{i}=0$ and hence the sum over neighbors $\vec{\mu}_j$ at positions ${\bf r}_{j} = {\bf r}_{ji}$ in Eq.~(\ref{Eq:Gk2}) excludes the set $(n_a,n_b,n_c) = (0,0,0)$ in Eq.~(\ref{Eq:rji}).  With our formulation, $\widehat{{\bf G}}_i({\bf k})$ does not explicitly contain the lattice parameters $a$ or $c$, and for tetragonal Bravais lattices just the dimensionless $c/a$ ratio appears as in Eqs.~(\ref{Eqs:spinPos}).

The reciprocal-lattice vectors in reciprocal-lattice units are
\be
{\bf k} = m_1 {\bf a}^\ast + m_2 {\bf b}^\ast + m_3 {\bf c}^\ast,
\ee
where the $m_i$ satisfy $0\leq m_i\leq 1$ and the reciprocal-lattice translation vectors are
\be
{\bf a}^\ast = \frac{2\pi}{a}\,\hat{\bf a}, \qquad {\bf b}^\ast = \frac{2\pi}{a}\,\hat{\bf b}, \qquad {\bf c}^\ast = \frac{2\pi}{c}\,\hat{\bf c},
\ee
and {\bf a}, {\bf b} and {\bf c} are the corresponding direct-lattice translation vectors.  We normalize {\bf k} by $1/a$, yielding
\be
{\bf k}a = 2\pi\Big(m_1 \hat{\bf a} + m_2 \hat{\bf b} + \frac{1}{c/a} m_3 \hat{\bf c}\Big).
\label{Eq:ka}
\ee
Using Eqs.~(\ref{Eq:ka}) and~(\ref{Eq:rji}), for the unit cell origins one has 
\bse
\be
{\bf k}\cdot {\bf r}_{ji} =  2\pi(m_1n_a + m_2n_b + m_3n_c)
\label{Eq:kdotrji}
\ee
and for the body-center positions
\be
{\bf k}\cdot {\bf r}_{ji} = 2\pi\bigg[m_1\bigg(n_a+\frac{1}{2}\bigg) + m_2\left(n_b+\frac{1}{2}\right) + m_3\left(n_c+\frac{1}{2}\right)\bigg],\label{Eq:kdotr1/2}
\ee
\ese
where the $c/a$ ratio has canceled out of both expressions.

The sum in Eq.~(\ref{Eq:Gk2}) gives an ``extinction condition'' for the contribution  to the sum in Eq.~(\ref{Eq:Gk2}) of the body-centered spins in the bcc lattice in Eq.~(\ref{Eq:kdotr1/2}), where the contribution is zero if ${\bf k}\cdot {\bf r}_{ji}$ is an odd multiple of $\pi/2$~rad.  This extinction occurs, for example, for AFM wavevectors
\be
{\bf k} = \bigg(\frac{1}{2},\ 0,\ 0\bigg),\ \ \bigg(0,\ 0,\ \frac{1}{2}\bigg),\ \ \bigg(\frac{1}{2},\ \frac{1}{2},\ \frac{1}{2}\bigg).
\label{Eq:kExtinction}
\ee
For such cases, according to Eq.~(\ref{Eq:Extinction}) which assumes a collinear magnetic structure, the spins at the body centers of the unit cells have zero ordered moment and they make no contribution to the dipolar interaction tensor in Eq.~(\ref{Eq:Gk2}).  The interaction tensor is then the same as for a simple tetragonal lattice of moments with the same $c/a$ ratio and {\bf k} value.

For the fcc lattice the lattice points are at the positions in Eq.~(\ref{Eq:rji}) and at
\bea
\frac{{\bf r}}{a} &=& \left(n_a+\frac{1}{2}\right)\hat{\bf a} + \left(n_b+\frac{1}{2}\right)\hat{\bf b} +0,\\*
\frac{{\bf r}}{a} &=& \left(n_a+\frac{1}{2}\right)\hat{\bf a}  + 0+\left(n_c+\frac{1}{2}\right)\hat{\bf c},\\*
\frac{{\bf r}}{a} &=& 0+\left(n_b+\frac{1}{2}\right)\hat{\bf b} + \left(n_c+\frac{1}{2}\right)\hat{\bf c},
\label{Eq:fccpositions}
\eea
with corresponding changes to the expressions for ${\bf k}\cdot {\bf r}_{ji}$.

\subsection{Simple Hexagonal (Triangular) Bravais Lattice}

\begin{figure}
\includegraphics[width=1.5in]{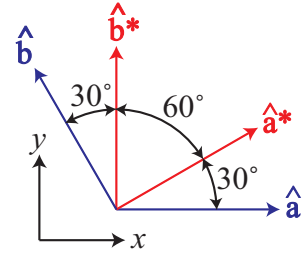}
\caption {(Color online) In-plane hexagonal lattice translation unit vectors $\hat{\bf a}$ and $\hat{\bf b}$ of the direct lattice and $\hat{\bf a}^\ast$ and $\hat{\bf b}^\ast$ of the reciprocal lattice, respectively.}
\label{Fig:Hexagonal_Recip_Latt}
\end{figure}

The normalized vectors ${\bf r}_{ji}$ for the simple hexagonal lattice with $a=b$ are given by 
\be
\frac{{\bf r}_{ji}}{a} = n_a\hat{\bf a} + n_b\hat{\bf b} + \frac{c}{a}n_c\hat{\bf c},
\label{Eq:rjiHex}
\ee
where here the {\bf b} axis is at an angle of 120$^\circ$ with respect to the positive $x$~axis as shown in Fig.~\ref{Fig:Hexagonal_Recip_Latt} and the $n_i$ are again positive or negative integers or zero.  In two dimensions one sets $n_c=0$.  In Cartesian coordinates the translation unit vectors are
\be
\hat{\bf a} = \hat{\bf i},\qquad\hat{\bf b} = -\frac{1}{2}\hat{\bf i} + \frac{\sqrt{3}}{2}\hat{\bf j},\qquad \hat{\bf c} = \hat{\bf k}.
\ee

A magnetic ordering wavevector {\bf k} is written in terms of the respective simple hexagonal reciprocal lattice vectors as
\bse
\label{Eqs:kHex}
\be
{\bf k} = m_1 {\bf a}^{\ast} + m_2 {\bf b}^{\ast} + m_3{\bf c}^{\ast},
\label{Eq:kmi}
\ee
where the $m_i$ are chosen to satisfy $0\leq m_i\leq 1$ and the reciprocal lattice translation vectors are given by
\be
{\bf a}^{\ast} = \frac{2\pi}{a}\left(\hat{\bf i} + \frac{1}{\sqrt{3}}\hat{\bf j}\right), \quad {\bf b}^{\ast} =\frac{4\pi}{a\sqrt{3}}\hat{\bf j}, \quad {\bf c}^{\ast} =\frac{2\pi}{c}\hat{\bf k},
\ee
\bea
|{\bf a}^{\ast}| &=& |{\bf b}^{\ast}| = \frac{4\pi}{\sqrt{3}a}\equiv 1~a,b{\textrm -}{\rm axis\ r.l.u.},\\*
|{\bf c}^{\ast}| &=& \frac{2\pi}{c}\equiv 1~c\textrm{-axis\ r.l.u.}
\eea
\ese
In terms of $\hat{\bf a}^{\ast}$ and $\hat{\bf b}^{\ast}$, the direct lattice unit vectors are
\be
\hat{\bf a} = \frac{1}{\sqrt{3}}(2\hat{\bf a}^{\ast}-\hat{\bf b}^{\ast}), \quad \hat{\bf b} =  \frac{1}{\sqrt{3}}(2\hat{\bf b}^{\ast}-\hat{\bf a}^{\ast}),\quad \hat{\bf c} = \hat{\bf c}^\ast.
\ee
The expression for ${\bf k}\cdot {\bf r}_{ji}$ is the same as in Eq.~(\ref{Eq:kdotrji}).

\clearpage
\section{\label{App:RtoInfty} Figures Showing the Approach to the Large-Radius Asymptotic Eigenvalues for Magnetic Ordering on 2D and 3D Spin Lattices}

\begin{figure}[h]
\includegraphics[width=3.3in]{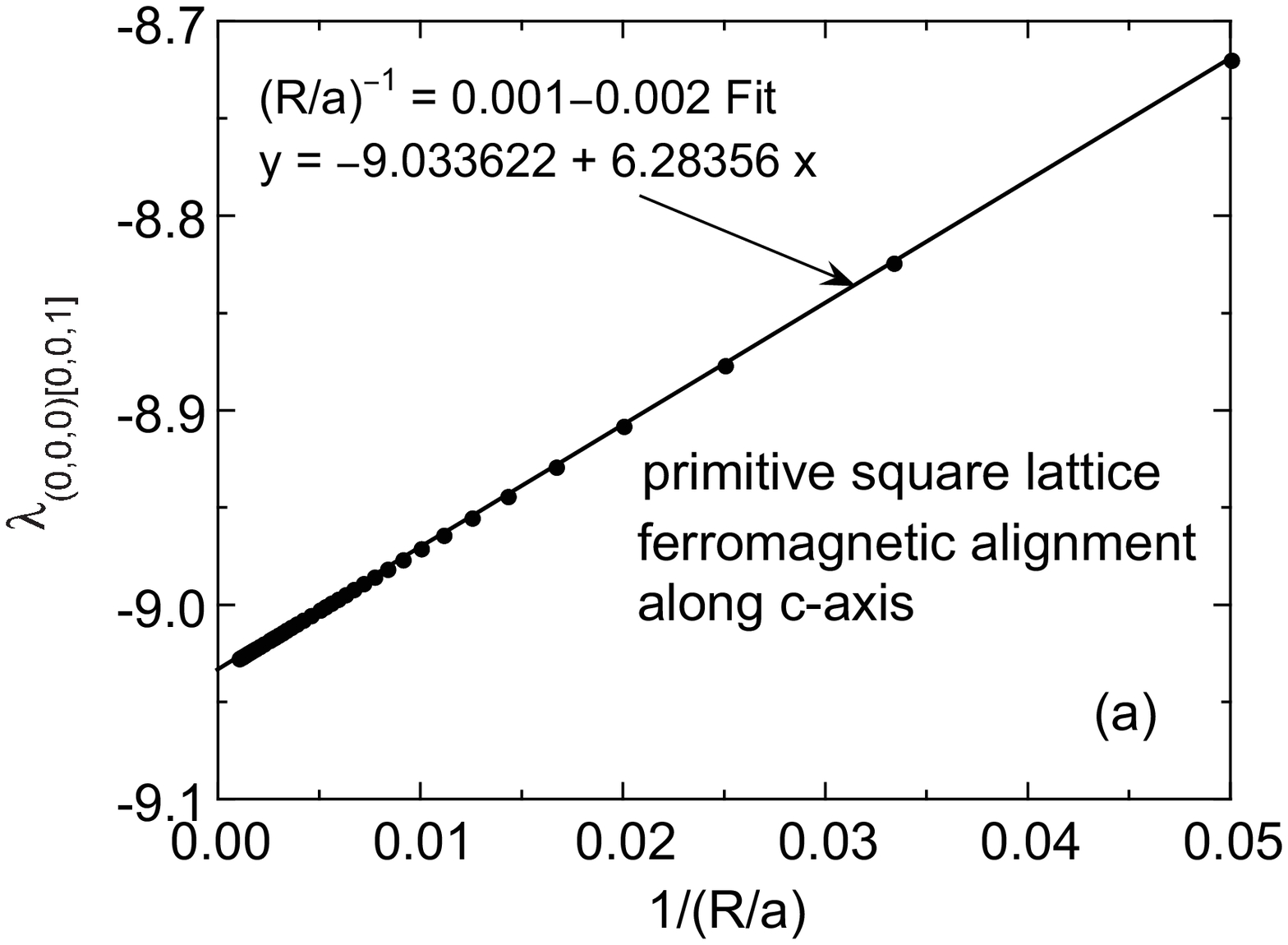}\vspace{0.1in}
\includegraphics[width=3.3in]{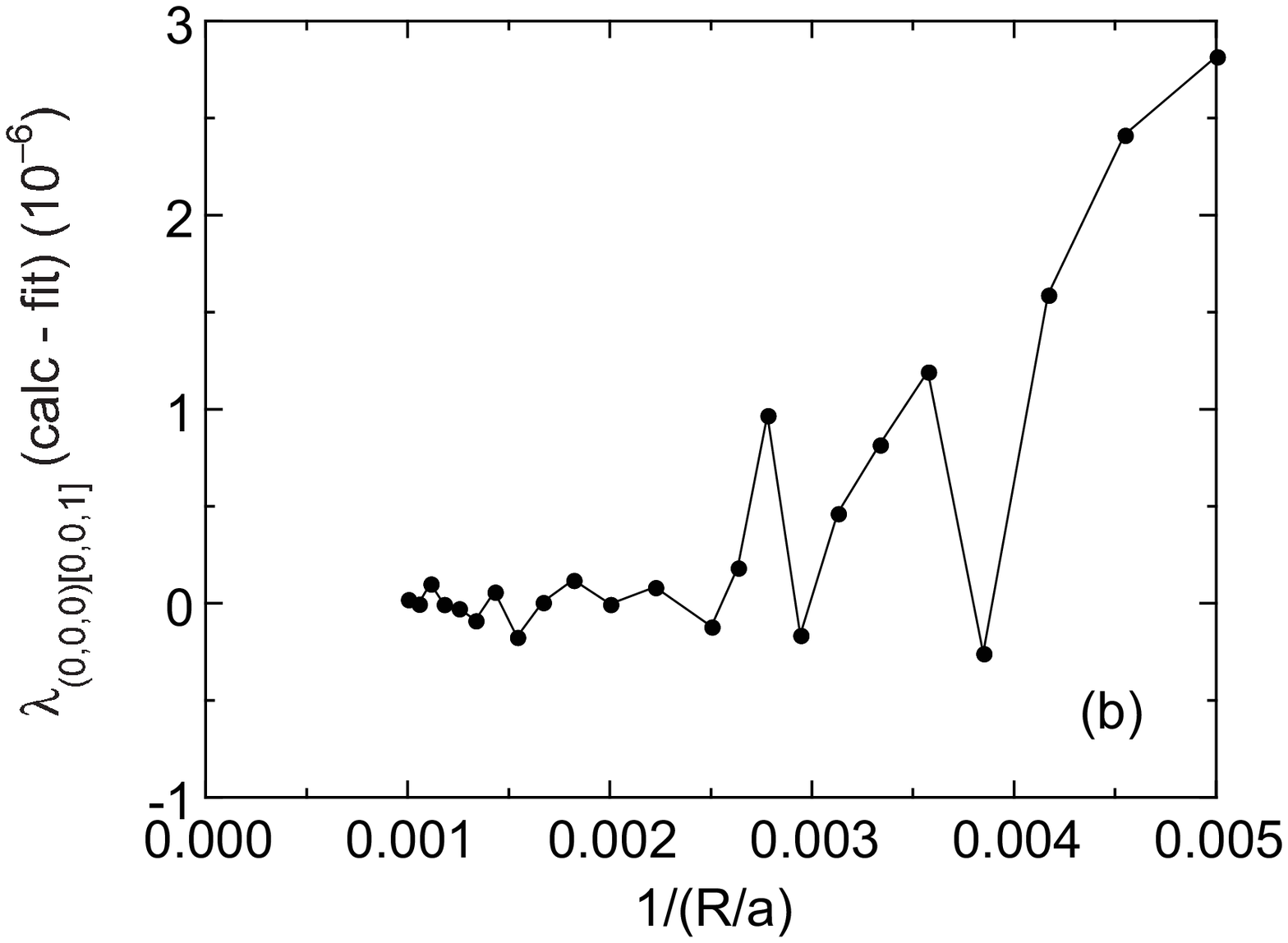}
\caption {(a) Eigenvalue $\lambda_{(0,0,0)[0,0,1]}$ for FM spin alignment along the $c$~axis versus the inverse of the circle radius~$R$ around the central moment in units of the square lattice parameter $a$ for the 2D simple square lattice. The $a$ and $b$-axis eigenvalues are each equal to $-\lambda_{(0,0,0)[0,0,1]}/2$. (b)~Deviation of the data from the fit. The ``noise'' is due to the discrete nature of the lattice, not to numerical inaccuracy.  The lines in (b) are guides to the eye.}
\label{Fig:SqLattk00}
\end{figure}

\begin{figure}[h]
\includegraphics[width=3.3in]{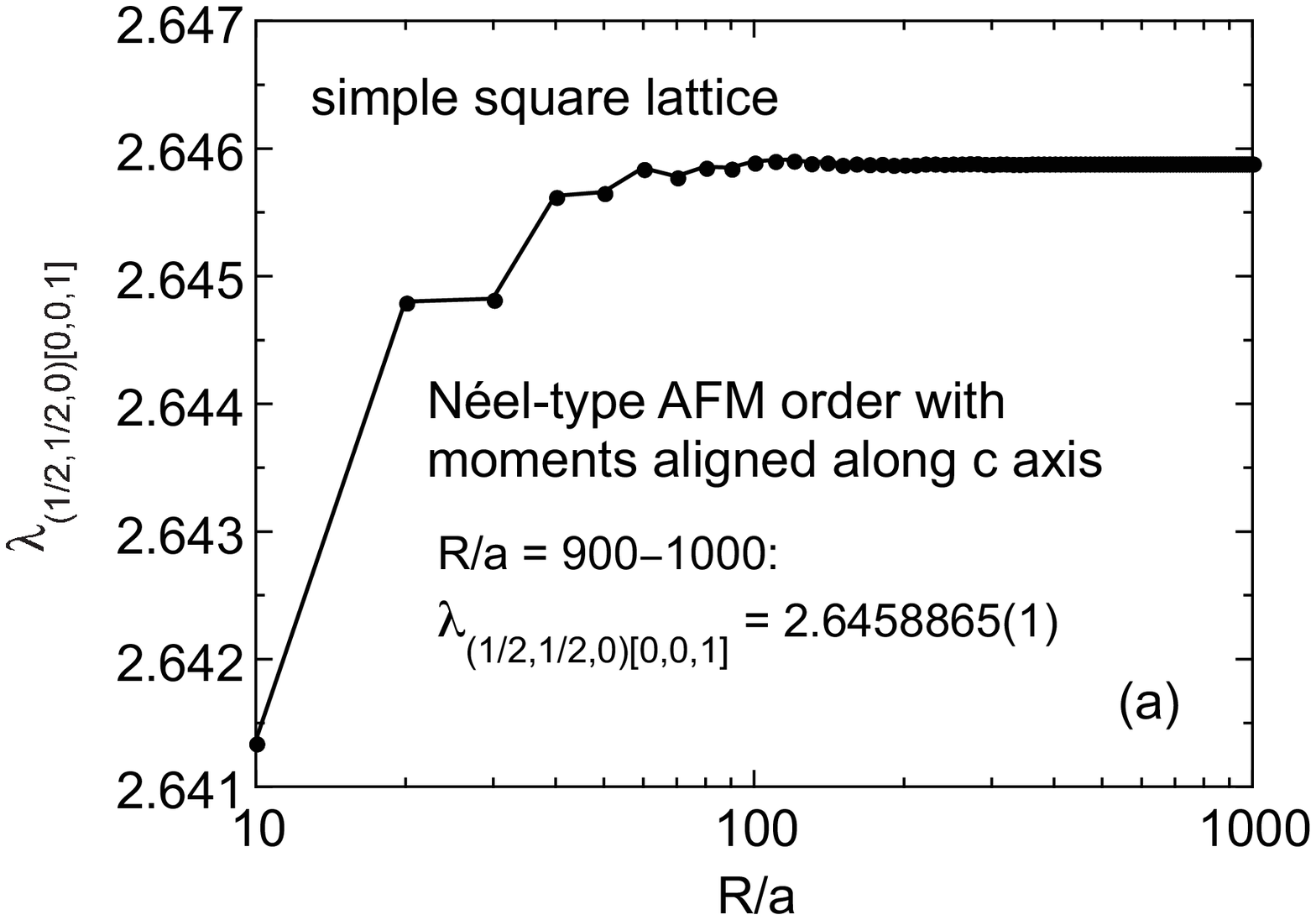}\vspace{0.1in}
\includegraphics[width=3.3in]{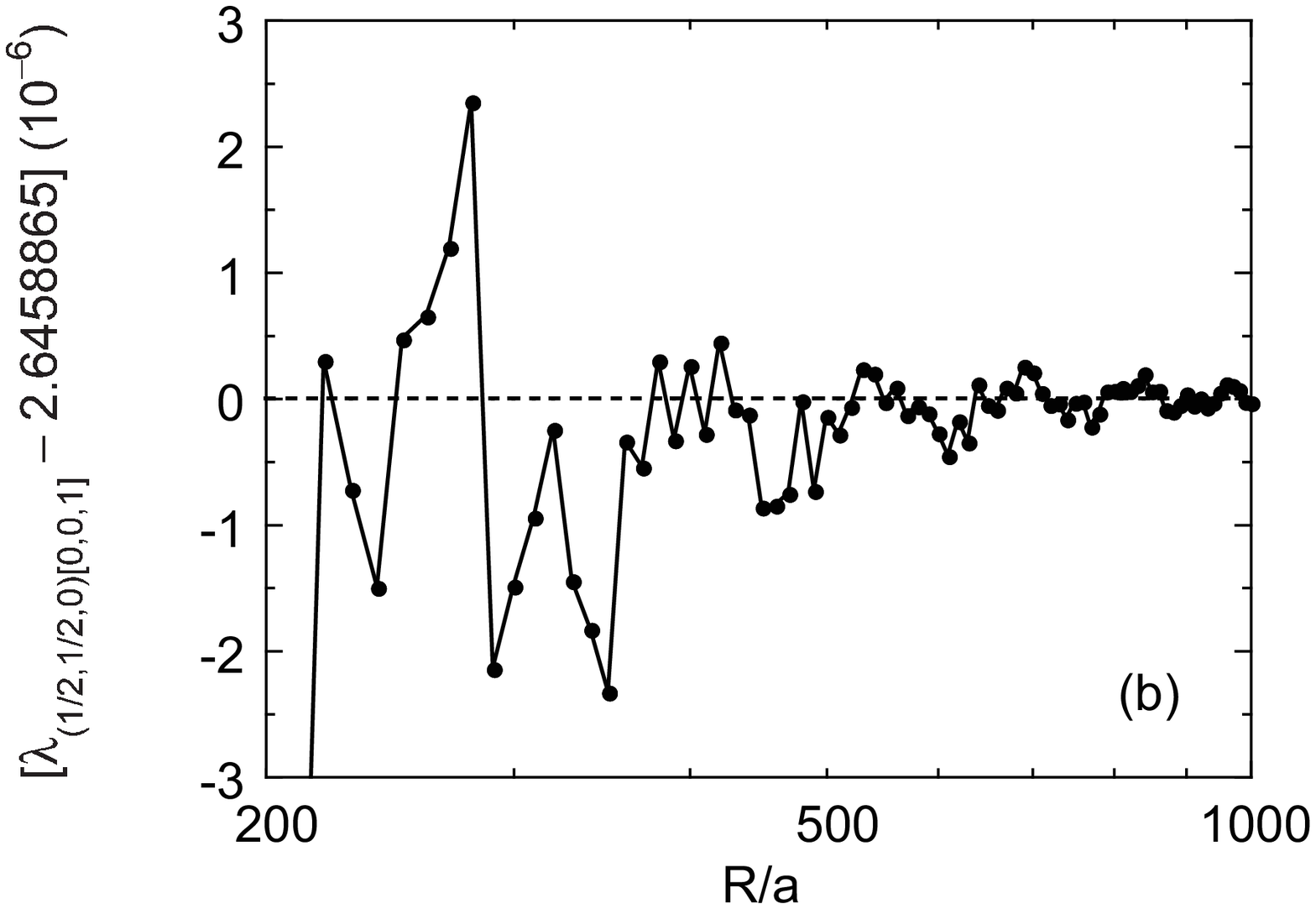}
\caption {(a) Eigenvalue $\lambda_{(1/2,1/2,0)[0,0,1]}$ for the N\'eel-type AFM moment alignment along the $c$~axis versus the circle radius~$R$ around the central moment in units of the square lattice parameter $a$ for the 2D simple square lattice. The $a$- and $b$-axis eigenvalues are each equal to $-\lambda_{(1/2,1/2,0)[0,0,1]}/2$. (b) Deviation of the data from the fit. The ``noise'' is due to the discrete nature of the lattice, not to numerical inaccuracy.  The lines in (b) are guides to the eye.}
\label{Fig:SqLattk11}
\end{figure}

\begin{figure}[h]
\includegraphics[width=3.3in]{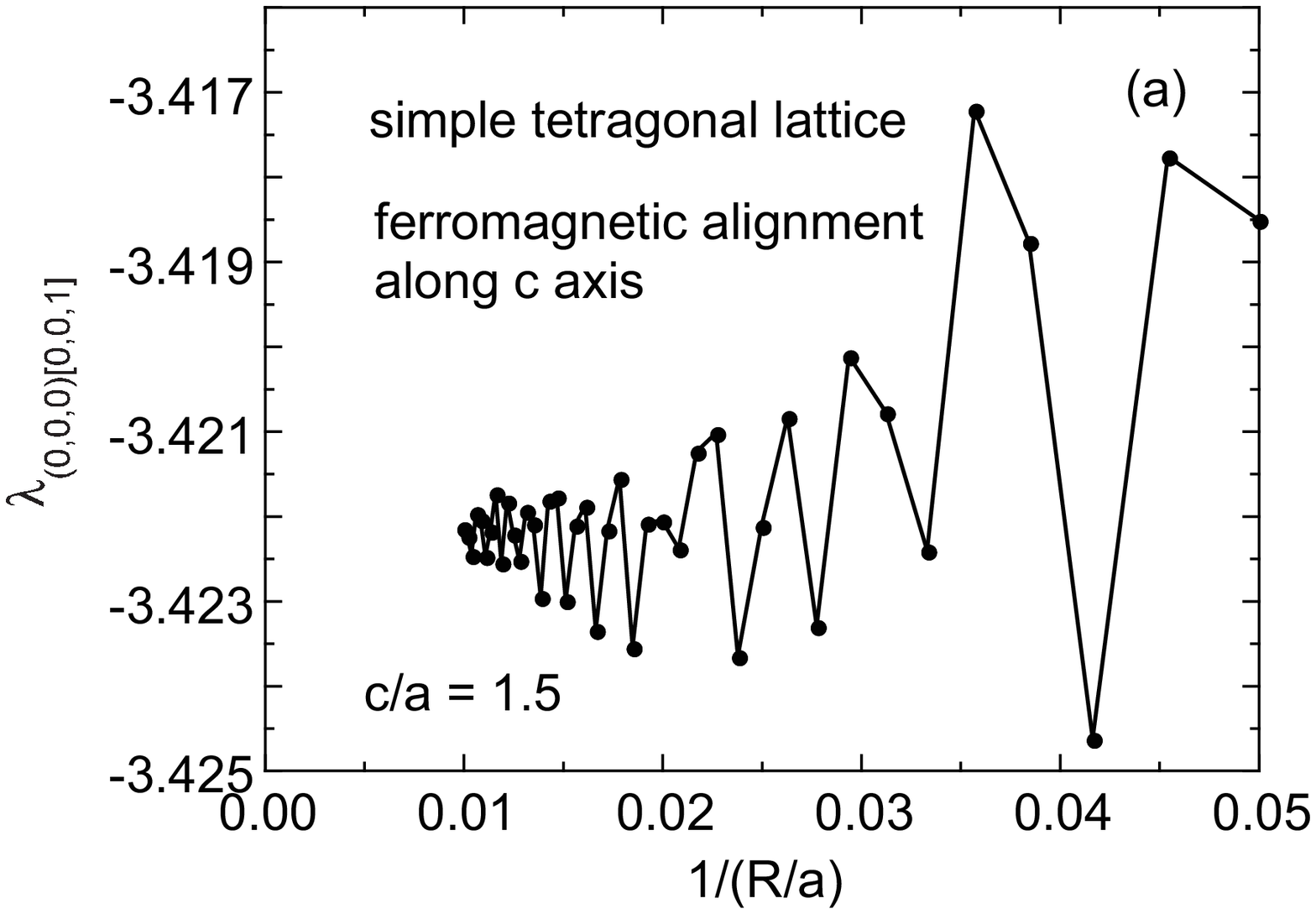}
\includegraphics[width=3.3in]{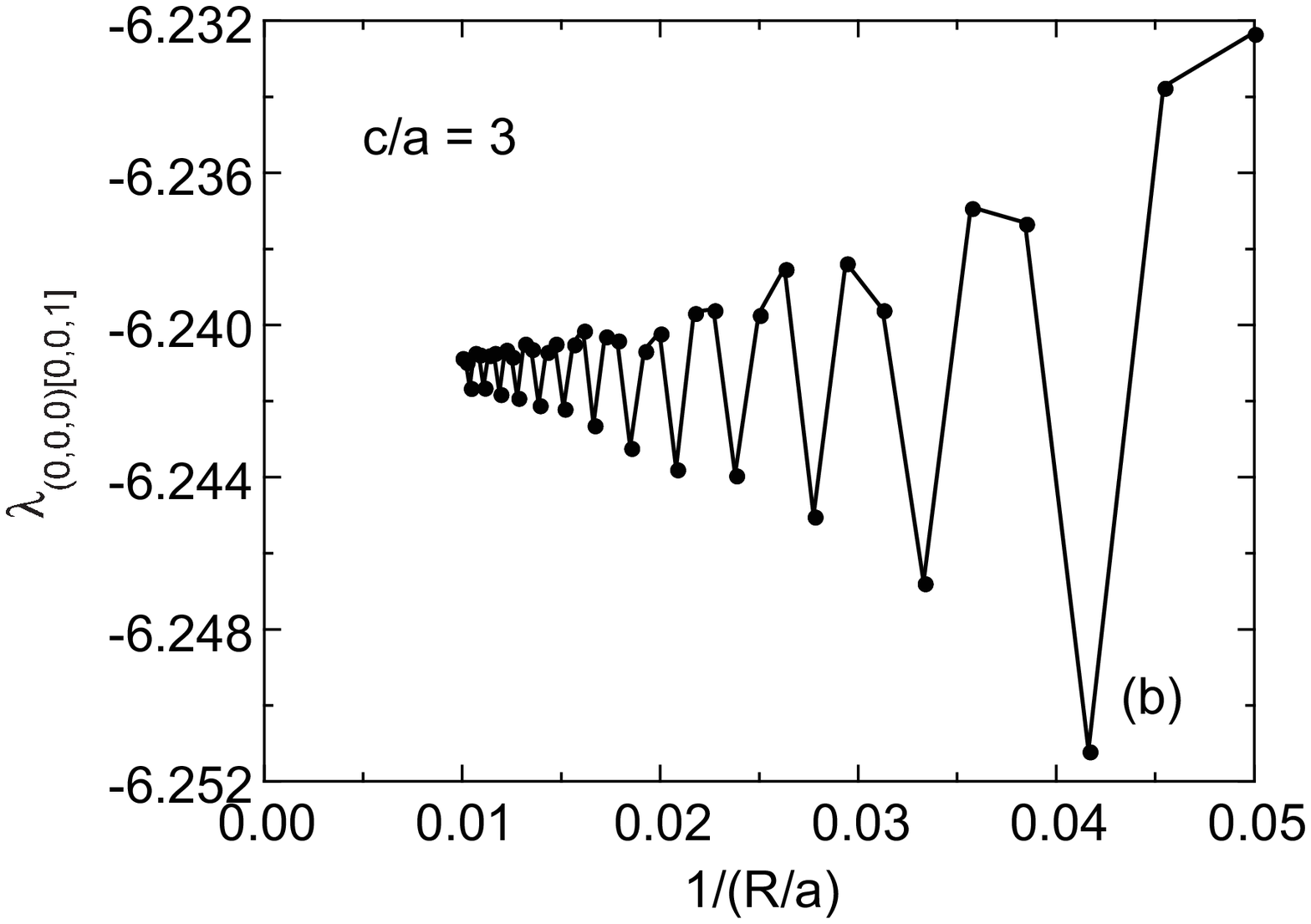}
\caption {Dependences of the eigenvalue $\lambda_{(0,0,0)[0,0,1]}$ on the inverse radius $(R/a)^{-1}$ of the Lorentz sphere for FM moment alignments [{\bf k} = (0,0,0)] along the $c$~axis in 3D simple tetragonal spin lattices with (a) $c/a = 1.5$ and (b)~$c/a = 3$.  The lines are guides to the eye.}
\label{Fig:PTCA1_5_100Convergence}
\end{figure}

\begin{figure}[h]
\includegraphics[width=3.3in]{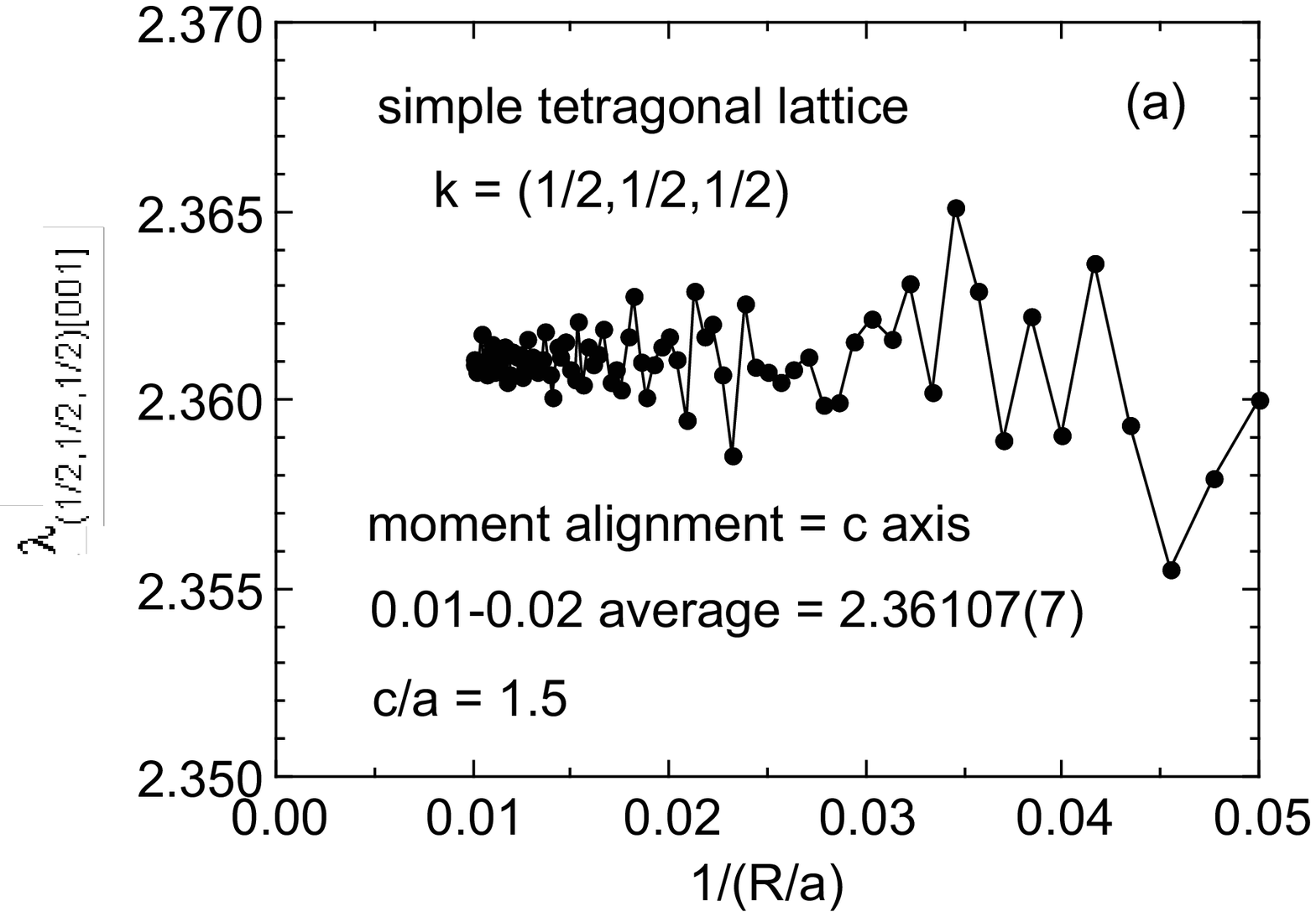}
\includegraphics[width=3.3in]{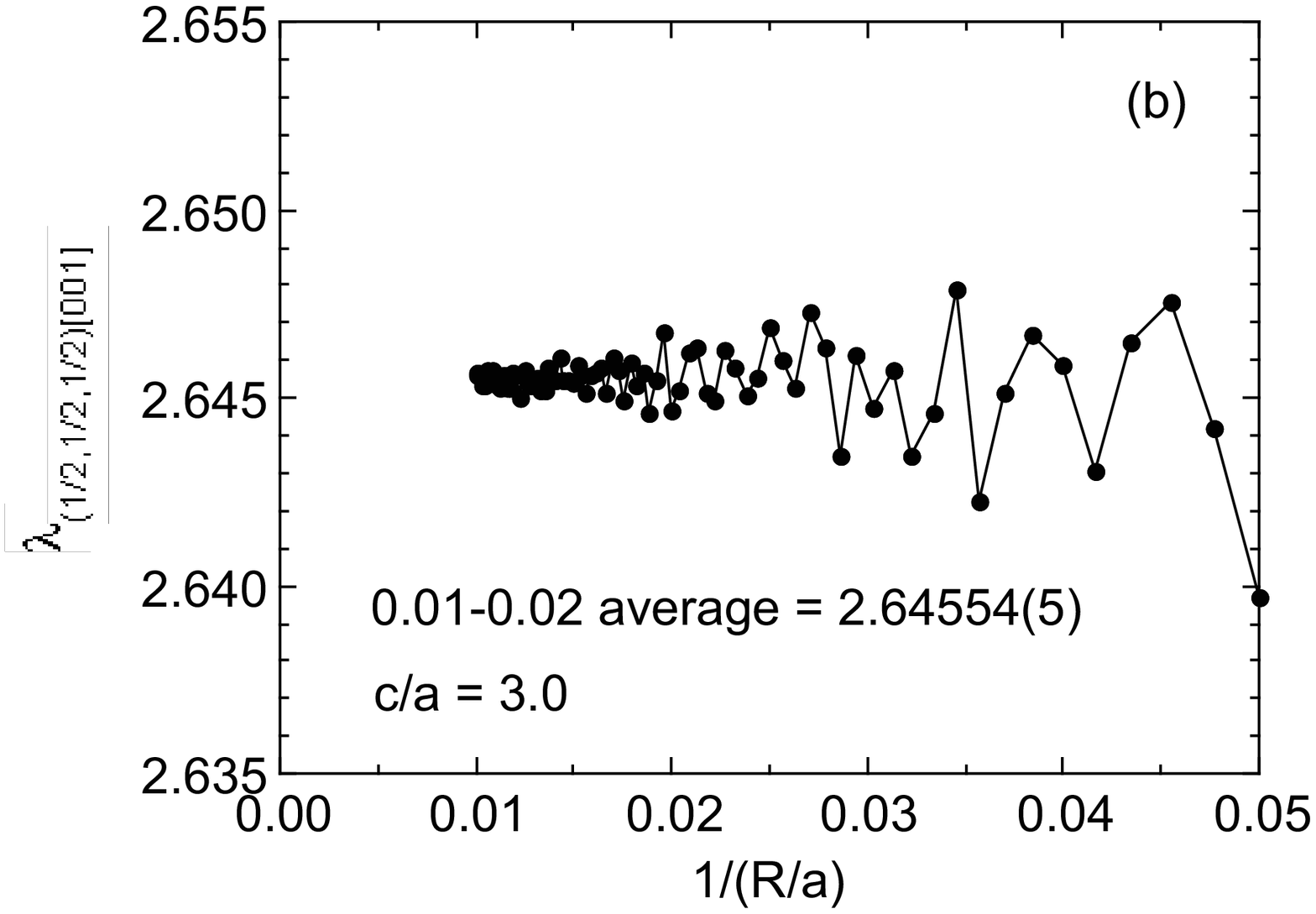}
\caption {Dependences of the eigenvalue $\lambda_{(1/2,1/2,1/2)[0,0,1]}$ for N\'eel-type ordering with {\bf k} = (1/2,1/2,1/2) on the inverse radius $(R/a)^{-1}$ of the Lorentz sphere for AFM moment alignments along the $c$~axis in 3D simple tetragonal spin lattices with (a) $c/a = 1.5$ and (b)~$c/a = 3$.  The lines are guides to the eye.  The averages for $R/a$ = 51--100 are shown.  With increasing $c/a$, the averages of $\lambda_{(1/2,1/2,1/2)[0,0,1]}$ for $R/a$ = 51--100 approach the 2D square-lattice limit $\lambda_{(1/2,1/2,0)[0,0,1]}= 2.645\,887$ in Table~\ref{Tab:EvecsEvals}, as shown in Fig.~\ref{Fig:AllPTDataCAk101Ave}(b).}
\label{Fig:AllPTk111ConvergCA1_5}
\end{figure}

\clearpage

\section{\label{App:lambdaFigs} Figures Showing Dipolar Eigenvectors and Eigenvalues versus the $c/a$ Ratio for Tetragonal and Hexagonal Bravais Spin Lattices and for the Honeycomb Lattice}

\begin{figure}[h]
\includegraphics[width=3.3in]{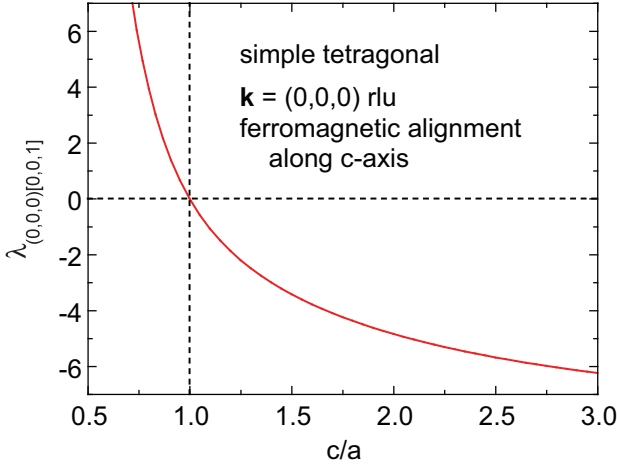}
\caption {(Color online) Dependence of the eigenvalue $\lambda_{(0,0,0)[0,0,1]}$ on the $c/a$ ratio for a simple tetragonal lattice with a FM alignment of the magnetic moments along the $c$~axis.  From the figure, one sees that FM alignment along the $c$~axis is the most stable for $c/a < 1$, but for $c/a > 1$ FM alignment along the $a$ or $b$~axis is energetically favorable.}
\label{Fig:SummaryPTDataCA}
\end{figure}

\begin{figure}[h]
\includegraphics[width=3.3in]{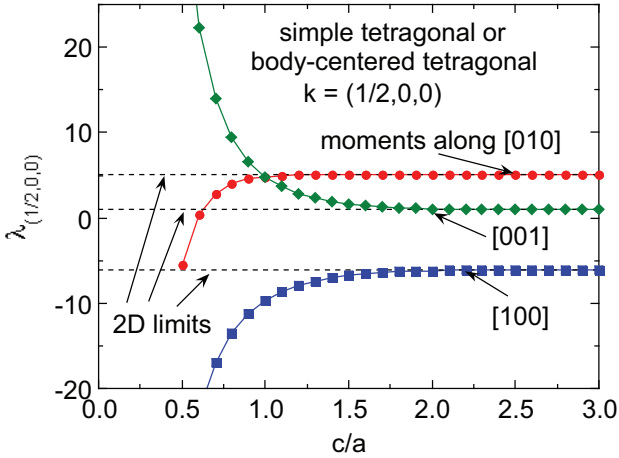}
\caption {(Color online) Eigenvalues for wavevector {\bf k} = (1/2,0,0)\,r.l.u.\ versus the $c/a$ ratio for a simple tetragonal or  bct lattice with the moments aligned along  [010] ($b$~axis, solid red circles), [001] ($c$~axis, solid green diamonds) or [100] ($a$~axis, solid blue squares). The 2D limits for $c/a\to\infty$ are shown as horizontal dashed lines.}
\label{Fig:AllPTDataCAk100Ave}
\end{figure}

\begin{figure}[h]
\includegraphics[width=3.3in]{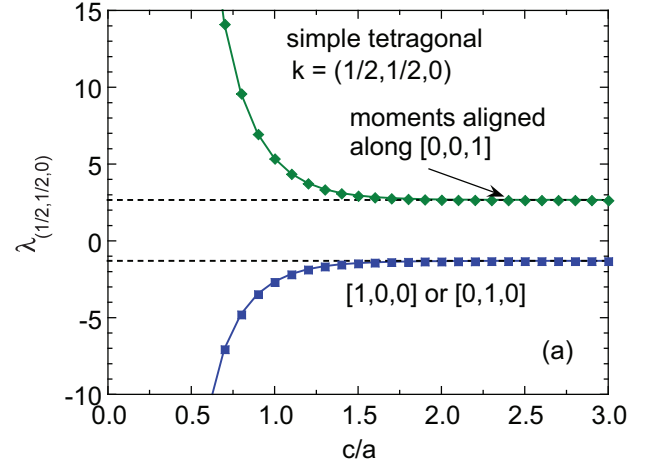}
\includegraphics[width=3.3in]{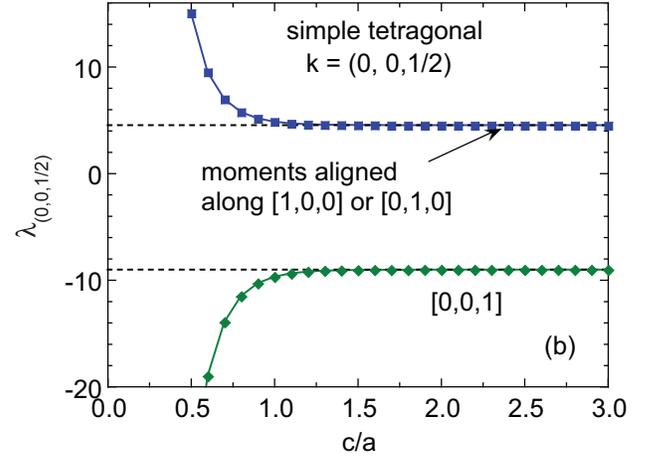}
\caption {(Color online) Eigenvalues (a) $\lambda_{(1/2,1/2,0)}$ for AFM wavevector {\bf k} = (1/2,1/2,0)\,r.l.u.\ and~(b) $\lambda_{(0,0,1/2)}$ for AFM wavevector {\bf k} = (0,0,1/2)\,r.l.u.\ versus the $c/a$ ratio for a simple tetragonal lattice with the moments aligned along [1,~0,~0] or [0,~1,~0] ($a$ or $b$~axis, solid blue squares) or [0,~0,~1] ($c$~axis, solid green diamonds).}
\label{Fig:AllPTDataCAk11Ave}
\end{figure}

\begin{figure}[h]
\includegraphics[width=3.3in]{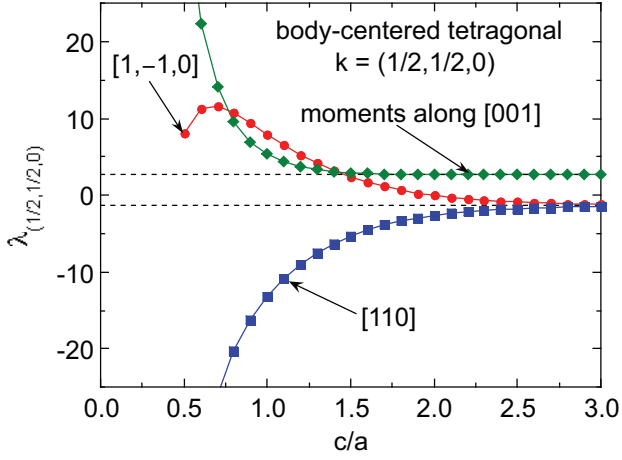}
\caption {(Color online) Eigenvalues for wavevector {\bf k} = (1/2,1/2,0)\,r.l.u.\ versus the $c/a$ ratio for a bct spin lattice with the moments aligned along [1,~$-1$,~0] (solid red circles), [001] ($c$~axis, solid green diamonds) or [110] (solid blue squares). }
\label{Fig:AllBCTDataCAk110Ave}
\end{figure}

\begin{figure}[h]
\includegraphics[width=3.3in]{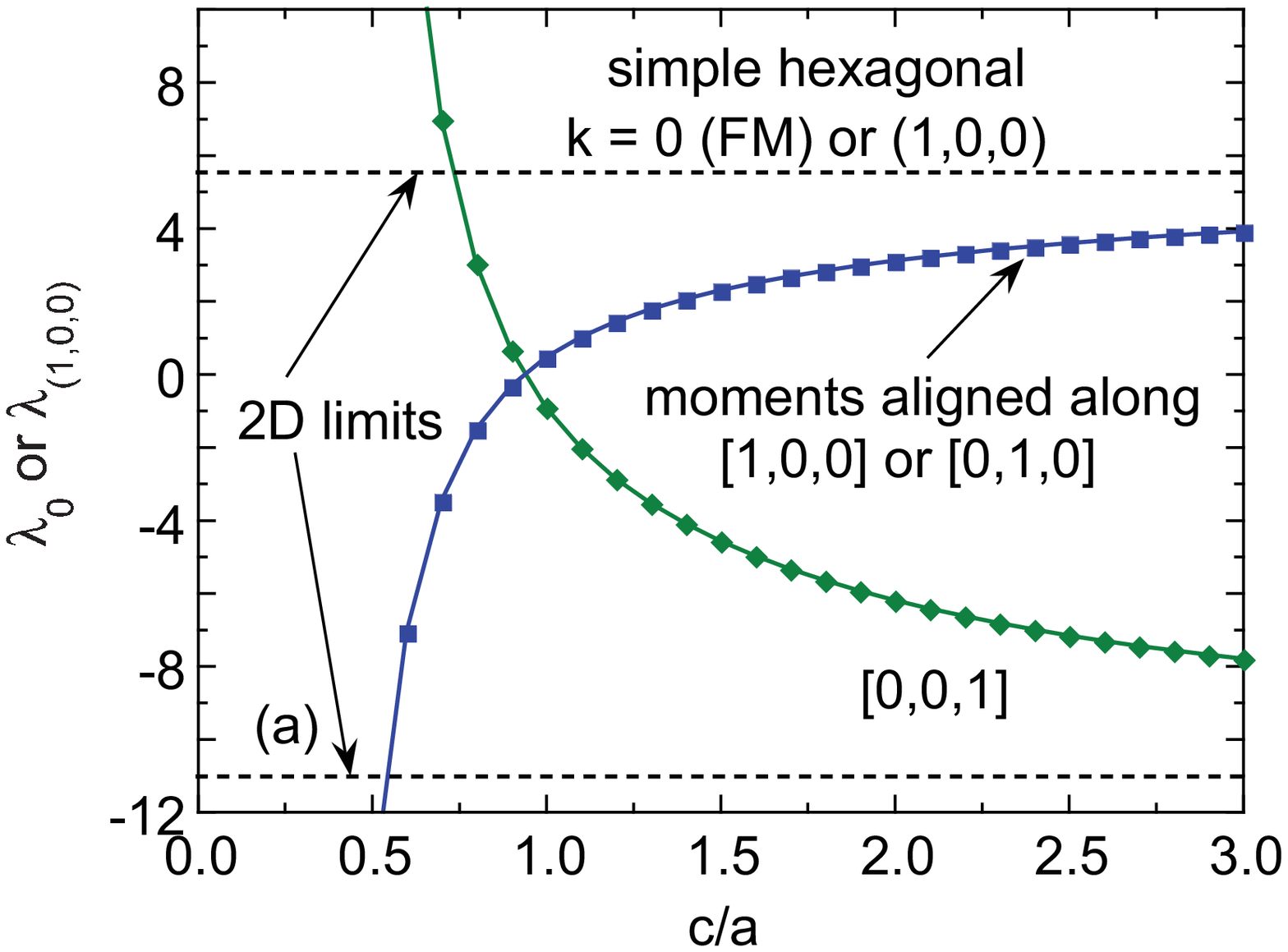}
\includegraphics[width=3.3in]{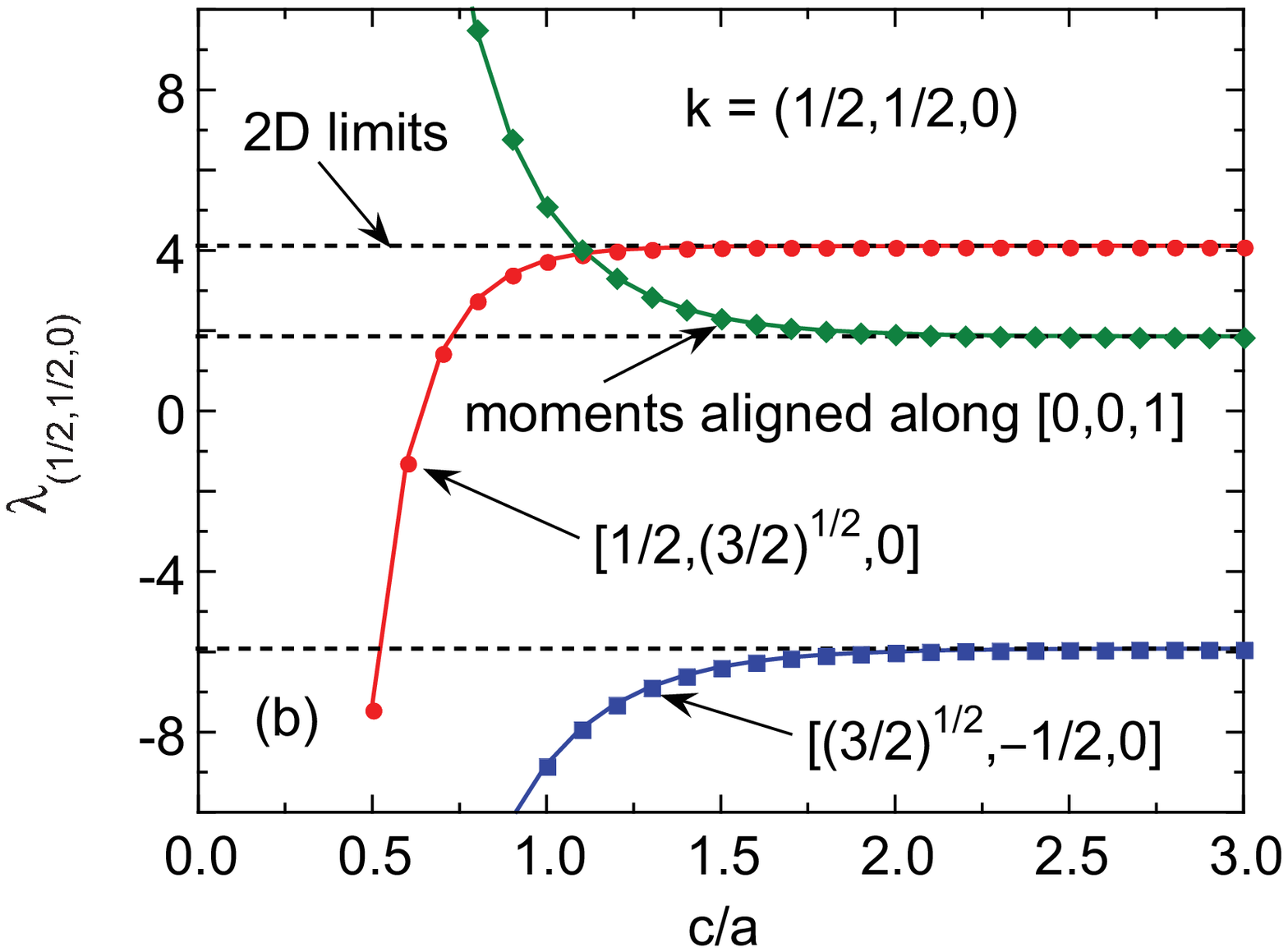}
\caption {(Color online) Eigenvalues for wavevectors (a) {\bf k} = 0 (FM) or (1,0,0) and (b) {\bf k} = (1/2,1/2,0)\,r.l.u.\ versus the $c/a$ ratio for a simple hexagonal (stacked triangular) spin lattice with the moments aligned along the indicated principal axes. The 2D limits of the respective eigenvalues for $c/a\to\infty$ are shown by horizontal dashed lines.}
\label{Fig:AllHexk000001DataAve}
\end{figure}

\begin{figure}[h]
\includegraphics[width=3.3in]{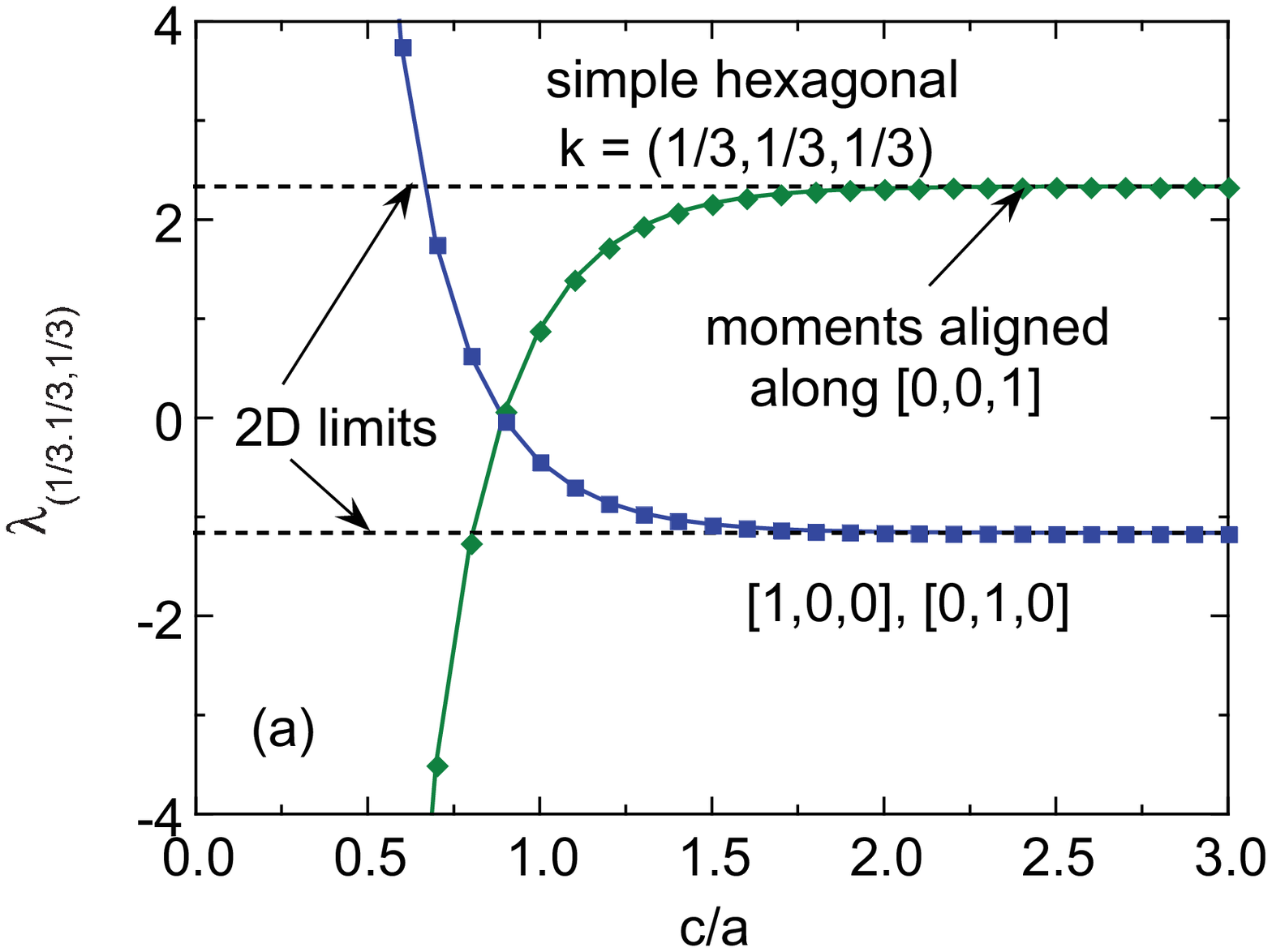}
\includegraphics[width=3.3in]{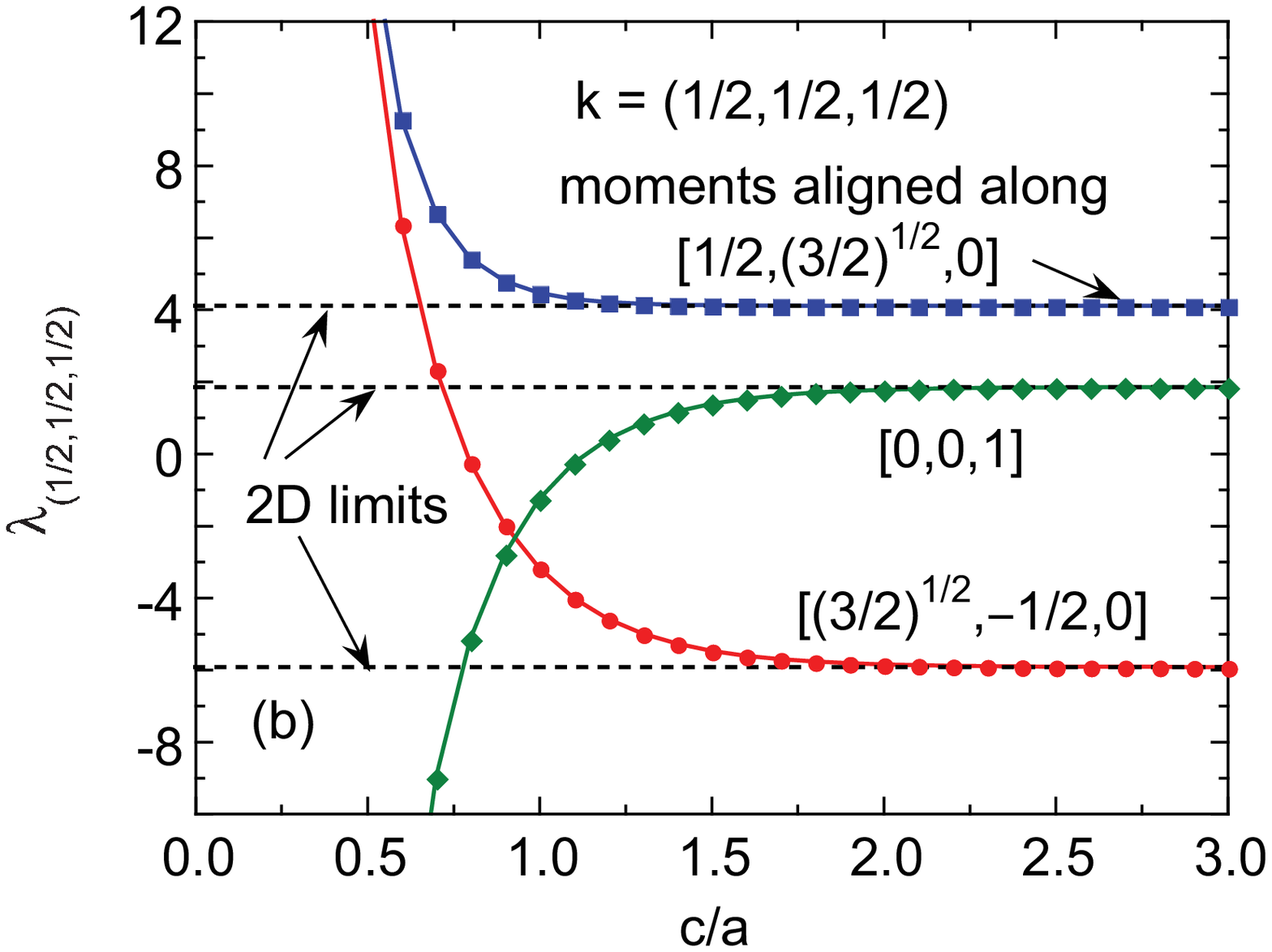}
\caption {(Color online) Eigenvalues for wavevectors (a) {\bf k} = (1/3,1/3,1/3) and (b) {\bf k} = (1/2,1/2,1/2)\,r.l.u.\ versus the $c/a$ ratio for a simple hexagonal (stacked triangular) spin lattice with the moments aligned along the indicated principal axes. The 2D limits of the respective eigenvalues for $c/a\to\infty$ are shown by horizontal dashed lines.}
\label{Fig:AllHexk131313DataAve}
\end{figure}

\begin{figure}[h]
\includegraphics[width=3.3in]{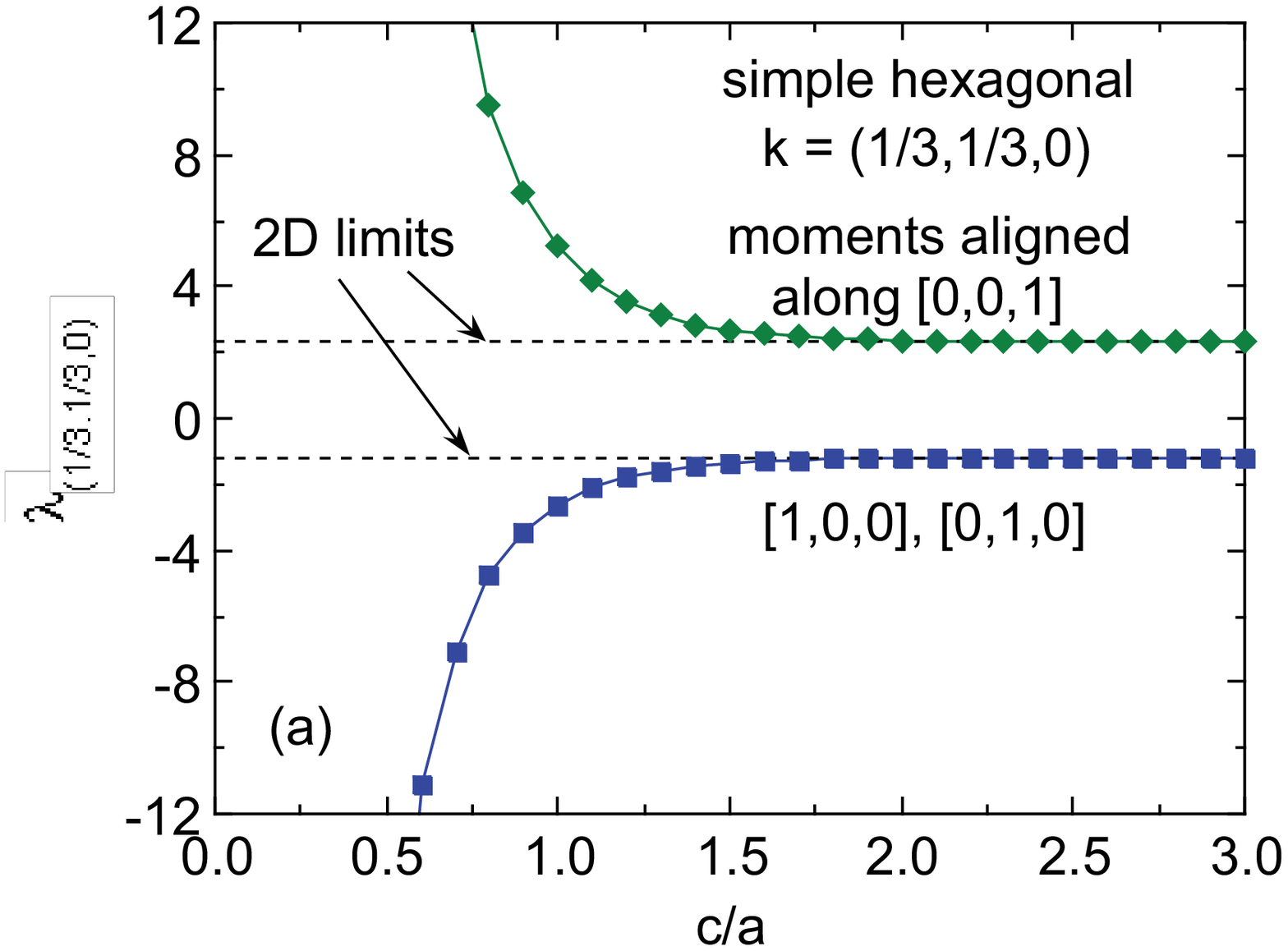}
\includegraphics[width=3.3in]{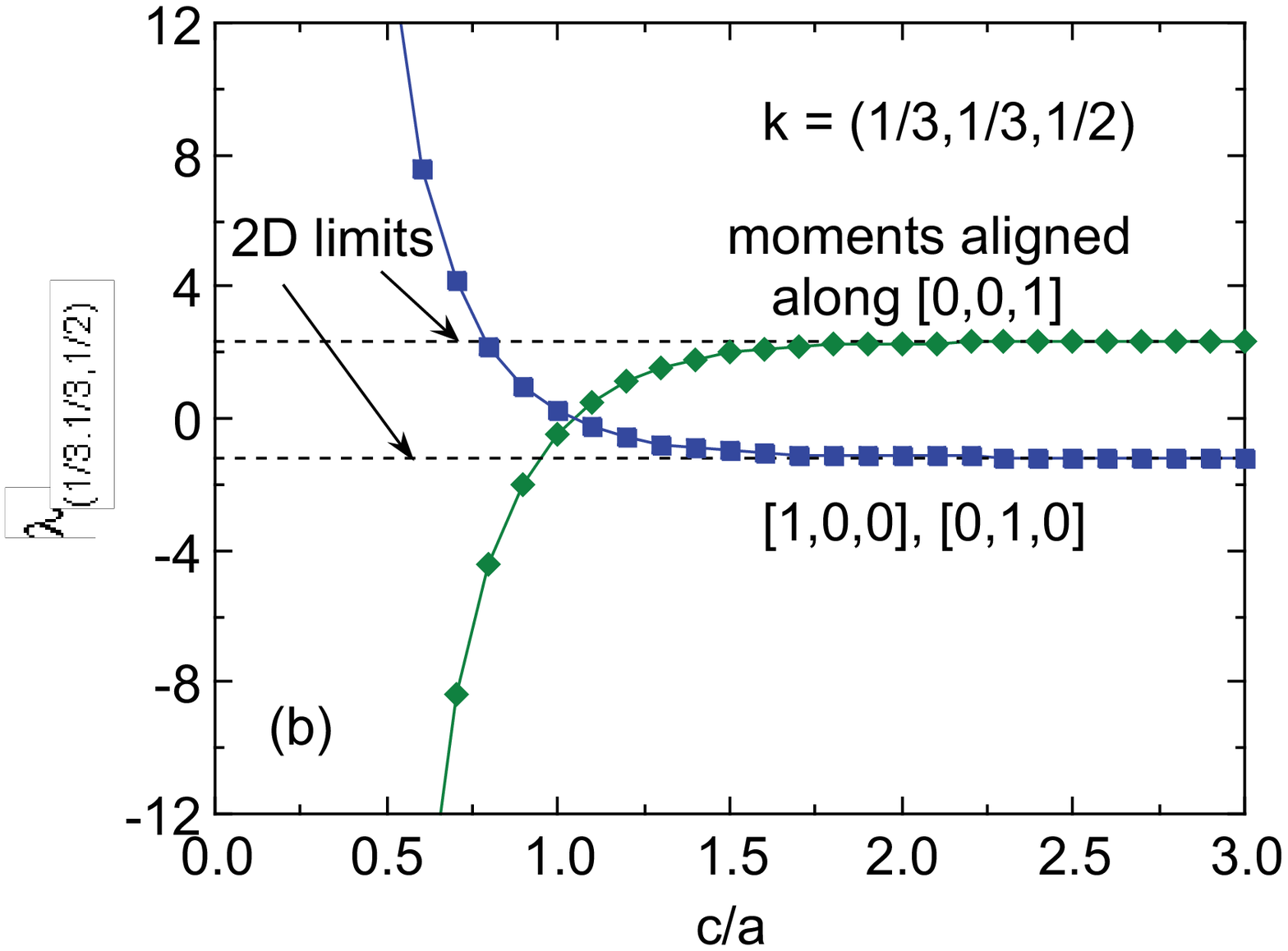}
\caption {(Color online) Eigenvalues for wavevectors (a) {\bf k} = (1/3,1/3,0) and (b) {\bf k} = (1/3,1/3,1/2)\,r.l.u.\ versus the $c/a$ ratio for a simple hexagonal (stacked triangular) spin lattice with the moments aligned along the indicated principal axes. The 2D limits of the respective eigenvalues for $c/a\to\infty$ are shown by horizontal dashed lines.}
\label{Fig:AllHexk13130DataAve}
\end{figure}

\begin{figure}[h]
\includegraphics[width=3.3in]{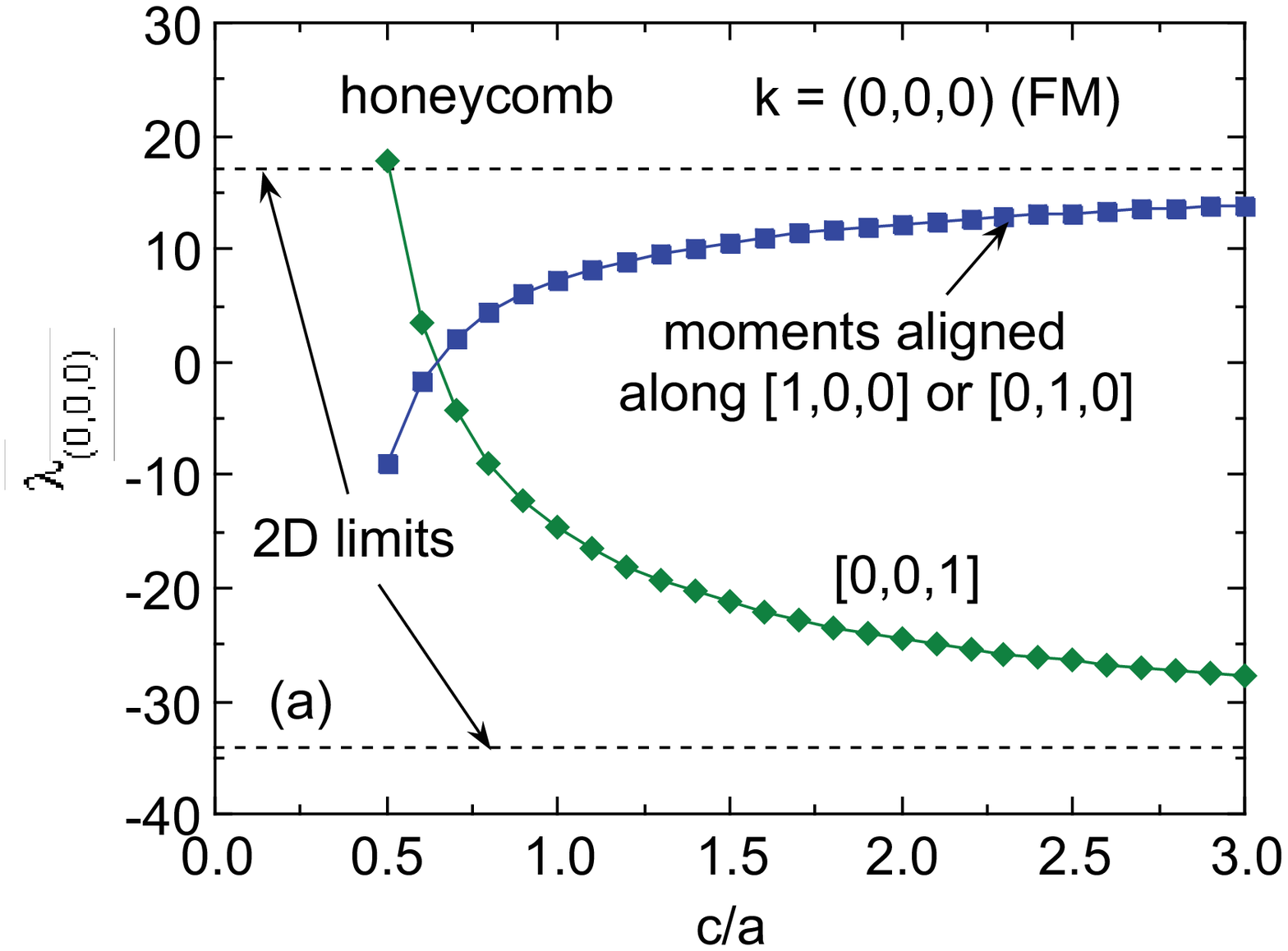}
\includegraphics[width=3.3in]{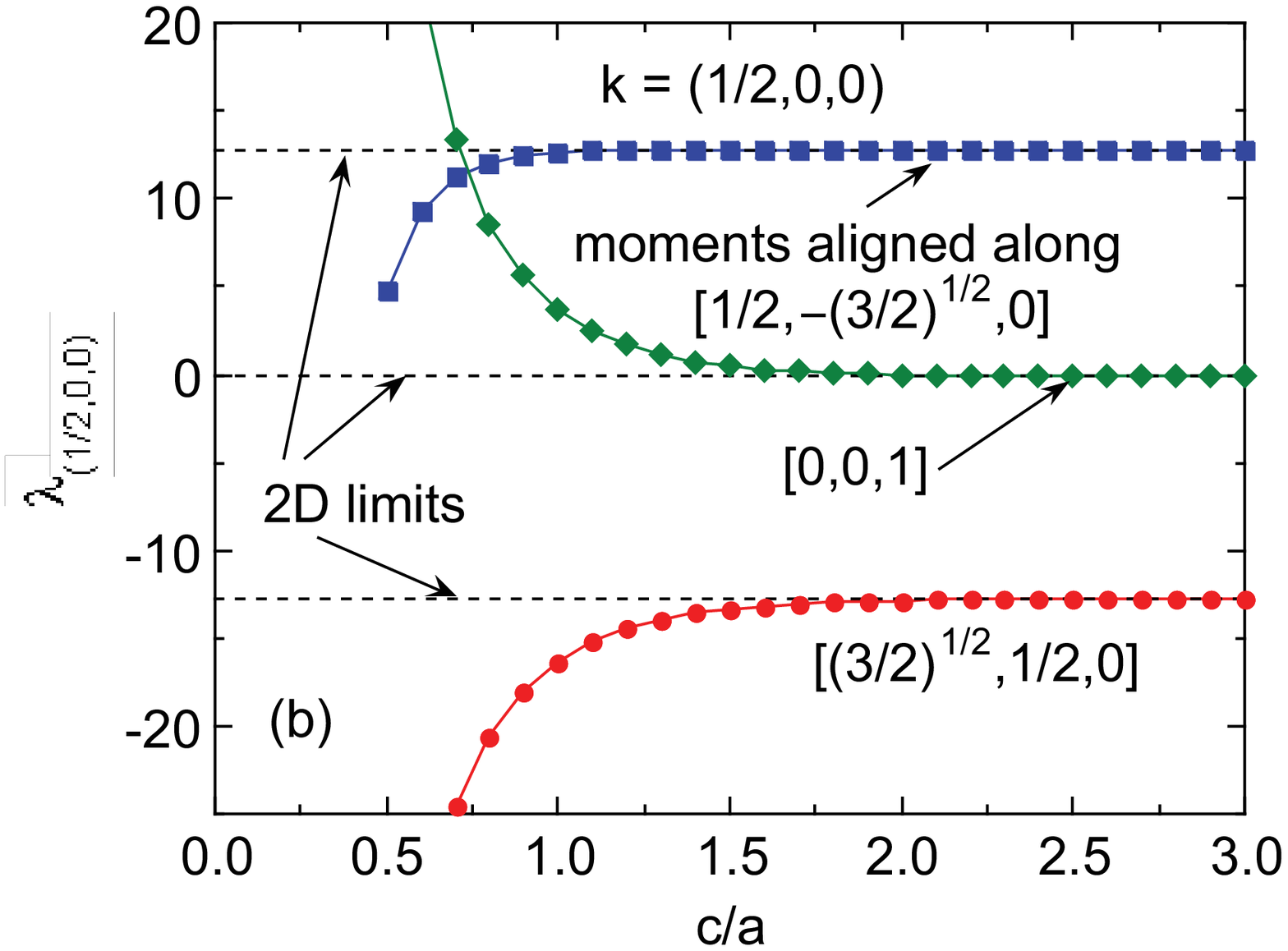}
\caption {(Color online) Eigenvalues for propagation vectors (a) {\bf k} = (0,0,0) (FM) and (b) {\bf k} = (1/2,0,0)\,r.l.u.\ versus the $c/a$ ratio for a honeycomb spin lattice with the moments aligned along the indicated principal axes. The 2D limits of the respective eigenvalues for $c/a\to\infty$ are shown by horizontal dashed lines.}
\label{Fig:All3DHoneyCo000_100AveLaTeX}
\end{figure} 

\begin{figure}[h]
\includegraphics[width=3.3in]{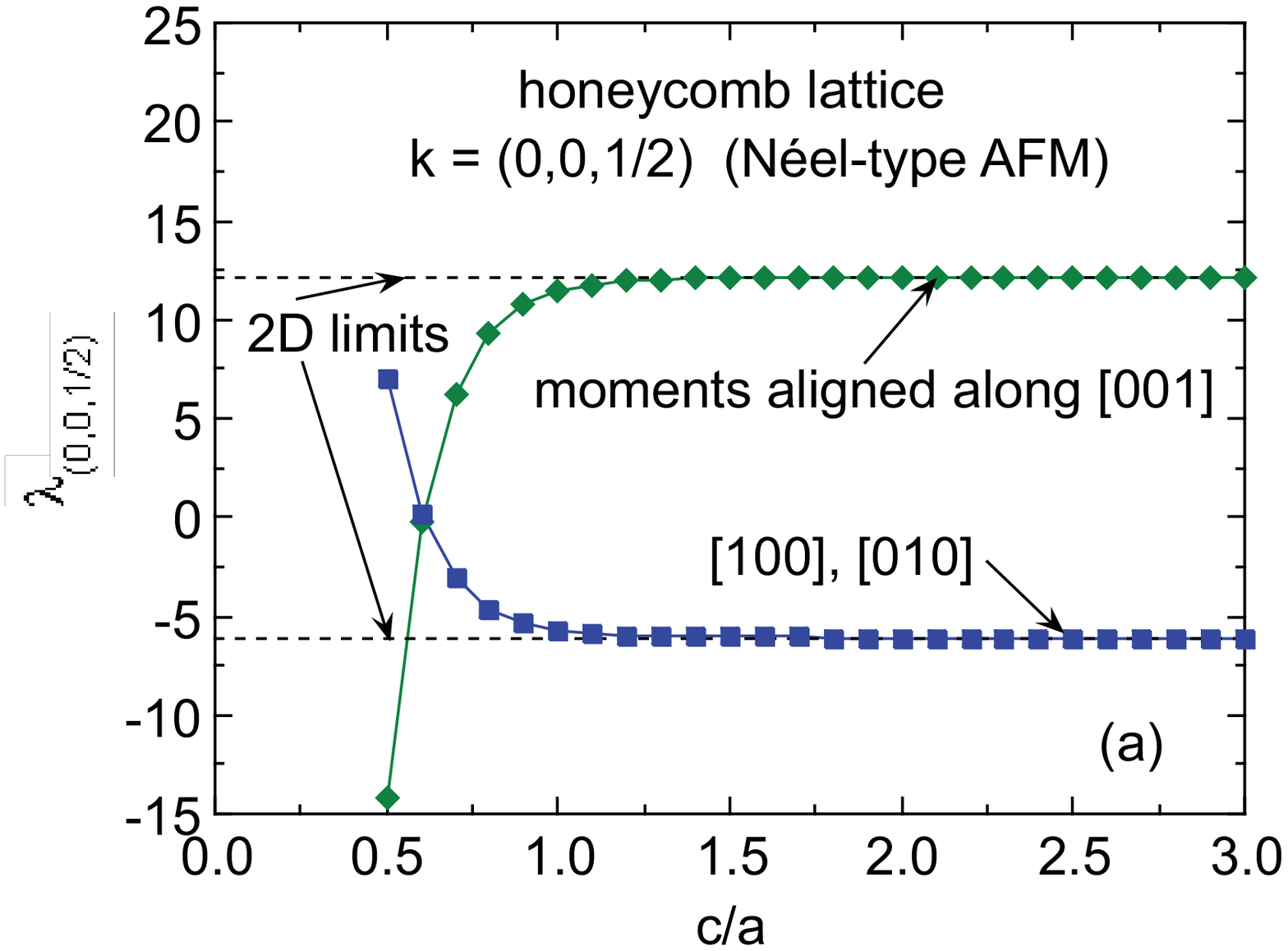}
\includegraphics[width=3.3in]{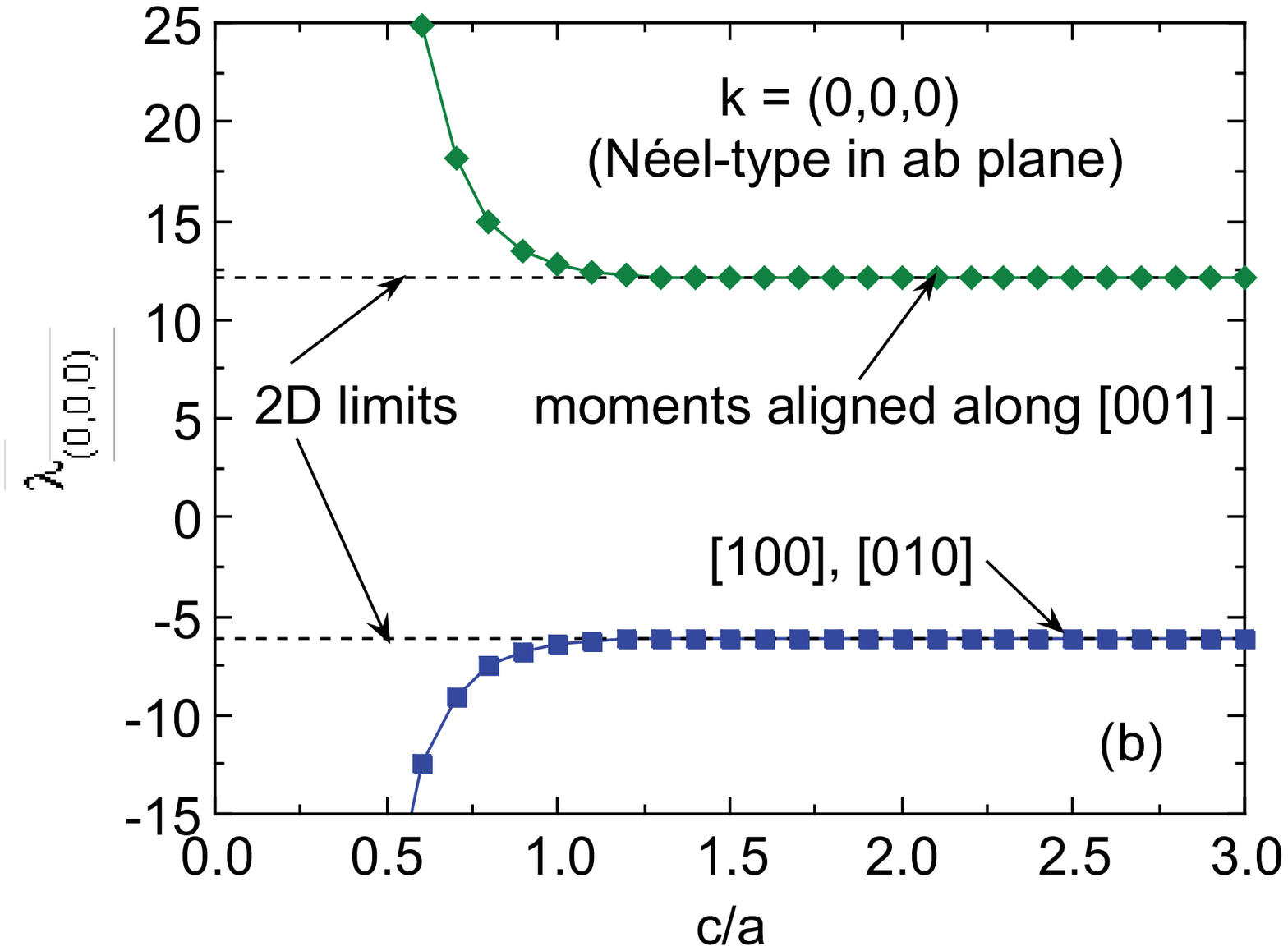}
\caption {(Color online) Eigenvalues for AFM propagation vectors (a) {\bf k} = (0,0,1/2) (N\'eel-type in all directions) and (b) {\bf k} = (0,0,0)\,r.l.u.\ (N\'eel-type in $ab$~plane and FM-alignment along $c$~axis) versus the $c/a$ ratio for a honeycomb spin lattice with the moments aligned along the indicated principal axes. The 2D limits of the respective eigenvalues for $c/a\to\infty$ are shown by horizontal dashed lines.}
\label{Fig:All3DHoneyComb3DNeelAve}
\end{figure}

\begin{figure}[h]
\includegraphics[width=3.3in]{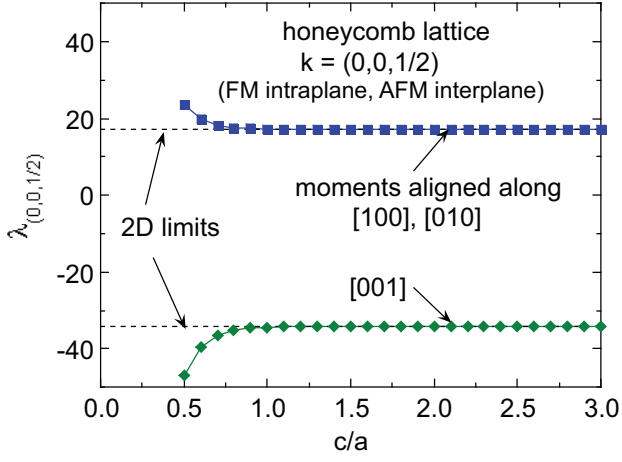}
\caption {(Color online) Eigenvalues for AFM propagation vectors {\bf k} = (0,0,1/2) (FM alignment intraplane and AFM alignment interplane) versus the $c/a$ ratio for a honeycomb spin lattice with the moments aligned along the indicated principal axes. The 2D  limits of the respective eigenvalues for $c/a\to\infty$ are shown by horizontal dashed lines.}
\label{Fig:All3DHoneyCombk0012Ave}
\end{figure}

\section*{\label{App:lambdaTabs} {\bf Supplementary Information}. Tables of Dipolar Eigenvectors and Eigenvalues versus the $c/a$ Ratio for Tetragaonal and Hexagonal Bravais Spin Lattices and for the Non-Bravais Honeycomb Lattice}

\begin{table}[h]
\caption{\label{Tab:FMSCEvecsEvals} {\bf Simple Tetragonal Spin Lattices with Ferromagnetic Alignment and k = (0,0,0).}  Eigenvalues $\lambda_{{\bf k}\alpha}$ and eigenvectors $\hat{\mu} = [\mu_x,\mu_y,\mu_z]$ in Cartesian coordinates of the magnetic dipole interaction tensor $\widehat{{\bf G}}_i({\bf k})$ in Eq.~(16c) for various values of the magnetic wavevector ${\bf k}$ in reciprocal lattice units (r.l.u.) for simple tetragonal spin lattices with {\bf collinear} magnetic moment alignments.  The most positive $\lambda_{{\bf k}\alpha}$ value(s) corresponds to the lowest energy value according to Eq.~(16d).  The Cartesian $x$, $y$ and $z$ axes are along the $a$, $b$ and $c$ axes of the tetragonal lattices, respectively. The accuracy of the values is estimated to be $\lesssim\pm 0.001$.  Also shown are the differences between the eigenvalues for different ordering axes for a given {\bf k}, which determine the anisotropy energies via Eq.~(16d).}
\begin{ruledtabular}
\begin{tabular}{cccc}
$c/a$ &		&	$\lambda_{(0,0,0)\alpha}$\\
	& [001]	&	[100], [010] & $[100]-[001]$ \\
\hline
0.5  &  30.0834  &  $-$15.0417  &  $-$45.1250  \\
0.6  &  15.2488  &  $-$7.6244  &  $-$22.8732  \\
0.7  &  7.9376  &  $-$3.9688  &  $-$11.9064  \\
0.8  &  3.9382  &  $-$1.9691  &  $-$5.9073  \\
0.9  &  1.5482  &  $-$0.7741  &  $-$2.3222  \\
1  &  0  &  0  &  0  \\
1.1  &  $-$1.0757  &  0.5378  &  1.6135  \\
1.2  &  $-$1.8728  &  0.9364  &  2.8091  \\
1.3  &  $-$2.4947  &  1.2474  &  3.7421  \\
1.4  &  $-$2.9994  &  1.4997  &  4.4991  \\
1.5  &  $-$3.4217  &  1.7109  &  5.1326  \\
1.6  &  $-$3.7830  &  1.8915  &  5.6745  \\
1.7  &  $-$4.0975  &  2.0487  &  6.1462  \\
1.8  &  $-$4.3750  &  2.1875  &  6.5625  \\
1.9  &  $-$4.6214  &  2.3107  &  6.9321  \\
2  &  $-$4.8436  &  2.4218  &  7.2653  \\
2.1  &  $-$5.0434  &  2.5217  &  7.5651  \\
2.2  &  $-$5.2249  &  2.6125  &  7.8374  \\
2.3  &  $-$5.3900  &  2.6950  &  8.0850  \\
2.4  &  $-$5.5425  &  2.7712  &  8.3137  \\
2.5  &  $-$5.6823  &  2.8411  &  8.5234  \\
2.6  &  $-$5.8110  &  2.9055  &  8.7165  \\
2.7  &  $-$5.9303  &  2.9652  &  8.8955  \\
2.8  &  $-$6.0413  &  3.0207  &  9.0620  \\
2.9  &  $-$6.1447  &  3.0724  &  9.2171  \\
3  &  $-$6.2410  &  3.1205  &  9.3616  \\
\end{tabular}
\end{ruledtabular}
\end{table}

\begin{table*}[h]
\caption{\label{Tab:AFMEvecsEvals1} {\bf Simple Tetragonal Antiferromagnetic Spin Lattices with k = (1/2,0,1/2) and (1/2,1/2,0)~r.l.u.}  The symbol descriptions are the same as in the caption to Table~\ref{Tab:FMSCEvecsEvals}. }
\begin{ruledtabular}
\begin{tabular}{ccccccccc}
 $\lambda_{{\bf k}\alpha}$	& $\left(\frac{1}{2},0,\frac{1}{2}\right)$ &   &  &  &	& $\left(\frac{1}{2},\frac{1}{2},0\right)$  \\
c/a	& [100] & [010] & [001] & [010]$-$[001] & [010]$-$[100] & [100], [010] & [001] & [001]$-$[100] \\
\hline
0.5  &  13.6119  &  15.1664  &  $-$28.7783  &  43.9447  &  1.5545  &  $-$19.236  &  38.4719  &  57.7079  \\
0.6  &  6.7759  &  9.7270  &  $-$16.5029  &  26.2299  &  2.9511  &  $-$11.1459  &  22.2918  &  33.4377  \\
0.7  &  2.8478  &  7.2967  &  $-$10.1445  &  17.4412  &  4.4489  &  $-$7.0563  &  14.1126  &  21.1689  \\
0.8  &  0.3134  &  6.1612  &  $-$6.4746  &  12.6358  &  5.8478  &  $-$4.7977  &  9.5955  &  14.3932  \\
0.9  &  $-$1.4328  &  5.6177  &  $-$4.1849  &  9.8026  &  7.0505  &  $-$3.4771  &  6.9543  &  10.4314  \\
1  &  $-$2.6767  &  5.3535  &  $-$2.6768  &  8.0303  &  8.0302  &  $-$2.6767  &  5.3535  &  8.0302  \\
1.1  &  $-$3.5784  &  5.2246  &  $-$1.6462  &  6.8708  &  8.8030  &  $-$2.1807  &  4.3614  &  6.5421  \\
1.2  &  $-$4.2360  &  5.1611  &  $-$0.9251  &  6.0862  &  9.3971  &  $-$1.8691  &  3.7381  &  5.6072  \\
1.3  &  $-$4.7177  &  5.1295  &  $-$0.4118  &  5.5413  &  9.8472  &  $-$1.6716  &  3.3431  &  5.0147  \\
1.4  &  $-$5.0709  &  5.1143  &  $-$0.0434  &  5.1577  &  10.1852  &  $-$1.5462  &  3.0925  &  4.6388  \\
1.5  &  $-$5.3294  &  5.1064  &  0.2230  &  4.8834  &  10.4358  &  $-$1.4655  &  2.9311  &  4.3966  \\
1.6  &  $-$5.5185  &  5.1026  &  0.4159  &  4.6867  &  10.6211  &  $-$1.4143  &  2.8286  &  4.2429  \\
1.7  &  $-$5.6572  &  5.1002  &  0.5570  &  4.5432  &  10.7574  &  $-$1.3815  &  2.7631  &  4.1446  \\
1.8  &  $-$5.7585  &  5.1001  &  0.6584  &  4.4417  &  10.8586  &  $-$1.3605  &  2.7209  &  4.0814  \\
1.9  &  $-$5.8328  &  5.0996  &  0.7332  &  4.3664  &  10.9324  &  $-$1.3473  &  2.6946  &  4.0419  \\
2  &  $-$5.8871  &  5.0991  &  0.7880  &  4.3111  &  10.9862  &  $-$1.3384  &  2.6767  &  4.0150  \\
2.1  &  $-$5.9269  &  5.0993  &  0.8276  &  4.2717  &  11.0262  &  $-$1.3328  &  2.6656  &  3.9984  \\
2.2  &  $-$5.9559  &  5.0992  &  0.8567  &  4.2425  &  11.0551  &  $-$1.3292  &  2.6585  &  3.9878  \\
2.3  &  $-$5.9771  &  5.0991  &  0.8780  &  4.2211  &  11.0762  &  $-$1.3269  &  2.6537  &  3.9806  \\
2.4  &  $-$5.9924  &  5.0992  &  0.8932  &  4.2060  &  11.0916  &  $-$1.3255  &  2.6511  &  3.9766  \\
2.5  &  $-$6.0038  &  5.0992  &  0.9046  &  4.1946  &  11.1030  &  $-$1.3247  &  2.6494  &  3.9741  \\
2.6  &  $-$6.0118  &  5.0988  &  0.9130  &  4.1858  &  11.1106  &  $-$1.3239  &  2.6478  &  3.9717  \\
2.7  &  $-$6.0180  &  5.0989  &  0.9191  &  4.1798  &  11.1169  &  $-$1.3235  &  2.6471  &  3.9706  \\
2.8  &  $-$6.0222  &  5.0990  &  0.9232  &  4.1758  &  11.1212  &  $-$1.3235  &  2.6471  &  3.9706  \\
2.9  &  $-$6.0256  &  5.0988  &  0.9268  &  4.1720  &  11.1244  &  $-$1.3234  &  2.6469  &  3.9703  \\
3  &  $-$6.0278  &  5.0989  &  0.9289  &  4.1700  &  11.1267  &  $-$1.3231  &  2.6461  &  3.9692  \\
\end{tabular}
\end{ruledtabular}
\end{table*}

\begin{table}[h]
\caption{\label{Tab:EvecsEvals2} {\bf Simple Tetragonal and Body-Centered Tetragonal Spin Lattice with k = (0,0,1/2)~r.l.u.}  The symbol descriptions are the same as in the caption to Table~\ref{Tab:FMSCEvecsEvals}.}
\begin{ruledtabular}
\begin{tabular}{cccc}
&& $\lambda_{{\bf k}\alpha}$	 \\
c/a	& [100], [010] & [001] & [100]$-$[001]  \\
\hline
0.5  &  15.0417  &  $-$30.0834  &  45.1251  \\
0.6  &  9.4972  &  $-$18.9943  &  28.4915  \\
0.7  &  6.9682  &  $-$13.9364  &  20.9046  \\
0.8  &  5.7521  &  $-$11.5042  &  17.2563  \\
0.9  &  5.1491  &  $-$10.2982  &  15.4473  \\
1  &  4.8436  &  $-$9.6871  &  14.5307  \\
1.1  &  4.6871  &  $-$9.3742  &  14.0613  \\
1.2  &  4.6061  &  $-$9.2122  &  13.8183  \\
1.3  &  4.5637  &  $-$9.1273  &  13.6910  \\
1.4  &  4.5416  &  $-$9.0833  &  13.6249  \\
1.5  &  4.5302  &  $-$9.0604  &  13.5906  \\
1.6  &  4.5238  &  $-$9.0476  &  13.5714  \\
1.7  &  4.5205  &  $-$9.0411  &  13.5616  \\
1.8  &  4.5187  &  $-$9.0373  &  13.5560  \\
1.9  &  4.5184  &  $-$9.0367  &  13.5550  \\
2  &  4.5173  &  $-$9.0346  &  13.5519  \\
2.1  &  4.5174  &  $-$9.0347  &  13.5521  \\
2.2  &  4.5168  &  $-$9.0337  &  13.5506  \\
2.3  &  4.5171  &  $-$9.0343  &  13.5514  \\
2.4  &  4.5163  &  $-$9.0325  &  13.5487  \\
2.5  &  4.5167  &  $-$9.0333  &  13.5500  \\
2.6  &  4.5168  &  $-$9.0337  &  13.5506  \\
2.7  &  4.5169  &  $-$9.0338  &  13.5507  \\
2.8  &  4.5167  &  $-$9.0333  &  13.5500  \\
2.9  &  4.5168  &  $-$9.0336  &  13.5504  \\
3  &  4.5175  &  $-$9.0351  &  13.5527  \\
\end{tabular}
\end{ruledtabular}
\end{table}

\begin{table*}[h]
\caption{\label{Tab:EvecsEvals3} {\bf Simple Tetragonal and Body-Centered Tetragonal Spin Lattices with k = (1/2,0,0) and (1/2,1/2,1/2)~r.l.u.}  The symbol descriptions are the same as in the caption to Table~\ref{Tab:FMSCEvecsEvals}.}
\begin{ruledtabular}
\begin{tabular}{ccccccccc}
 $\lambda_{{\bf k}\alpha}$	& $\left(\frac{1}{2},0,0\right)$ &   &  &  &	& $\left(\frac{1}{2},\frac{1}{2},\frac{1}{2}\right)$  \\
c/a	& [100] & [010] & [001] & [010]$-$[001] & [010]$-$[100] & [100], [010] & [001] & [001]$-$[100] \\
\hline
0.5  &  $-$32.9866  &  $-$5.4794  &  38.4660  &  $-$43.9454  &  27.5072  &  13.8825  &  $-$27.7650  &  $-$41.6475  \\
0.6  &  $-$22.6079  &  0.3470  &  22.2609  &  $-$21.9139  &  22.9549  &  7.4078  &  $-$14.8156  &  $-$22.2234  \\
0.7  &  $-$16.8911  &  2.8706  &  14.0205  &  $-$11.1499  &  19.7617  &  3.9638  &  $-$7.9276  &  $-$11.8914  \\
0.8  &  $-$13.4266  &  4.0289  &  9.3977  &  $-$5.3688  &  17.4555  &  1.9691  &  $-$3.9383  &  $-$5.9074  \\
0.9  &  $-$11.1910  &  4.5782  &  6.6128  &  $-$2.0346  &  15.7692  &  0.7562  &  $-$1.5124  &  $-$2.2686  \\
1  &  $-$9.6874  &  4.8437  &  4.8437  &  0.0000  &  14.5311  &  0.0000  &  0.0000  &  0.0000  \\
1.1  &  $-$8.6477  &  4.9729  &  3.6748  &  1.2981  &  13.6206  &  $-$0.4778  &  0.9556  &  1.4334  \\
1.2  &  $-$7.9160  &  5.0370  &  2.8790  &  2.1580  &  12.9530  &  $-$0.7821  &  1.5641  &  2.3461  \\
1.3  &  $-$7.3954  &  5.0683  &  2.3271  &  2.7412  &  12.4637  &  $-$0.9763  &  1.9526  &  2.9289  \\
1.4  &  $-$7.0217  &  5.0837  &  1.9380  &  3.1457  &  12.1054  &  $-$1.1009  &  2.2018  &  3.3027  \\
1.5  &  $-$6.7521  &  5.0915  &  1.6606  &  3.4309  &  11.8436  &  $-$1.1805  &  2.3611  &  3.5416  \\
1.6  &  $-$6.5566  &  5.0950  &  1.4616  &  3.6334  &  11.6516  &  $-$1.2315  &  2.4630  &  3.6945  \\
1.7  &  $-$6.4150  &  5.0969  &  1.3181  &  3.7788  &  11.5119  &  $-$1.2644  &  2.5287  &  3.7931  \\
1.8  &  $-$6.3119  &  5.0980  &  1.2139  &  3.8841  &  11.4099  &  $-$1.2852  &  2.5704  &  3.8556  \\
1.9  &  $-$6.2367  &  5.0986  &  1.1381  &  3.9605  &  11.3353  &  $-$1.2987  &  2.5974  &  3.8961  \\
2  &  $-$6.1821  &  5.0986  &  1.0835  &  4.0151  &  11.2807  &  $-$1.3074  &  2.6147  &  3.9220  \\
2.1  &  $-$6.1421  &  5.0985  &  1.0436  &  4.0549  &  11.2406  &  $-$1.3132  &  2.6263  &  3.9395  \\
2.2  &  $-$6.1131  &  5.0988  &  1.0143  &  4.0845  &  11.2119  &  $-$1.3169  &  2.6339  &  3.9509  \\
2.3  &  $-$6.0917  &  5.0988  &  0.9929  &  4.1059  &  11.1905  &  $-$1.3188  &  2.6376  &  3.9564  \\
2.4  &  $-$6.0760  &  5.0990  &  0.9770  &  4.1220  &  11.1750  &  $-$1.3201  &  2.6402  &  3.9603  \\
2.5  &  $-$6.0652  &  5.0992  &  0.9660  &  4.1332  &  11.1644  &  $-$1.3213  &  2.6426  &  3.9639  \\
2.6  &  $-$6.0566  &  5.0989  &  0.9577  &  4.1412  &  11.1555  &  $-$1.3218  &  2.6437  &  3.9655  \\
2.7  &  $-$6.0507  &  5.0988  &  0.9519  &  4.1469  &  11.1495  &  $-$1.3222  &  2.6444  &  3.9666  \\
2.8  &  $-$6.0463  &  5.0990  &  0.9473  &  4.1517  &  11.1453  &  $-$1.3223  &  2.6445  &  3.9668  \\
2.9  &  $-$6.0430  &  5.0988  &  0.9442  &  4.1546  &  11.1418  &  $-$1.3226  &  2.6452  &  3.9678  \\
3  &  $-$6.0408  &  5.0990  &  0.9418  &  4.1572  &  11.1398  &  $-$1.3227  &  2.6455  &  3.9682  \\
\end{tabular}
\end{ruledtabular}
\end{table*}

\begin{table}[h]
\caption{\label{Tab:EvecsEvals2B} {\bf Body-Centered Tetragonal Spin Lattice with ferromagnetic moment alignment and {\bf k} = (0,0,0)~r.l.u.}  The symbol descriptions are the same as in the caption to Table~\ref{Tab:FMSCEvecsEvals}.}
\begin{ruledtabular}
\begin{tabular}{crrr}
&& $\lambda_{{\bf k}\alpha}$	 \\
c/a	& [001] & [100], [010] & [100]$-$[001]  \\
\hline
0.50  &  21.9948  &  $-$10.9974  &  $-$32.9923  \\
0.60  &  9.0718  &  $-$4.5359  &  $-$13.6077  \\
0.70  &  3.5213  &  $-$1.7607  &  $-$5.2820  \\
0.80  &  1.1666  &  $-$0.5833  &  $-$1.7499  \\
0.90  &  0.2635  &  $-$0.1317  &  $-$0.3952  \\
0.93  &  0.1405  &  $-$0.0702  &  $-$0.2107  \\
0.95  &  0.0825  &  $-$0.0412  &  $-$0.1237  \\
0.97  &  0.0404  &  $-$0.0202  &  $-$0.0606  \\
0.99  &  0.0105  &  $-$0.0052  &  $-$0.0157  \\
1  &  0  		&  0		 &  0  \\
1.01  &  $-$0.0083  &  0.0042  &  0.0125  \\
1.03  &  $-$0.0182  &  0.0091  &  0.0273  \\
1.05  &  $-$0.0231  &  0.0115  &  0.0346  \\
1.07  &  $-$0.0219  &  0.0109  &  0.0328  \\
1.09  &  $-$0.0170  &  0.0085  &  0.0255  \\
1.10  &  $-$0.0143  &  0.0071  &  0.0214  \\
1.11  &  $-$0.0098  &  0.0049  &  0.0147  \\
1.13  &  $-$0.0015  &  0.0007  &  0.0022  \\
1.15  &  0.0084  &  $-$0.0042  &  $-$0.0127  \\
1.17  &  0.0179  &  $-$0.0089  &  $-$0.0268  \\
1.19  &  0.0269  &  $-$0.0135  &  $-$0.0404  \\
1.20  &  0.0307  &  $-$0.0154  &  $-$0.0461  \\
1.21  &  0.0346  &  $-$0.0173  &  $-$0.0520  \\
1.23  &  0.0413  &  $-$0.0206  &  $-$0.0619  \\
1.25  &  0.0463  &  $-$0.0232  &  $-$0.0695  \\
1.27  &  0.0493  &  $-$0.0246  &  $-$0.0739  \\
1.29  &  0.0497  &  $-$0.0248  &  $-$0.0745  \\
1.30  &  0.0491  &  $-$0.0246  &  $-$0.0737  \\
1.31  &  0.0489  &  $-$0.0244  &  $-$0.0733  \\
1.33  &  0.0448  &  $-$0.0224  &  $-$0.0672  \\
1.35  &  0.0379  &  $-$0.0190  &  $-$0.0569  \\
1.37  &  0.0294  &  $-$0.0147  &  $-$0.0440  \\
1.39  &  0.0175  &  $-$0.0088  &  $-$0.0263  \\
1.40  &  0.0110  &  $-$0.0055  &  $-$0.0165  \\
1.41  &  0.0039  &  $-$0.0019  &  $-$0.0058  \\
$\sqrt{2}$  &  0  &  0  &  0  \\
1.43  &  $-$0.0122  &  0.0061  &  0.0183  \\
1.45  &  $-$0.0324  &  0.0162  &  0.0486  \\
1.47  &  $-$0.0535  &  0.0268  &  0.0803  \\
1.49  &  $-$0.0764  &  0.0382  &  0.1146  \\
1.50  &  $-$0.0896  &  0.0448  &  0.1344  \\
1.60  &  $-$0.2452  &  0.1226  &  0.3678  \\
1.70  &  $-$0.4435  &  0.2217  &  0.6652  \\
1.80  &  $-$0.6731  &  0.3366  &  1.0097  \\
1.90  &  $-$0.9220  &  0.4610  &  1.3830  \\
2.00  &  $-$1.1823  &  0.5911  &  1.7734  \\
2.10  &  $-$1.4452  &  0.7226  &  2.1678  \\
2.20  &  $-$1.7061  &  0.8530  &  2.5591  \\
2.30  &  $-$1.9611  &  0.9806  &  2.9417  \\
2.40  &  $-$2.2097  &  1.1048  &  3.3145  \\
2.50  &  $-$2.4475  &  1.2238  &  3.6713  \\
2.60  &  $-$2.6739  &  1.3369  &  4.0108  \\
2.70  &  $-$2.8900  &  1.4450  &  4.3349  \\
2.80  &  $-$3.0954  &  1.5477  &  4.6431  \\
2.90  &  $-$3.2898  &  1.6449  &  4.9348  \\
3.00  &  $-$3.4735  &  1.7367  &  5.2102  \\
\end{tabular}
\end{ruledtabular}
\end{table}

\begin{table*}[h]
\caption{\label{Tab:BCTEvecsEvals1} {\bf Body-Centered Tetragonal Spin Lattices with {\bf k} = (1/2,0,1/2)~r.l.u.}   Eigenvalues $\lambda_{{\bf k}\alpha}$ and eigenvectors $\hat{\mu} = [\mu_x,\mu_y,\mu_z]$ in Cartesian coordinates of the magnetic dipole interaction tensor $\widehat{{\bf G}}_i({\bf k})$ in Eq.~(16c)   with {\bf collinear} magnetic moment alignments.  The accuracy of the eigenvalues is estimated to be $\lesssim\pm 0.001$.  Also shown are two differences between the eigenvalues for the three different ordering axes.}
\begin{ruledtabular}
\begin{tabular}{ccccccc}
 $\hat{\mu}$	&   & [010] & $[\sqrt{1-x^2},0,-x]$  & $[x,0,\sqrt{1-x^2}]$ & $[010]-[\sqrt{1-x^2},0,-x]$ & $[010]-[x,0,\sqrt{1-x^2}]$	\\
c/a	& $x$ &  &  & &  &  \\
\hline
0.5  &  0.15297  &  15.1665  &  14.6525  &  $-$29.8189  &  0.5140  &  44.9854  \\
0.6  &  0.32332  &  9.7270  &  9.8527  &  $-$19.5797  &  $-$0.1257  &  29.3067  \\
0.7  &  0.48513  &  7.2967  &  8.6251  &  $-$15.9218  &  $-$1.3284  &  23.2185  \\
0.8  &  0.59407  &  6.1612  &  8.4577  &  $-$14.6189  &  $-$2.2965  &  20.7801  \\
0.9  &  0.66190  &  5.6177  &  8.3111  &  $-$13.9288  &  $-$2.6934  &  19.5465  \\
1  &  0.70711  &  5.3535  &  7.9437  &  $-$13.2971  &  $-$2.5902  &  18.6506  \\
1.1  &  0.74051  &  5.2246  &  7.3770  &  $-$12.6016  &  $-$2.1524  &  17.8262  \\
1.2  &  0.76766  &  5.1611  &  6.6884  &  $-$11.8495  &  $-$1.5273  &  17.0106  \\
1.3  &  0.79145  &  5.1295  &  5.9492  &  $-$11.0787  &  $-$0.8197  &  16.2082  \\
1.4  &  0.81346  &  5.1143  &  5.2144  &  $-$10.3287  &  $-$0.1001  &  15.4430  \\
$\sqrt{2}$ & 0.81650 & 5.1127 & 5.1127 & $-10.2254$  & 0 & 15.3381\\
1.5  &  0.83442  &  5.1064  &  4.5196  &  $-$9.6260  &  0.5868  &  14.7324  \\
1.6  &  0.85465  &  5.1026  &  3.8873  &  $-$8.9899  &  1.2153  &  14.0925  \\
1.7  &  0.87419  &  5.1002  &  3.3301  &  $-$8.4303  &  1.7701  &  13.5305  \\
1.8  &  0.89285  &  5.1001  &  2.8480  &  $-$7.9481  &  2.2521  &  13.0482  \\
1.9  &  0.91035  &  5.0996  &  2.4435  &  $-$7.5431  &  2.6561  &  12.6427  \\
2  &  0.92634  &  5.0991  &  2.1105  &  $-$7.2096  &  2.9886  &  12.3087  \\
2.1  &  0.94057  &  5.0993  &  1.8402  &  $-$6.9395  &  3.2591  &  12.0388  \\
2.2  &  0.95284  &  5.0992  &  1.6258  &  $-$6.7250  &  3.4734  &  11.8242  \\
2.3  &  0.96316  &  5.0991  &  1.4576  &  $-$6.5567  &  3.6415  &  11.6558  \\
2.4  &  0.97160  &  5.0992  &  1.3273  &  $-$6.4266  &  3.7719  &  11.5258  \\
2.5  &  0.97836  &  5.0992  &  1.2281  &  $-$6.3273  &  3.8711  &  11.4265  \\
2.6  &  0.98367  &  5.0988  &  1.1528  &  $-$6.2517  &  3.9460  &  11.3505  \\
2.7  &  0.98776  &  5.0989  &  1.0965  &  $-$6.1954  &  4.0024  &  11.2943  \\
2.8  &  0.99088  &  5.0990  &  1.0540  &  $-$6.1530  &  4.0450  &  11.2520  \\
2.9  &  0.99324  &  5.0988  &  1.0231  &  $-$6.1220  &  4.0757  &  11.2208  \\
3  &  0.99501  &  5.0989  &  0.9996  &  $-$6.0985  &  4.0993  &  11.1974  \\
\end{tabular}
\end{ruledtabular}
\end{table*}

\begin{table*}[h]
\caption{\label{BCTEvecsEvals2} {\bf Body-Centered Tetragonal Spin Lattices with ${\bf k} = $ (1/2,1/2,0) and (0,0,1)~r.l.u.} Eigenvalues $\lambda_{{\bf k}\alpha}$ and eigenvectors $\hat{\mu} = [\mu_x,\mu_y,\mu_z]$ in Cartesian coordinates of the magnetic dipole interaction tensor $\widehat{{\bf G}}_i({\bf k})$ in Eq.~(16c) for two values of the magnetic wavevector ${\bf k}$ in reciprocal lattice units (r.l.u.) and {\bf collinear} magnetic moment alignments.  The accuracy of the eigenvalues is estimated to be $\lesssim\pm 0.001$.  Also shown are the differences between the eigenvalues for different ordering axes for a given {\bf k}.}
\begin{ruledtabular}
\begin{tabular}{ccccccccc}
 $\lambda_{{\bf k}\alpha}$	& $\left(\frac{1}{2},\frac{1}{2},0\right)$ &   &  &  &	& (0,0,1) &  &  \\
c/a	& [001] & $[1\bar{1}0]$ & [110] & [001]$-[1\bar{1}0]$ & [001]$-$[110] & [100],[010] & [001] & [100]$-$[001] \\
\hline
0.5  &  38.4705  &  8.0868  &  $-$46.5573  &  30.3837  &  85.0278  &  $-$19.086  &  38.172  &  $-$57.258  \\
0.6  &  22.2918  &  11.2533  &  $-$33.5451  &  11.0385  &  55.8369  &  $-$10.713  &  21.426  &  $-$32.139  \\
0.7  &  14.1126  &  11.5369  &  $-$25.6495  &  2.5757  &  39.7621  &  $-$6.177  &  12.354  &  $-$18.531  \\
0.8  &  9.5955  &  10.6756  &  $-$20.2711  &  $-$1.0801  &  29.8666  &  $-$3.355  &  6.710  &  $-$10.065  \\
0.9  &  6.9543  &  9.3705  &  $-$16.3248  &  $-$2.4162  &  23.2791  &  $-$1.416  &  2.833  &  $-$4.249  \\
1  &  5.3534  &  7.9437  &  $-$13.2971  &  $-$2.5903  &  18.6505  &  0.000  &  0.000  &  0.000  \\
1.1  &  4.3614  &  6.5552  &  $-$10.9166  &  $-$2.1938  &  15.2780  &  1.069  &  $-$2.137  &  3.206  \\
1.2  &  3.7381  &  5.2829  &  $-$9.0210  &  $-$1.5448  &  12.7591  &  1.888  &  $-$3.776  &  5.664  \\
1.3  &  3.3431  &  4.1586  &  $-$7.5017  &  $-$0.8155  &  10.8448  &  2.519  &  $-$5.039  &  7.558  \\
1.4  &  3.0925  &  3.1897  &  $-$6.2822  &  $-$0.0972  &  9.3747  &  3.005  &  $-$6.010  &  9.015  \\
$\sqrt{2}$  &  3.0640  &  3.0640  &  $-$6.1280  &  0  &  9.1920  &  3.064  &  $-$6.128  &  9.192  \\
1.5  &  2.9311  &  2.3693  &  $-$5.3004  &  0.5618  &  8.2315  &  3.377  &  $-$6.754  &  10.131  \\
1.6  &  2.8286  &  1.6843  &  $-$4.5129  &  1.1443  &  7.3415  &  3.660  &  $-$7.321  &  10.981  \\
1.7  &  2.7627  &  1.1176  &  $-$3.8803  &  1.6451  &  6.6430  &  3.876  &  $-$7.752  &  11.627  \\
1.8  &  2.7209  &  0.6513  &  $-$3.3722  &  2.0696  &  6.0931  &  4.038  &  $-$8.077  &  12.115  \\
1.9  &  2.6946  &  0.2700  &  $-$2.9646  &  2.4246  &  5.6592  &  4.160  &  $-$8.321  &  12.481  \\
2  &  2.6765  &  $-$0.0384  &  $-$2.6381  &  2.7149  &  5.3146  &  4.252  &  $-$8.505  &  12.757  \\
2.1  &  2.6656  &  $-$0.2896  &  $-$2.3760  &  2.9552  &  5.0416  &  4.321  &  $-$8.642  &  12.962  \\
2.2  &  2.6585  &  $-$0.4921  &  $-$2.1664  &  3.1506  &  4.8249  &  4.372  &  $-$8.744  &  13.116  \\
2.3  &  2.6537  &  $-$0.6555  &  $-$1.9982  &  3.3092  &  4.6519  &  4.409  &  $-$8.819  &  13.228  \\
2.4  &  2.6511  &  $-$0.7876  &  $-$1.8635  &  3.4387  &  4.5146  &  4.438  &  $-$8.875  &  13.313  \\
2.5  &  2.6494  &  $-$0.8938  &  $-$1.7556  &  3.5432  &  4.4050  &  4.459  &  $-$8.917  &  13.376  \\
2.6  &  2.6478  &  $-$0.9783  &  $-$1.6695  &  3.6261  &  4.3173  &  4.474  &  $-$8.948  &  13.422  \\
2.7  &  2.6471  &  $-$1.0462  &  $-$1.6009  &  3.6933  &  4.2480  &  4.485  &  $-$8.971  &  13.456  \\
2.8  &  2.6471  &  $-$1.1014  &  $-$1.5457  &  3.7485  &  4.1928  &  4.494  &  $-$8.987  &  13.481  \\
2.9  &  2.6469  &  $-$1.1461  &  $-$1.5008  &  3.7930  &  4.1477  &  4.500  &  $-$9.000  &  13.499  \\
3  &  2.6461  &  $-$1.1803  &  $-$1.4658  &  3.8264  &  4.1119  &  4.504  &  $-$9.009  &  13.513  \\
\end{tabular}
\end{ruledtabular}
\end{table*}
 
\begin{table*}[h]
\caption{\label{HexEvecsEvals1} {\bf Simple 3D Hexagonal (Stacked Triangular) Spin Lattices with {\bf k} = (1/2,1/2,0) and (0,0,0) or (1,0,0)~r.l.u.} Eigenvalues $\lambda_{{\bf k}\alpha}$ and eigenvectors $\hat{\mu} = [\mu_x,\mu_y,\mu_z]$ in Cartesian coordinates of the magnetic dipole interaction tensor $\widehat{{\bf G}}_i({\bf k})$ in Eq.~(16c) for two values of the magnetic wavevector ${\bf k}$ in reciprocal lattice units (r.l.u.) and {\bf collinear} magnetic moment alignments.  The $x$~axis is in the direction of the hexagonal $a$~axis.  The accuracy of the eigenvalues is estimated to be $\lesssim\pm 0.001$.  Also shown are the differences between the eigenvalues for different ordering axes for a given {\bf k}.}
\begin{ruledtabular}
\begin{tabular}{ccccccccc}
 $\lambda_{{\bf k}\alpha}$	& $\left(\frac{1}{2},\frac{1}{2},0\right)$ &   &   & $\left[\frac{1}{2},\frac{\sqrt{3}}{2},0\right]$ &	$\left[\frac{1}{2},\frac{\sqrt{3}}{2},0\right]$	& (0,0,0), (1,0,0) &  &  \\
c/a	& [001] & $\left[\frac{\sqrt{3}}{2},-\frac{1}{2},0\right]$ & $\left[\frac{1}{2},\frac{\sqrt{3}}{2},0\right]$ & $-[001]$ & $-\left[\frac{\sqrt{3}}{2},-\frac{1}{2},0\right]$ & [001] & [100], [010]  & $[100]-[001]$ \\
\hline
0.5  &  38.4698  &  $-$31.0341  &  $-$7.4357  &  $-$45.9055  &  23.5983  &  28.7829  &  $-$14.3914  &  $-$43.1743  \\
0.6  &  22.2769  &  $-$20.9871  &  $-$1.2899  &  $-$23.5668  &  19.6972  &  14.1513  &  $-$7.0757  &  $-$21.2270  \\
0.7  &  14.0669  &  $-$15.5182  &  1.4513  &  $-$12.6156  &  16.9695  &  6.9640  &  $-$3.4820  &  $-$10.4460  \\
0.8  &  9.4972  &  $-$12.2545  &  2.7573  &  $-$6.7399  &  15.0118  &  3.0273  &  $-$1.5137  &  $-$4.5410  \\
0.9  &  6.7848  &  $-$10.1901  &  3.4053  &  $-$3.3795  &  13.5953  &  0.6542  &  $-$0.3271  &  $-$0.9813  \\
1  &  5.1009  &  $-$8.8361  &  3.7353  &  $-$1.3656  &  12.5714  &  $-$0.9104  &  0.4552  &  1.3656  \\
1.1  &  4.0221  &  $-$7.9278  &  3.9056  &  $-$0.1165  &  11.8334  &  $-$2.0227  &  1.0113  &  3.0340  \\
1.2  &  3.3153  &  $-$7.3096  &  3.9943  &  0.6789  &  11.3038  &  $-$2.8679  &  1.4339  &  4.3018  \\
1.3  &  2.8443  &  $-$6.8859  &  4.0416  &  1.1972  &  10.9275  &  $-$3.5418  &  1.7709  &  5.3128  \\
1.4  &  2.5278  &  $-$6.5942  &  4.0664  &  1.5386  &  10.6606  &  $-$4.1002  &  2.0501  &  6.1503  \\
1.5  &  2.3125  &  $-$6.3921  &  4.0797  &  1.7672  &  10.4718  &  $-$4.5732  &  2.2866  &  6.8599  \\
1.6  &  2.1656  &  $-$6.2521  &  4.0865  &  1.9209  &  10.3386  &  $-$4.9823  &  2.4911  &  7.4734  \\
1.7  &  2.0650  &  $-$6.1551  &  4.0901  &  2.0251  &  10.2452  &  $-$5.3406  &  2.6703  &  8.0110  \\
1.8  &  1.9952  &  $-$6.0874  &  4.0922  &  2.0970  &  10.1796  &  $-$5.6583  &  2.8291  &  8.4874  \\
1.9  &  1.9476  &  $-$6.0410  &  4.0934  &  2.1458  &  10.1343  &  $-$5.9423  &  2.9712  &  8.9135  \\
2  &  1.9144  &  $-$6.0082  &  4.0938  &  2.1794  &  10.1019  &  $-$6.1973  &  3.0986  &  9.2959  \\
2.1  &  1.8915  &  $-$5.9856  &  4.0941  &  2.2026  &  10.0798  &  $-$6.4280  &  3.2140  &  9.6420  \\
2.2  &  1.8757  &  $-$5.9700  &  4.0944  &  2.2187  &  10.0644  &  $-$6.6369  &  3.3184  &  9.9553  \\
2.3  &  1.8641  &  $-$5.9587  &  4.0946  &  2.2305  &  10.0533  &  $-$6.8282  &  3.4141  &  10.2423  \\
2.4  &  1.8569  &  $-$5.9512  &  4.0943  &  2.2375  &  10.0456  &  $-$7.0036  &  3.5018  &  10.5055  \\
2.5  &  1.8515  &  $-$5.9458  &  4.0943  &  2.2428  &  10.0402  &  $-$7.1647  &  3.5824  &  10.7471  \\
2.6  &  1.8477  &  $-$5.9421  &  4.0944  &  2.2467  &  10.0365  &  $-$7.3135  &  3.6567  &  10.9702  \\
2.7  &  1.8452  &  $-$5.9396  &  4.0944  &  2.2493  &  10.0340  &  $-$7.4513  &  3.7256  &  11.1769  \\
2.8  &  1.8432  &  $-$5.9378  &  4.0946  &  2.2513  &  10.0324  &  $-$7.5797  &  3.7898  &  11.3695  \\
2.9  &  1.8419  &  $-$5.9366  &  4.0947  &  2.2528  &  10.0313  &  $-$7.6983  &  3.8491  &  11.5474  \\
3  &  1.8410  &  $-$5.9360  &  4.0950  &  2.2540  &  10.0311  &  $-$7.8098  &  3.9049  &  11.7147  \\
\end{tabular}
\end{ruledtabular}
\end{table*}
 
\begin{table*}[h]
\caption{\label{HexEvecsEvals2} {\bf Simple 3D Hexagonal (Stacked Triangular) Spin Lattices with {\bf k} = (1/2,1/2,1/2) and (1/3,1/3,1/3).} Eigenvalues $\lambda_{{\bf k}\alpha}$ and eigenvectors $\hat{\mu} = [\mu_x,\mu_y,\mu_z]$ in Cartesian coordinates of the magnetic dipole interaction tensor $\widehat{{\bf G}}_i({\bf k})$ in Eq.~(16c) for two values of the magnetic wavevector ${\bf k}$ in reciprocal lattice units (r.l.u.) and {\bf collinear} magnetic moment alignments.  The $x$~axis is in the direction of the hexagonal $a$~axis.  The accuracy of the eigenvalues is estimated to be $\lesssim\pm 0.001$.  Also shown are the differences between the eigenvalues for different ordering axes for a given {\bf k}.}
\begin{ruledtabular}
\begin{tabular}{ccccccccc}
 $\lambda_{{\bf k}\alpha}$	& $\left(\frac{1}{2},\frac{1}{2},\frac{1}{2}\right)$ &   &   & $\left[\frac{1}{2},\frac{\sqrt{3}}{2},0\right]$ &	$\left[\frac{1}{2},\frac{\sqrt{3}}{2},0\right]$	& $\left(\frac{1}{3},\frac{1}{3},\frac{1}{3}\right)$ &  &  \\
c/a	& [001] & $\left[\frac{1}{2},\frac{\sqrt{3}}{2},0\right]$ & $\left[\frac{\sqrt{3}}{2},-\frac{1}{2},0\right]$ & $-[001]$ & $-\left[\frac{\sqrt{3}}{2},-\frac{1}{2},0\right]$ & [001] & [100], [010]  & $[001]-[100]$ \\
\hline
0.5  &  $-$28.2681  &  14.9063  &  13.3618  &  43.1743  &  1.5444  &  $-$15.2234  &  7.6117  &  $-$22.8352  \\
0.6  &  $-$15.6463  &  9.2784  &  6.3679  &  24.9247  &  2.9105  &  $-$7.4986  &  3.7493  &  $-$11.2479  \\
0.7  &  $-$9.0113  &  6.6815  &  2.3298  &  15.6928  &  4.3517  &  $-$3.5057  &  1.7528  &  $-$5.2585  \\
0.8  &  $-$5.1660  &  5.4157  &  $-$0.2497  &  10.5817  &  5.6653  &  $-$1.2648  &  0.6324  &  $-$1.8972  \\
0.9  &  $-$2.7943  &  4.7791  &  $-$1.9848  &  7.57341  &  6.7638  &  0.0638  &  $-$0.0319  &  0.0957  \\
1  &  $-$1.2721  &  4.4523  &  $-$3.1802  &  5.72436  &  7.6325  &  0.8795  &  $-$0.4397  &  1.3192  \\
1.1  &  $-$0.2708  &  4.2831  &  $-$4.0123  &  4.55388  &  8.2954  &  1.3929  &  $-$0.6965  &  2.0894  \\
1.2  &  0.3984  &  4.1944  &  $-$4.5928  &  3.79602  &  8.7873  &  1.7210  &  $-$0.8605  &  2.5816  \\
1.3  &  0.8511  &  4.1472  &  $-$4.9984  &  3.29611  &  9.1456  &  1.9333  &  $-$0.9666  &  2.8999  \\
1.4  &  1.1587  &  4.1227  &  $-$5.2814  &  2.96403  &  9.4042  &  2.0710  &  $-$1.0355  &  3.1065  \\
1.5  &  1.3696  &  4.1103  &  $-$5.4799  &  2.74075  &  9.5903  &  2.1607  &  $-$1.0803  &  3.2410  \\
1.6  &  1.5145  &  4.1027  &  $-$5.6172  &  2.58825  &  9.7199  &  2.2194  &  $-$1.1097  &  3.3291  \\
1.7  &  1.6147  &  4.0985  &  $-$5.7132  &  2.48379  &  9.8116  &  2.2580  &  $-$1.1290  &  3.3870  \\
1.8  &  1.6835  &  4.0972  &  $-$5.7807  &  2.41363  &  9.8779  &  2.2834  &  $-$1.1417  &  3.4252  \\
1.9  &  1.7305  &  4.0971  &  $-$5.8276  &  2.36665  &  9.9247  &  2.2998  &  $-$1.1499  &  3.4496  \\
2  &  1.7641  &  4.0958  &  $-$5.8599  &  2.33175  &  9.9557  &  2.3108  &  $-$1.1554  &  3.4662  \\
2.1  &  1.7870  &  4.0953  &  $-$5.8823  &  2.30833  &  9.9776  &  2.3182  &  $-$1.1591  &  3.4773  \\
2.2  &  1.8030  &  4.0947  &  $-$5.8977  &  2.29180  &  9.9924  &  2.3230  &  $-$1.1615  &  3.4845  \\
2.3  &  1.8141  &  4.0945  &  $-$5.9086  &  2.28032  &  10.0031  &  2.3261  &  $-$1.1631  &  3.4892  \\
2.4  &  1.8219  &  4.0946  &  $-$5.9165  &  2.27277  &  10.0111  &  2.3280  &  $-$1.1640  &  3.4919  \\
2.5  &  1.8273  &  4.0945  &  $-$5.9217  &  2.26724  &  10.0162  &  2.3294  &  $-$1.1647  &  3.4940  \\
2.6  &  1.8306  &  4.0949  &  $-$5.9255  &  2.26425  &  10.0203  &  2.3303  &  $-$1.1652  &  3.4955  \\
2.7  &  1.8331  &  4.0947  &  $-$5.9278  &  2.26155  &  10.0225  &  2.3307  &  $-$1.1653  &  3.4960  \\
2.8  &  1.8346  &  4.0952  &  $-$5.9298  &  2.26055  &  10.0249  &  2.3311  &  $-$1.1655  &  3.4966  \\
2.9  &  1.8365  &  4.0949  &  $-$5.9314  &  2.25844  &  10.0263  &  2.3312  &  $-$1.1656  &  3.4968  \\
3  &  1.8372  &  4.0949  &  $-$5.9321  &  2.25775  &  10.0270  &  2.3314  &  $-$1.1657  &  3.4971  \\
\end{tabular}
\end{ruledtabular}
\end{table*}
 
\begin{table*}[h]
\caption{\label{HexEvecsEvals3} {\bf Simple 3D Hexagonal (Stacked Triangular) Spin Lattices with {\bf k} = (1/3,1/3,0) and (1/3,1/3,1/2).} Eigenvalues $\lambda_{{\bf k}\alpha}$ and eigenvectors $\hat{\mu} = [\mu_x,\mu_y,\mu_z]$ in Cartesian coordinates of the magnetic dipole interaction tensor $\widehat{{\bf G}}_i({\bf k})$ in Eq.~(16c) for two values of the magnetic wavevector ${\bf k}$ in reciprocal lattice units (r.l.u.) and {\bf collinear} magnetic moment alignments.  The $x$~axis is in the direction of the hexagonal $a$~axis.  The accuracy of the eigenvalues is estimated to be $\lesssim\pm 0.001$.  Also shown are the differences between the eigenvalues for different ordering axes for a given {\bf k}.}
\begin{ruledtabular}
\begin{tabular}{ccccccc}
 $\lambda_{\bf k\alpha}$	& $\left(\frac{1}{3},\frac{1}{3},0\right)$ &&&	$\left(\frac{1}{3},\frac{1}{3},\frac{1}{2}\right)$\\
c/a	& [001] & [100], [010] & $[001]-[100]$ & [001] & [100], [010] & $[001]-[100]$  \\
\hline
0.5  &  38.4703  &  $-$19.2351  &  57.7054  &  $-$27.9927  &  13.9964  &  $-$41.9891  \\
0.6  &  22.2841  &  $-$11.1420  &  33.4261  &  $-$15.1774  &  7.5887  &  $-$22.7661  \\
0.7  &  14.0899  &  $-$7.0450  &  21.1349  &  $-$8.3821  &  4.1910  &  $-$12.5731  \\
0.8  &  9.5488  &  $-$4.7744  &  14.3231  &  $-$4.4347  &  2.2173  &  $-$6.6520  \\
0.9  &  6.8755  &  $-$3.4378  &  10.3133  &  $-$2.0181  &  1.0090  &  $-$3.0271  \\
1  &  5.2388  &  $-$2.6194  &  7.8582  &  $-$0.4912  &  0.2456  &  $-$0.7368  \\
1.1  &  4.2100  &  $-$2.1050  &  6.3150  &  0.4898  &  $-$0.2449  &  0.7346  \\
1.2  &  3.5533  &  $-$1.7767  &  5.3300  &  1.1260  &  $-$0.5630  &  1.6889  \\
1.3  &  3.1295  &  $-$1.5648  &  4.6943  &  1.5405  &  $-$0.7702  &  2.3107  \\
1.4  &  2.8539  &  $-$1.4270  &  4.2809  &  1.8130  &  $-$0.9065  &  2.7195  \\
1.5  &  2.6742  &  $-$1.3371  &  4.0113  &  1.9907  &  $-$0.9954  &  2.9861  \\
1.6  &  2.5570  &  $-$1.2785  &  3.8356  &  2.1078  &  $-$1.0539  &  3.1617  \\
1.7  &  2.4793  &  $-$1.2397  &  3.7190  &  2.1846  &  $-$1.0923  &  3.2770  \\
1.8  &  2.4287  &  $-$1.2144  &  3.6431  &  2.2347  &  $-$1.1174  &  3.3521  \\
1.9  &  2.3953  &  $-$1.1977  &  3.5930  &  2.2681  &  $-$1.1340  &  3.4021  \\
2  &  2.3738  &  $-$1.1869  &  3.5607  &  2.2898  &  $-$1.1449  &  3.4346  \\
2.1  &  2.3597  &  $-$1.1798  &  3.5395  &  2.3044  &  $-$1.1522  &  3.4566  \\
2.2  &  2.3499  &  $-$1.1749  &  3.5248  &  2.3139  &  $-$1.1569  &  3.4708  \\
2.3  &  2.3435  &  $-$1.1717  &  3.5152  &  2.3196  &  $-$1.1598  &  3.4794  \\
2.4  &  2.3396  &  $-$1.1698  &  3.5095  &  2.3244  &  $-$1.1622  &  3.4865  \\
2.5  &  2.3371  &  $-$1.1686  &  3.5057  &  2.3266  &  $-$1.1633  &  3.4899  \\
2.6  &  2.3350  &  $-$1.1675  &  3.5025  &  2.3283  &  $-$1.1642  &  3.4925  \\
2.7  &  2.3341  &  $-$1.1670  &  3.5011  &  2.3296  &  $-$1.1648  &  3.4945  \\
2.8  &  2.3335  &  $-$1.1667  &  3.5002  &  2.3304  &  $-$1.1652  &  3.4956  \\
2.9  &  2.3329  &  $-$1.1665  &  3.4994  &  2.3308  &  $-$1.1654  &  3.4962  \\
3  &  2.3325  &  $-$1.1662  &  3.4987  &  2.3311  &  $-$1.1656  &  3.4967  \\
\end{tabular}
\end{ruledtabular}
\end{table*}

\begin{table*}[h]
\caption{\label{HoneycombEvecsEvals} {\bf 3D Honeycomb Spin Lattices with {\bf k} = (0,0,0) (ferromagnetic) and (1/2,0,0).} Eigenvalues $\lambda_{{\bf k}\alpha}$ and eigenvectors $\hat{\mu} = [\mu_x,\mu_y,\mu_z]$ in Cartesian coordinates of the magnetic dipole interaction tensor $\widehat{{\bf G}}_i({\bf k})$ in Eq.~(16c) for two values of the magnetic wavevector ${\bf k}$ in reciprocal lattice units (r.l.u.) and {\bf collinear} magnetic moment alignments.  The $x$~axis is in the direction of the hexagonal $a$~axis.  The accuracy of the eigenvalues is estimated to be $\lesssim\pm 0.001$.  Also shown are the differences between the eigenvalues for different ordering axes for a given {\bf k}.}
\begin{ruledtabular}
\begin{tabular}{ccccccccc}
 $\lambda_{{\bf k}\alpha}$	& $\left(0,0,0\right)$ &   & [100]  & $\left(\frac{1}{2},0,0\right)$ & && $\left[\frac{1}{2},-\frac{\sqrt{3}}{2},0\right]$ &	$\left[\frac{1}{2},-\frac{\sqrt{3}}{2},0\right]$\\
c/a	& [001] & [100], [010] & $-[001]$ & 	$\left[\frac{\sqrt{3}}{2},\frac{1}{2},0\right]$	& 	[001]	& 	$\left[\frac{1}{2},-\frac{\sqrt{3}}{2},0\right]$	&	$-[001]$ & $-\left[\frac{\sqrt{3}}{2},\frac{1}{2},0\right]$  \\
\hline
0.5  &  17.8809  &  $-$8.9405  &  $-$26.8214  &  $-$42.9938  &  38.2651  &  4.7287  &  $-$33.5364  &  47.7226  \\
0.6  &  3.4722  &  $-$1.7361  &  $-$5.2083  &  $-$31.1917  &  21.8433  &  9.3484  &  $-$12.4948  &  40.5401  \\
0.7  &  $-$4.2063  &  2.1032  &  6.3095  &  $-$24.5981  &  13.3674  &  11.2307  &  $-$2.1367  &  35.8289  \\
0.8  &  $-$8.9442  &  4.4721  &  13.4163  &  $-$20.5983  &  8.5385  &  12.0598  &  3.5214  &  32.6581  \\
0.9  &  $-$12.1969  &  6.0984  &  18.2953  &  $-$18.0445  &  5.5981  &  12.4464  &  6.8483  &  30.4909  \\
1  &  $-$14.6077  &  7.3038  &  21.9115  &  $-$16.3605  &  3.7263  &  12.6342  &  8.9080  &  28.9947  \\
1.1  &  $-$16.4856  &  8.2428  &  24.7284  &  $-$15.2276  &  2.5001  &  12.7275  &  10.2274  &  27.9552  \\
1.2  &  $-$18.0090  &  9.0045  &  27.0134  &  $-$14.4562  &  1.6822  &  12.7740  &  11.0918  &  27.2302  \\
1.3  &  $-$19.2757  &  9.6378  &  28.9135  &  $-$13.9270  &  1.1274  &  12.7996  &  11.6723  &  26.7266  \\
1.4  &  $-$20.3528  &  10.1764  &  30.5292  &  $-$13.5625  &  0.7498  &  12.8127  &  12.0629  &  26.3753  \\
1.5  &  $-$21.2806  &  10.6403  &  31.9209  &  $-$13.3097  &  0.4905  &  12.8193  &  12.3288  &  26.1290  \\
1.6  &  $-$22.0891  &  11.0445  &  33.1336  &  $-$13.1347  &  0.3120  &  12.8227  &  12.5107  &  25.9575  \\
1.7  &  $-$22.8019  &  11.4009  &  34.2028  &  $-$13.0134  &  0.1889  &  12.8245  &  12.6356  &  25.8379  \\
1.8  &  $-$23.4352  &  11.7176  &  35.1528  &  $-$12.9291  &  0.1034  &  12.8257  &  12.7223  &  25.7548  \\
1.9  &  $-$24.0010  &  12.0005  &  36.0016  &  $-$12.8708  &  0.0446  &  12.8262  &  12.7816  &  25.6971  \\
2  &  $-$24.5114  &  12.2557  &  36.7671  &  $-$12.8298  &  0.0033  &  12.8265  &  12.8233  &  25.6564  \\
2.1  &  $-$24.9722  &  12.4861  &  37.4583  &  $-$12.8016  &  $-$0.0251  &  12.8267  &  12.8518  &  25.6283  \\
2.2  &  $-$25.3902  &  12.6951  &  38.0853  &  $-$12.7818  &  $-$0.0449  &  12.8267  &  12.8716  &  25.6085  \\
2.3  &  $-$25.7725  &  12.8863  &  38.6588  &  $-$12.7680  &  $-$0.0586  &  12.8266  &  12.8853  &  25.5946  \\
2.4  &  $-$26.1232  &  13.0616  &  39.1848  &  $-$12.7585  &  $-$0.0679  &  12.8264  &  12.8944  &  25.5850  \\
2.5  &  $-$26.4457  &  13.2228  &  39.6685  &  $-$12.7519  &  $-$0.0745  &  12.8264  &  12.9009  &  25.5783  \\
2.6  &  $-$26.7431  &  13.3716  &  40.1147  &  $-$12.7475  &  $-$0.0795  &  12.8270  &  12.9064  &  25.5745  \\
2.7  &  $-$27.0190  &  13.5095  &  40.5285  &  $-$12.7443  &  $-$0.0827  &  12.8271  &  12.9098  &  25.5714  \\
2.8  &  $-$27.2751  &  13.6376  &  40.9127  &  $-$12.7421  &  $-$0.0850  &  12.8271  &  12.9120  &  25.5692  \\
2.9  &  $-$27.5129  &  13.7565  &  41.2694  &  $-$12.7402  &  $-$0.0867  &  12.8269  &  12.9136  &  25.5672  \\
3  &  $-$27.7360  &  13.8680  &  41.6040  &  $-$12.7392  &  $-$0.0877  &  12.8270  &  12.9147  &  25.5662  \\
\end{tabular}
\end{ruledtabular}
\end{table*}

\begin{table*}[h]
\caption{\label{HoneycombEvecsEvalsNeel} {\bf 3D Honeycomb Spin Lattices with {\bf k} = (0,0,0) and (0,0,1/2) (N\'eel-type in the $ab$~plane and then also along the $c$~axis, respectively.} Eigenvalues $\lambda_{{\bf k}\alpha}$ and eigenvectors $\hat{\mu} = [\mu_x,\mu_y,\mu_z]$ in Cartesian coordinates of the magnetic dipole interaction tensor $\widehat{{\bf G}}_i({\bf k})$ in Eq.~(16c) for two values of the magnetic wavevector ${\bf k}$ in reciprocal lattice units (r.l.u.) and {\bf collinear} magnetic moment alignments.  The $x$~axis is in the direction of the hexagonal $a$~axis.  The accuracy of the eigenvalues is estimated to be $\lesssim\pm 0.001$.  Also shown are the differences between the eigenvalues for different ordering axes for a given {\bf k}.}
\begin{ruledtabular}
\begin{tabular}{ccccccc}
 $\lambda_{\bf k\alpha}$	& $\left(0,0,0\right)$ &&&	$\left(0,0,\frac{1}{2}\right)$\\
c/a	& [001] & [100], [010] & $[001]-[100]$ & [001] & [100], [010] & $[001]-[100]$  \\
\hline
0.5    &    39.6849    &    $-$19.8424    &    59.5273    &    $-$14.1112    &    7.0556    &    $-$21.1667    \\
0.6    &    24.8305    &    $-$12.4152    &    37.2457    &    $-$0.283961    &    0.1420    &    $-$0.425942    \\
0.7    &    18.1343    &    $-$9.0671    &    27.2014    &    6.17048    &    $-$3.0852    &    9.25572    \\
0.8    &    14.9989    &    $-$7.4994    &    22.4983    &    9.25024    &    $-$4.6251    &    13.8754    \\
0.9    &    13.5052    &    $-$6.7526    &    20.2579    &    10.7321    &    $-$5.3660    &    16.0981    \\
1    &    12.7869    &    $-$6.3934    &    19.1803    &    11.4468    &    $-$5.7234    &    17.1702    \\
1.1    &    12.4402    &    $-$6.2201    &    18.6603    &    11.7927    &    $-$5.8963    &    17.6890    \\
1.2    &    12.2732    &    $-$6.1366    &    18.4099    &    11.9585    &    $-$5.9793    &    17.9378    \\
1.3    &    12.1920    &    $-$6.0960    &    18.2880    &    12.0404    &    $-$6.0202    &    18.0606    \\
1.4    &    12.1524    &    $-$6.0762    &    18.2286    &    12.0792    &    $-$6.0396    &    18.1189    \\
1.5    &    12.1341    &    $-$6.0671    &    18.2012    &    12.0987    &    $-$6.0493    &    18.1480    \\
1.6    &    12.1246    &    $-$6.0623    &    18.1868    &    12.1075    &    $-$6.0537    &    18.1612    \\
1.7    &    12.1206    &    $-$6.0603    &    18.1809    &    12.1119    &    $-$6.0560    &    18.1679    \\
1.8    &    12.1187    &    $-$6.0593    &    18.1780    &    12.1146    &    $-$6.0573    &    18.1718    \\
1.9    &    12.1164    &    $-$6.0582    &    18.1746    &    12.1161    &    $-$6.0580    &    18.1741    \\
2    &    12.1168    &    $-$6.0584    &    18.1752    &    12.1160    &    $-$6.0580    &    18.1739    \\
2.1    &    12.1162    &    $-$6.0581    &    18.1743    &    12.1159    &    $-$6.0579    &    18.1738    \\
2.2    &    12.1164    &    $-$6.0582    &    18.1746    &    12.1166    &    $-$6.0583    &    18.1749    \\
2.3    &    12.1161    &    $-$6.0581    &    18.1742    &    12.1161    &    $-$6.0581    &    18.1742    \\
2.4    &    12.1159    &    $-$6.0579    &    18.1738    &    12.1164    &    $-$6.0582    &    18.1746    \\
2.5    &    12.1162    &    $-$6.0581    &    18.1743    &    12.1165    &    $-$6.0583    &    18.1748    \\
2.6    &    12.1162    &    $-$6.0581    &    18.1743    &    12.1167    &    $-$6.0584    &    18.1751    \\
2.7    &    12.1165    &    $-$6.0583    &    18.1748    &    12.1164    &    $-$6.0582    &    18.1745    \\
2.8    &    12.1158    &    $-$6.0579    &    18.1737    &    12.1156    &    $-$6.0578    &    18.1734    \\
2.9    &    12.1164    &    $-$6.0582    &    18.1745    &    12.1165    &    $-$6.0582    &    18.1747    \\
3    &    12.1164    &    $-$6.0582    &    18.1746    &    12.1164    &    $-$6.0582    &    18.1746    \\
\end{tabular}
\end{ruledtabular}
\end{table*}

\begin{table}[h]
\caption{\label{Tab:EvecsEvalsHC0012} {\bf 3D Honeycomb Spin Lattices with {\bf k} = (0,0,1/2) (FM intraplane and AFM interplane).} Eigenvalues $\lambda_{{\bf k}\alpha}$ and eigenvectors $\hat{\mu} = [\mu_x,\mu_y,\mu_z]$ in Cartesian coordinates of the magnetic dipole interaction tensor $\widehat{{\bf G}}_i({\bf k})$ in Eq.~(16c) for the magnetic wavevector ${\bf k}$ in reciprocal lattice units (r.l.u.) and {\bf collinear} magnetic moment alignments.  The $x$~axis is in the direction of the hexagonal $a$~axis.  The accuracy of the eigenvalues is estimated to be $\approx\pm 0.001$.  Also shown are the differences between the eigenvalues for different ordering axes for a given {\bf k}.}
\begin{ruledtabular}
\begin{tabular}{cccc}
&& $\lambda_{{\bf k}\alpha}$	 \\
c/a	& [001] &  [100], [010] & [100]$-$[001]  \\
\hline
0.5  &  $-$47.097  &  23.549  &  70.646  \\
0.6  &  $-$39.490  &  19.745  &  59.235  \\
0.7  &  $-$36.499  &  18.249  &  54.748  \\
0.8  &  $-$35.235  &  17.618  &  52.853  \\
0.9  &  $-$34.674  &  17.337  &  52.011  \\
1  &  $-$34.415  &  17.208  &  51.623  \\
1.1  &  $-$34.295  &  17.148  &  51.443  \\
1.2  &  $-$34.238  &  17.119  &  51.358  \\
1.3  &  $-$34.211  &  17.105  &  51.316  \\
1.4  &  $-$34.198  &  17.099  &  51.296  \\
1.5  &  $-$34.191  &  17.096  &  51.287  \\
1.6  &  $-$34.188  &  17.094  &  51.282  \\
1.7  &  $-$34.186  &  17.093  &  51.279  \\
1.8  &  $-$34.185  &  17.093  &  51.278  \\
1.9  &  $-$34.185  &  17.093  &  51.278  \\
2  &  $-$34.186  &  17.093  &  51.279  \\
2.1  &  $-$34.185  &  17.092  &  51.277  \\
2.2  &  $-$34.186  &  17.093  &  51.279  \\
2.3  &  $-$34.186  &  17.093  &  51.279  \\
2.4  &  $-$34.184  &  17.092  &  51.276  \\
2.5  &  $-$34.185  &  17.093  &  51.278  \\
2.6  &  $-$34.185  &  17.093  &  51.278  \\
2.7  &  $-$34.186  &  17.093  &  51.280  \\
2.8  &  $-$34.185  &  17.092  &  51.277  \\
2.9  &  $-$34.182  &  17.091  &  51.274  \\
3  &  $-$34.184  &  17.092  &  51.275  \\
\end{tabular}
\end{ruledtabular}
\end{table}

\end{document}